\documentclass[twocolumn,twocolappendix, tighten]{aastex63}
\usepackage{graphicx}
\usepackage{url}
\usepackage{epstopdf}
\usepackage{color}
\usepackage{textcomp}
\usepackage{gensymb}
\usepackage{multirow}
\usepackage{ragged2e}
\usepackage{hyperref}
\usepackage{url}
\usepackage{amsmath} 
\usepackage{booktabs}
\usepackage{comment}
\usepackage{afterpage}
\usepackage{mathtools}
\usepackage{tabularx}
\usepackage{bold-extra}
\usepackage{xspace}
\usepackage{relsize}
\usepackage{colordvi}
\usepackage{braket}

\usepackage[nolist,nohyperlinks]{acronym}

\usepackage[flushleft]{threeparttable}

\usepackage{eht}

\usepackage[T1]{fontenc} 

\usepackage{comment}
\usepackage[encapsulated]{CJK}
\usepackage{ucs}
\usepackage[utf8x]{inputenc}
\newcommand{\cntext}[1]{\begin{CJK}{UTF8}{gbsn}#1\end{CJK}}

\begin{document}
\shorttitle{EHT \sgra Paper VI}
\shortauthors{EHTC et al.}
\newcounter{iPap}\setcounter{iPap}{1}
\newcommand{\ehtsubtitle}{This is just the GAL for now}

\ifnum\value{iPap}=1 \renewcommand{\ehtsubtitle}{The Shadow of the Supermassive Black Hole in the Center of the Milky Way}\fi
\ifnum\value{iPap}=2 \renewcommand{\ehtsubtitle}{EHT and Multi-wavelength Observations, Data Processing, and Calibration}\fi
\ifnum\value{iPap}=3 \renewcommand{\ehtsubtitle}{Imaging of the Galactic Centre Supermassive Black Hole}\fi
\ifnum\value{iPap}=4 \renewcommand{\ehtsubtitle}{Variability, morphology, and black hole mass}\fi
\ifnum\value{iPap}=5 \renewcommand{\ehtsubtitle}{Testing Astrophysical Models of the Galactic Center Black Hole}\fi
\ifnum\value{iPap}=6 \renewcommand{\ehtsubtitle}{Testing the Black Hole Metric}\fi

\title{First Sagittarius A$^*$ Event Horizon Telescope Results VI: Testing the Black Hole Metric}

\author[0000-0002-9475-4254]{Kazunori Akiyama}
\affiliation{Massachusetts Institute of Technology Haystack Observatory, 99 Millstone Road, Westford, MA 01886, USA}
\affiliation{National Astronomical Observatory of Japan, 2-21-1 Osawa, Mitaka, Tokyo 181-8588, Japan}
\affiliation{Black Hole Initiative at Harvard University, 20 Garden Street, Cambridge, MA 02138, USA}

\author[0000-0002-9371-1033]{Antxon Alberdi}
\affiliation{Instituto de Astrof\'{\i}sica de Andaluc\'{\i}a-CSIC, Glorieta de la Astronom\'{\i}a s/n, E-18008 Granada, Spain}

\author{Walter Alef}
\affiliation{Max-Planck-Institut f\"ur Radioastronomie, Auf dem H\"ugel 69, D-53121 Bonn, Germany}

\author[0000-0001-6993-1696]{Juan Carlos Algaba}
\affiliation{Department of Physics, Faculty of Science, Universiti Malaya, 50603 Kuala Lumpur, Malaysia}

\author[0000-0003-3457-7660]{Richard Anantua}
\affiliation{Black Hole Initiative at Harvard University, 20 Garden Street, Cambridge, MA 02138, USA}
\affiliation{Center for Astrophysics $|$ Harvard \& Smithsonian, 60 Garden Street, Cambridge, MA 02138, USA}
\affiliation{Department of Physics \& Astronomy, The University of Texas at San Antonio, One UTSA Circle, San Antonio, TX 78249, USA}

\author[0000-0001-6988-8763]{Keiichi Asada}
\affiliation{Institute of Astronomy and Astrophysics, Academia Sinica, 11F of Astronomy-Mathematics Building, AS/NTU No. 1, Sec. 4, Roosevelt Rd, Taipei 10617, Taiwan, R.O.C.}

\author[0000-0002-2200-5393]{Rebecca Azulay}
\affiliation{Departament d'Astronomia i Astrof\'{\i}sica, Universitat de Val\`encia, C. Dr. Moliner 50, E-46100 Burjassot, Val\`encia, Spain}
\affiliation{Observatori Astronòmic, Universitat de Val\`encia, C. Catedr\'atico Jos\'e Beltr\'an 2, E-46980 Paterna, Val\`encia, Spain}
\affiliation{Max-Planck-Institut f\"ur Radioastronomie, Auf dem H\"ugel 69, D-53121 Bonn, Germany}

\author[0000-0002-7722-8412]{Uwe Bach}
\affiliation{Max-Planck-Institut f\"ur Radioastronomie, Auf dem H\"ugel 69, D-53121 Bonn, Germany}

\author[0000-0003-3090-3975]{Anne-Kathrin Baczko}
\affiliation{Max-Planck-Institut f\"ur Radioastronomie, Auf dem H\"ugel 69, D-53121 Bonn, Germany}

\author{David Ball}
\affiliation{Steward Observatory and Department of Astronomy, University of Arizona, 933 N. Cherry Ave., Tucson, AZ 85721, USA}

% ...

\collaboration{0}{The Event Horizon Telescope Collaboration}

\ifnum\value{iPap}=1 \include{./SAL1}\fi 
\ifnum\value{iPap}=2 \include{./SAL2}\fi
\ifnum\value{iPap}=3 \include{./SAL3}\fi
\ifnum\value{iPap}=4 \include{./SAL4}\fi
\ifnum\value{iPap}=5 \include{./SAL5}\fi
\ifnum\value{iPap}=6 \include{./SAL6}\fi


\begin{abstract}
Astrophysical black holes are expected to be described by the Kerr metric. This is the only stationary, vacuum, axisymmetric metric, without electromagnetic charge, that satisfies Einstein’s equations and does not have pathologies outside of the event horizon. We present new constraints on potential deviations from the Kerr prediction based on 2017 EHT observations of Sagittarius~A$^*$ (\sgra). We calibrate the relationship between the geometrically defined black hole shadow and the observed size of the ring-like images using a library that includes both Kerr and non-Kerr simulations. We use the exquisite prior constraints on the mass-to-distance ratio for \sgra to show that the observed image size is within $\sim 10\%$ of the Kerr predictions. We use these bounds to constrain metrics that are parametrically different from Kerr as well as the charges of several known spacetimes. To consider alternatives to the presence of an event horizon we explore the possibility that \sgra is a compact object with a surface that either absorbs and thermally re-emits incident radiation or partially reflects it. Using the observed image size and the broadband spectrum of \sgra, we conclude that a thermal surface can be ruled out and a fully reflective one is unlikely. We compare our results to the broader landscape of gravitational tests. Together with the bounds found for stellar mass black holes and the M87 black hole, our observations provide further support that the external spacetimes of all black holes are described by the Kerr metric, independent of their mass.
 \end{abstract}
 \keywords{galaxies: individual: \sgra -- Galaxy: center -- black hole physics }
\clearpage

\section{Introduction}\label{sec:intro}
\sbox0{\phantom{\cite{IV_EHT2019_M87,V_EHT2019_M87,VI_EHT2019_M87, PaperII, PaperIII, PaperIV, PaperV}}}

Horizon-scale images of supermassive black holes provide a conceptually new avenue for testing the theory of General Relativity. These images are formed by photons that originate in the deep gravitational fields of black holes, and, therefore, carry imprints of the spacetime properties in the strong-field regime \citep{Jaroszynski1997,Falcke2000}. In this series of papers, we report the first horizon-scale images of \sgra, the black hole at the center of our Galaxy, obtained with the Event Horizon Telescope (EHT), a global interferometric array observing at 1.3\,mm wavelength (\citealt{PaperII,PaperIII},
hereafter papers II and III). This paper in the series explores new constraints on the potential deviations from General Relativity imposed by these images. 

General Relativity has been tested in numerous settings with different observational tools and with different astrophysical systems~(see the review by \citealt{Will2014} and references therein). Traditionally, tests have been carried out in the solar system, with the periastron precession of Mercury~\citep{Verma2014}, the deflection of light observed during solar eclipses~\citep{Lambert2011}, and the detection of Shapiro delays in photons grazing the solar surface~\citep{Bertotti2003}. Radio observations of pulsars in binary systems expanded these tests, probing the radiative aspects of the theory and the strong-field coupling of the matter to the gravitational field~(see \citealt{Stairs2003} for a review and \citealt{Wex2020,Kramer2021} for some recent examples). Cosmological observations of the accelerated expansion of the universe probed gravity at the largest scales in the cosmos and gave evidence for the presence of dark matter and dark energy~(see \citealt{Ferreira2019} for a review).

As is clear from this overview, each test probes a different aspect of the theory of General Relativity. First, different astrophysical objects possess widely different mass and length scales and hence map to a very broad range of gravitational potentials and curvatures~\citep{Baker2015}. Second, some tests probe the stationary spacetimes of objects while others probe the dynamic and radiative aspects of the theory. Third, some settings involve vacuum spacetimes while others are affected by the coupling of matter and radiation to gravity~(see, e.g.,~\citealt{Damour1993}). Because modifications to the theory of gravity can be introduced independently in each of these aspects, without necessarily affecting the others, each of these tests brings a unique ability to constrain such modifications. 

Although general relativistic predictions have shown a high degree of consistency with the aforementioned tests, there remain unresolved questions at the fundamental level; e.g., whether curvature singularities are generally covered by event horizons (cosmic censorship conjecture) or can be naked. These become most urgent for black holes as those objects have the strongest gravitational fields in the universe and possess a curvature singularity in their center. The combination with quantum theory could tame curvature singularities but at the same time predicts inherent randomness for quantum particles at the event horizon leading to the black hole information loss paradox (see, e.g., \citealt{Harlow2014} for a review). All the concerns involve the presence of event horizons and are, therefore, accessible only to tests with black holes. Until recently, however, precision tests of gravity with black holes have not been possible. This situation has changed dramatically in the recent years with the detection of gravitational waves from coalescing stellar mass black holes with LIGO/Virgo~\citep{Abbott2016,Abbott2021}, the detection of relativistic effects in the orbits of stars around \sgra~\citep{2018A&A...615L..15G,2019Sci...365..664D,2020A&A...636L...5G}, and the imaging observations of the black hole in the center of the M87 galaxy~\citep{VI_EHT2019_M87,Psaltis2020,Kocherlakota2021}.

Tests of gravity with black holes benefit from a remarkable General Relativistic prediction encapsulated in the so-called no-hair theorem: the only vacuum spacetime that is stationary, axisymmetric, asymptotically flat, contains a horizon and is free of pathologies is the one described by the Kerr metric~\citep{Kerr1963,Israel1967,Israel1968,Carter1968,Carter1971,Hawking1972,Price1972a,Price1972b,Robinson1975}. Testing this prediction involves using spacetimes that introduce deviations from this metric and applying observational constraints to place bounds on the magnitudes of the deviations. In order for these spacetimes to evade the no-hair theorem while remaining free of pathologies, they cannot be solutions to the vacuum General Relativistic field equations but instead involve additional fields or parametric deviations that are agnostic to the underlying theory of gravity. In either case, measuring conclusively a deviation from the Kerr metric, while demonstrating that the compact object has a horizon, will constitute a demonstration of a violation of the no-hair theorem and, therefore, of the General Relativistic field equations.

Horizon-scale images of \sgra\ offer a distinct set of advantages in testing General Relativistic predictions with black holes~\citep{Psaltis2011,Psaltis2016,Goddi2017,Cunha2018,Psaltis2019}. At $4 \times 10^6 M_\odot$, this black hole has a mass that bridges those of the stellar-mass black holes observed with LIGO/Virgo ($\sim 10^1-10^2 M_\odot$) and that of the M87 black hole ($\sim 6.5 \times 10^9 M_\odot$), and, therefore, probes a curvature scale that is different from those of other tests. Perhaps more importantly, it enables an approach that is different from other tests in its methodology. Because of the detection of relativistic effects in the stellar orbits around this black hole, its mass and distance are accurately known resulting in precise predictions of its space time properties \citep{2019Sci...365..664D,2021arXiv211207478G}. As a result, contrary to other tests, where the mass of the black hole is measured from the same data simultaneously with the other spacetime properties (or possess significant astrophysical uncertainties as in M87), tests with \sgra\ rely on mass priors with completely orthogonal systematics and potential biases. In addition, the very small uncertainties in the prior mass measurement lead to a parameter-free prediction on the gravitational effects in the images, which can be tested precisely with the EHT observations. 

The most prominent gravitational effect on black hole images is the black hole shadow \citep{Falcke2000}. The boundary of the shadow on the image plane of a distant observer is marked by the impact parameters of photons, which, when traced back towards the black hole, become tangent to the spherical photon orbits close to the horizon~\citep{Bardeen1973,Luminet1979}. Although we define the shadow as a purely geometric feature that does not depend on astrophysical effects, we relate this feature to the brightness depression in observed images. Photons with impact parameters smaller than this critical value have paths that cross the horizon and, hence, have small optical paths through this spacetime. These reduced optical paths lead to much smaller radiation intensities compared to photons at larger impact parameters and, therefore, to the brightness depression~\citep{Jaroszynski1997,Johannsen2010,Narayan2019,Ozel2021,Bronzwaer2021,Kocherlakota+2022}.

In the Kerr metric, because of a cancellation between the effects of frame dragging and the quadrupole moment of the spacetime, the shape and size of the shadow boundary have a very weak dependence on black hole spin and the observer's inclination (i.e., the radius ranges from $\sim 4.8GMc^{-2}$ to $\sim 5.2GMc^{-2}$, see \citealt{Johannsen2010} for a detailed study of the dependence on spin). Instead, they are determined predominantly by the mass-to-distance ratio of the black hole, which are known precisely for \sgra, making the shadow a direct probe of the metric properties~(see e.g., \citealt{Psaltis2015}). 
For the black hole shadow to become observable, two conditions need to be satisfied. First, a sufficiently bright source of photons needs to be present close to the horizon such that these photons experience strong gravitational lensing. Second, this source needs to be optically thin (i.e., transparent) at the observing wavelength such that the shadow is not enshrouded by the material generating this radiation. 
Both of these conditions are satisfied at 1.3\,mm in the radiatively inefficient accretion flow around \sgra\  
(\citealt{Ozel2000}; see also \citealt[][ hereafter \citetalias{PaperV}]{PaperV}).

For such a configuration, the predicted image of the black hole is a bright ring of emission surrounding the shadow. The imaging observations with the EHT capture this ring and allow us to measure its properties, such as its diameter and fractional width. Earlier work has shown that, when this ring of emission is observed, the ring diameter can be used, with proper calibration, as a proxy for the shadow diameter itself~\citep{V_EHT2019_M87,VI_EHT2019_M87,Narayan2019,Ozel2021,Younsi2021,Kocherlakota+2022}. 
This is the approach that we follow in this paper to compare the predictions of General Relativity for the size of the black hole shadow to the observed measurement of the ring in the images of \sgra. 

The presence of a brightness depression also allows us to explore different possibilities for the nature of the compact object itself. In particular, if \sgra\ contained a reflecting surface at 1.3\,mm instead of a horizon or a naked singularity, we would have observed a less pronounced brightness depression. Alternatively, if it contained a surface that was fully absorptive and reemitting thermally the accreting energy, it would still create a depression in the EHT image but would generate bright emission at wavelengths shorter than 1.3\;mm. We use the EHT images in conjunction with the broadband spectrum of \sgra\ to place strong constraints on such alternatives. 

In Section \ref{sec:priors}, we summarize the prior information on the mass-to-distance ratio and the spectrum of \sgra. In Section \ref{sec:error}, we quantify the measurements of the image diameter as well as the relationship between image and shadow diameters using extensive simulations and synthetic data. We combine these to place bounds on potential deviations between the predicted and inferred shadow size for \sgra. In Section \ref{sec:eh}, we constrain alternatives to the black hole nature of the compact object that involve reflecting or absorbing surfaces. In Section \ref{sec:tests}, we impose constraints on the potential metric deviations from Kerr as well as address the possibility that \sgra\ contains a naked singularity. In Section \ref{sec:compare}, we leverage our gravity tests with those that involve other compact objects and solar system bodies in order to draw general conclusions about the theory of gravity. We summarize our findings in Section \ref{sec:conclusions}. 

\section{Priors}\label{sec:priors}

\subsection{Priors on $\theta_g$}

 The mass and distance of \sgra have been extensively studied by analyzing the dynamics of the central stellar cluster in the innermost 10 arcsec of the Galactic Center \citep{1997MNRAS.291..219G,1998ApJ...509..678G,1999A&A...352L..22E,2000MNRAS.317..348G,2000Natur.407..349G,2002MNRAS.331..917E,2002ApJ...576..790G,2003ApJ...594..812G,2007A&A...469..125S,2008ApJ...672L.119M,2010RvMP...82.3121G,2012RAA....12..995M,2013ApJ...779L...6D,2019ApJ...873....9J}. Near-infrared observation with 8-m to 10-m class telescopes supported by adaptive optics (AO) revealed the orbits of individual stars in the innermost arcsec (the so-called S-stars), in particular the star S0-2\footnote{S0-2 is called S2 in the VLTI naming convention.}
For this star, the combined fit for orbital elements and black hole parameters (mass, distance, projected position in the sky, proper motions, and radial velocity) has provided the most precise estimates for \sgra's mass and distance so far (\citealt{2003ApJ...596.1015S,2005ApJ...620..744G,2005ApJ...628..246E,2008ApJ...689.1044G,2009ApJ...692.1075G,2009ApJ...707L.114G,2012Sci...338...84M,2016ApJ...830...17B,2017ApJ...837...30G,2018ApJ...854...12C,2019AJ....158....4O,2019ApJ...880...87H,2021A&A...645A.127G}).

S0-2 is a star with an apparent $2.2$~$\mu$m (NIR K-band) magnitude of $m_K = 14$, an orbital period of $P\approx16$ years, a semi-major axis of $a = 125$~mas (or $\sim 10^3$~AU at an 8\,kpc distance), and, thus, is the brightest star with a comparatively close orbit and short period at the Galactic Center. The study of S0-2's orbit has predominantly been conducted with two sets of instruments, the two 10-m telescopes of the Keck Observatory, and the Very Large Telescope (VLT) of the European Southern Observatory (ESO), using its individual telescopes as well as GRAVITY, an interferometer combining all four 8.2-m telescopes of the VLT (VLTI). The orbit of S0-2 provides some of the best evidence for the existence of a black hole. S0-2 has concluded an entire revolution between 2002 and 2018 covered by observations and has allowed the Keck and VLTI teams to test relativistic effects like the gravitational redshift or the Schwarzschild precession (\citealt{2020A&A...636L...5G,2019Sci...365..664D,2018A&A...615L..15G,2019A&A...625L..10G,PhysRevLett.122.101102}) and to constrain alternative theories of gravity (\citealt{2017PhRvL.118u1101H,DeMartinoI,DellaMonica}) and variations of the fine structure constant (\citealt{2020PhRvL.124h1101H}). 

\begin{figure}
\centering
\includegraphics[width=.45\textwidth]{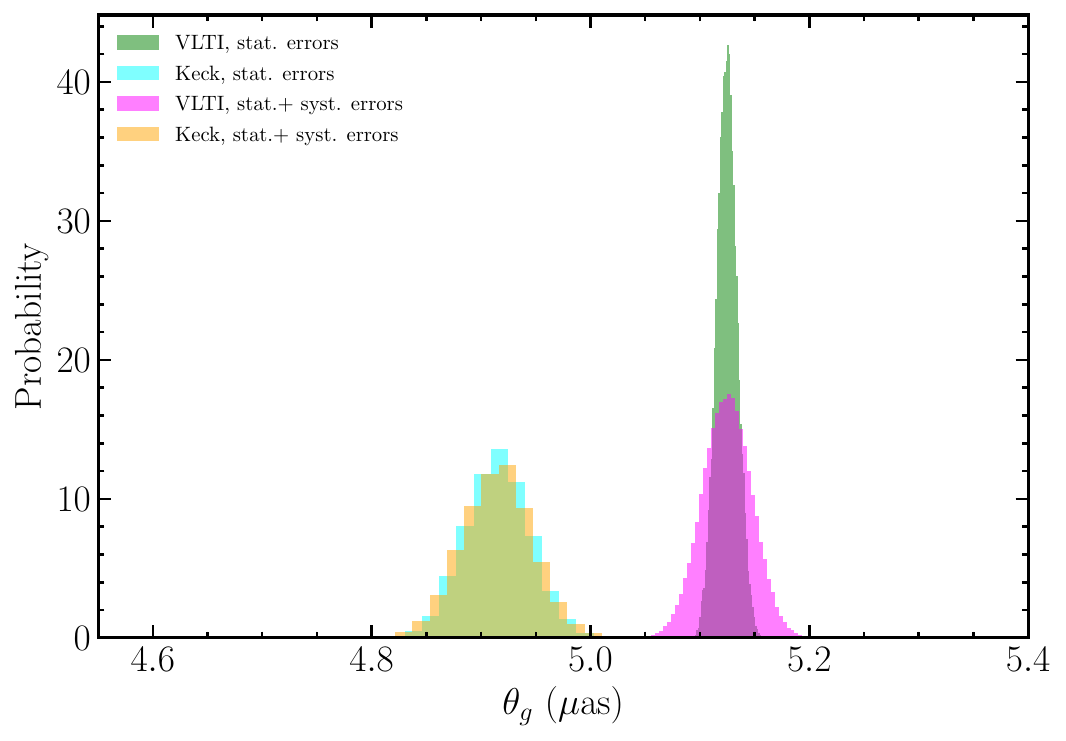}
\caption{Posteriors on $\theta_{\rm g}=GM/Dc^2$ as derived from the dynamical analyses of the orbit of the Galactic Center star  S0-2 by the Keck and the VLTI teams, respectively.}
\label{fig:Prior_MoverD}
\end{figure}

In order to measure S0-2's projected position in the sky, an astrometric reference frame has to be established. \cite{1997ApJ...475L.111M} proposed the idea to use a group of SiO maser stars at the Galactic Center --- visible both in the radio as well as in the NIR --- with positions and proper motions determined by interferometric astrometry at radio wavelengths. These masers allow it to establish a reference frame in the NIR. For both imaging instruments (KeckII/NIRC2, VLT/NACO) the field of view (10~arcsec and 14~arcsec, respectively) is not large enough to capture the S-stars and the seven masers in the same pointing. Instead, a dither pattern of pointings overlapping with one another is observed. Astrometric measurements in the central field are then executed via secondary astrometric standards, either in the form of matching coordinate lists or by generating mosaic images. In this process, systematic astrometric errors occur due to the geometric distortions of the camera optics and field dependence of the point spread function (PSF) caused by anisoplanatism of the AO-correction and higher order aberrations of the optics (\citealt{2010ApJ...725..331Y,2015MNRAS.453.3234P,2019ApJ...873...65S}). 

The VLTI team included interferometric data in their analysis starting in 2016. VLTI/GRAVITY provided high precision distance measurements to \sgra during  the S0-2's closest approach with $\sim$1~mas resolution and $\sim 40 \,\mu$as astrometric precision.
 
This subset of the interferometric data is not affected by the systematic uncertainties of the reference frame because the projected position of S2 is directly referenced to the projected (center of light) position of Sgr A*.
However, also VLTI/GRAVITY data have systematic uncertainties, mainly due to aberrations of the optical trains of the individual telescopes (\citealt{2021A&A...647A..59G}).

Information on the third dimension of the stellar orbits is obtained in the form of radial velocities, which in the case of  S0-2 can be determined by observing the 2.167~$\mu$m  HI (Br$\gamma$)
and the 2.11~$\mu$m HeI lines with integral field spectrographs like VLT/SINFONI or Keck/OSIRIS.

In their latest publication on the measurement of the gravitational redshift (\citealt{2019Sci...365..664D}, Table~1) the Keck team found for the distance a value of $(R_0 = 7959 \pm 59 \pm 32)$\,pc (for the fit that leaves the redshift parameter free). They also published a posterior version with the assumption that General Relativity is true (redshift parameter set to unity), $R_0 = (7935 \pm 50)$~pc, which is practically equivalent within the uncertainties. Their estimates for the black hole mass are $M = (3.975 \pm 0.058 \pm 0.026)  \times 10^6 M_\odot$ and $M = (3.951 \pm 0.047)  \times 10^6 M_\odot$, respectively.

In the publication on the detection of the Schwarzschild precession (\citealt{2020A&A...636L...5G}), the VLTI team found $R_0 = (8246.7 \pm 9.3)$\,pc and $M = (4.261 \pm 0.012)  \times 10^6 M_\odot$. Their latest paper on the mass distribution in the Galactic Center (\citealt{2021arXiv211207478G}) changes these values slightly.
Their table B.1 is an overview of recently published VLTI values for the black hole mass and distance; they also give an estimate of their systematics due to aberrations in GRAVITY's optics: $R_0 = (8277 \pm 9 \pm 33)$\,pc (\citealt{2021A&A...647A..59G}). For the mass they find: $M = (4.297 \pm 0.012 \pm 0.040)  \times 10^6 M_\odot$.
Additionally, the team provided a file with the posterior chains of their Bayesian analysis (Gillessen, priv. communication), assuming General Relativity to be true, which has median values of $R_0 = (8278 \pm 10)$\,pc and $M = (4.298 \pm 0.013)  \times 10^6 M_\odot$. 

It is interesting to point out that a third, independent estimate for the distance to \sgra has been provided by the Bessel project, a study of the Milky Way structure with VLBI astrometry: $R_0 = (8.15 \pm 0.15)$\,kpc (\citealt{2009ApJ...700..137R,2014ApJ...783..130R,2019ApJ...885..131R}). This value for the distance is in marginally better agreement with the VLTI results.

Here, the two values considered for the distance are $R_0 = (7935 \pm 50 \pm 32)$\,pc and  $R_0 = (8277 \pm 9 \pm 33)$~pc. Mass and distance set a characteristic scale of the orbit in its projection on the sky and are highly correlated. The values for $\theta_g\equiv GM/Dc^2$ as derived from the posterior distributions are:  $\theta_g = (5.125\pm 0.009 \pm 0.020) \,\mu$as (VLTI) and $\theta_g = (4.92 \pm 0.03 \pm 0.01) \,\mu$as (Keck). For the VLTI value, the systematics were derived by error propagation according to $M \propto R_0^2$ (\citealt{2021arXiv211207478G}). For the Keck value, a dedicated jackknife analysis was conducted to quantify the systematics stemming from the reference system (Do, priv. communication). We show the posteriors in Figure~\ref{fig:Prior_MoverD}. The discrepancy between the values of the two studies is about 4\%.

\begin{figure*}
\centering
\includegraphics[width=\textwidth]{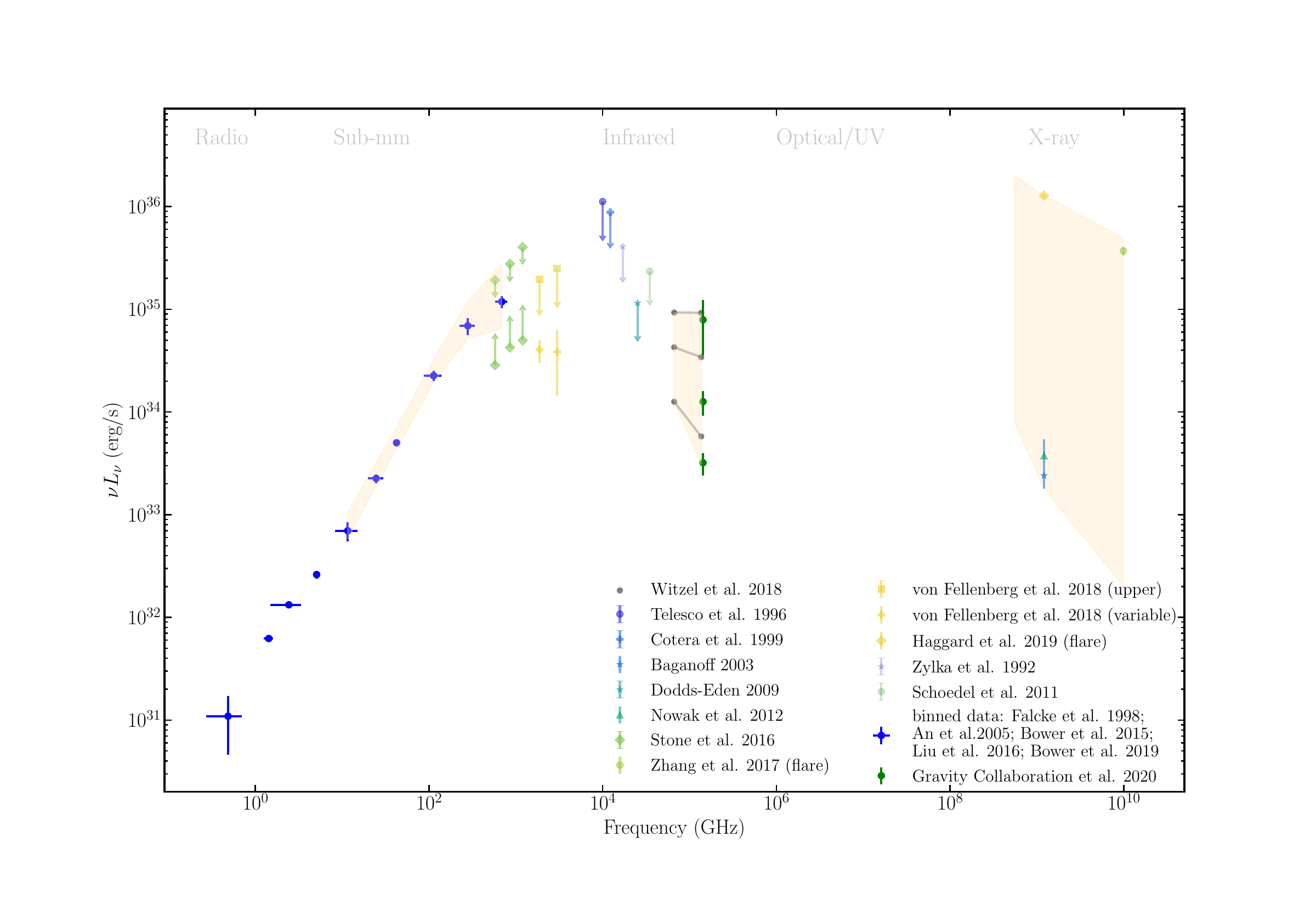}
\caption{\sgra radio to X-ray Spectral Energy Distribution (SED). Blue points show re-binned observed flux densities (\citealt{1998ApJ...499..731F,2005ApJ...634L..49A,2015ApJ...802...69B,2016A&A...593A.107L,2019ApJ...881L...2B}), the faint colored points show upper limits for the median flux density from the papers listed on the right. Solid gray lines from 4.5 to 2.2 $\mu$m show model SEDs for the 5th, 50th, and 95th percentile and demonstrate the predicted slope change as a function of flux density. The green points at 2.2 $\mu$m represent the most recent analysis of the 5th, 50th, and 95th percentile of NIR flux density distribution based on VLTI/GRAVITY data. The orange-shaded envelope represents an estimate of the range of flux densities in 90\% of the observed time.
}
\label{fig:Prior_SED}
\end{figure*}

\subsection{Priors on the spectral energy distribution of \sgra} \label{sec:sed}

The spectral energy distribution (SED) of \sgra is shown in Figure~\ref{fig:Prior_SED}. It has been compiled from the large body of literature, starting as early as 1992. Points show SED values taken from
\cite{1992A&A...261..119Z,1996ApJ...456..541T,1998ApJ...499..731F,1999ASPC..186..240C,2005ApJ...634L..49A,2009ApJ...698..676D,2011A&A...532A..83S,2011ApJ...728...37D,2012ApJ...759...95N,2015ApJ...802...69B,2016A&A...593A.107L,2016ApJ...825...32S,2017ApJ...843...96Z,2018ApJ...862..129V,2018ApJ...863...15W,2019ApJ...881L...2B,2019ApJ...886...96H,2020A&A...638A...2G}. We present the radio part of the SED (\citealt{1998ApJ...499..731F,2005ApJ...634L..49A,2015ApJ...802...69B,2016A&A...593A.107L,2019ApJ...881L...2B}) in a binned version (for a more detailed version showing all historic literature values in the radio to sub-millimeter regime, including some epochs of heightened variability, see \citetalias{PaperII}). The steepening of the SED slope at cm wavelengths  (\citealt{1998ApJ...499..731F}), is clearly visible between 10-20\,GHz and the sub-millimeter. From THz-frequencies to the mid IR \sgra has not be detected, and we have included lower and upper limits.

The SED shows variability in all observable parts. Especially in the NIR and X-ray regime \sgra is strongly variable with regular flux density changes of factors of tens and hundreds, respectively, within $10-20$ min (e.g.,  \citealt{2001Natur.413...45B,2003Natur.425..934G,2004A&A...427....1E,2005ApJ...635.1087G,2009ApJ...691.1021D,2011ApJ...728...37D,2012ApJS..203...18W,2013ApJ...774...42N,2015ApJ...799..199N,2017MNRAS.468.2447P,2018ApJ...864...58F}). However, at radio frequencies and in the mm/sub-millimeter regime the variability is comparatively minor with typical excursions of about 50\% or less of the mean flux density. (\citealt{1999ASPC..186..113F,2004AJ....127.3399H,2008ApJ...682..373M,Dexter2014,2015A&A...576A..41B, 2017A&A...601A..80S, 2018ApJ...864...58F, 2021ApJ...920L...7M,Goddi2021}).

Here we are focusing on the NIR properties of \sgra, in particular on limits for a steady component that is not varying on timescales of minutes and hours. Figure~\ref{fig:Prior_SED} shows the percentiles of the observed flux density distributions at 2.2~$\mu$m (VLT/NACO and KECK/NIRC2) and 4.5~$\mu$m (SPITZER/IRAC) as well as the corresponding spectral indices that change with flux density\footnote{Note that for several brighter flares at the 95th precentile level and above even flatter spectral indices have been observed that correspond to positive slopes in this plot \citep{2007ApJ...667..900H,2021A&A...654A..22G}.} \citep{2018ApJ...863...15W}. Additionally, we present the same percentiles for the flux density distribution measured with VLTI/GRAVITY at 2.2~$\mu$m \citep{2020A&A...638A...2G}. While the VLT and KECK data are confusion limited and noise dominated at the low end of flux density distribution resulting in non-detections of the source against the background, \cite{2020A&A...638A...2G} report a clear detection of \sgra at all times. Because this detected source is variable at all times, their 5th percentile of the variable flux density distribution represents a conservative upper limit for any steady source component that may lie underneath.

\mbox{}

\section{EHT Observations and Error budget}\label{sec:error}

The EHT observations of \sgra\ show a bright ring of emission surrounding a brightness depression that we have identified with the black hole shadow \citep{PaperIII}. In principle, the diameter of this ring, $\hat{d}_m$, can be used to measure the properties of the black hole metric and to assess its compatibility with the Kerr solution in General Relativity for a black hole of given angular size $\theta_{g}$. In practice, this comparison first requires establishing a quantitative relation (i.e., a calibration factor) between the diameter of a bright ring feature and that of the corresponding shadow. We can then use this relationship, in combination with the measured ring diameter, to infer any potential deviations from the General Relativistic predictions. 

To accomplish this, we write 
\begin{eqnarray}
\hat{d}_{m} &=& \frac{\hat{d}_m} {d_{\rm sh}} d_{\rm sh}= \alpha_{\rm c} \; d_{\rm sh} = \alpha_{\rm c} \; (1 + \delta) \; d_{\rm sh,Sch} \nonumber\\
&=& \alpha_{\rm c} \; (1+\delta) \; 6\sqrt{3} \; \theta_g\;. 
\label{eq:delta}
\end{eqnarray}
In this expression, $\hat{d}_m$ is the ring diameter measured from imaging and model-fitting to the \sgra\ data, where the hat signifies the fact that this is a measured quantity that may differ from the true value because of measurement biases. The quantity $\alpha_{\rm c} \equiv \hat{d}_m / d_{\rm sh}$ is the calibration factor, defined as the ratio of the measured diameter of the image to that of the shadow, which addresses the extent to which the ring diameter can be used as a proxy for the shadow diameter. The shadow diameter depends on the metric and its properties, such as the black-hole spin and potential charges, as well as on the observer inclination. 

The calibration factor $\alpha_{\rm c}$ is determined primarily by the physics of image formation near the horizon and quantifies the degree to which the image diameter tracks that of the shadow, for any underlying metric and for different realistic models of the accreting plasma. For example, the calibration factor would be $\alpha_{\rm c} =1.1$ whether the image diameter is $11\, GM/c^2$ and the shadow diameter is $10\, GM/c^2$ or, for some non-Kerr black hole, the image diameter is $110\, GM/c^2$ and the shadow diameter is $100\, GM/c^2$. 

The quantity $\delta \equiv (d_{\rm sh} / d_{\rm sh, Sch}) -1$, on the other hand, quantifies any deviation between the inferred shadow diameter and that of a Schwarzschild black hole of angular size $\theta_g$, given by $d_{\rm sh, Sch} = 6 \sqrt{3} \theta_g$. Note that, for the Kerr metric, the Schwarzschild limit provides the largest possible value for the shadow diameter. Black holes with non-zero spin observed at different inclinations can have shadow sizes that are smaller by up to $\sim 7.5\%$ from this limit \citep{Takahashi2004,Chan2013}. As a result, values of $\delta$ in the range $[-0.075, 0]$ are consistent with the Kerr predictions, while values outside this range can be considered to be in tension with it. We also note the small differences in the definitions of these quantities with respect to earlier work (see, e.g., \citealt{VI_EHT2019_M87,Psaltis2021}), which simply scaled the image diameter to $\theta_g$, and hence, did not cleanly separate the effects of different spacetimes from other astrophysical effects. We will use equation~(\ref{eq:delta}) to infer the posterior on the deviation parameter $\delta$ given the EHT measurements and prior information. 

\begin{figure}
\centering
\includegraphics[width=.45\textwidth]{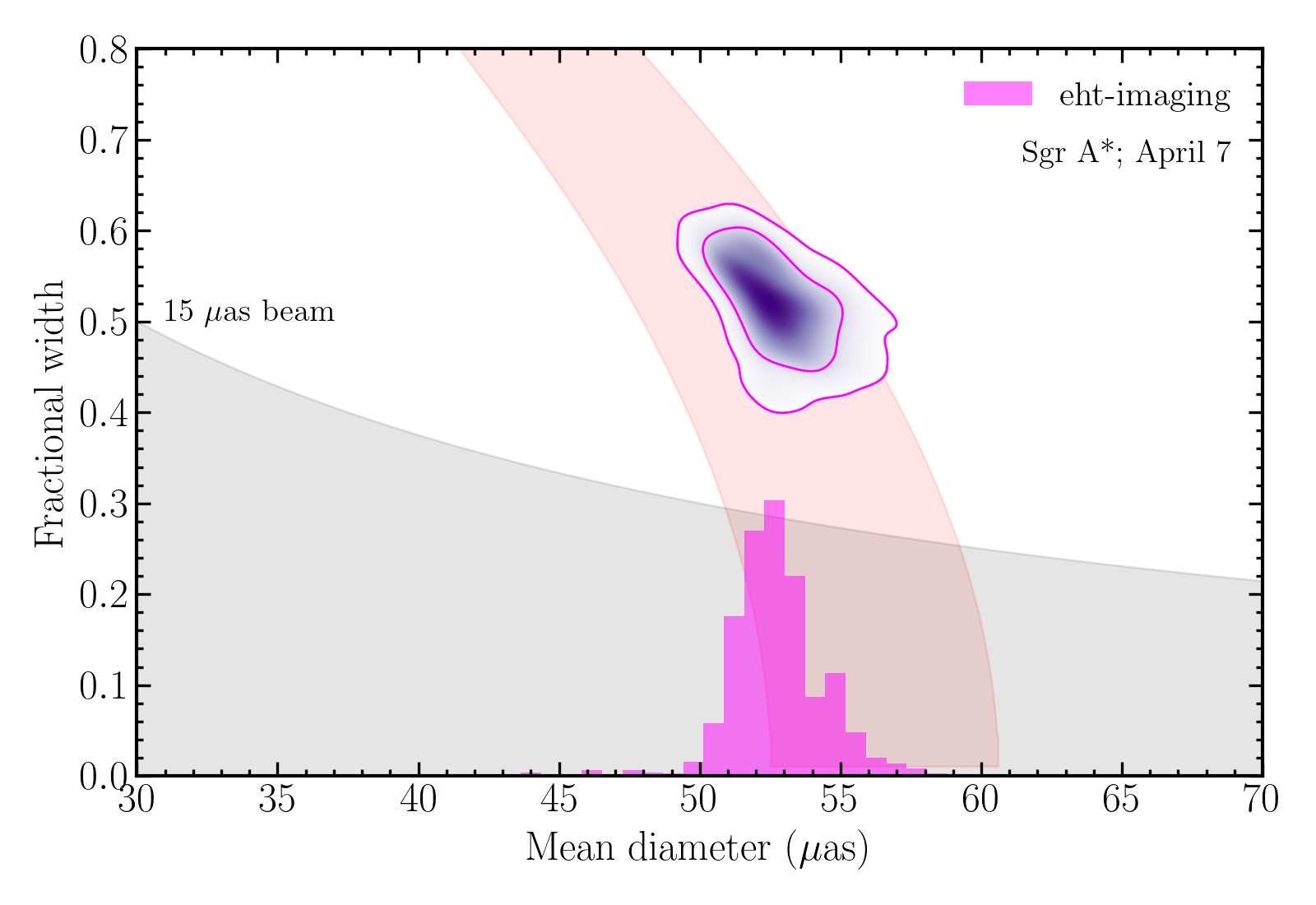}
\includegraphics[width=.45\textwidth]{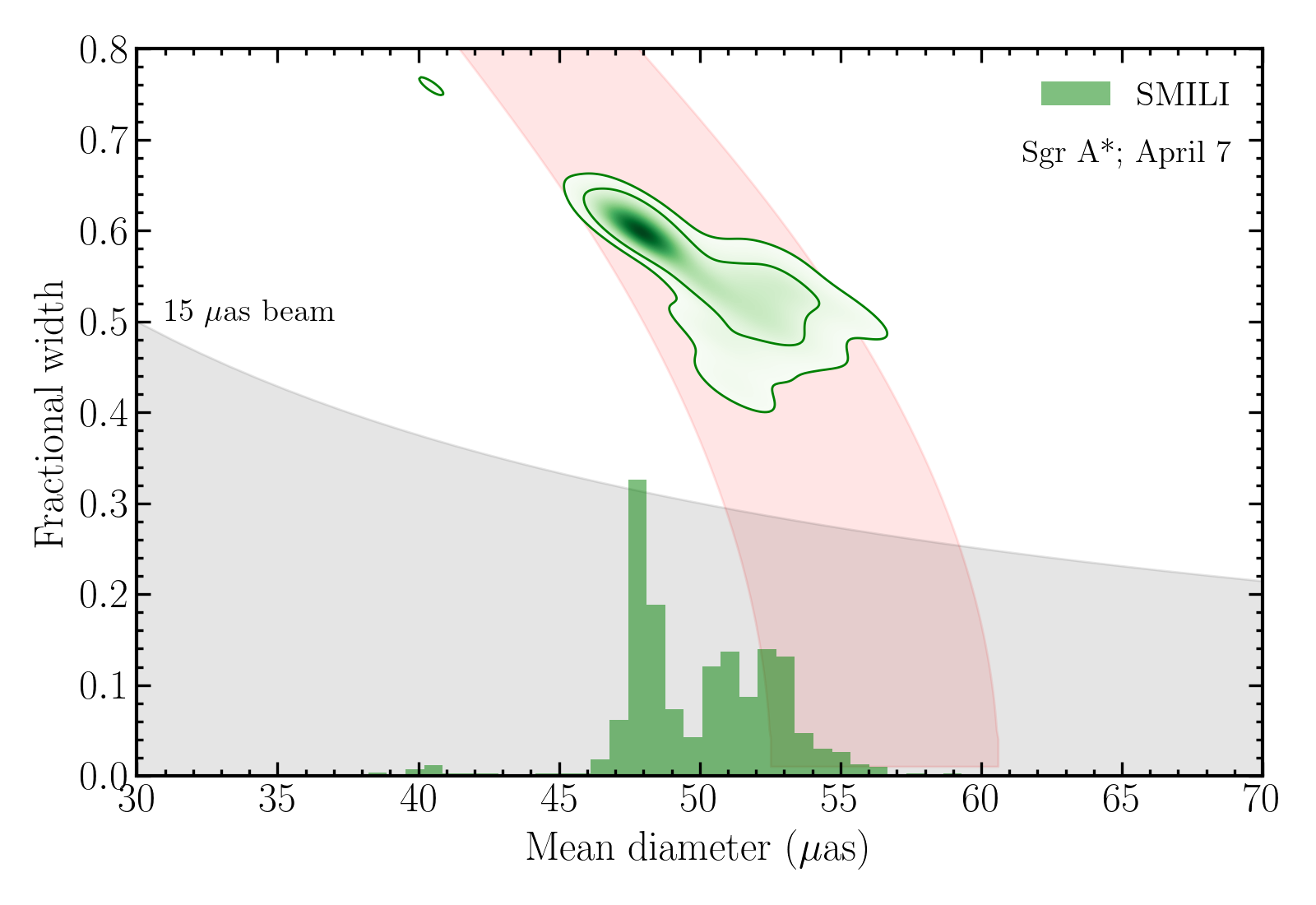}
\includegraphics[width=.45\textwidth]{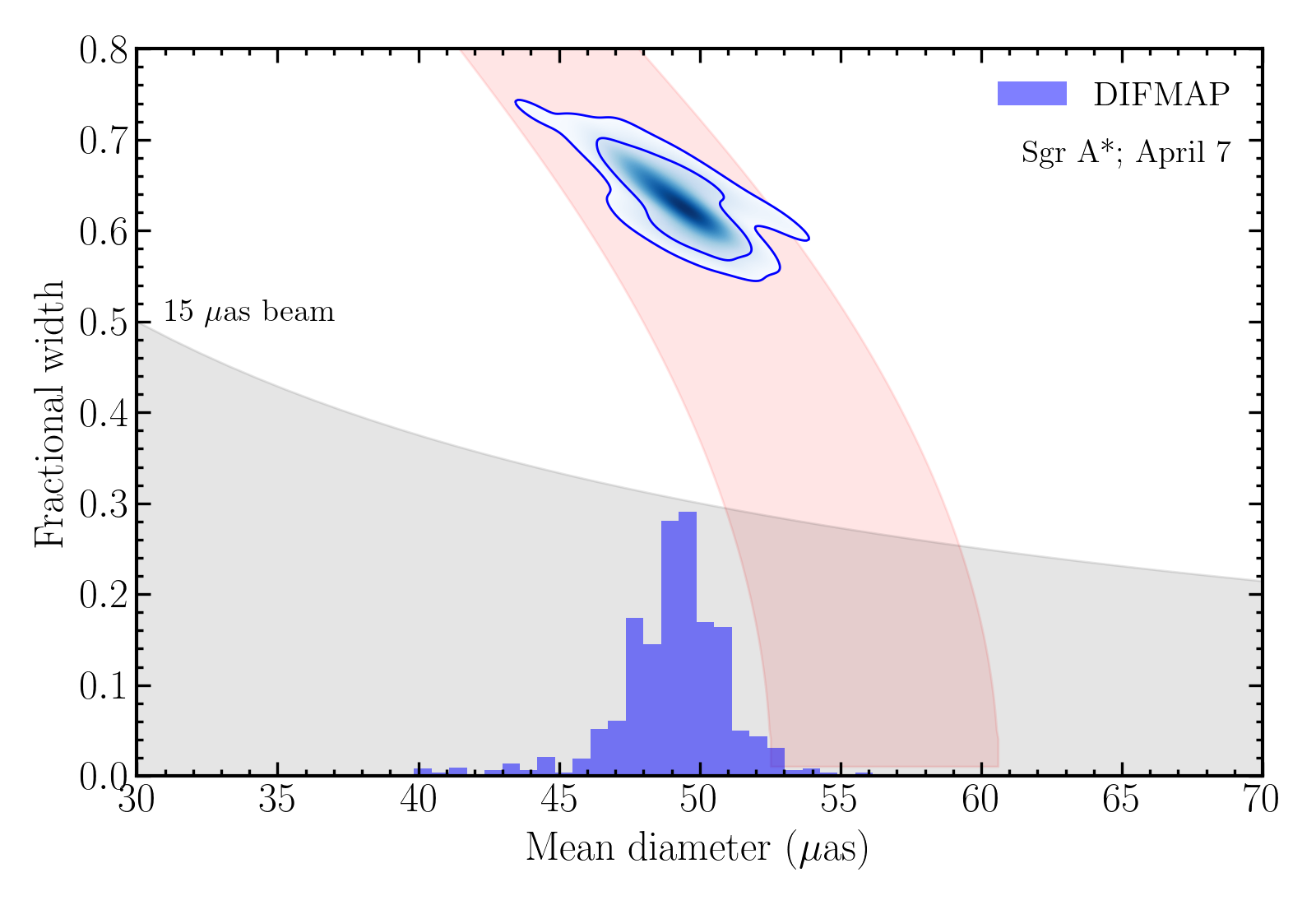}
\caption{The distributions of fractional ring widths and diameters for the top set images for the \sgra\ April 7 data obtained using {\it (top)} \ehtim, {\it (middle)} \smili, and {\it (bottom)} \difmap algorithms (see \citetalias{PaperIII}). Each image in these top sets is characterized using \charm. The two contour levels correspond to 68th and 95th percentiles of the images. The histograms are the projections of the distributions on the mean diameter axis. The grey shaded area corresponds to a nominal 15$\;\mu$as resolution of the telescope array. The pink shaded area shows the expected anticorrelation between diameter and width that is caused by Gaussian broadening of a thin ring.}
\label{fig:imaging_data}
\end{figure}

Even though we used, for simplicity, a single calibration factor in writing equation~(\ref{eq:delta}), in reality, this factor has two components that are multiplicative in nature, i.e., $\alpha_{\rm c}=\alpha_1 \times \alpha_2$. This is because the calibration factor encompasses both a theoretical bias ($\alpha_1$) as well as potential measurement biases ($\alpha_2$), which are generally independent of each other and need to be quantified separately. As a result, there are four sources of uncertainty in total that contribute to the error budget in the measurement of the deviation parameter $\delta$. These are:
\begin{enumerate}
    \item the uncertainty in the measurement of $\theta_g$ from stellar dynamics, as described in the previous section,
    \item the formal uncertainties obtained from measuring the diameter $\hat{d}_m$ of the bright ring from the data (see Section~\ref{sec:msmt}), 
    \item the theoretical uncertainties in the ratio $\alpha_1\equiv d_m/d_{\rm sh}$ between the true diameter $d_m$ of the bright ring of emission and the diameter of the shadow $d_{\rm sh}$, given a model for the black-hole spacetime and emissivity in the surrounding plasma (see Section~\ref{sec:alpha1}), and 
    \item the uncertainties in the ratio $\alpha_2\equiv \hat{d}_m/d_m$ between the measured ring diameter $\hat{d}_m$ and its true value $d_m$ that result from fitting analytic or pixel-based models to EHT data and arises, e.g., from the limited $u-v$ coverage, model complexity, and incomplete prior knowledge of telescope gains (see Section~\ref{sec:alpha2}). 
 \end{enumerate}   
We present below our quantitative inference of the formal measurement uncertainties as well as of the various calibration factors.

\subsection{Measurement Uncertainties}
\label{sec:msmt}
 We focus here on the 2017 April 7 data set because it satisfies three important criteria: The ALMA array, which leads to the highest SNR data, participated in the observation; there is no evidence for an X-ray flare or large excursion in the 1.3 mm flux; and the interferometric coverage samples the visibility amplitude minima in the $u-v$ plane, which are critical for establishing an accurate image size measurement. We also note that the analysis presented in \citetalias{PaperIII} for the 2017 April 6 data provide consistent results. The measurement uncertainties are obtained from modeling these data with imaging and model fitting tools, as discussed in \citetalias{PaperIII} and \citet[][hereafter \citetalias{PaperV}]{PaperIV}, respectively. Here, we quantify these results using characterization tools, as we describe below. 

\begin{figure}[t]
\centering
\includegraphics[width=.47\textwidth]{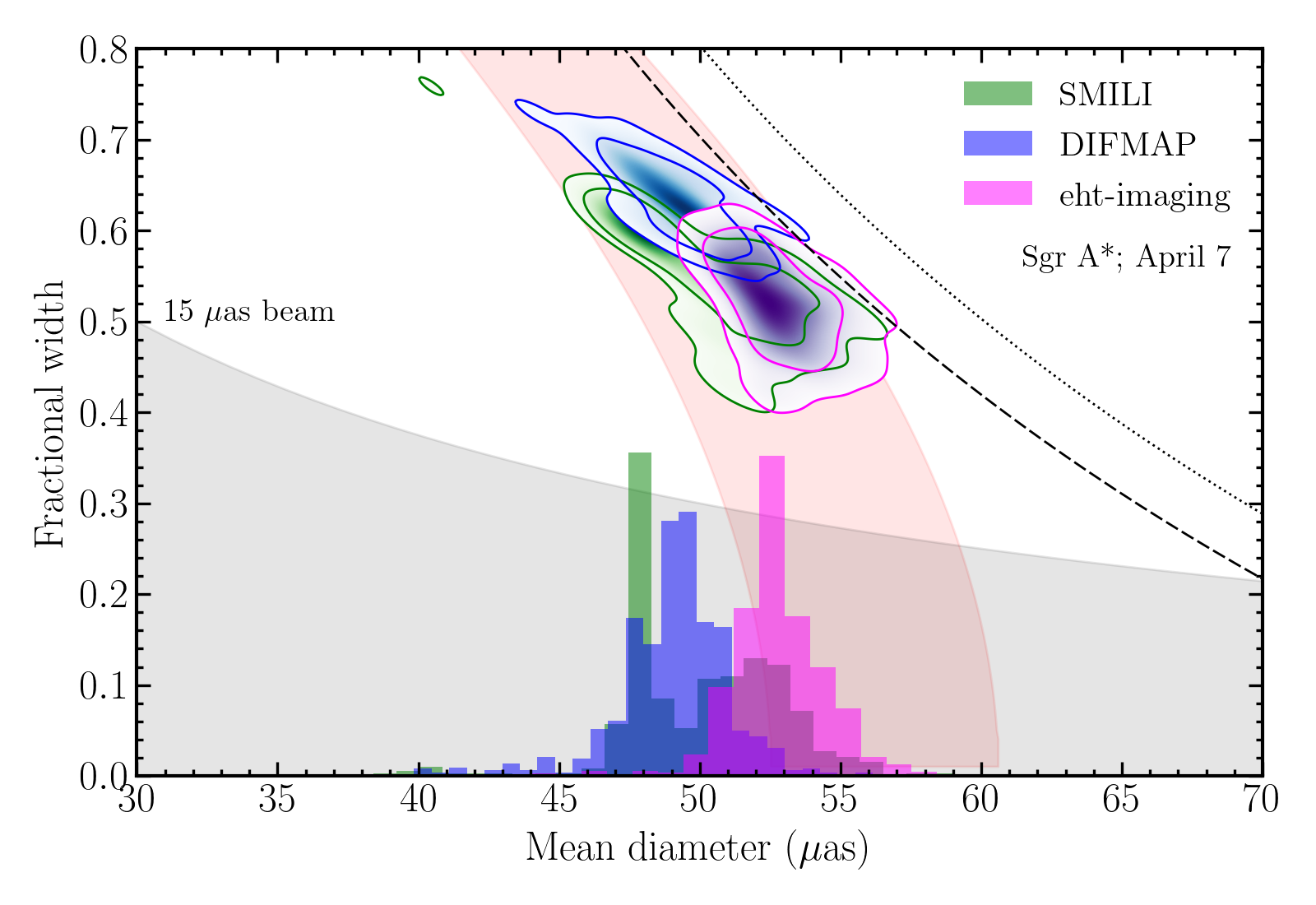}
\caption{A comparison of the fractional widths and mean diameters measured in the top sets of the three imaging algorithms for the reconstruction of the \sgra\ April 7 data. The contours show the 68th and 95th percentile of the top set images, as before. The pink shaded area shows the expected anticorrelation as in Figure~\ref{fig:imaging_data}. The small differences in the inferred parameters from each algorithm lie along this expected anticorrelation. The dashed and dotted lines correspond to (ring diameter $+$ ring width) $=90\,\mu$as and $80\,\mu$as, respectively (see the discussion in section~\ref{sec:surface_radius}). }
\label{fig:imaging_all}
\end{figure}

We use the CHaracterization Algorithm for Radius Measurements (\charm) that is based on the feature extraction algorithm that was employed in \citet{VI_EHT2019_M87} and improved further in \cite{Ozel2021}. Briefly, the algorithm {\it (i)} chooses a trial center for a potential ring-like feature; {\it (ii)} uses a rectangular bivariate spline interpolation to obtain radial cross sections of the filtered image brightness at 128 equidistant azimuthal orientations starting from the trial center; {\it (iii)} measures, in each radial cross section, the distance of the location of peak brightness from the trial center and identifies the ring diameter as two times the median value of this distance; {\it (iv)} iterates the location of the trial center and steps {\it (i)}$-${\it (iii)} such that the variance in the diameter along different cross sections is minimized; and {\it (v)} measures a median FWHM of the ring by fitting an equivalent asymmetric Gaussian to each radial cross section such that the corresponding integrated brightness of the cross section of the filtered image is equal to that of the Gaussian. We then define the fractional width as the FWHM of the ring in units of the ring diameter.

We show in Figure~\ref{fig:imaging_data} the fractional width and diameter measurements obtained for \ehtim, \smili, and \difmap top-set images for the April 7 \sgra data (see \citetalias{PaperIV} for the details of these three imaging algorithms). Even though we apply \charm to all of the topset images, without employing clustering filters (e.g., to select only ring-like images), we find that the 68th and 95th percentile contours for the ring parameters form compact regions for each algorithm. This indicates that there is a discernible brightness depression in each image that is surrounded by a bright region that has a robust characteristic size. 

\begin{figure}
\centering
\includegraphics[width=.47\textwidth]{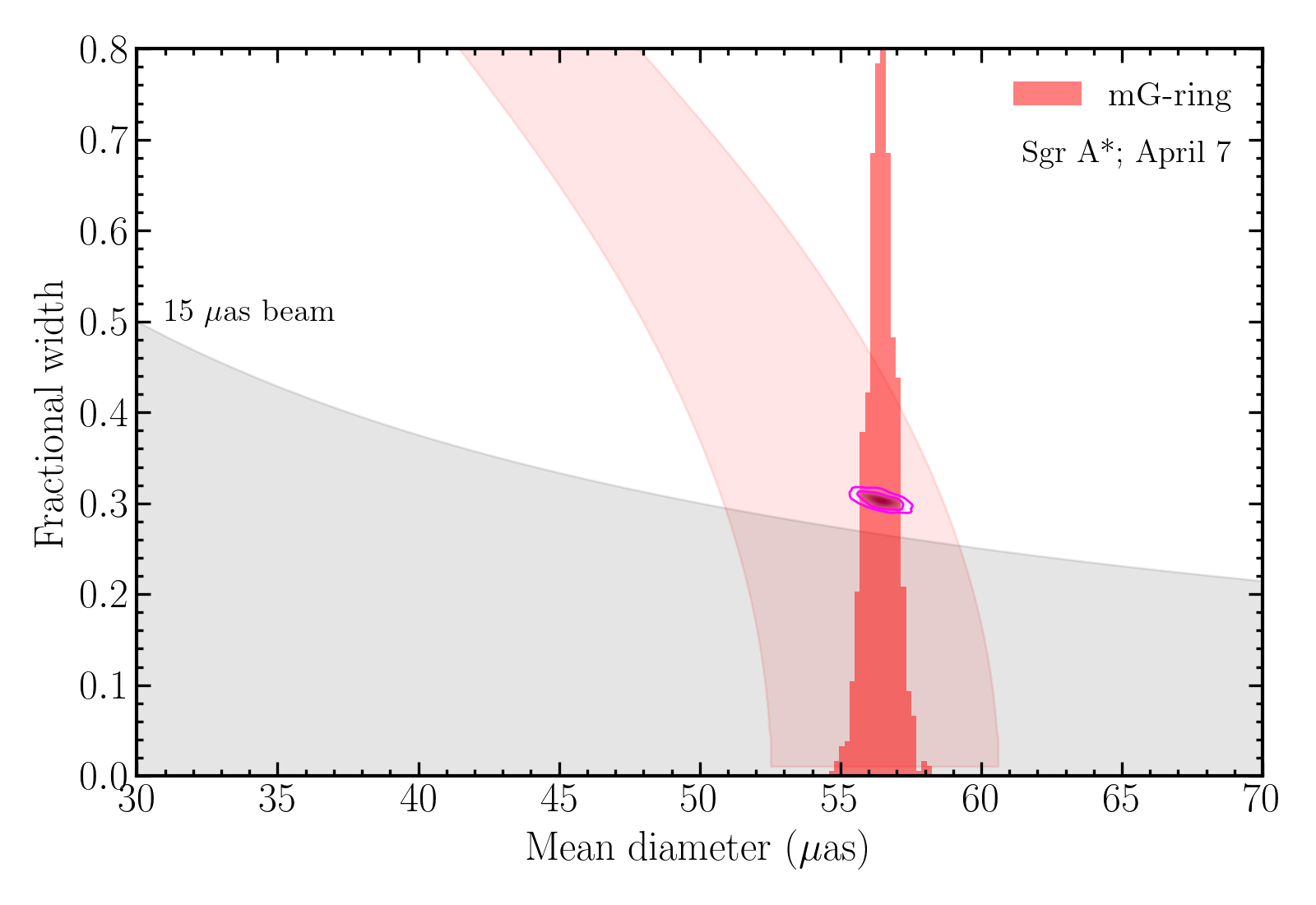}
\caption{The fractional ring width and diameter measurements obtained from fitting mG-ring models to the April 7 visibility domain data for \sgra. The shaded areas are the same as in Figure~\ref{fig:imaging_all}.} \label{fig:mrings_data}
\end{figure}

The grey bands in Figure~\ref{fig:imaging_data} mark the effective limit of the fractional width that can be measured with imaging methods because of the finite resolution of the EHT array. The pink bands show the expected anticorrelation between the ring FWHM and the measured diameter $\hat{d}_{\rm m}$ that arises from the Gaussian broadening of an infinitesimally thin ring of diameter $d$. To first order in the fractional ring width, this anticorrelation follows (see Appendix~G of \citealt{IV_EHT2019_M87}) 
\begin{equation}
    \hat{d}_{\rm m} = d - \frac{1}{4 \ln 2} \frac{{\rm FWHM}^2}{d}. 
\end{equation}
Because some of the inferred fractional widths are relatively large, in calculating the actual shaded areas in Figure~\ref{fig:imaging_data} we do not make this first-order approximation but rather employ a numerical evaluation of the complete expression.  

\begin{figure*}
\centering
\includegraphics[width=0.99\textwidth]{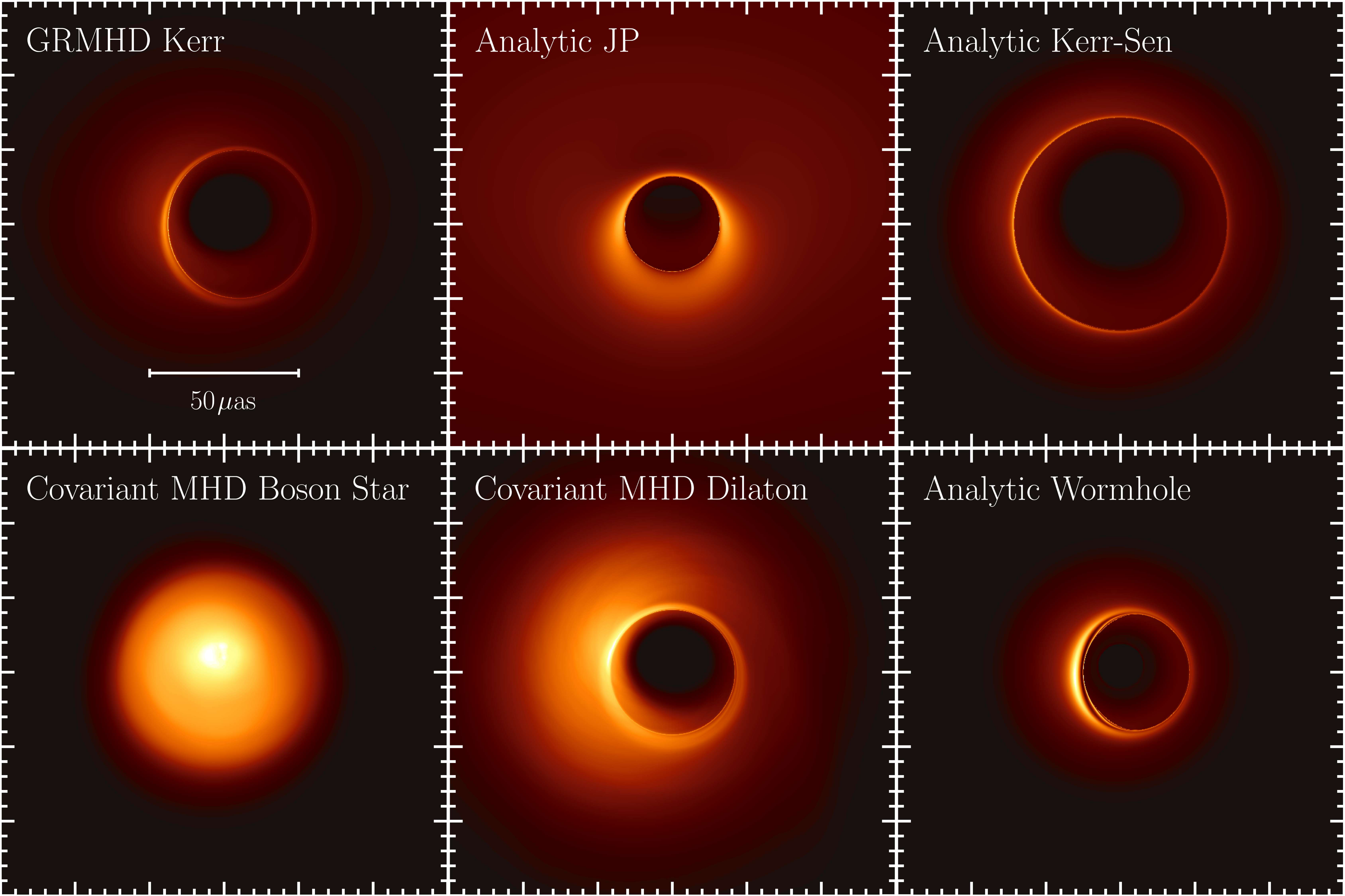}
\caption{Selection of simulated $1.3$~mm Sgr A* images for different spacetime geometries and plasma models. From left to right, the upper row presents time-averaged images of a Kerr MAD GRMHD simulation and semi-analytic accretion flow models with the background spacetime given by the JP metric and the Kerr-Sen spacetime, respectively.
The bottom row presents images from a covariant MHD simulation of accretion onto a boson star, a covariant MHD simulation of a dilaton black hole, and a semi-analytic accretion flow model from a traversable wormhole spacetime, respectively. Images from top row models were used in the analysis in this study, whereas bottom row models were not used and are shown to provide examples of images from non-Kerr spacetimes and horizonless compact objects. In all cases for which a central brightness depression is present, the size of the ring-like image scales with the boundary of the black-hole shadow.}
\label{fig:image_gallery}
\end{figure*}

In Figure~\ref{fig:imaging_all}, we compare the fractional widths and mean diameters inferred for \sgra\ with the three imaging algorithms. Even though there appear to be small differences in the mean diameter, all of the contours lie along the expected anticorrelation. This suggests that the differences are simply caused by the various algorithmic choices and do not reflect inconsistencies between them. 

We also use the image diameter and fractional width obtained from fitting analytic models to the visibility data (see \citetalias{PaperIV}). In particular, we focus on the mG-ring model described in \citetalias{PaperIV}, which comprises a Gaussian broadened ring with flux enhancements on the ring with m-fold azimuthal symmetry and an additional central Gaussian floor component. We use the posteriors obtained from the fitting algorithm \texttt{Comrade} \citep{tiede2022}. In Figure~\ref{fig:mrings_data}, we show the posterior over the diameter and the fractional width obtained from fitting the mG-ring model to the April~7 data. The narrow posterior in diameter for this model reflects primarily the insufficient degree of model complexity in the model, as can be seen in the synthetic data analysis below (see also discussion in~\citealt{Psaltis2020b}). Nevertheless, the inferred diameter is consistent with those of the imaging methods, given the expected anticorrelations.

\begin{figure}
\centering
\includegraphics[width=.47\textwidth]{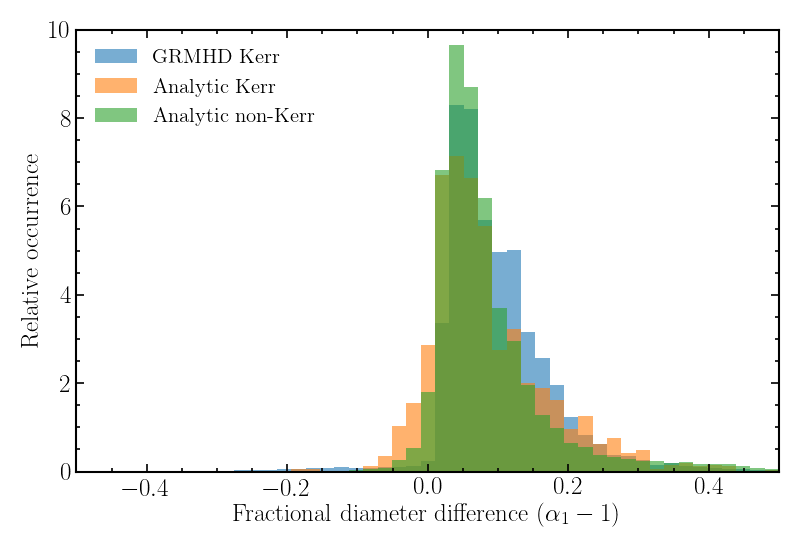}
\caption{The fractional diameter difference between the diameter of peak emission in the image of a black hole and that of its shadow obtained from three different types of simulations. The blue histogram shows the result from 180,000 snapshots from time-dependent GRMHD simulations in the Kerr metric, spanning a broad range of spins, inclinations, and plasma parameters. The orange histogram shows the same for analytic plasma models in the Kerr metric that relaxes some of the assumptions of the GRMHD simulations, while the green histogram shows the results for analytic plasma models in metrics that deviate from Kerr either parametrically (JP) or through different solutions to the field equations (EMDA). All distributions peak at small positive values. }
\label{fig:a1_calib}
\end{figure}

\subsection{The $\alpha_1$ Calibration Factor and its Uncertainties}
\label{sec:alpha1}

In this section, we use simulated black-hole images to quantify the correction factor $\alpha_1$, which is the ratio between the diameter of the peak brightness of the image and the diameter of the black-hole shadow. We employ three different types of models to explore a range of effects releated to the plasma properties, spacetime characteristics, and different numerical realizations of the turbulent flow. 

The first category of images comprises $\sim 180,000$ snapshots of GRMHD accretion-flow simulations discussed in \citetalias{PaperV}. The simulations cover a range of 
black-hole spins ($a=-0.94,-0.5,0.,0.5,0.94$), observer inclinations ($i=$10, 30, 50, 70, and 90 degrees), MAD and SANE magnetic field configurations, and thermal electron distributions with temperature prescriptions characterized by $R_{\rm high}=$10, 40, and 160. For each combination in these sets of parameters, we also considered snapshots calculated with two different GRMHD simulation algorithms: \texttt{KHARMA}~\citep{Prather2021} and \texttt{BHAC}~\citep{Porth2017}, and corresponding images calculated using two different covariant radiation transport schemes: \texttt{ipole} \citep{Moscibrodzka+2018} and \texttt{BHOSS} \citep{Younsi+2012,Younsi+2016}. 

The second set comprises $\sim 4000$ images from covariant plasma models in the Kerr metric that go beyond some assumptions of GRMHD. These employ analytic calculations that are agnostic to the particular microprocesses responsible for angular momentum transport and particle heating. The particular parameters of these models are discussed in detail in  \citet{Ozel2021}. 

The third category includes $\sim 200,000$ images from analytic models that explore a range of black hole metrics that are either parametrically different from the Kerr metric or represent other known solutions to the field equations \citep{Younsi2021}. For the former, we employ the Johannsen-Psaltis (JP) metric \citep{Johannsen2011,Johannsen2013b}, which enables parametric deviations from Kerr and recovers the Kerr spacetime when its deviation parameters vanish, whilst still guaranteeing many of the basic properties of the Kerr metric (i.e., it is Petrov Type-D, free of pathologies, etc). For the latter, we utilize the EMDA (Kerr-Sen) metric \citep{Garcia+1995}, which is a solution to the field equations of a modified gravity theory with additional scalar degrees of freedom. The plasma model is the same covariant analytic model of \citet{Ozel2021} and the model library spans different black-hole spins, observer inclinations, magnetic field configurations, plasma parameters and, where appropriate, the metric parameters, as discussed in \citet{Younsi2021}. We refer to these models as analytic non-Kerr.

 Using the covariant radiation transport code \texttt{BHOSS} \citep{Younsi+2012,Younsi+2016}, Figure~\ref{fig:image_gallery} presents a selection of illustrative simulated 1.3 mm \sgra\ images from five different non-Kerr spacetimes, together with an image from a GRMHD simulation of a Kerr black hole. The field of view in all panels is $150~\mu \mathrm{as}$ in both directions, with the brightest pixel value in each panel normalized to unity. We show in the top row mean images from covariant MHD simulations averaged over a time window of $5000~GM/c^{3}$, with snapshots every $10~GM/c^{3}$ ($\sim 3.5$ minutes for \sgra).
The Kerr GRMHD simulation parameters are: MAD magnetic field configuration, $a=0.9375,\, i=30^{\circ},\, R_{\rm low}=1,\, \mathrm{and} \ R_{\rm high}=10$ (see~\citetalias{PaperV} for further details of the modeling). The upper middle panel shows an image of accretion onto a non-rotating dilaton black hole \citep{Mizuno+2018}. The upper right panel presents the image from a simulation of accretion onto a boson star \citep{Olivares+2020,Fromm2021}.
The boson star image represents one example of a compact object without an event horizon or an unstable photon orbit, thereby lacking a central brightness depression or a photon ring in its image. We do not consider such configurations in the calibration procedure discussed here but explore them  in detail in Section~\ref{sec:eh}. 

We present in the bottom row of Figure~\ref{fig:image_gallery} images from non-Kerr spacetimes with the background semi-analytic accretion flow model as specified in \cite{Ozel2021} and \cite{Younsi2021}. These spacetimes are the JP and the Kerr-Sen (EMDA) metrics, as well as a spinning traversable wormhole spacetime \citep{Teo1998,Harko+2009}. The JP metric for this example is non-spinning, with deformation parameters chosen to push the unstable photon orbit very close to the event horizon (hence the smaller central brightness depression). The Kerr-Sen spacetime parameters (axion and dilaton field couplings) have been chosen to produce an image with a photon ring larger than is possible with a Kerr black hole. Finally, the rotating wormhole spacetime is chosen to have a throat radius equal to the event horizon radius of a Kerr black hole with the same spin ($a=0.9375$). In all of the examples with a central brightness depression, the size of the ring-like image scales with that of the black-hole shadow.

We convolve all of the images in the three categories with an $n=2$, 15\;G$\lambda$ Butterworth filter to mimic the resolution of the EHT array. We then apply the characterization algorithm \charm to all of these images to measure the median diameter $D_{\rm im}$ of the bright ring of emission, with respect to the analytically calculated center of the black hole shadow. We also calculate the shadow diameter in each spacetime; for Kerr, we use the analytic approximation derived in \citet{Chan2013}. We then define the calibration factor $\alpha_1$ as the ratio of the median diameter to the diameter of the shadow. We will refer to the difference $\alpha_1 - 1$ as the fractional diameter difference. If the peak emission in the bright ring coincides with the shadow boundary, then the fractional diameter difference would be equal to zero. 

We show in Figure~\ref{fig:a1_calib} the distribution over the fractional diameter difference for the three types of images. As discussed in \citet{Ozel2021} and \citet{Younsi2021}, the distribution peaks at small positive values of $\alpha_1-1$, indicating that the peak of the bright ring is slightly larger than the boundary of the black hole shadow. 

\begin{figure}
\centering
\includegraphics[width=.47\textwidth]{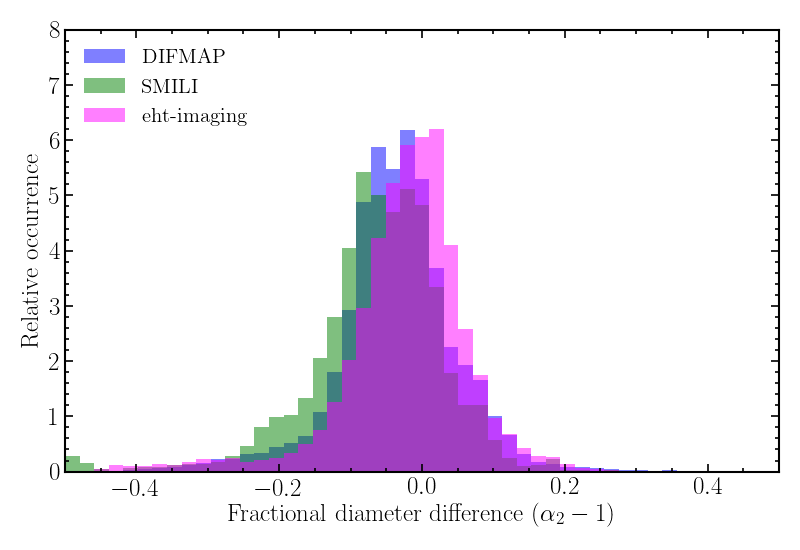}
\caption{The fractional diameter difference between the diameter of peak emission in ground truth images and the those reconstructed through the three different imaging methods used for EHT analyses. Synthetic data cover 145 sets selected from numerical and analytic Kerr and non-Kerr models, while the image diameters were inferred for all of the top-set images for the image reconstructions of each data set. The small offsets in the calibration parameter mimic those seen in the analysis of the actual \sgra\ data.}
\label{fig:a2_calib}
\end{figure}

\begin{figure}
\centering
\includegraphics[width=.47\textwidth]{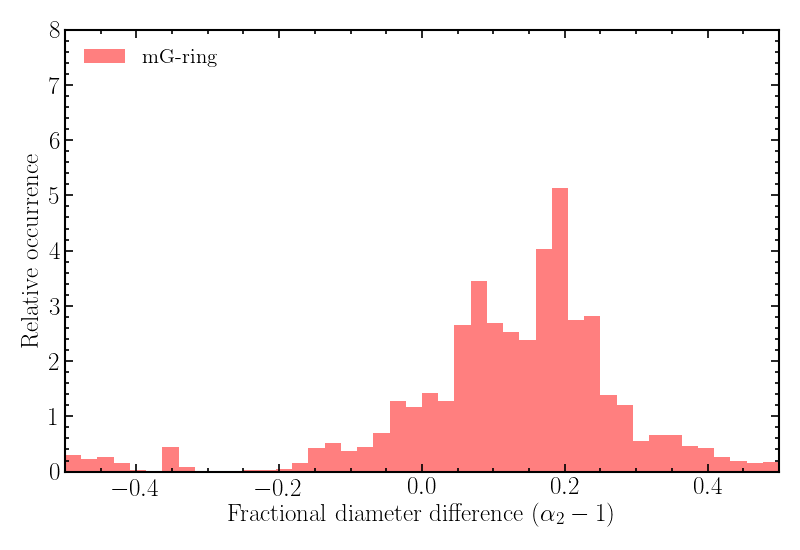}
\caption{The fractional diameter difference between the diameter of peak emission in ground truth images and those reconstructed through fitting the mG-ring model to the visibility domain synthetic data used for EHT analyses. }
\label{fig:a2_calib_mring}
\end{figure}

\subsection{The $\alpha_2$ Calibration Factor and its Uncertainties}
\label{sec:alpha2}

We turn to quantifying the correction factor $\alpha_2$ and its uncertainty that arises from applying imaging and model fitting tools to infer the size of a ring-like image. To this end, we first characterized all simulations discussed in Section \ref{sec:alpha1} based on image morphology and size, degree of variability, spacetime metric, and plasma model. We then randomly selected segments and snapshots from each category. We assigned a random position angle in the sky to each image and generated synthetic EHT data from them using the VLBI synthetic data generation pipeline SYMBA (\citealt{2019Janssen,Roelofs2020,natarajan2022}). SYMBA accounts for the effects of interstellar scattering through the Galactic disk, as well as several realistic atmospheric, instrumental, and calibration effects. In addition, we designated a last category in which synthetic data were generated from a small number of snapshots but with several different realizations of all the measurement uncertainties. This yielded a total of 145 synthetic data sets. 

We carried out blind image reconstructions and mG-ring fits to all the synthetic data using the same EHT imaging pipelines as those applied to \sgra\ data, separating into teams who did not have any prior knowledge of the synthetic data characteristics. As for the case of the real data, imaging teams generated a {\it top set} of reconstructions for each synthetic data set, using the exact set of algorithmic parameters as those used for the real \sgra\ data. We applied \charm to the entire top-set image reconstructions (for a total of 145 data sets $\times$ 2000 top-set parameters $\times$ 3 algorithms) as well as to the ground-truth images to measure the calibration factor $\alpha_2$. Modeling teams applied the snapshot fitting procedure with an mG-ring model and returned their posteriors for the model diameter, which we used to calculate the $\alpha_2$ calibration factor. 

In the majority of cases, the set of reconstructions that correspond to the full range of top-set parameters or posteriors yielded a narrow range of diameters and widths for the ring features, indicating a robust inference of the prevalent features with little sensitivity to the choice of regularizers. However, in $< 30\%$ of the data sets, the features of the images varied significantly within the topset parameters, leading to an uncertainty in the ring diameter that is $\sim 3-8$ times larger than what is measured in the \sgra\ data (see Fig.~\ref{fig:imaging_data}). This primarily happens when the image size, position angle, and asymmetry of the ground truth image that led to the particular synthetic data set conspire in a way to remove any prominent salient features in the visibility domain and the image reconstruction is dominated by the priors rather than any unique features in the data that an imaging or model fitting algorithm can pick up on. More quantitatively, we define the spread in the diameter for all the reconstructions of a given synthetic data set by using the metric 
\begin{equation}
{\rm diameter\ spread} = \frac{d_{85}-d_{15}}{d_{50}},
\end{equation}
where $d_{85}, d_{50}$, and $d_{15}$ refer to the 85th, 50th, and 15th percentile of diameters in a given distribution, respectively. The spread in the top-set reconstructions of the actual \sgra\ data using this metric is $0.06-0.1$ (see Fig~\ref{fig:imaging_data}). We place a conservative limit of diameter spread less than 0.2 for the synthetic data reconstructions and include only the data sets that fulfill this criterion in our derivation of the $\alpha_2$ calibration parameters. 

Figures~\ref{fig:a2_calib} and \ref{fig:a2_calib_mring} show the distributions of the fractional diameter difference $\alpha_2-1$ for the imaging reconstructions and mG-ring model fits, respectively, of the synthetic data sets discussed above. The trend in Figure~\ref{fig:a1_calib}, i.e., the slight offset between the peaks of the distributions calculated for the different imaging methods, follows the one we see in the reconstruction of the actual EHT \sgra\ data very closely (see Fig.~\ref{fig:imaging_data}). This result reinforces our conclusion that the small differences in the inferred diameters between different algorithms are primarily caused by the different methodologies, prior, and regularizer choices in those methods (see \citetalias{PaperIII}). The same is true for the trend in the mG-ring results, albeit corresponding to more marked differences. 

\begin{figure}
\centering
\includegraphics[width=.47\textwidth]{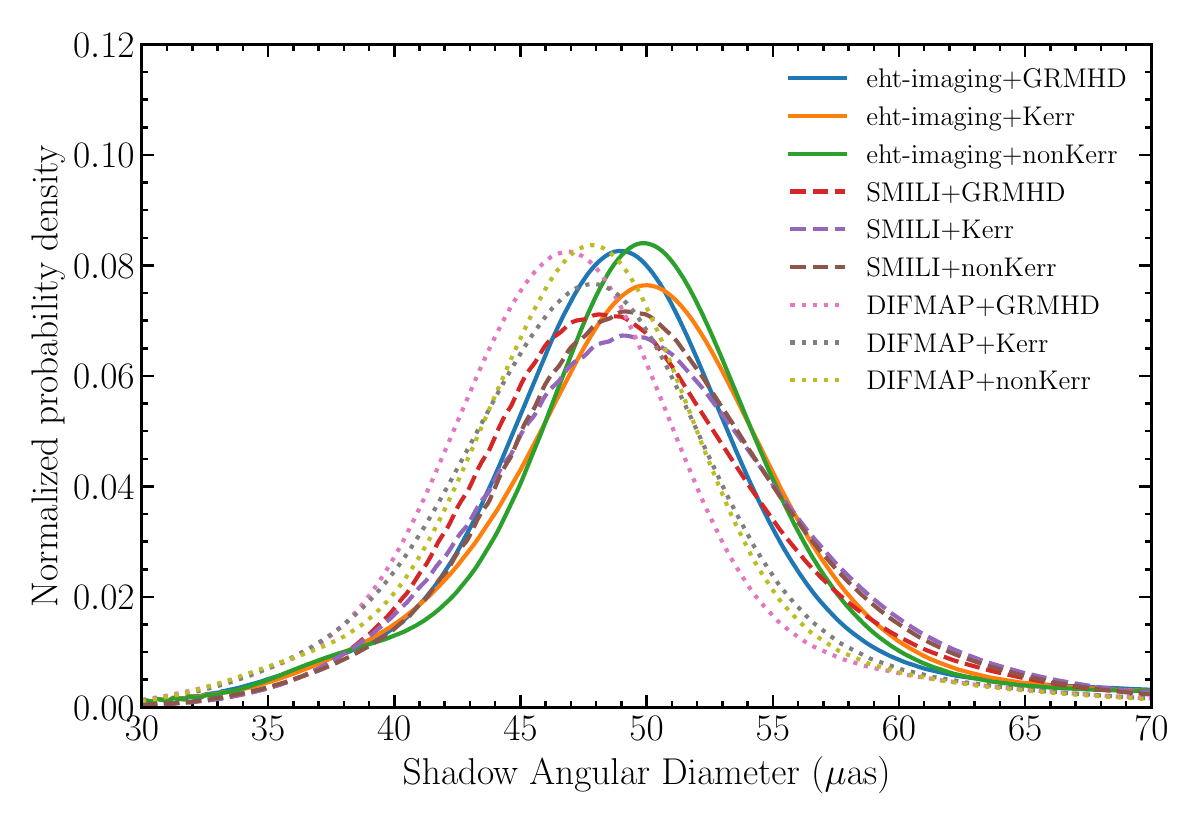}
\caption{Posteriors over the shadow diameter inferred using the measurements of the ring diameter size $\hat{d}_{\rm m}$ based on three image-domain algorithms, as well as the two  factors $\alpha_1$ and $\alpha_2$ that quantify the theoretical and measurement calibrations.}
\label{fig:dsh_post}
\end{figure}

\begin{table}[t]
\caption{The Inferred Shadow Diameter of \sgra\ in $\mu$as.}
\label{tab:delta_constr}
\begin{ruledtabular}
\begin{tabular}{lccc}
\textrm{~} & GRMHD & Analytic & Analytic \\
& & Kerr & non-Kerr \\
\colrule
\texttt{eht-imaging}  & $48.9_{-5.1}^{+5.2}$  & $50.0_{-5.6}^{+5.6}$ & $49.9_{-4.9}^{+5.2}$ \\
\smili & $48.1_{-5.2}^{+6.3}$  & $49.1_{-5.7}^{+6.5}$  & $49.2_{-5.4}^{+6.2}$ \\ 
\difmap & $46.9_{-5.2}^{+4.9}$  & $47.8_{-5.7}^{+5.1}$  & $47.8_{-5.2}^{+4.9}$ \\ 
\end{tabular}
\end{ruledtabular}
\end{table}

\subsection{The Diameter of the Black-Hole Shadow}
\label{sec:dsh}
We use the combination of the measurements and calibrations discussed in the previous sections to infer the diameter of the boundary of the black hole shadow, $d_{\rm sh} = \hat{d}_{\rm m}/(\alpha_1 \; \alpha_2)$. The posterior over the shadow diameter is given by
\begin{eqnarray}
    P(d_{\rm sh} \vert \hat{d})&=&C \int d\alpha_1 \int d\alpha_2  \;{\cal L}[\hat{d}|d_{\rm sh},\alpha_1,\alpha_2] \nonumber\\
    &&\qquad\qquad\times P(d_{\rm sh})  P(\alpha_1) P(\alpha_2)\;, 
    \label{eq:d_posterior}
\end{eqnarray}
where ${\cal L}[\hat{d}|d_{\rm sh},\alpha_1,\alpha_2]$ is the likelihood of measuring a ring diameter $\hat{d}$ given the model parameters, $P(\alpha_1)$ and $P(\alpha_2)$ are the distributions of the calibration parameters, and $C$ is an appropriate normalization constant. $P(d_{\rm sh})$ is the prior over the shadow diameter, which we assume to be flat over a range that is much broader than that of the posteriors. 

\begin{figure}
\centering
\includegraphics[width=.4\textwidth]{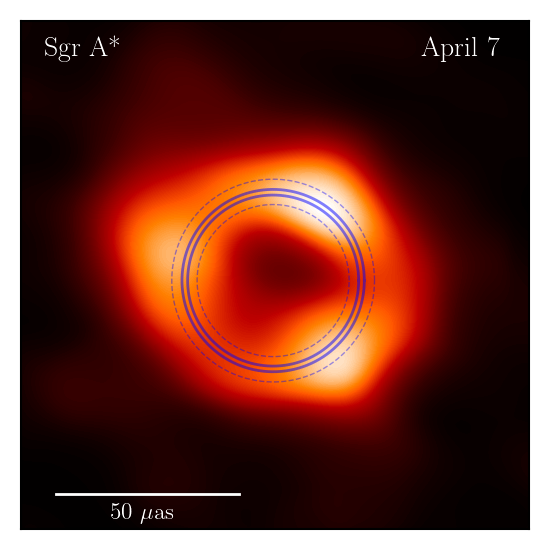}
\caption{Inferred diameter of the black hole shadow boundary overlaid on the average EHT image of \sgra\ obtained from the 2017 April 7 data. Solid lines show the range of most likely values while the dashed lines show the envelope of the 68th percentile credible intervals for all methods. }
\label{fig:shadow_image}
\end{figure}

We show in Figure~\ref{fig:dsh_post} the posteriors over the shadow diameter as inferred from the three image-domain algorithms and for the different theoretical calibrations discussed in Section~\ref{sec:alpha1}. In Table~2, we report the most likely values of the black hole shadow diameter for \sgra\ as well as the 68th percentile credible levels. Finally, in Figure~\ref{fig:shadow_image}, we overlay the inferred shadow boundaries on the average EHT image of \sgra\ obtained from the 2017 April 7 data \citep{PaperIII}. In this plot, the solid lines show the range $46.9-50.0 \; \mu$as of the most likely values and the dashed lines show the envelope of the 68th percentile credible intervals across the different methods, spanning $41.7-55.6 \; \mu$as.

\subsection{Constraints on the Deviation Parameter}
\label{sec:delta}

Using the uncertainties discussed above, we obtain the posterior over the deviation parameter $\delta$ by
\begin{eqnarray}
    P(\delta\vert \hat{d})&=&C \int d\alpha_1 \int d\alpha_2 \int d\theta_{\rm g} \;{\cal L}[\hat{d}|\theta_{\rm g},\alpha_1,\alpha_2,\delta] \nonumber\\
    &&\qquad\qquad\times P(\delta) P(\theta_{\rm g}) P(\alpha_1) P(\alpha_2)\;.
    \label{eq:d_posterior}
\end{eqnarray}
Here $C$ is an appropriate normalization constant, ${\cal L}[\hat{d}|\theta_{\rm g},\alpha_1,\alpha_2,\delta]$ is the likelihood of measuring a ring diameter $\hat{d}$ given the model parameters, which we identify with the distributions of measurements from the imaging and visibility domain methods, $P(\theta_{\rm g})$ denotes the prior in $\theta_{\rm g}$ given by stellar dynamics measurements, and $P(\alpha_1)$ and $P(\alpha_2)$ are obtained from the calibration procedures outlined in Secs. \S\ref{sec:alpha1} and \S\ref{sec:alpha2}. 

\begin{figure}
\centering
\includegraphics[width=.47\textwidth]{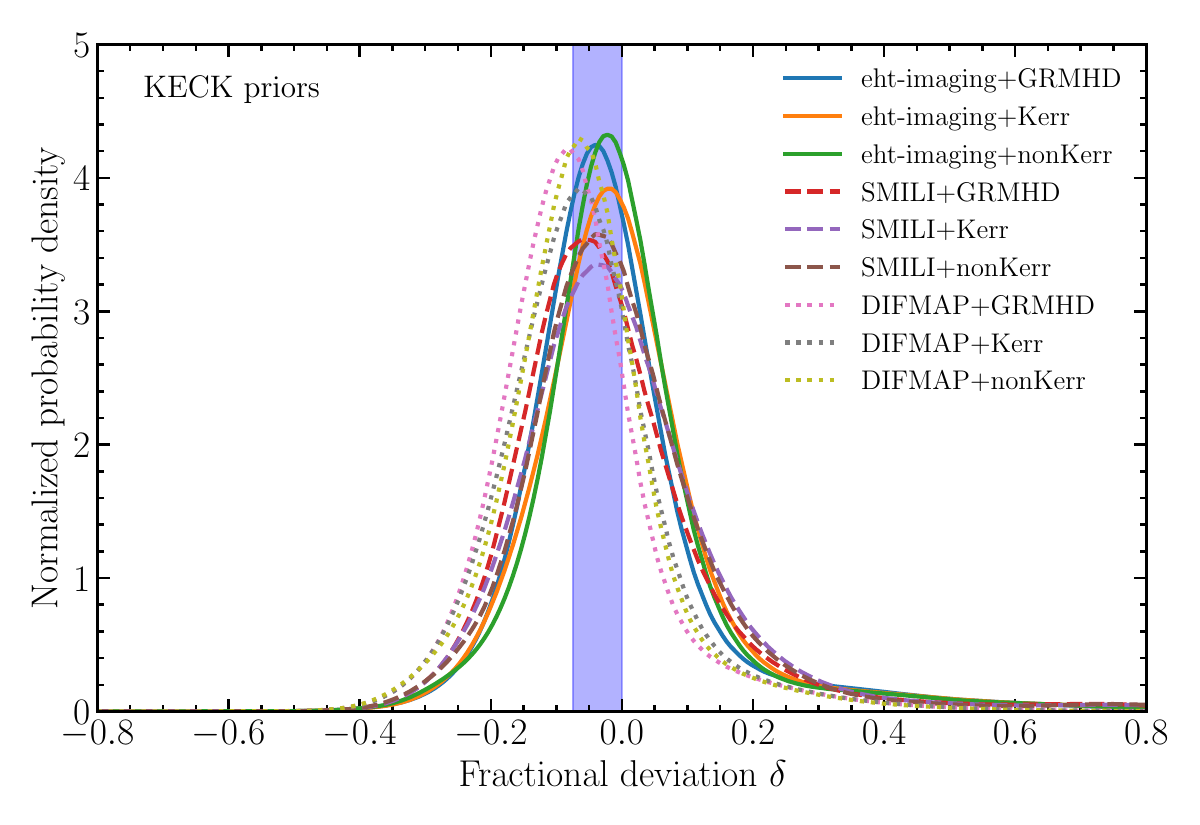}
\includegraphics[width=.47\textwidth]{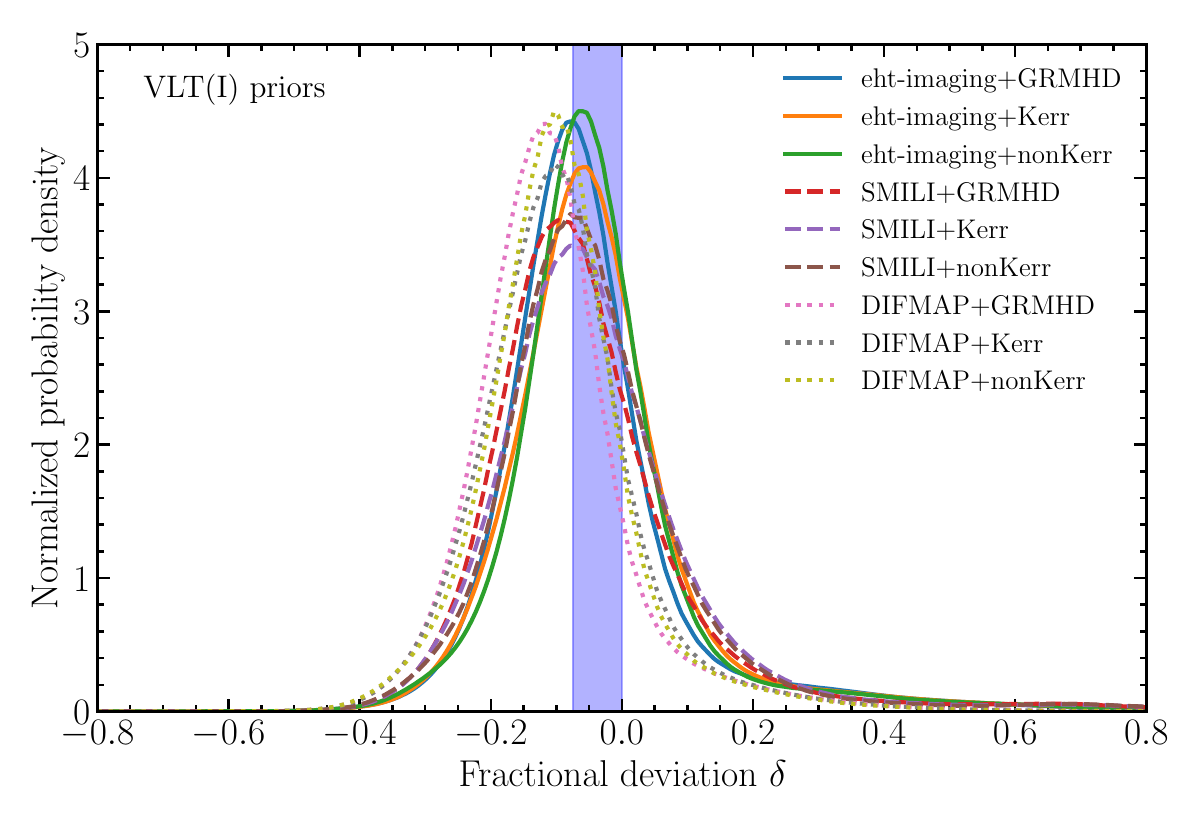}
\caption{Posteriors on the parameter $\delta$ that measures the deviation of the black hole shadow size obtained for \sgra from the Schwarzschild predictions. The top panel uses as a prior the angular size $\theta_g$ obtained with Keck observations, while the bottom uses the same quantity from VLT(I). The various curves correspond to different sets of theoretical models used for calibration and the measurements obtained with the various imaging methods. The purple-shaded area shows the $\sim 8$\% range predicted for the Kerr metric, depending on the black-hole spin and observer inclination.}
\label{fig:delta_post}
\end{figure}

\begin{table}[t]
\caption{Schwarzschild Deviation Parameter $\delta$ for \sgra.}
\label{tab:delta_constr}
\begin{ruledtabular}
\begin{tabular}{llccc}
\textrm{~}& $\theta_g$  Prior& GRMHD & Analytic & Analytic \\
& & & Kerr & non-Kerr \\
\colrule
\texttt{eht-} & VLT(I) & $-0.08_{-0.09}^{+0.09}$  & $-0.05_{-0.11}^{+0.09}$ & $-0.07_{-0.09}^{+0.10}$ \\ \texttt{imaging}
& Keck & $-0.04_{-0.10}^{+0.09}$ & $-0.02_{-0.11}^{+0.11}$ & $-0.02_{-0.09}^{+0.10}$ \\
\hline
\smili & VLT(I) & $-0.10_{-0.10}^{+0.12}$  & $-0.08_{-0.11}^{+0.13}$  & $-0.08_{-0.10}^{+0.12}$ \\ 
& Keck & $-0.06_{-0.10}^{+0.13}$ & $-0.04_{-0.11}^{+0.13}$ & 
$-0.04_{-0.10}^{+0.13}$ \\
\hline
\difmap & VLT(I) & $-0.12_{-0.08}^{+0.10}$  & $-0.10_{-0.10}^{+0.09}$  & $-0.10_{-0.09}^{+0.09}$ \\ 
& Keck & $-0.08_{-0.09}^{+0.09}$ & $-0.07_{-0.11}^{+0.10}$ & $-0.07_{-0.09}^{+0.09}$ \\
\hline
mG-ring & VLT(I) & $-0.17_{-0.10}^{+0.11}$  & $-0.14_{-0.13}^{+0.10}$  & 
$-0.17_{-0.09}^{+0.13}$ \\ 
& Keck & $-0.13_{-0.11}^{+0.11}$ & $-0.12_{-0.11}^{+0.11}$ & 
$-0.12_{-0.11}^{+0.10}$ \\
\end{tabular}
\end{ruledtabular}
\end{table}

\begin{figure}
\centering
\includegraphics[width=.47\textwidth]{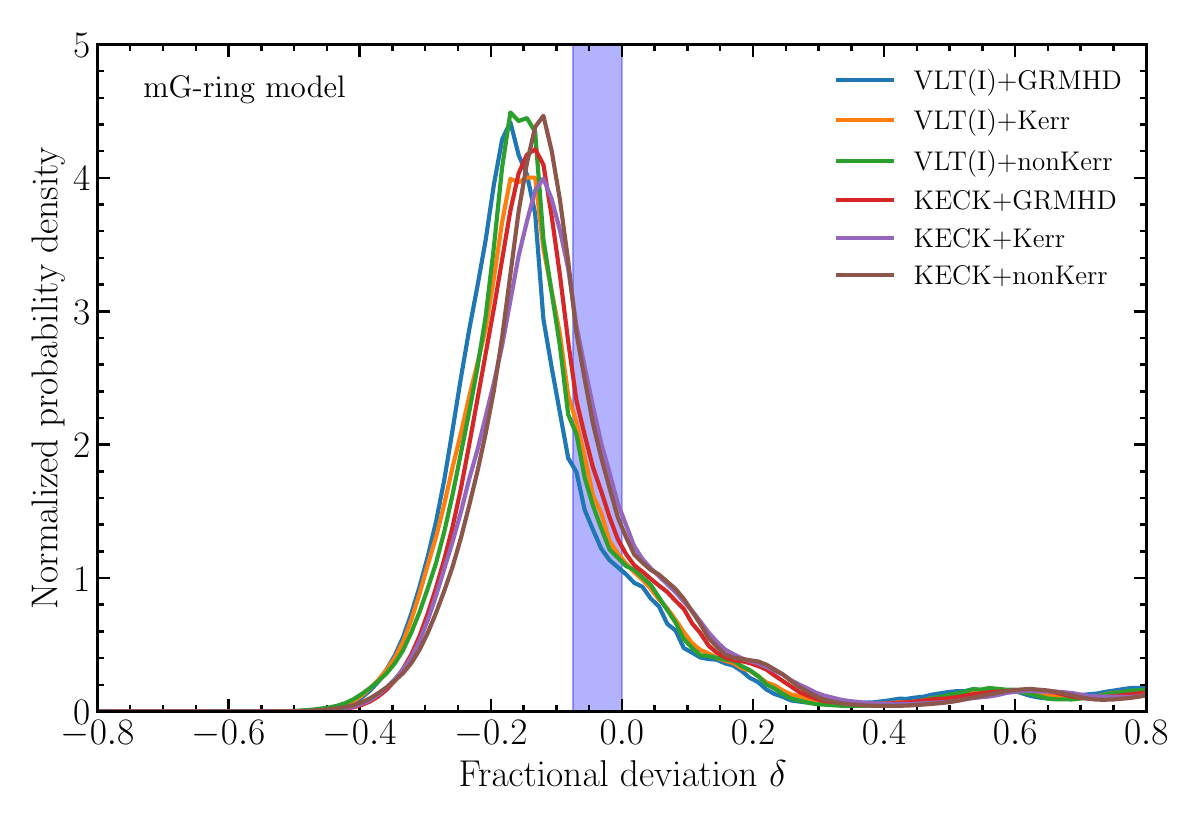}
\caption{Same as Fig.~\ref{fig:delta_post} but for the measurements obtained from fitting mG-ring to the visibility domain data.}
\label{fig:delta_post_mrings}
\end{figure}

As discussed earlier, we consider two separate priors for $\theta_{\rm g}$ denoted by Keck and VLTI, three different measurements of the ring diameter from imaging methods (together with their corresponding $\alpha_2$ calibrations) denoted by \ehtim, \smili, and \difmap, as well as three different sets of snapshot images for the $\alpha_1$ calibration denoted by GRMHD, Analytic Kerr, and Analytic JP. We assume a flat prior in the fractional deviation $\delta$, with limits that cover a range that is sufficiently broad not to affect the posteriors. We perform the two integrals in eq.~(\ref{eq:d_posterior}) numerically and show the resulting posteriors in the deviation parameter $\delta$ in Figure~\ref{fig:delta_post}.

We repeat the same procedure for the measurements obtained from mG-ring fits to the \sgra\ data. We show the corresponding result for the deviation parameter $\delta$ in Figure~\ref{fig:delta_post_mrings}. 

We present in Table~\ref{tab:delta_constr} the means and 68th percentile credible levels for the posteriors we obtain for the deviation parameter $\delta$ using different combinations of black hole mass priors, theoretical models used for calibration, as well as the imaging and model-fitting methods used on the \sgra data. All of the posteriors are consistent with each other and with no deviation from the General Relativistic predictions, we choose the \ehtim+Keck+GRMHD and \ehtim+VLTI+GRMHD combinations as the two fiducial cases to calculate constraints on the individual metric parameters in the remainder of this paper.

\section{Are there Viable Alternatives to an Event Horizon?}\label{sec:eh}

While there is overwhelming evidence that \sgra contains a large amount of mass confined within a very small volume, the question of whether it is a true black hole remains unresolved. The defining characteristic of a black hole is the presence of an event horizon. While it is relatively easy to show that observations of \sgra are consistent with the presence of an event horizon (e.g., the many black hole based models discussed in \citetalias{PaperV}), proving that all alternatives are ruled out is well-nigh impossible. Here we discuss what EHT observations of \sgra are able to add to this question.

If \sgra\ does not have an event horizon, it is likely to have some kind of a surface. Alternatively, the object might be a  boson star, naked singularity, or some other exotic solution of gravitational physics (see \citealt{Cardoso2019} for a review of exotic compact object models). If we could rule out some of these possibilities using observational data, then the case for \sgra having an event horizon would become significantly stronger. We discuss below two arguments against \sgra possessing a {\rm radiating} surface. One argument (Section~\ref{sec:thermal}) is well-developed in the literature \citep{Narayan+1998,Narayan_2002,Broderick_Narayan_2006,Broderick_Narayan_2007,Narayan_McClintock_2008,Broderick+2009}, while the other (Section~\ref{sec:reflect}) is new. Models involving boson stars and certain kinds of naked singularities are considered in Section~\ref{sec:SHCOs}, and other exotic possibilities, including wormholes, are discussed in Section~\ref{sec:known}.

\subsection{Thermalizing Surface}\label{sec:thermal}

Accretion in \sgra is believed to occur via a hot accretion flow\footnote{In this Section, by a "hot accretion flow" we mean hot gas with near-virial temperature that is located external to the central gravitating object, as distinct from whatever gas may be present at rest on the surface of the object. The external gas could be accreting toward the center, or flowing out in a jet. Generically, both types of motion are expected to be present in a hot accretion flow \citep[see][for reviews]{Falcke2013,Yuan_Narayan_2014}. Suggestions that \sgra may have hot inflowing gas and/or an outflowing jet go back to \citet{Rees_1982}, \citet{Falcke1993}, \citet{Narayan+1995}, and \citet{Falcke_Markoff2000}.} \citep{Yuan_Narayan_2014}. 
Now that  the EHT image of \sgra (\citetalias{PaperIII}) has revealed a brightness temperature well in excess of $10^{9}$\,K, the evidence for the presence of very hot gas is particularly compelling.

The radiative luminosities of hot accretion flows are generally far less than $\dot{M}c^2$ \citep{Narayan+1995,Yuan_Narayan_2014}, where $\dot{M}$ is the mass accretion rate. Therefore, the accreting gas in these systems reaches the compact object at the center with a considerable amount of thermal and kinetic energy. If the compact object is a black hole, this energy simply disappears through the  event horizon. On the other hand, if the object has a surface, the energy will be thermalized and re-radiated (once the system reaches steady state), giving a large surface luminosity that should be visible to a distant observer. Observations can thus tell the difference between an event horizon and a thermalizing surface.

In the previous paragraph, and also in the rest of Section 4, we assume that (i) matter in the compact object at the center of \sgra satisfies energy conservation, (ii) that it obeys the laws of thermodynamics, in particular, that it approaches statistical equilibrium in steady state, and (iii) that it couples to and radiates in all electromagnetic modes.
These assumptions can be considered "natural" minimal principles, but they can be violated in extreme models. For example, the shell-like black hole mimicker described in \citet{Danielsson2021} can be designed either not to produce any electromagnetic radiation, or to radiate only in a handful of modes, thereby violating assumption (iii). 
It is not possible to constrain such models using astronomical observations in electromagnetic bands, though in certain cases it may be possible to distinguish them via gravitational waves \citep{Abbott2021} \citep[e.g.,][for the case of gravastars]{Chirenti+2007}. Note that even very exotic objects would satisfy our assumptions, including (iii), if only a small fraction of the accreted baryonic gas survives on their surface as normal matter. To be optically thick in the electromagnetic bands of interest to us, the skin of normal matter should have a surface density as little as $1\,{\rm g\,cm^{-2}}$, which corresponds to just $10^{-14}$ of the total mass of \sgra. An exotic object would need to convert {\it all} accreted gas on its surface to electromagnetically-inactive material if it is to escape detection by electromagnetic observations.

For a spherically symmetric spacetime, matter that starts from rest at infinity and then accretes via a radiatively inefficient mode to come to rest on a surface at radius $R_*$, will release thermal energy as measured at infinity equal to a fraction $\eta$ of the rest mass energy of the gas, where (the following expression is obtained for the Schwarzschild metric, \citealt{Broderick_Narayan_2006}),
\begin{equation}
\eta = 1-\left(1-\frac{2M}{R_*}\right)^{1/2} \gtrsim \frac{M}{R_*}, \label{eq:eta}
\end{equation}
and we use geometrized units: $G=c=1$. If the released thermal energy is radiated back to infinity -- we emphasize that this is unavoidable once the object reaches steady state -- the extra luminosity from the thermalizing surface will be typically much larger than the luminosity radiated by the hot accretion flow itself. This feature can be exploited to distinguish black holes, which by definition have an event horizon, from other kinds of compact object that have a surface. In the context of stellar-mass black holes, this argument provides a convenient way of distinguishing black holes from neutron stars \citep[][see \citealt{Narayan_McClintock_2008} for a review]{Narayan_Yi_1995,Narayan+1997,Garcia+2001,Done_Gierlinski_2003,McClintock+2004}.

In the case of \sgra, the argument proceeds differently. In essence, the observed sub-millimeter radiation provides a lower limit on the mass accretion rate, $\dot{M}_{\rm min}$, regardless of whether the radiation is produced by inflowing hot gas or an outflowing jet. Therefore, given an assumed radius $R_*$ of the surface, we can estimate the minimum surface luminosity that should be observed at infinity,
\begin{equation}
L_\infty > \eta\dot{M}_{\rm min}c^2. \label{eq:Linf}
\end{equation}

As we show below, the surface radiation should appear in the infrared, where observations provide strong upper limits on the luminosity of \sgra. These limits lie far below the predicted minimum surface luminosity, implying that \sgra does not have a radiating surface. Versions of this argument have been made in previous papers in the context of \sgra \citep[see][for a review]{Narayan_McClintock_2008}. A similar argument also applies to the supermassive black hole in M87 \citep{Broderick+2015,VI_EHT2019_M87}. In related work, \citet{Lu+2017} argued that the absence of flashes of radiation from stars crashing on supermassive black hole candidates in galactic nuclei requires these candidates to be true black holes with event horizons.

\subsubsection{EHT Limit on the Radius of the Surface}\label{sec:surface_radius}

In the case of \sgra, a somewhat weak link in the argument outlined above was the hitherto lack of a strong upper limit on the radius $R_*$ of a putative surface in \sgra. Since the surface luminosity for a given $\dot{M}$ scales as $L_\infty \propto \dot{M}(M/R_*)$ (Equations \ref{eq:eta} and \ref{eq:Linf}), one could make the predicted luminosity small by arbitrarily increasing $R_*/M$, thereby evading observational limits. This loophole has now been closed by EHT observations. 

Using a maximally conservative analysis of EHT 2017 visibility data, and without any model assumptions, \citetalias{PaperII} estimates the FWHM of the image of \sgra to lie in the range $39-87\,\mu$as. With a conversion factor, $GM/c^2D\approx 5\,\mu$as (see Figure~\ref{fig:Prior_MoverD}), this corresponds to an image diameter $<18M$. 

The observed 230\,GHz radiation in \sgra is from the hot accretion flow, not from the surface (which should radiate in the infrared). Any surface must lie interior to the 230\,GHz-emitting hot accretion flow and should have an apparent diameter smaller than $18M$. Thus, from the analysis in \citetalias{PaperII}, we set the following upper limit on the apparent radius of the surface as viewed by a distant observer: $R_{\rm app} < 9M$.

\citetalias{PaperIII} presents image reconstructions of \sgra based on the EHT 2017 data. Table 7 in that paper summarizes the results of fitting a ring model to image reconstructions based on several methods. Using the imaging results from \difmap, \ehtim and \smili, and combining the ring analyses with \rex and \vida (see \citetalias{PaperIII} for details), the average ring diameter estimate is $d=51.3\pm2.0\,\mu$as, and the ring width estimate is $w=29.6\pm3.6\,\mu$as (these results correspond to descattered images from April 7 data). We take $(d+w) = 80.9 \pm 4.1\,\mu$as as a reasonable proxy for the apparent outer diameter of the source. Using the 95\% confidence upper limit, $(d+w) < 88\,\mu$as, we obtain $R_{\rm app} < 8.8M$~(95\%cl).
\citetalias{PaperIII} obtains a tighter constraint using the Bayesian imaging method {\themis}, while \citetalias{PaperIV} similarly reports tighter constraints by fitting mG-ring models (based on \citealt{Johnson+2020})  directly to visibility data. To be conservative, we do not use these limits.

The analyses described in the previous paragraph treat $d$ and $w$ as uncorrelated quantities. However, as the careful analysis in  Section~\ref{sec:msmt} of the present paper shows, there is a strong anti-correlation between the estimated values of $d$ and $w$, such that their sum $(d+w)$ is quite tightly constrained. The dotted and dashed curves in Figure~\ref{fig:imaging_all} correspond to $(d+w) = 90\,\mu$as and $80\,\mu$as, or equivalently, $R_{\rm app}=9M$ and $8M$, respectively. Clearly, from this analysis, $R_{\rm app}<8M$ is a safe upper limit (at about 95\% confidence).

To be very safe, we choose as a  conservative upper limit on the apparent radius of a surface in \sgra, $R_{\rm app} < 9M$. For a Schwarzschild spacetime, gravitational deflection of rays causes the apparent radius of a spherical surface as viewed by an observer at infinity to be larger than the true areal radius $R_*$. The relation between the two is
\begin{eqnarray}
R_{\rm app} &=& 3\sqrt{3} M, \qquad\qquad\qquad  R_* \leq 3M, \nonumber\\
&=& R_*\left(1-\frac{2M}{R_*}\right)^{-1/2}, ~ R_*>3M.
\label{eq:Rapp}
\end{eqnarray}
Our upper limit, $R_{\rm app}<9M$, then corresponds to
$R_*< 8M$. In the discussion below, we consider the full range of allowed $R_*$ values, from the event horizon radius $R_H=2M$ to the upper limit, namely, $2M < R_* \leq 8M$. 

\subsubsection{Predicted Spectrum of Surface Radiation}\label{sec:spectrum}

\citetalias{PaperV} discusses hot accretion flow models of \sgra based on extensive GRMHD simulations. The models indicate that the mass accretion rate in \sgra is typically $\dot{M} \sim 10^{-8} M_\odot {\rm yr^{-1}}$ (similar to estimates reported in, e.g., \citealt{Falcke1993,Yuan+2003,Chael+2018,Ressler+2020}), but with a broad distribution that extends from $\dot{M}_{\rm min} \sim 10^{-9} M_\odot {\rm yr^{-1}}$ to $\dot{M}_{\rm max} \sim 10^{-6} M_\odot {\rm yr^{-1}}$. The models at the lower end of this range are actually ruled out by various constraints (see \citetalias{PaperV});  nevertheless, we stick to $\dot{M}_{\rm min} = 10^{-9} M_\odot {\rm yr^{-1}}$ as a safe and conservative lower limit on the mass accretion rate. Equations (\ref{eq:eta}) and (\ref{eq:Linf}), combined with our upper limit on $R_*$, then show that the surface luminosity measured at infinity must be $\gtrsim 10^{37} ~{\rm erg\,s^{-1}}$.

\begin{figure*}
\centering
\includegraphics[width=0.95\columnwidth]{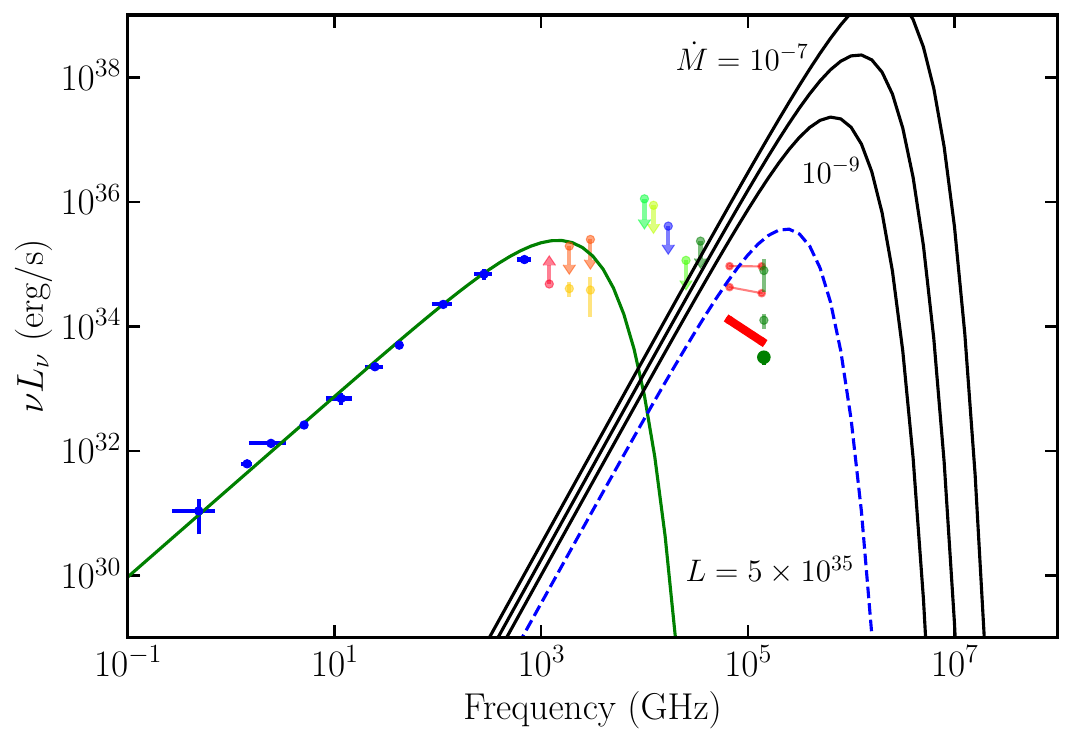}
\hspace{0.5cm}
\includegraphics[width=0.95\columnwidth]{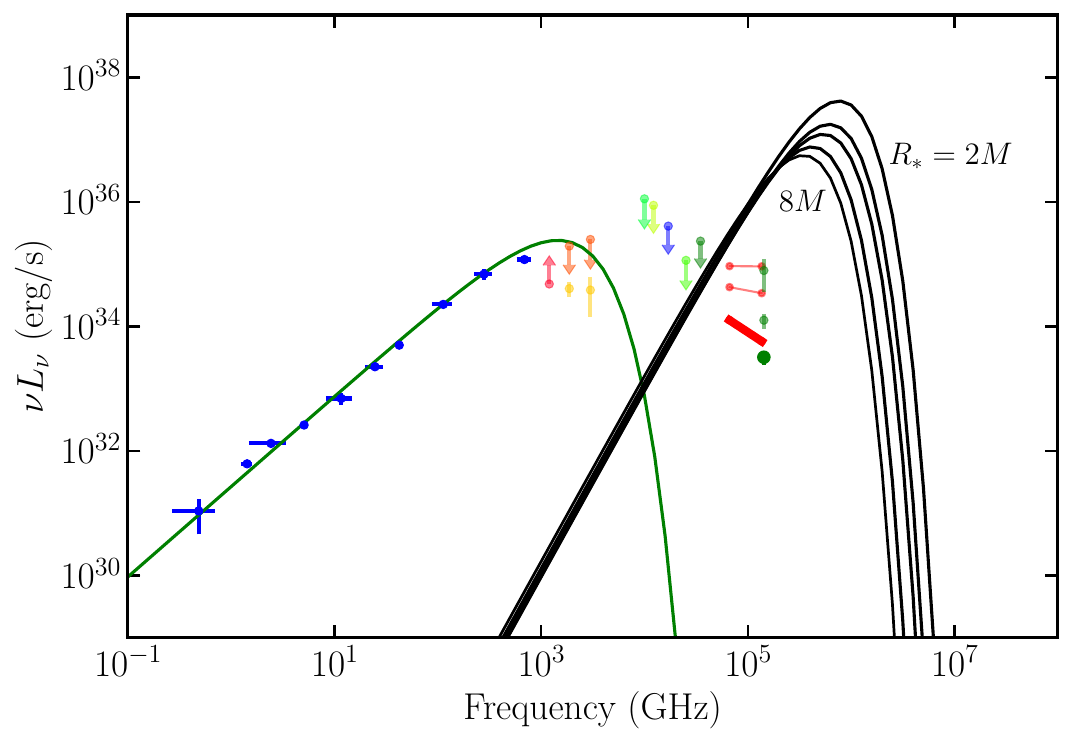}
\caption{\textit{Left:} Predicted  spectral energy distributions (SEDs), $\nu L_\nu$ vs $\nu$, of thermal radiation from {\sgra}, if the object has a thermalizing surface with radius $R_*=2.5M$ (selected as a fiducial model). Three choices of the mass accretion rate on the surface, $\dot{M} = [10^{-7}, ~10^{-8}, ~10^{-9}]\,M_\odot{\rm yr^{-1}}$, are shown as solid black curves. The dashed blue curve corresponds to a model with a surface luminosity of $5\times10^{35} {\rm erg\,s^{-1}}$, a conservative lower bound.  Observational data in various wavebands are plotted as solid circles. The thick red line and large green filled circle correspond to the 5th percentile of the variable infrared emission \citep{2018ApJ...863...15W,2020A&A...638A...2G}, 
which we treat as an upper limit on any quiescent infrared luminosity from a potential surface in \sgra. This upper limit from the observations lies well below the theoretical SEDs, and therefore rules out these models. The green curve on the left is an empirical fit to the radio and sub-millimeter part of the observed SED; the radiation in this region of the spectrum is produced by synchrotron emission from the hot accretion flow, and the corresponding luminosity is $\sim5\times10^{35}{\rm erg\,s^{-1}}$.  \textit{Right:} Theoretical SEDs of surface emission from an accretion flow in \sgra with $\dot{M} = 10^{-9}M_\odot{\rm yr^{-1}}$  falling on surfaces of radii, $R_*=[2M, ~3M, ~4M, ~6M, ~8M]$, respectively (solid black lines). All the SEDs are inconsistent with infrared observations (thick red line, large green filled circle). Hence these models are ruled out. }
\label{fig:thermal}
\end{figure*}

Meanwhile, we know that the hot accretion flow in \sgra produces synchrotron radiation at sub-millimeter wavelengths with a luminosity $\sim 5\times10^{35}{\rm erg\,s^{-1}}$, shown by the green curves in Figure~\ref{fig:thermal}. Even in the absence of any independent estimate of $\dot{M}$, just the fact that accretion results in this much radiation implies a certain minimum energy flow on to the surface. Since the accreting gas generally moves radially inward, relativistic beaming causes more radiation to impinge on the central object compared to what escapes to infinity. Thermalization of this infalling radiation would then give a surface luminosity greater than\footnote{This argument fails of course if the radiating gas does not accrete inward but moves away from the surface, and especially if its emission is beamed preferentially toward Earth.  \citet{Brinkerink+2021} propose a model of this kind for \sgra in which relativistically outflowing gas with $\beta\gamma\sim 1.5$ moves nearly directly toward us. In such a model, reprocessing of jet radiation by the surface could be negligible.} $5\times10^{35}{\rm erg\,s^{-1}}$. Any additional energy released by the mechanical and thermal energy of the infalling gas (this is expected to dominate in most scenarios) would further increase the surface luminosity. We therefore treat $5\times10^{35}{\rm erg\,s^{-1}}$ as an even more conservative lower bound on the surface luminosity of {\sgra} than that discussed in the previous paragraph.

A key feature of radiation emitted from a central surface in a hot accretion flow is that it will appear in a different region of the electromagnetic spectrum than the emission from the hot accreting gas and jet. The latter dominates in the radio and sub-millimeter bands (Figure~\ref{fig:thermal}, green curves). Meanwhile, the radiating gas at the surface, being optically thick, will radiate like a blackbody to a very good approximation (McClintock et al. 2004, Broderick \& Narayan 2006, 2007).

The temperature of this radiation, measured at infinity, is given by
\begin{eqnarray}
T_\infty &=& \left( \frac{L_\infty}{4\pi R_{\rm app}^2 \sigma_{\rm SB}}\right)^{1/4} \nonumber \\ &\approx& 2100\,{\rm K} \left( \frac{L_\infty}{10^{36}\,{\rm erg\,s^{-1}}}\right)^{1/4}\left( \frac{9M}{R_{\rm app}} \right)^{1/2}, \label{eq:Tinf}
\end{eqnarray}
where $\sigma_{\rm SB}$ is the Stefan-Boltzmann constant. For the estimates of $L_\infty$ and $R_{\rm app}$ derived earlier, the predicted radiation should be in the near-infrared and optical bands. If we define the characteristic frequency $\nu_*$ of the blackbody radiation by $h\nu_*=kT_\infty$, the  spectral energy distribution (SED) at infinity takes the form
\begin{equation}
    \nu L_{\nu} = \frac{15}{\pi^4} \,L_\infty \frac{(\nu/\nu_*)^4}{\exp(\nu/\nu_*)-1}, \quad \nu_* \equiv \frac{kT_\infty}{h}.
\end{equation}

The left panel in Figure~\ref{fig:thermal} shows predicted SEDs of surface radiation from {\sgra}, if the object has a surface with a radius $R_*=2.5M$; we choose this radius as a fiducial model for illustration. The three solid black curves correspond to mass accretion rates $\dot{M} = [10^{-7}, ~10^{-8}, ~10^{-9}]~M_\odot{\rm yr^{-1}}$, respectively, the last of which is the conservative lower limit from \citetalias{PaperV} mentioned earlier. The dashed blue curve corresponds to the absolute lower limit on the surface luminosity, $L_\infty = 5\times10^{35}{\rm erg\,s^{-1}}$, discussed above.

The right panel in Figure~\ref{fig:thermal} shows another sequence of models in which we vary the surface radius $R_*$. Taking the previously mentioned conservative mass accretion rate estimate of $10^{-9}M_\odot{\rm yr^{-1}}$, we consider surface radii $R_* = [2M, ~3M, ~4M, ~6M, ~8M]$, respectively.

In all the models shown in the two panels in Figure~\ref{fig:thermal}, the predicted surface emission (black and blue curves) is spectrally well separated from the synchrotron emission of the hot accretion flow (green curve).

Therefore, this predicted signature of surface emission is easy to identify via observations, making it possible to develop a robust test for the presence of a thermalizing surface.

\subsubsection{Observational Limit on Surface Luminosity}\label{sec:obse_limit}

Observations of \sgra have improved substantially in recent years. The current status is summarized in  Section~\ref{sec:sed} and Figure~\ref{fig:Prior_SED}, and the data are shown again in Figure \ref{fig:thermal}. The infrared data are of most interest to us and are highlighted by the red line segments and green filled circles, which correspond to the 5th (thick red line and large green filled circle at the bottom), 50th (thin line, small circle), and 95th (thin line, small circle) percentiles, respectively, of the variable infrared luminosity. \sgra exhibits frequent flares in its infrared light curve \citep{Eckart+2004,2005ApJ...628..246E,Hora+2014,2018ApJ...863...15W}, which are interpreted as transient electron heating events in the hot accretion flow or jet. A few bright flares have been shown to come from gas orbiting the central object at a {\rm projected} radius $R_{\rm flare}\sim6-10M$ \citep{2018A&A...618L..10G}. This location is not very different from the region of the flow that produces the sub-millimeter radiation observed by the EHT. 

If \sgra were an object with a thermalizing surface, then given its large mass we would expect it to have an enormous thermal capacity. Consequently, thermal emission from its surface is not expected to show violent flaring activity. The observed infrared flares are thus much more likely to be produced by the hot accretion flow, possibly in transient turbulent heating or magnetic reconnection events \citep{Markoff+2001,Yuan+2004,Ball+2016,Ressler+2017,Davelaar2018,Dexter+2020,Nathanail+2020,Nathanail+2021,Chatterjee+2021,Porth+2021,Ripperda+2021,Ball+2021}.

Since any surface infrared emission in \sgra must be steady, we ignore the fluctuating flare emission and treat the 5th percentile (the thick red line and large filled green circle in Figure \ref{fig:thermal}) as the maximum steady infrared emission from a surface\footnote{Because it is hard to tell how much time variability is present below the 5th percentile, we take this as a conservative estimate of the maximum level of steady surface emission. Note that we expect some of the radiation below the 5th percentile to be produced by synchrotron emission from nonthermal electrons in the hot accretion flow and Compton scattering of synchrotron radiation by the same electrons. By ignoring these possibilities and counting all the radiation below the 5th percentile as surface emission, we are being conservative}. Note that \citetalias{PaperV} uses an upper limit of $\nu L_\nu < 10^{34}{\rm erg\,s^{-1}}$ in infrared (50th percentile) when evaluating their GRMHD-based accretion-jet models. As Figure \ref{fig:thermal} shows, this upper limit (especially the large green filled circle) lies nearly two orders of magnitude below the strict lower bound on the predicted surface luminosity discussed earlier (dashed blue line), and three orders of magnitude below predictions of more realistic models (solid black lines). We thus conclude that \sgra cannot have a thermalizing surface with characteristics similar to any of the models considered in Figure \ref{fig:thermal}, ergo the case for an event horizon is much strengthened.

\subsubsection{Discussion and Caveats}\label{sec:horizon_caveats}

Compared to previous discussions of this topic, what has improved is that, thanks to the EHT image of \sgra, we are now able to limit ourselves safely to surface radii $R_*<8M$, whereas in earlier works much larger radii were considered (as large as $1000M$ in \citealt{Narayan_2002}, and $100M$ in \citealt{Broderick+2009}).  Moreover, the infrared constraints are also now very much stronger (Figure~\ref{fig:thermal}). Correspondingly, the argument for the absence of a thermalizing surface is substantially strengthened. 

The discrepancy between the maximum steady infrared luminosity that \sgra can possibly have (the 5th percentile thick red line and large green circle in Figure \ref{fig:thermal}) and the minimum possible luminosity it could theoretically have and still possess a thermalizing surface (the dashed blue line) is too large to be circumvented with small fixes to model details. This statement is true even if we use the 50th percentile of the infrared observations (the middle red line and middle green circle in Figure~\ref{fig:thermal}), which would be equivalent to counting all the observed infrared radiation, including the flares, as surface emission. If we wish to consider models of \sgra with a surface, we have to find a weakness in one of the links in the underlying logic of the argument. An easy way out is to give up one of the basic physics assumptions listed in the third paragraph of Section~\ref{sec:thermal}. Here we consider other less-drastic possibilities.

Could the predicted infrared radiation from a hypothetical surface in \sgra be obscured by foreground matter such as dust? This is highly unlikely since the radiation from the infrared flares is clearly visible, and that radiation comes from hot external gas (not from the surface) at radii within $10M$ \citep{2018A&A...618L..10G}. It is hard to imagine an obscuring medium that allows flare emission to make it through but blocks radiation from the surface.

Another minor worry may be quickly dealt with. Because of spacetime curvature, radiation from a surface at areal radius $R_*$ in a Schwarzschild spacetime takes longer to reach a distant observer compared to a ray that travels in flat spacetime. Could this delay be so large that surface radiation has not yet reached us? Let us write
\begin{equation}
R_* = (1+\mu) R_H,
\end{equation}
where $R_H=2M$ is the radius of the event horizon. For $\mu\ll1$, the additional time delay is $\sim 2M\ln(1/\mu)$, which is $\sim 40\,\ln(1/\mu)$\,s in the case of \sgra. Even if the logarithm is as large as $100$ (corresponding to $R_*$ being located a Planck length above the event horizon), the extra time delay is only about an hour. 

A related worry is that the gravitational redshift, $(1+z) \sim \mu^{-1/2}$, between the surface and infinity might dilute the observed luminosity sufficiently to make the surface radiation invisible. \citet{Abramowicz+2002} noted that this effect causes the radiation luminosity that reaches the observer to be reduced by a factor of $1/(1+z)^4 \sim \mu^2$ compared to what is emitted at the surface. They claimed that, if $(1+z)$ were large enough, no detectable radiation would reach the observer and it would be impossible to distinguish an event horizon from a surface. 

However, gravitational redshift is not an issue for the line of argument we have presented in this paper because we expressed everything in terms of energy and luminosity as measured at infinity; in such a framework, all redshift factors drop out. For instance, if the radiation observed at infinity has a temperature $T_\infty$, then the radiation emitted by the surface will have a temperature $T_{\rm local} = (1+z)T_\infty$ in the local rest frame. The radiation emerging from the surface will have a flux equal to $\sigma_{\rm SB}T_\infty^4(1+z)^4$, and the corresponding luminosity is larger than what reaches infinity by precisely the factor of $(1+z)^4$ noted by \citet{Abramowicz+2002}. The only question is whether the system has enough time to heat up to such a high local temperature. We discuss this important issue next.

\sgra is presumably as old as the Milky Way Galaxy, i.e., several billion years old. Over much of that time, it must have accreted gas at a rate equal at least\footnote{ There is clear evidence that \sgra went through episodes of much larger $\dot{M}$ a few hundred years ago \citep{Ponti+2010,Clavel+2013}, and there are suggestive arguments for enhanced accretion over the last millions of years \citep[e.g.,][]{Mou+2014}. On the time scale of the age of the Galaxy, if \sgra acquired much of its mass by accretion, it would have had to accrete at an average rate of more than $10^{-4}M_\odot{\rm yr^{-1}}$, i.e., orders of magnitude larger than the conservative rates we have been assuming. Such large average accretion rates are routinely predicted by cosmological models of galaxy and supermassive black hole evolution.} to the present $\dot{M}$. The time needed to achieve the steady state condition implicit in equation (\ref{eq:Tinf}), namely, $\eta\dot{M}c^2 = L_\infty = (\sigma T_\infty^4)(4\pi R_{\rm app}^2)$, or equivalently $T_{\rm local}=(1+z)T_\infty$, is far shorter than the age of \sgra for almost any model. 

The one exception is if $\mu\ll1$, i.e., if the surface $R_*$ is extremely close to the event horizon. In this limit, as \citet{Lu+2017} argued, the time required to achieve steady state scales as $\mu^{-1}$ and can become arbitrarily long. The physical reason is that the region between $R=R_*$ and the photon sphere, $R_{\rm ph}=3M$, traps radiation. This volume has a large thermal capacity, and therefore takes a long time to reach steady state. Applying this logic to \sgra, \citet{Lu+2017} concluded that the absence of infrared radiation in \sgra rules out a thermalizing surface only if $\mu \gtrsim 10^{-14}$. If the surface is even closer to the horizon radius than this limit, i.e., if $R_* \lesssim R_H + 10^{-2}$\,cm, then the steady state condition will be invalid. Their argument thus provides an {\it upper} limit on $R_*$.

Using completely different reasoning, \citet{Carballo-Rubio+2018} set a {\it lower} limit on $\mu$. The argument goes as follows. Because \sgra accretes mass continuously, its horizon radius $2M$ increases with time. In order to maintain $R_*=(1+\mu)R_H$, the surface also needs to expand. However, if $\mu$ is too small, the required expansion speed is greater than the speed of light in the local frame, which is unphysical. Using a conservative estimate of $\dot{M} \sim 10^{-11} M_\odot {\rm yr^{-1}}$ (which is two orders of magnitude smaller than the lower limit given in \citetalias{PaperV} and more than seven orders of magnitude less than the likely average accretion rate over the life of \sgra), \citet{Carballo-Rubio+2018} conclude that \sgra can avoid the faster-than-light conflict only if $\mu \gtrsim 10^{-23}$, i.e., if $R_* \gtrsim R_H + 10^{-11}$\,cm. Note that this rules out models in which the surface lies a Planck length ($10^{-33}$\,cm) above the event horizon, as gravastar models \citep{Mazur_Mottola_2001,Chapline_2003}
often implicitly assume. 

Combining the arguments in the previous two paragraphs, we are left with an interesting class of models with $\mu$  in the range $10^{-23} < \mu < 10^{-14}$ for which \sgra is presently allowed to have a thermalizing surface and yet not be ruled out by infrared constraints. This gap in model space merits further investigation.

Another issue worth serious discussion is the assumption that the surface will radiate like a blackbody. Since we are considering an object which (i) is in steady state and therefore in thermal equilibrium (by our assumptions), (ii) is likely nearly isothermal in the sense that the redshifted temperature $T_\infty$ is independent of radius inside the object, and (iii) has an enormous optical depth, it seems unavoidable that the emission must be close to a blackbody. (For instance, stars radiate roughly like blackbodies because of their large optical depths, and would be perfect blackbodies if they were isothermal.) Any deviations from a perfect blackbody in the putative surface radiation in \sgra might thus be expected to be minor. However, the specific case of radiation produced by energy release from matter falling on the surface of a compact supermassive ($>10^6M_\odot$) object has not been studied and merits further attention. Models of spherical accretion on neutron stars ($M=1.4M_\odot$) studied by \citet{Shapiro_Salpeter_1975} suggest that modest deviations from a perfect blackbody are expected in that case; their models show some hardening of the thermal spectrum plus the appearance of a power-law spectral component extending to higher frequencies. If the corresponding effects in the case of a surface in \sgra ($M=4\times10^6M_\odot$) are similarly modest, then our blackbody assumption is quite safe. Note that \citet{Shapiro_Salpeter_1975} did not include ray deflections and strong lensing in their model.

The argument for a blackbody spectrum is very strong in one particular limit. When the surface has a radius $R_*$ close to $R_H$, i.e., $\mu \ll 1$, the volume between $R_*$ and $R_{\rm ph} = 3M$ acts like an enclosed cavity, with radiation allowed to escape only over a small solid angle $\sim\mu$. The cavity then behaves like a textbook isothermal ``furnace'' with a tiny pinhole for escaping radiation. In this limit, the radiation that reaches a distant observer will be indistinguishable from a perfect blackbody \citep{Broderick_Narayan_2006}. 

If the quiescent infrared radiation in \sgra corresponds to blackbody emission from a surface, it should be completely unpolarized. On the other hand, if the radiation is produced by synchrotron emission in optically thin (weak) flares, we might expect a certain degree of linear polarization. Bright infrared flares in \sgra show clear evidence for strong linear polarization \citep{Eckart+2006,GRAVITY+2020}, but there is presently no information on the degree of polarization of the weak emission below the 5th percentile. Sensitive polarimetry could be used in the future to explore this regime, and might help to reduce even further the maximum level of blackbody emission allowed in \sgra.

\subsection{Reflecting Surface}\label{sec:reflect}

In this Subsection, we focus again on the possibility that \sgra may have a surface, but now we explore models in which the surface {\it reflects} incident radiation. We assume that, in the rest frame of the surface at some fixed areal radius $R_*$, the following properties hold: (i) Any inward-moving ray that is incident with wave-vector $k^\mu$ becomes an outward-moving ray with $k^r$ reversed and the other components of $k^\mu$ unchanged. (ii) If the intensity of the incoming ray is $I_\nu$, the outgoing ray has an intensity $AI_\nu$, where $A\leq1$ is the albedo of the surface. The motivation for considering such a model is that it makes interesting predictions that an interferometer like the EHT might be able to observe.

\subsubsection{Synthetic Images Based on GRMHD Simulations}\label{sec:GRMHD}

As an illustration of the effects we expect from a reflecting surface, we use a long-duration simulation of a hot accretion flow in the magnetically arrested disk (MAD) state around a black hole of spin $a_*=0$ \citep{Narayan+2021}. We take the profiles of density, pressure, four-velocity and magnetic field in the poloidal $(r,\theta)$ plane, time-averaged over the simulation period $t=50,000-100,000M$. We set the electron temperature using the prescription given in \citet[][which is based on \citealt{Moscibrodzka+2016}]{V_EHT2019_M87} with parameter values, $R_{\rm high}=20$, $R_{\rm low}=1$. We scale the density, and proportionately the gas pressure and magnetic energy density, such that the observed 230\,GHz flux density is equal to 2.4\,Jy, as  measured during the 2017 EHT observations of \sgra \citep{Wielgus2022}. We then compute a synthetic 230\,GHz image for an observer at an inclination angle of $i=60^o$.

\begin{figure*}
\centering
\includegraphics[width=0.74\textwidth]{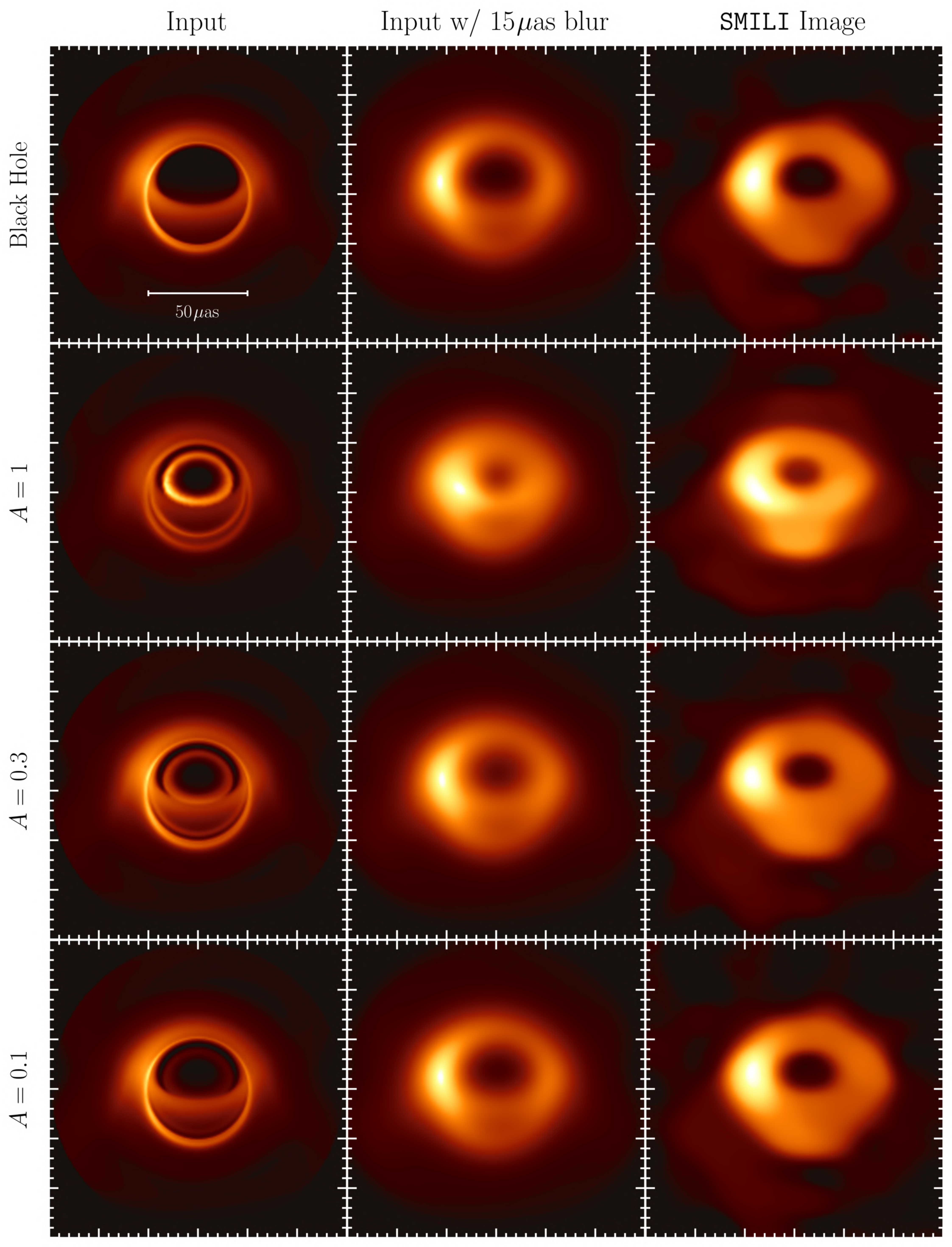}

\caption{\textit{Comparison of a synthetic image of a hot accretion flow around a black hole with corresponding images when the central object has a reflecting surface. Top-Left}: Model image of \sgra at 230\,GHz as seen by an observer at inclination angle of $60^o$, computed by ray-tracing a GRMHD simulation of a hot accretion flow around a non-spinning black hole. The color scale corresponds to brightness temperature (K). The image shows a relatively diffuse elliptical feature corresponding to equatorial emission from the accretion flow, a sharp circular lensing ring, and a dark shadow region in the middle. \textit{Top-Middle}: Shows the image on the left blurred with a gaussian beam with FWHM $=15\,\mu$as.
\textit{Top-Right}: Image reconstruction with {\smili} top set (see \citetalias{PaperIII} for details) using synthetic visibility data generated from the model shown in Top-Left.
\textit{Second Row Left}: Model image when the black hole horizon is replaced by a reflecting surface with radius $R_*=2.5M$ and albedo $A=1$. Additional features appear in the image, especially a thick bright secondary ring inside the shadow region. \textit{Second-Middle}: Blurred (FWHM $=15\,\mu$as) version of the image on the left. \textit{Second-Right}: \smili reconstruction using synthetic visibilities generated from the model in Second-Left. Comparison with the Top-Right image shows that the shadow region, which is prominent in the upper image (black hole case), is largely filled in by surface reflection. This change is potentially distinguishable with EHT observations. 
\textit{Third Row}: Similar to the Second Row, but for albedo $A=0.3$. Compared to the $A=1$ model, in this case the difference from the black hole image due to the presence of a surface is only marginally detectable. \textit{Bottom Row}: Model with surface albedo $A=0.1$. Here the image is indistinguishable from the black hole image (Top row) at the resolution and sensitivity of the EHT 2017 data.}
\label{fig:reflect}
\end{figure*}

The top left panel in Figure \ref{fig:reflect} shows the 230\,GHz image of the above model, assuming that the object at the center is a Schwarzschild black hole. The image is computed using the ray-tracing code HEROIC \citep{Zhu+2015,Narayan+2016}. The top middle panel shows the same image blurred with a Gaussian beam of FWHM equal to $15\,\mu$as; this beam size corresponds to the typical resolution that is achieved by the EHT using super-resolution image reconstruction techniques. 

The unblurred image in the top left panel in Figure \ref{fig:reflect} shows the usual features. The sharp circular ring is the photon ring produced by strong gravitational lensing by the black hole. The diffuse elliptical feature is the image of equatorial emission from the accretion flow, flattened in the vertical direction because of the $60^o$ inclination of the observer. These two features are visible even in the $15\,\mu$as blurred image in the top middle panel (the features merge if we blur with a $20\,\mu$as beam, the nominal resolution of the EHT). Most importantly, a dark shadow region is clearly seen in the middle of even the blurred image.

The second row in Figure \ref{fig:reflect} shows the effect of including a reflecting surface with albedo $A=1$ (100\% reflection) at a radius $R_*=2.5M$ (selected as an example). In addition to the diffuse disk emission and sharp photon ring already described in the top left image, we find additional components that are caused by reflection. The thick bright ring at the center of the image corresponds to radiation from the equatorial accretion flow that is reflected from the side of the surface facing the observer. The thin ring (close to the original photon ring) is from rays that reflect off the far side of the surface and are then lensed around the compact object. Interestingly, the new features from reflection, especially the first one, appear in the the shadow region of the original black hole image. When blurred, the resulting image, shown in the second row middle panel, has much of the shadow region filled in. This fairly dramatic effect is potentially distinguishable by the EHT.

The third and fourth rows in Figure \ref{fig:reflect} correspond to models with albedos $A=0.3$ and $0.1$, respectively. The image of the $A=0.3$ model, when blurred, is only marginally different from the black hole image (top middle panel), while the blurred $A=0.1$ model is indistinguishable from the black hole image.

The implication of these test images is that models in which \sgra has a reflecting surface with perfect albedo, $A=1$, could potentially be distinguished by the EHT 2017 observations, but models with only partial albedo, e.g., $A=0.3, ~0.1$, are harder to distinguish from the case of a black hole. Interestingly, in the latter models, a fraction $(1-A)$ of the radiation that falls on the surface must be absorbed, and will presumably be re-radiated as part of the thermalized emission discussed in Section~\ref{sec:thermal}. For any value of $(1-A) \gtrsim 0.1$, this thermally reprocessed emission will lie well above the infrared limits discussed in Section~\ref{sec:obse_limit} and shown in Figure~\ref{fig:thermal}. These models could thus be ruled out by that argument.

Note that several arbitrary choices were made in the above models: spin $a_*=0$, temperature ratios, $R_{\rm high}=20$, $R_{\rm low}=1$, observer inclination $i=60^o$, and surface radius $R_*=2.5M$. The values of $R_{\rm high}$ and $i$ were chosen to lie near the center of the corresponding ranges considered in \citetalias{PaperV}. As it happens, for spin $a_*=0$, GRMHD-based models with these parameter values are fairly consistent with observations (see \citetalias{PaperV}). Varying the parameters will certainly affect the predictions for the effect of surface reflection. The results may not change excessively since we pin the 230\,GHz flux to 2.4\,Jy. Nevertheless, we caution that the results presented in Section~\ref{sec:ehtimage} below are for a preliminary toy model, and are in the nature of a proof of concept. More detailed investigations are needed before we can draw firm conclusions.

An additional caveat is that, in this toy model, we have taken the flow solution to be the same as in a simulation that was run with a black hole event horizon at the center \citep{Narayan+2021}. We simply truncated that solution at $R=R_*$. As mentioned in Section~\ref{sec:horizon_caveats}, the problem of self-consistently solving the gas dynamics and radiation field for a supermassive object with a surface has not yet been studied. 

Another caveat is that we have considered only the case of specular reflection. Diffuse reflection, where radiation incident on the surface is reflected isotropically (or with a more complicated angular distribution), is also worth exploring. In that case, the surface reflected intensity will not be restricted to a few narrow features in the image, but will be spread more uniformly over the entire shadow region.  This would eliminate any truly dark regions in the center of the  image, conceivably making it easier to constrain such models.

Additionally, we considered a time-averaged steady image whereas in reality we expect the image to fluctuate, which can lead to interesting time correlations of features. Also, we have focused here on a spherically symmetric spacetime around a non-spinning object. Once we allow the central object to rotate we will need to solve for the corresponding spacetime, which in general will not belong to the Kerr family of solutions.

\subsubsection{Constraints from EHT Images}\label{sec:ehtimage}

Figures 13, 14 and 17 in \citetalias{PaperIII} show a range of images of \sgra obtained by applying different image reconstruction techniques to the EHT 2017 data. The vast majority of images show a ring-like morphology with a pronounced dark shadow region at the center. These images are visibly different from the synthetic blurred image shown in the second row middle panel in our Figure~\ref{fig:reflect}. We can thus exclude this particular model using EHT observations.

The right four panels in Figure~\ref{fig:reflect} show image reconstructions of the four synthetic models shown in the left panels using one of the image reconstruction methods described in \citetalias{PaperIII}. For each model, synthetic visibility data were generated with the same $(u,v)$-coverage as in the April 7 EHT observations of \sgra, and the appropriate amount of noise was added to match the noise present in the real data. These synthetic visibilities were then analysed using the {\smili} top set (see \citetalias{PaperIII} for details) and the resulting images are shown in the panels on the right in Figure~\ref{fig:reflect}. The \smili reconstructions are fairly similar to the $15\,\mu$as-blurred images shown in the corresponding central panels, though the \smili images appear to be slightly more blurred. More interestingly, the \smili image of the $A=1$ reflecting surface model (second row right panel) is quite different from the reconstruction of the black hole model (top row right panel). This confirms our expectation that a surface which reflects in-falling radiation with 100\% efficiency can potentially be ruled out by EHT 2017 observations of \sgra (modulo the many caveats mentioned in Section~\ref{sec:GRMHD}). In the case of the $A=0.3$ model, and especially the $A=0.1$ model, the \smili reconstructions do not differ much from the black hole image, hence it would be hard to distinguish such models using the current EHT data.

Another way of comparing models is to measure the brightness depression in the shadow region of the image. For instance, Table 13 in \citetalias{PaperIII} presents estimates of a parameter $f_c$, which measures the ratio of the brightness at the center of the ring image to the mean brightness around the ring. This quantity is a measure of the brightness depression in the central shadow region of the image. From Figure~22 in \citetalias{PaperIII}, 
the estimate based on {\smili} is $f_c \approx 0.2$, with not much probability that $f_c>0.3$. This implies that the image intensity in the shadow region in \sgra is very likely $< 30\%$ of the mean intensity around the ring.\footnote{The upper limit on $f_c$ is significantly lower when estimated via {\themis} (see Table 13 in \citetalias{PaperIII}) or by the modeling methods described in \citetalias{PaperIV}. To be consistent with Figure~\ref{fig:reflect}, here we focus on only the {\smili} results.}
 
Such a degree of flux depression is inconsistent with the blurred synthetic image and \smili reconstructed image in the second row of  Figure~\ref{fig:reflect}. That particular model could be potentially ruled out. However, the dynamic range of images based on the current EHT 2017 data is too low, and its angular resolution is too poor, to constrain a weakly reflecting surface in \sgra, such as the models in the third and fourth rows of Figure ~\ref{fig:reflect}.

An array with more stations and with larger bandwidth (e.g., the proposed Next Generation Event Horizon Telescope, \citealt{Doeleman+2019,Raymond+2021}) would solve the sensitivity problem. However, to improve the angular resolution substantially, it will be necessary to expand the telescope array into space by sending one or more radio dishes into large-radius orbits around the Earth \citep{Palumbo+2019,Roelofs+2020,Fish+2020,Gurvits2021,Kudriashov+2021,Fromm2021}. With such an expanded array, we might be able to observe images with sufficient angular resolution to see some of the details revealed in the various panels in the left column of Figure~\ref{fig:reflect} and to check for the presence of a surface in \sgra. Furthermore, if a surface is present, we might be able to determine the albedo and the surface radius.

\citet{Chael+2021} show that, with better sensitivity and angular resolution than the present EHT is able to provide, the observed image of a hot accretion flow around a supermassive black hole could be used to delineate the inner edge of the accretion disk. Their interest is to use this technique to identify the edge of the event horizon. However, their method could equally well be used to measure the radius of a potential surface. This might provide a direct estimate of $R_*$.

Additionally, it may be possible to observe time-delayed echoes from a reflecting surface. In the case of time-variable emission from the accretion flow, e.g., sub-millimeter or infrared flares, the observer would see both the primary signal from the emitting gas element as well as a delayed reflected copy of the same radiation. This is the electromagnetic analog of gravitational wave echoes that have been searched for in LIGO/Virgo observations of merging stellar-mass black holes \citep{Abedi+2017,Westerweck+2018}.  However, here it could be done with spatially resolved images, with all the rich detail they can provide. We do not pursue electromagnetic echoes further, but note that their presence could potentially be explored already using existing image-integrated light curve data.

\subsection{Surfaceless Horizonless Compact Objects} \label{sec:SHCOs}

As examples of horizonless compact objects without surfaces, we consider black-hole mimickers such as (mini) boson stars, for which synthetic images from covariant MHD simulations of radiatively inefficient accretion flows  have recently been obtained in \cite{Olivares+2020}. It was shown there that a region with a central brightness depression could appear in the final observed image of an unstable boson star (model A there), despite the absence of an unstable photon orbit in that spacetime. These features are the result of an effective low-density region that appears in the center of the spacetime due to a centrifugal barrier.
The unstable boson star has a significantly smaller intrinsic source size than corresponding constraints of \sgra thereby ruling it out as a candidate alternative. 
Similarly, for the stable boson star configuration considered in \cite{Olivares+2020}, there is a complete absence of a central brightness depression with the inner image being extremely bright, akin to a radiating surface \citep[see, e.g.,][]{Fromm2021}.
Given the EHT constraints a mini boson star becomes an unlikely candidate as a black hole mimicker to describe \sgra, since their image morphologies are generally too compact and lack both a characteristic ring-like feature and a central brightness depression. However a more extensive study of a boson star spin, compactness and astrophysical setup should be considered to make this argument conclusive (e.g., \citealt{Vincent+2021}).

While the size of the bright emission ring/central brightness depression will be necessary to rule out or constrain the black hole and non-black hole models considered in Section \ref{sec:tests} below, the very presence of a central brightness depression in the 2017 image of \sgra (see\citetalias{PaperIII,PaperIV}, and \citetalias{PaperV}) is sufficient to rule out various models for compact objects that do not possess photon spheres. For example, these observations rule out the possibility that \sgra is a nonspinning Joshi-Malafarina-Narayan-2 (\citealt{Joshi+2014}; JMN-2) naked singularity since these exotic compact objects do not cast shadows \citep{Shaikh+2019}. We note that the JMN-2 spacetime is an exact solution of the Tolman-Oppenheimer-Volkoff equations of General Relativity, can form as the nonempty end-state of the gravitational collapse of a (non-thermal) perfect fluid from regular Cauchy data, and photons in the spacetime move on null geodesics of the metric tensor (see also the associated discussion in Section \ref{sec:known}). We assume the naked singularity present at $r=0$ does not interact with matter or radiation (classical gravity).

\section{Metric tests from shadow size}\label{sec:tests}

In Section~\ref{sec:delta}, we used the prior information on the mass-to-distance ratio of the \sgra\ black hole to calculate the predicted size of its shadow and compared the result to the size inferred from the EHT images and visibility-domain model fitting. We based this prediction on the Kerr metric and found that there is no evidence for any violations of the theory of General Relativity. Our goal in this section is to use these bounds on plausible deviations in the shadow size that are still consistent with the imaging data in order to place constraints on deviations of the parameters of the underlying black-hole metric.

We will follow two complementary approaches. First, in Section~\ref{sec:parametric}, we will constrain the parameters of stationary metrics that are agnostic to the underlying physical theory. These have been designed in a way that they reduce to the Kerr metric, when the deviation parameters are set to zero, but remain free of pathologies for a wide range of parameter values~\citep{Psaltis2020}. Although these metrics do not arise from any particular modification to gravity, they allow us to explore phenomenologically a very broad range of possibilities, which can be mapped afterwards to parameters of a fundamental theory.  Second,  in Section \ref{sec:known}, we will constrain the parameters of stationary metrics that are generated by various matter distributions and/or those that arise from specific modifications to the theory of gravity, and which depend on additional generalized charges~\citep{Kocherlakota2021}. Although the latter represent only particular types of deviations from General Relativity, they allow us to translate directly the constraints from the EHT images to bounds on physical parameters. Finally, in Section \ref{sec:PPN}, we will compare the constraints on the various metrics in terms of their asymptotic post-Newtonian parameters in order to demonstrate that, fundamentally, the bounds imposed by the EHT imaging observations of \sgra\ depend weakly on the particular metrics used to describe deviations from Kerr.

\begin{figure*}
\centering
\includegraphics[width=2\columnwidth]{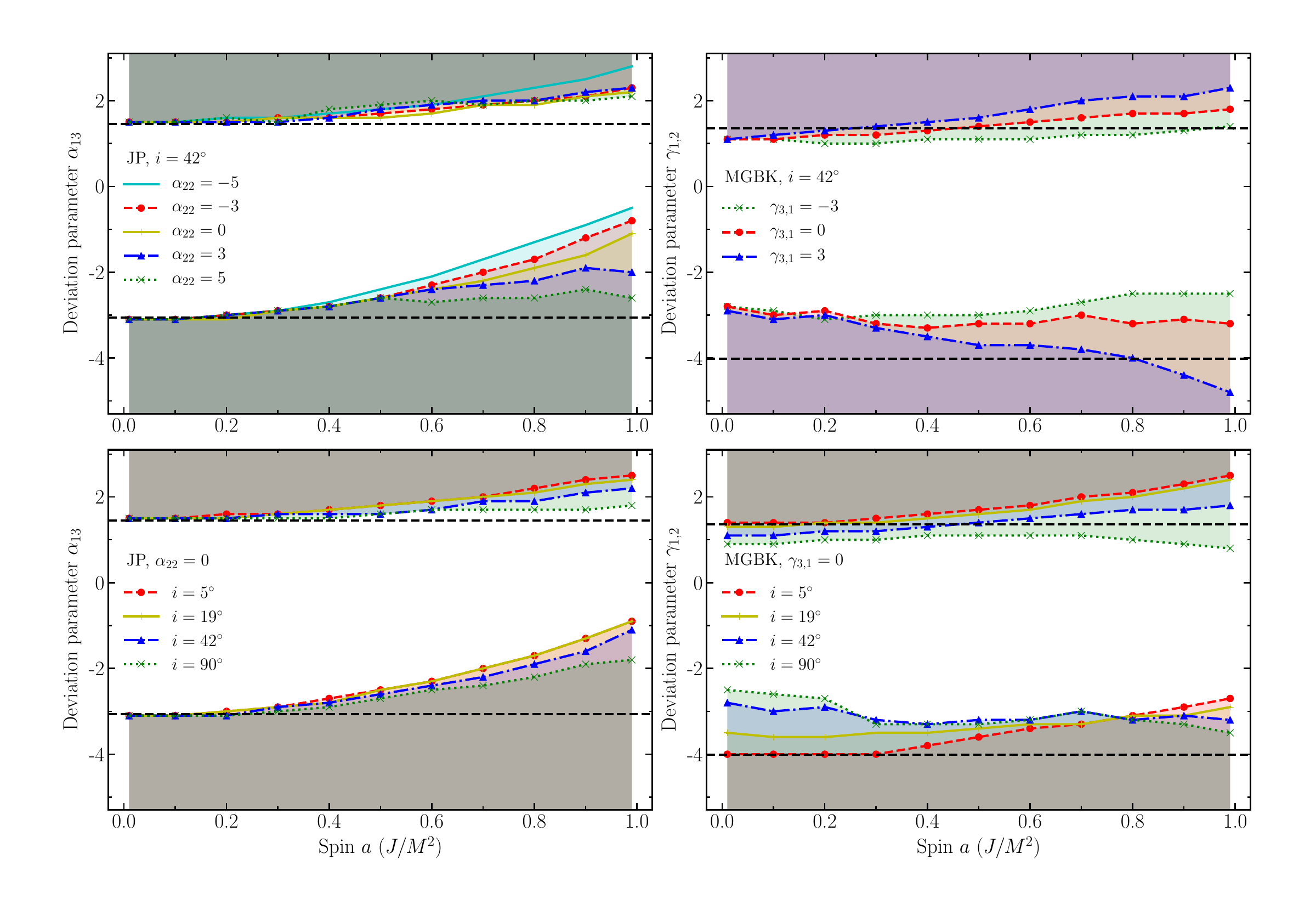}
\caption{Numerical constraints on various deviation parameters for the JP and MGBK metrics as a function of the dimensionless black-hole spin placed by the requirement that the predicted size of the black-hole shadow is consistent with the size inferred for \sgra. Here the shading shows the regions of the parameter space that are ruled out by the bound derived from the calibration based on the Keck mass measurement, the \ehtim algorithm, and the GRMHD simulation library, as an example. The left (right) panels show constraints for the JP (MGBK) metrics, while the different symbols/colors in the top and bottom panels show the effect of the secondary deviation parameter and the inclination of the observer, respectively. In the top panels we set the inclination angle to $i=42^{\circ}$ and set the secondary deviation parameter to zero in the bottom panels. In all panels the black dashed lines correspond to the analytic constraints for non-spinning black holes. The non-monotonic behavior of the constraints are due to the fact that both the size and the shape of the shadows are affected by the deviation parameters. The measurement of the shadow size constrains primarily the parameters that quantify deviations in the $tt-$components of the various metrics, as expressed in areal coordinates, and the resulting bounds depend only weakly on black-hole spin or observer inclination.}
\label{fig:numerical_param_constraints}
\end{figure*}

Throughout this section, for clarity of presentation, we will use the bounds $\delta = -0.04^{+0.09}_{-0.10}$ (Keck) and $\delta = -0.08^{+0.09}_{-0.09}$ (VLTI) on the fractional deviation inferred from the predicted shadow size, as calculated for the fiducial analyses that use the Keck and VLTI priors on the mass-to-distance ratio, the \ehtim imaging method, and the GRMHD library for quantifying the theoretical uncertainties. Where it is not possible to show both bounds, we will show the Keck bound as an example. This constraint depends weakly on the choice of priors and techniques. In particular, the fiducial bounds correspond to the following constraints on shadow size:
\begin{equation}\label{eq:constraintKeck}
\begin{split}
    3\sqrt{3}(1 - 0.14) M & \lesssim \tilde{r}_{\text{sh}} \lesssim 3\sqrt{3}(1 + 0.05) M,\\
    4.5 M & \lesssim \tilde{r}_{\text{sh}} \lesssim 5.5 M,
\end{split}
\end{equation}
for Keck and 
\begin{equation}\label{eq:constraintVLT}
\begin{split}
    3\sqrt{3}(1 - 0.17) M & \lesssim \tilde{r}_{\text{sh}} \lesssim 3\sqrt{3}(1 + 0.01) M,\\
    4.3 M & \lesssim \tilde{r}_{\text{sh}} \lesssim 5.3 M,
\end{split}
\end{equation}
for VLTI, where we have set $G=c=1$, as we will do throughout this section.
For the different metrics employed in this section, we calculate the fractional diameter deviation of the metric as
the relative difference between the analytic shadow diameter and the Schwarzschild diameter, namely:
\begin{equation}
\delta_{\rm metric} = \frac{\tilde{d}_{\mathrm {metric}}}{6\sqrt{3}} - 1 \,,
\end{equation}
where $\tilde{d}_{\rm metric}$ is the median shadow diameter, i.e., the locus of critical impact parameters.

Throughout this section we will not consider constraints on the circularity of the shadow. This is due to the sparse interferometric coverage of 2017 observations, which may lead to significant uncertainties in circularity measurements that we do not quantify here. In addition to measurement uncertainty, we also do not quantify the theoretical uncertainty between the circularity of the shadow and that of the observational feature. However, in future EHT observations with additional telescopes the circularity of the shadow may potentially be used for constraints on deviations from the Kerr metric.

\subsection{Constraints on Metrics with Parametric Deviations}\label{sec:parametric}

According to the no-hair theorem in General Relativity, the only stationary, asymptotically flat, Ricci-flat spacetime that is free of pathologies\footnote{By pathologies we mean closed timelike loops, naked singularities, and non-Lorentzian signatures outside of the event horizon.} is the one described by the Kerr metric. We do not consider here the astrophysically irrelevant case of black holes with a net electric charge (c.f. Section~\ref{sec:known}). As a result, introducing simple phenomenological deviations to the Kerr metric leads to pathologies that severely constrain our ability to make predictions for the size of the black-hole shadow, especially in the case of spinning black holes~(see \citealt{Johannsen2013a} for a detailed study of the pathologies of several parametrized metrics and \citealt{Kocherlakota+2022} for an analysis of theoretically-allowed parameter spaces of the RZ metric). For this reason, several parametrized metrics have been developed in the last decade that allow for general deviations from the Kerr metric while minimizing pathologies mostly by relaxing the assumption of Ricci flatness. These parametrized metrics are completely agnostic to an underlying physical theory and, therefore, significant assumptions must be made for stability tests (see e.g., \citealt{Suvorov2021} for a quasi-normal-mode analysis of the RZ metric). Among these metrics, we choose three representative ones: the so-called JP metric (\citealt{Johannsen2011}, which was further developed to ensure the presence of a Carter-like integral of motion in~\citealt{Johannsen2013b});  the Modified Gravity Bumpy Kerr (MGBK) metric \citep{Vigeland2011}; and the so-called RZ metric (\citealt{Rezzolla2014}, which was further developed to include the effects of spin in~\citealt{Konoplya2016}). We will derive analytic constraints for all three metrics and will use numerical calculations to derive spin-dependent constraints for both the JP and MGBK metrics. 

Earlier studies have demonstrated that, because of a near cancellation between the effects of frame dragging and of the quadrupole moment of the spacetime, the spin of the black hole affects the shadow size only marginally (see, e.g., \citealt{Johannsen2010,Psaltis2020}). Therefore, the bounds on the deviation parameters imposed by the measurement of the black-hole shadow in \sgra\ are also expected to depend weakly on black-hole spin. We demonstrate this in Figure~\ref{fig:numerical_param_constraints}, which shows the limits on different deviation parameters of the JP and MGBK metrics as a function of black-hole spin, for various observer inclinations, and for different values of the secondary deviation parameters. The horizontal dashed curves show the bounds for non-spinning black holes when the secondary deviation parameters are zero. For this figure, we have set the parameters that affect the $g_{tt}$--component of the metric at the $r^{-2}$ order to zero (i.e., $\alpha_{12}=0$ and $\gamma_{4,2}=-\frac{1}{2}\gamma_{1,2}$, see also Section~\ref{sec:known}) so we can focus on higher order effects. The resulting constraints are of order-unity and weakly dependent on spin, inclination angle, and the values of secondary deviation parameters. 

The simulations used for this figure are described in \citet{Medeiros2020}. In these simulations we assume that the geodesic equation holds for all metrics and solve for the trajectories of photons ignoring matter effects. We define the boundary of the black hole shadow as the critical impact parameter between photon trajectories that fall into the event horizon and those that escape to infinity. As was done in Section \ref{sec:error} we define the size of a shadow as its median radius and compare this measurement to the bounds on shadow size from the \ehtim algorithm, the GRMHD simulation library, and the prior mass and distance measurements from Keck. 

As found in~\cite{Psaltis2020}, the measurement of the size of the black-hole shadow places constraints of order unity primarily on parameters such as $\alpha_{13}$ and $\gamma_{1,2}$ that depend weakly on the magnitude of the black-hole spin. What is common between these parameters is that they describe deviations in the $tt-$component of the black-hole metric, as expressed in areal coordinates.

Since spin has a relatively small effect on the predicted shadow size and, hence, on the metric constraints, we now focus on a more detailed exploration of the constraints on these three metrics when we set the spin to zero (i.e., in the limit of spherical symmetry). The radius of the shadow in this limit is given by \citep{Psaltis2020}
\begin{equation}
r_{\mathrm{sh}}=\ \frac{r_{\mathrm{ph}}}{\sqrt{-g_{tt}(r_{\mathrm{ph}})}},
\label{eq:rsh}
\end{equation}
where 
\begin{equation}
r_{\mathrm{ph}} \equiv 2g_{tt}(r_{\mathrm{ph}})\left(\left. \frac{dg_{tt}}{dr}\right\vert_{r_{\mathrm{ph}}}\right)^{-1}\;,
\end{equation}
is the radius of the photon orbit.

For non-spinning black holes, the $tt-$component of the JP metric in areal coordinates is (see \citealt{Johannsen2013a})
\begin{equation}
g_{tt}^{\mathrm{JP}} = -\left( 1- \frac{2}{r_A}\right) \left( 1 + \sum^{\infty}_{i=2} \frac{\alpha_{1i}}{r_A^i}\right)^{-2}
\end{equation}
where $\alpha_{1i}$ are deviation parameters and the subscript $A$ denotes the fact that we use areal coordinates.

\begin{figure}[t!]
\centering
\includegraphics[width=\columnwidth]{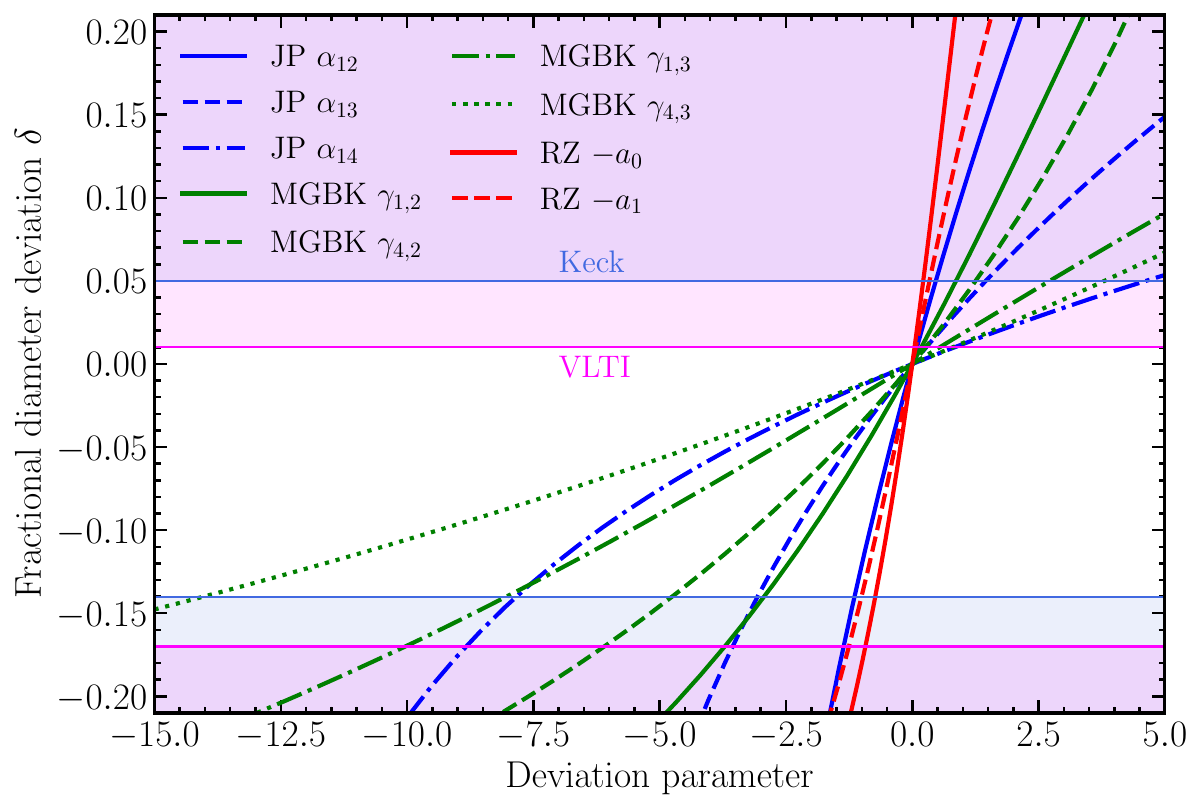}
\caption{Analytic constraints on deviation parameters for the JP, MGBK, and RZ metrics. The regions excluded by the \sgra constraints using Keck (VLTI) are shown in blue (magenta), see also Equations~\ref{eq:constraintKeck} and \ref{eq:constraintVLT}. For each curve we allow only one parameter to vary while setting the others to zero. The original $g_{tt}$ components of the metrics were used for this plot, not their expansions.}
\label{fig:analytic_param_constraints}
\end{figure}

The $tt-$component of the non-spinning MGBK metric in areal coordinates is (see \citealt{Vigeland2011,Gair2011})
\begin{equation}
g_{tt}^{\mathrm{MGBK}} = -\left( 1-\frac{2}{r_A}\right)\left[ 1-\gamma_{1}(r_A)-2 \gamma_4(r_A)\left( 1-\frac{2}{r_A}\right)\right]
\end{equation}
where $\gamma_1(r_A)$ and $\gamma_{4}(r_A)$ are defined by 
\begin{equation}
\gamma_A = \sum^{\infty}_{n=2}\frac{\gamma_{A,n}}{r_A^n},
\end{equation}
$A=1$ or 4, and $\gamma_{A,n}$ are the deviation parameters.  

Finally, the $tt-$component of the spherically symmetric RZ metric in areal coordinates is (see \citealt{Rezzolla2014})
\begin{equation}
g_{tt}^{\mathrm{RZ}} = -x\left[ 1-\epsilon(1-x)+ (a_0-\epsilon)(1-x)^2+\tilde{A}(x)(1-x)^3 \right]
\end{equation}
where
\begin{equation}
\begin{split}
x &\equiv 1-\frac{r_0}{r_A},\\
\tilde{A}(x) &= \frac{a_1}{1+\frac{a_2 x}{1+\frac{a_3 x}{...}}},
\end{split}
\end{equation}
and $r_0$ is the coordinate radius of the infinite redshift surface (identified with the horizon if no pathologies exist). The parameters $\epsilon, \, a_1,\, a_2,\,...$ are the deviation parameters. As done in \citet{Psaltis2021}, we write all radii in terms of the mass of the black hole at infinity which fixes one of the parameters
\begin{equation}
\epsilon = -\left( 1- \frac{2}{r_0}\right).
\end{equation}
For simplicity we assume $r_0=2$ throughout the rest of this section.

Using these analytic expressions, we calculate the size of the black-hole shadow as a function of the various deviation parameters using equation~(\ref{eq:rsh}). We then apply the bounds on the shadow size imposed by the \sgra\ images and obtain constraints on the deviation parameters of the various metrics. The dashed lines in Figure~\ref{fig:numerical_param_constraints} compare the analytic bounds with those obtained numerically for the spinning spacetimes.

Because we use only one measured quantity from the EHT image of \sgra, i.e., the size of the black-hole shadow, but the $tt-$components of each metric depend on a series of deviation parameters, it follows that the EHT observations place, in fact, correlated constraints on these parameters. These can be thought of as subspaces in the multidimensional parameter space of deviations and are very difficult to visualize in full generality. In Figure~\ref{fig:numerical_param_constraints}, we showed one particular cross section of this parameter space in which the various parameters were combined such that deviations from Kerr appear only at the second or higher post-Newtonian order. 

Figure~\ref{fig:analytic_param_constraints} shows a different cross section of these constraints, where we have varied only one parameter at a time, setting all others equal to zero. The constraints derived from the Keck bounds are also summarized in Table~\ref{table:Solutions_Summary_parametrized}. The deviation parameters that correspond to higher-order corrections (as denoted by the second integer in their subscripts) affect the size of the shadow less strongly that the lower-order parameters and are, therefore, less constrained. We will return to the magnitudes of these constraints in Section \ref{sec:PPN}, after we discuss the bounds on metrics that correspond to solutions of particular modified gravity theories.

\begin{table}[t]
\begin{center}
\caption{Constraints on Deviation Parameters of Various Metrics}
\label{table:Solutions_Summary_parametrized}
\begin{ruledtabular}
\begin{tabular}[t]{cc}
Parametrized metric & Constraints \\
\colrule
 JP & $\begin{array} {lcl} -1.1 \lesssim &\alpha_{12} \lesssim 0.5\\  -3.1 \lesssim &\alpha_{13}\lesssim 1.5\\ -7.8 \lesssim &\alpha_{14} \lesssim 4.6\\ \end{array}$ \\
\hline
 MGBK & $\begin{array} {lcl} -4.0 &\lesssim \gamma_{1,2}\lesssim 1.4\\  -4.8 &\lesssim \gamma_{4,2} \lesssim 1.3\\ -8.0 &\lesssim \gamma_{1,3}\lesssim 2.7\\ -14.0 &\lesssim \gamma_{4,3} \lesssim 3.8 \end{array}$ \\
\hline
RZ &  $\begin{array} {lcl} -0.2\lesssim &a_0 \lesssim0.7\\ -0.3 \lesssim &a_1 \lesssim 1.0\\ \end{array}$ \\
\end{tabular}
\begin{tablenotes}
\item Notes: Here and in Table~\ref{table:Solutions_Summary_parametrized_PPN} we use the bound derived from the calibration based on the Keck mass measurement, the eht-imaging algorithm, and the GRMHD simulation library, as an example.
\item JP, MGBK, and RZ are parametrized metrics that deviate from Kerr (see \citealt{Johannsen2011,Vigeland2011,Rezzolla2014} for details on these metrics).
\end{tablenotes}
\end{ruledtabular}
\end{center}
\end{table}

\begin{figure*}[t!]
\includegraphics[width=\columnwidth]
{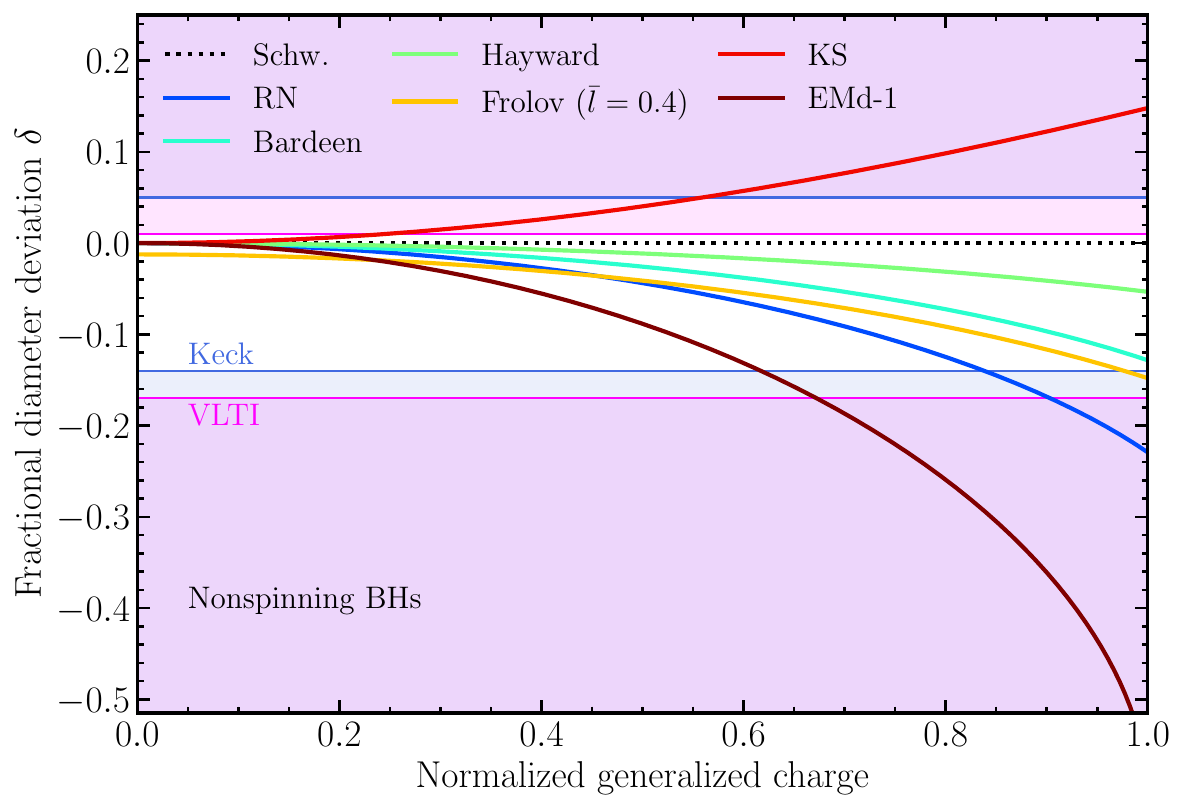}
\hfill
\includegraphics[width=\columnwidth]
{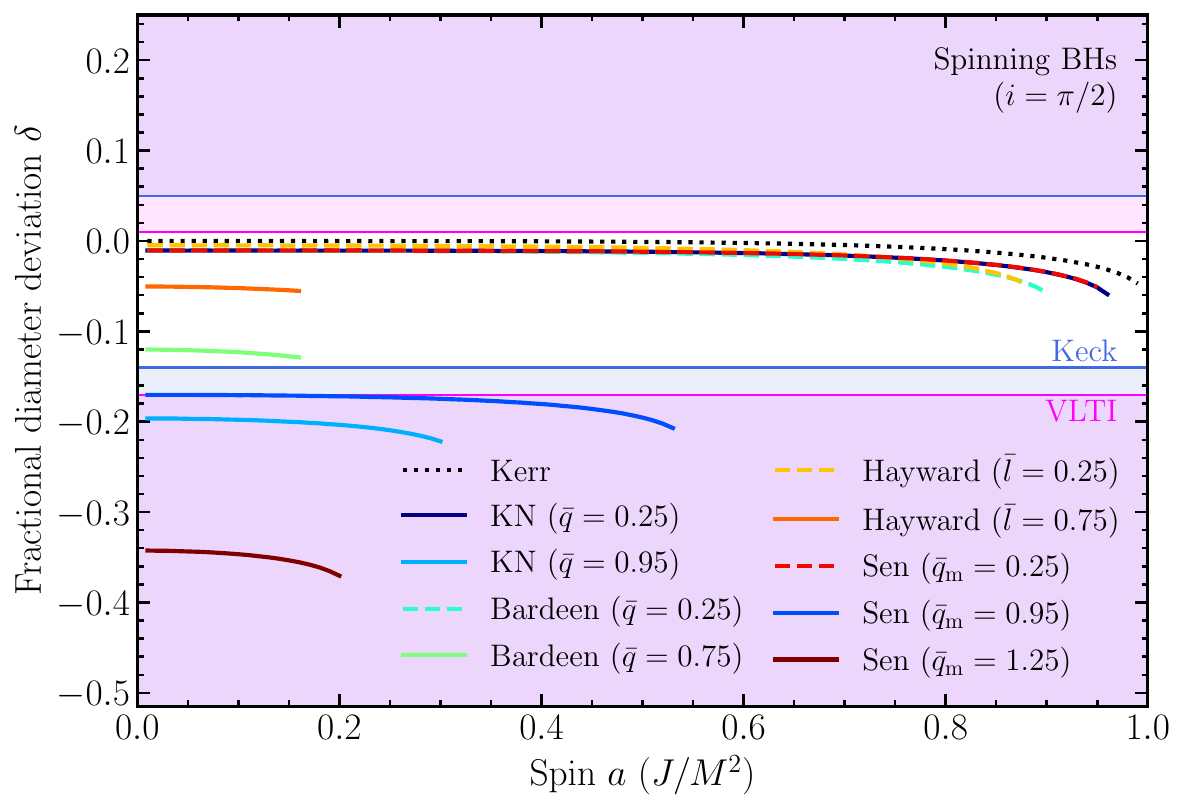}
\caption{The dependence of the fractional shadow diameter deviation from the Schwarzschild value {\em (left)\/} on the relevant physical ``charges''  of various nonspinning black hole metrics and {\em (right)\/} on the black-hole spins, for fixed values of the charges and for an observer inclination $i =\pi/2$. The white regions correspond to shadow sizes that are consistent at the 68\% level with the 2017 EHT observations for \sgra. As in the case of the Kerr and of the parametric metrics, the spin of the black hole introduces only minor corrections to the predicted shadow size and, hence, to the metric constraints. Current EHT imaging observations of \sgra\ are inconsistent with some metrics when their physical charges are comparable to their maximum theoretically-allowed values. We use the median shadow diameter to characterize the size of the noncircular shadows cast by the spinning black holes {\em (right)\/}, as done in Section \ref{sec:error}.} 
\label{fig:KnownSolutionsBHFig}
\end{figure*}

\subsection{Constraints on Specific Compact Object Spacetime Metrics}\label{sec:known}

Detecting possible deviations from the Kerr metric using the agnostic approach discussed above can be used to infer constraints on multiple asymptotic expansion coefficients of the spacetime, test the no-hair theorem, assess the Ricci-flatness of black-hole metrics, etc. In this Section, we follow a complementary approach to determining whether specific fundamental principles of the theory of gravity are violated by considering explicitly theories that incorporate such violations by design, finding (stationary) solutions to the associated field equations that describe supermassive compact objects such as \sgra, and determining whether their images, when undergoing similar accretion processes, are compatible with those observed with the EHT.

Adopting this approach helps us assess the necessity of including additional fundamental fields (such as dilatons or axions) in the description of the classical theory and yield quantitative constraints on the amount of build-up of various fields in the vicinity of supermassive compact objects (see, e.g., \citealt{Kocherlakota2021}). This could be instructive of the astrophysical processes that may have produced them. Studying the images of available solutions allows us also to address questions related to the type of object that \sgra\ is, e.g., whether it is a naked singularity, a boson star, or a black hole. Finally, working with a specific theory enables a comparison of its predictions for a variety of other physical scenarios with already existing or future observations, for its overall compatibility, as discussed in Section \ref{sec:compare}.

{\em Alternative Black Holes.---\/}As an example, the equivalence principle is a fundamental building block of the theory of gravity and has thus far been tested in various regimes by complementary experiments. It comprises three aspects~\citep{Dicke1964, Will2014}: the weak equivalence principle (WEP), local Lorentz invariance (LLI), and local positional invariance (LPI). To demonstrate the scope of testing theories that violate the WEP and LPI with the EHT, we will consider here black-hole solutions from two Einstein-Maxwell-dilaton-axion (EMda) theories (\citealt{Gibbons+1988, Garfinkle+1991, Sen1992, Garcia+1995, Kallosh+1992}; see also the discussion in, e.g., \citealt{Magueijo2003} and \citealt{Kocherlakota+2020}), which emerge as the low-energy effective descriptions of the heterotic string. This conservative choice allows us to be certain that (a) the form of the equations that describe the dynamics of accreting plasma flow around EMda black holes is identical to those in General Relativity due to the minimal coupling of matter to Einstein-Hilbert gravity via the metric tensor, and (b) photons move on null geodesics of the metric tensor since electromagnetism is described by the (linear) Maxwell Lagrangian (see, e.g., Sec. 4.3 of \citealt{Wald1984}).

Synthetic images of radiative inefficient accretion flows onto Gibbons-Maeda-Garfinkle-Horowitz-Strominger black holes \citep{Gibbons+1988, Garfinkle+1991}, which describe charged, static black holes in one of the EMda theories (henceforth the EMd-1 for brevity), have been constructed in \cite{Mizuno+2018}, using MHD simulations (see also Figure \ref{fig:image_gallery}). It was demonstrated there that the final images of these EMd-1 black holes are comparable to those of the Schwarzschild/Kerr black holes. More recently, properties of images of (Kerr-)Sen black holes~\citep{Sen1992}, which are the spinning generalizations of the EMd-1 black holes, have been calculated and characterized in \cite{Younsi2021}, when undergoing accretion that is described by the semi-analytic model of \cite{Ozel2021} (see also Figure \ref{fig:image_gallery}). To compare the features of Sen black holes against their general relativistic counterparts, we consider the Reissner-Nordstrom (RN; \citealt{Reissner1916, Nordstrom1918}) and the Kerr-Newman (KN; \citealt{Newman+1965}) solutions, which describe charged black holes with and without spin respectively. 

We also consider solutions arising from various attempts to regularize the central singularities of classical black holes within General Relativity. In particular, we consider solutions by \cite{Bardeen1968}, \cite{Hayward2006}, and \cite{Frolov2016}.\footnote{These spacetimes have also been obtained as solutions in other theories (see, e.g., \citealt{Ayon-Beato+1998, Ayon-Beato+2000, 
Held+2019}).}
Such solutions are typically nonempty and the matter present, in a stationary configuration, typically violates one or more energy conditions (\citealt{Hawking+1973, Curiel2014}). Studying the images of available solutions that can be used to model compact objects allows us to test for possible violations of components of the equivalence principle or of energy conditions. We also include here the static \cite{Kazakov+1994} (KS) solution which attempts to smear out the central singularity onto a surface. Additionally, we also consider the spinning counterparts of the Bardeen and Hayward metrics \citep{Abdujabbarov+2016}. To conduct the analysis below, we have implicitly assumed that the ``ordinary'' matter in the accretion flow does not interact with the background matter in the nonempty spacetimes we consider here. 

\begin{figure}[t!]
\includegraphics[width=\columnwidth]
{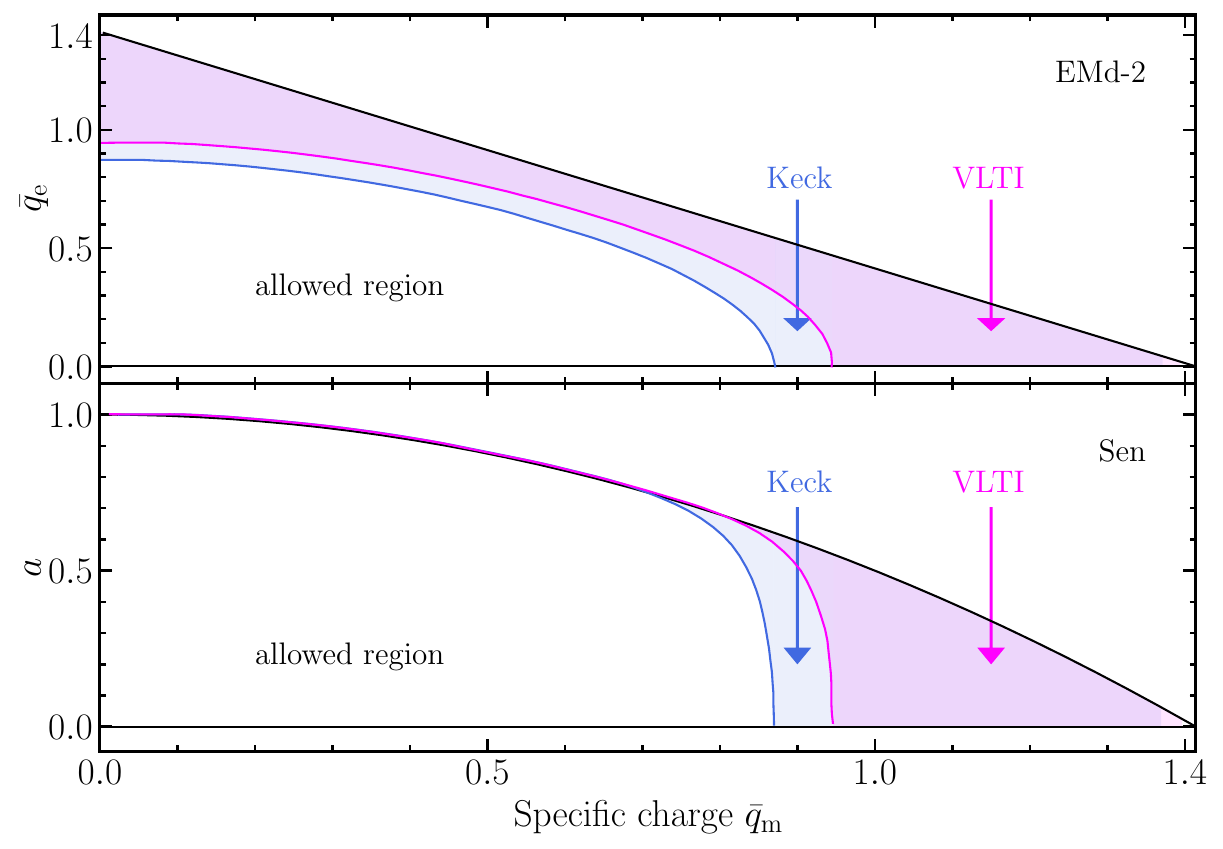}
\caption{We show here the constraints on two EMda solutions from different EMda theories. In the top plot we show the constraints on the parameter space of a nonspinning black hole from an EMda theory with two $U(1)$ gauge fields \citep{Kallosh+1992} whereas in the bottom plot we show the constraints on the parameter space of a spinning Sen black hole from an EMda theory with a single $U(1)$ gauge field \citep{Sen1992}.}
\label{fig:KnownSolutionsEMdaFig}
\end{figure} 

\begin{figure}[t!]
\includegraphics[width=\columnwidth]
{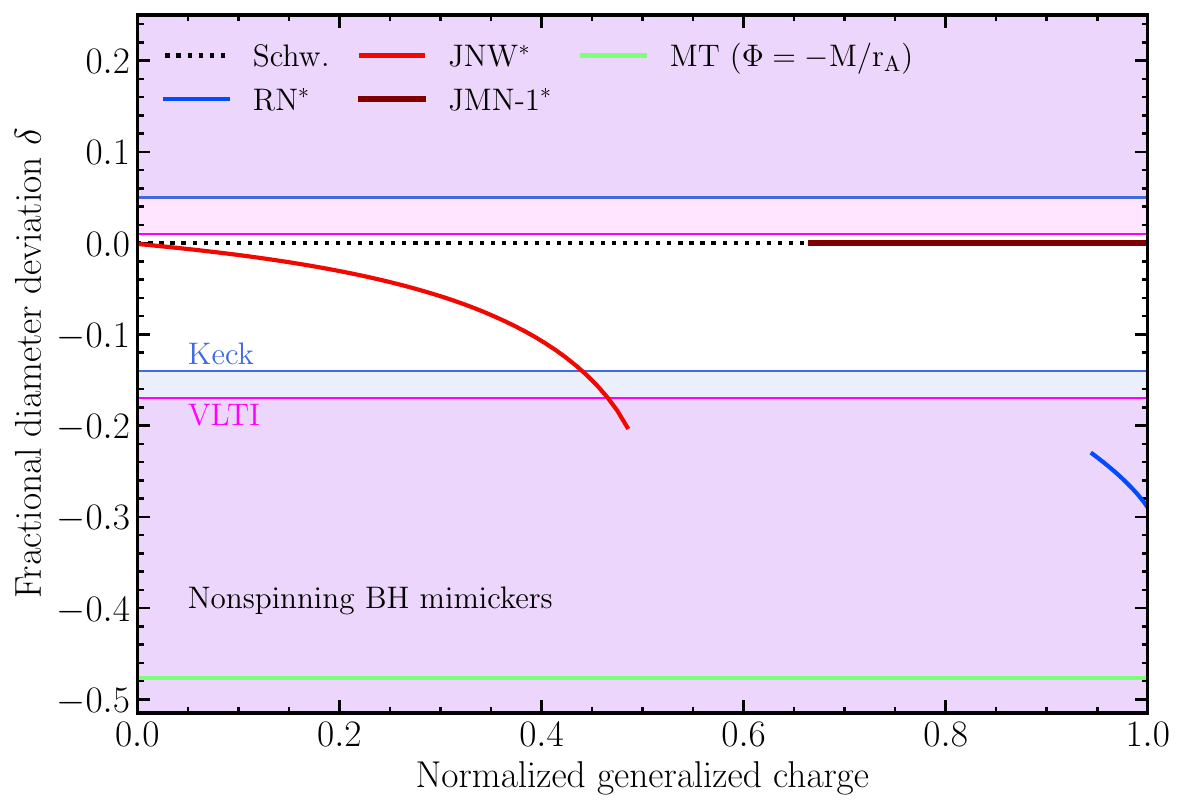}
\caption{Same as Figure~\ref{fig:KnownSolutionsBHFig} but for metrics that describe various naked singularities (denoted by a star) and a wormhole.}
\label{fig:KnownSolutionsNSFig}
\end{figure} 

Figure~\ref{fig:KnownSolutionsBHFig} shows the dependence of the deviation of the shadow size from the Schwarzschild prediction on the parameters (the ``generalized charges'' or simply charges henceforth) of the various metrics discussed above (see also Section IV of \citealt{Kocherlakota+2020} for further details). In this figure, each relevant physical parameter has been normalized to its maximum theoretically-allowed value (see also Table \ref{table:Solutions_Summary}).%
\footnote{For the Kazakov-Solodukhin (KS; \citealt{Kazakov+1994}) black hole, the theoretically-permitted range of the relevant parameter is noncompact, $\bar{\alpha} > 0$, and we show the range $0 < \bar{\alpha} \leq \sqrt{2}$ in Figure~\ref{fig:KnownSolutionsBHFig}.} %
Similarly to the case of the Kerr and of the parametric metrics, we find that the black-hole spin introduces minor corrections to the size of the shadow, which allows us to focus on nonspinning spacetimes. Moreover, the current bounds imposed by the EHT images of \sgra\ place constraints of order unity to the charges of several of the spacetimes, which are comparable to their maximum values, by construction. Figure~\ref{fig:KnownSolutionsEMdaFig} focuses on the constraints that we can set on the relevant parameter spaces of all the black holes from the two EMda theories considered here: the top panel shows the charged, nonspinning ``EMd-2'' solution from one theory \citep{Kallosh+1992}. From the other theory, we show in the bottom panel the parameter space for the spinning Sen black hole \citep{Sen1992} (the EMd-1 black hole corresponds to the $a=0$ line). As can be seen from these figures, we find no evidence of violations of the equivalence principle or of the presence of energy-conditions violating matter within the present context. 

Thus far we have considered in detail the possibility that \sgra\ is a supermassive black hole (described by different metrics) as well as the alternative that it possesses a material surface (Section \ref{sec:eh}). We have also considered the possibility that \sgra\ is a surfaceless horizonless compact object without a photon sphere, with focus on mini-boson stars and naked singularities, in Section \ref{sec:SHCOs}. Here, using specific solutions, we will address whether the spacetime in its vicinity can be well modeled by that of a naked singularity with a photon sphere, and, later on, by a wormhole. Since all of the naked singularity solutions we consider here arise from metric theories of gravity with the electromagnetic sector being governed by the linear Maxwell Lagrangian, photons move on null geodesics of the metric tensor, as discussed above. Since we consider exact solutions to the classical theory, we assume that the background spacetimes are static and that the naked singularities at $r=0$ do not interact with matter or radiation in any way.

{\em Naked Singularities.---\/}For an example of a naked singularity spacetime, we will consider the Reissner-Nordstr{\"o}m metric \citep{Reissner1916,Nordstrom1918}, characterized by specific electromagnetic charges of $\bar{q} > 1$, and denote it by RN$^*$. These spacetimes admit photon spheres only for $1 < \bar{q} \leq \sqrt{9/8}$, and Figure~\ref{fig:KnownSolutionsNSFig} shows only this range, normalized to the maximum. The Janis-Newman-Winicour (JNW; \citealt{Janis+1968}) naked singularity spacetime is a solution of the Einstein-Maxwell-scalar theory with a theoretically-allowed scalar charge parameter range of $0 < \hat{\bar{\nu}} < 1$. However, the JNW naked singularities only cast shadows when $0 < \hat{\bar{\nu}} \leq 0.5$, as indicated in Figure~\ref{fig:KnownSolutionsNSFig}. We will also consider a new class of naked singularities within General Relativity, namely the Joshi-Malafarina-Narayan-1 (JMN-1; \citealt{Joshi+2011}) naked singularities. This class of solutions describes a one-parameter $M_0$ family of static spacetimes containing a compact region $r_A < r_{A, \text{b}} \equiv 2M/M_0$ filled by an anisotropic fluid, where $M$ is the ADM mass of the spacetime. This spacetime can be attained at asymptotically late times as a result of gravitational collapse from regular initial data (\citealt{Joshi+2011}; see also Section IV of \citealt{Dey+2019}) and contains a photon sphere when $r_{A, \text{b}} < 3M$ or equivalently when $M_0 \geq 2/3$ contributed by the exterior Schwarzschild spacetime. 

Spherical Bondi-Michel accretion onto JMN-1 naked singularities has been studied in \cite{Shaikh+2019}, where it was found that the final images of JMN-1 naked singularities with photon spheres are indistinguishable from those of a Schwarzschild black hole. More strikingly, for the same approximate luminosity as \sgra at 200~GHz, accretion flows onto these singularities have spectra nearly identical to that of a Schwarzschild black hole (see Fig.~6 therein), indicating that a JMN-1 naked singularity with a photon sphere may be one of the best possible black-hole mimickers for \sgra\ (cf. Section \ref{sec:eh}).  

Figure~\ref{fig:KnownSolutionsNSFig} shows the bounds imposed by the EHT images of \sgra\ on the the physical charges of these metrics that describe naked singularities. With the exception of the Reissner-Nordstr{\"o}m metric, which predicts shadow sizes that are significantly smaller than what is observed, the possibility that \sgra\ is a naked singularity cannot be ruled out based on the metric tests we describe in this section.

{\em Wormholes.---\/}As an example of a wormhole, we consider nonspinning, traversible Morris-Thorne (MT) metrics \citep{Morris+1988} in General Relativity, for which the $tt$-components of the metric are determined by the ``redshift function'' $\Phi$ as $g_{tt} = -\exp{(2\Phi)}$. For wormholes in General Relativity to be traversible, the spacetime must necessarily contain energy-condition-violating matter and lack event horizons. The location(s) of the circular null geodesic(s) in this spacetime can be obtained by solving $-1 + r_A~d\Phi/dr_A = 0$. If we restrict to the simplest case of an MT wormhole with a single unstable circular null geodesic,\footnote{We note that there exist wormhole spacetimes that could be particularly difficult to distinguish from a black hole using the present considerations (see, e.g., \citealt{Morris+1988b}).} which can then be identified as the location of the photon sphere, e.g., by setting $\Phi = -r_{A, \text{t}}/r_A$ as in \cite{Bambi2013}, we find that the (Keplerian/ADM) mass definition forces $r_{A, \text{t}} = M$. This implies a shadow radius of $r_{\text{sh}} = e~M \approx 2.72~M$ or equivalently $\delta \approx -0.48$, which is immediately ruled out by the present considerations (see Figure~\ref{fig:KnownSolutionsNSFig}).

Finally, as noted above in Section \ref{sec:eh}, improved angular resolution with space-VLBI would greatly help constrain possible metric-deviations from the Kerr geometry. Notably, it would be possible to infer the spin of \sgra, when modelled as a Kerr black hole, if we are able to achieve a precision of $|\delta| < 0.07$ (\citealt{Johannsen2010}; see also Fig. \ref{fig:KnownSolutionsBHFig}). Shadows of spinning MT wormholes \citep{Teo1998} have recently been considered in \cite{Shaikh2018}, where it was shown that their shape can vary considerably from that of a Kerr black hole, and possibly be detected with future EHT or ngEHT measurements. Spacetimes admitting multiple circular null geodesics were considered by \citet{Wielgus+2020}, an example of which is given by black holes in a non-minimal Einstein-Maxwell-scalar theory \citep{Gan+2021}. Presence of a persistent multi-ring structure in an EHT image would constitute a robust topological discriminant of this family of spacetimes, particularly with future observations at higher resolution and flux-sensitivity.

\begin{table*}[t]
\begin{center}
\caption{Post-Newtonian Coefficients of Parametrized Metrics}
\label{table:Solutions_Summary_parametrized_PPN}
\begin{ruledtabular}
\begin{tabular}[t]{ccccc}
Parametrized metric & $\kappa_1$ & $\kappa_2$ & Constraints\\
\colrule
 JP &  $-\alpha_{12}$ &  $-2\alpha_{12}+\alpha_{13}$ &$\begin{array} {lcl} -0.3 &\lesssim \kappa_1 \lesssim 0.7\\   -3.1 &\lesssim \kappa_2 \lesssim 1.5\\ \end{array}$ \\
\hline
 MGBK  & $-\frac{1}{2}\gamma_{1,2}-\gamma_{4,2}$ & $-\gamma_{1,2}+\frac{1}{2}\gamma_{1,3}-4\gamma_{4,2}+\gamma_{4,3}$ &$\begin{array} {lcl}  -0.3 &\lesssim \kappa_1 \lesssim 0.9 \\   -4.0 &\lesssim \kappa_2 \lesssim 1.3\\ \end{array}$\\
\hline
RZ &   $2a_0$ & $4\left(a_0-\frac{a_1}{1+\frac{a_2}{1+\frac{a_3}{...}}}\right)$ &$\begin{array} {lcl}  -0.3 &\lesssim \kappa_1 \lesssim 0.9\\   -4.0 & \lesssim \kappa_2 \lesssim 1.3\\ \end{array}$\\
\end{tabular}
\end{ruledtabular}
\end{center}
\end{table*}

\begin{table*}
\begin{center}
\caption{Post-Newtonian Coefficients of Specific Compact Object Spacetime Metrics}
\label{table:Solutions_Summary}
\begin{ruledtabular}
\begin{tabular}[t]{lcccc}
Spacetime & Charge Range & Constraints & $\kappa_1$ & $\kappa_2$ \\
\colrule
RN & $0 < \bar{q} \leq 1$ & $0 < \bar{q}\leq 0.84$ & $\bar{q}^2/2$ & $0$ \\
RN* & $1 < \bar{q}$ & $\times$ & $\bar{q}^2/2$ & $0$ \\
Schwarzschild & $-$ & $-$ & $0$ & $0$ \\
\hline
Bardeen & $0 < \bar{q} < \sqrt{16/27}$ & $-$ & $0$ & $-3\bar{q}^2/2$ \\
\hline
Frolov  & $0 < \bar{l} < \sqrt{16/27}, 0 < \bar{q} \leq 1$ & see, e.g., Figure \ref {fig:KnownSolutionsBHFig} & $\bar{q}^2/2$ & 0 \\
Hayward & $0 < \bar{l} < \sqrt{16/27}$ & $-$ & $0$ & $0$ \\
\hline 
MT Wormhole & $-$ & $\times$ & $-1$ & $-2/3$ \\
\hline 
KS & $0 < \bar{\alpha}$ & $0 < \bar{\alpha} \leq 0.79$ & $-\bar{\alpha}^2/4$ & $0$ \\
\hline
EMd-1$^\dagger$ & $0 < \bar{q} < \sqrt{2}$ & $0 < \bar{q} < 0.87$  & $\bar{q}^2/2$ & $\bar{q}^4/8$ \\
\hline
EMd-2$^\dagger$ & $0 < \bar{q}_{\text{e}} \leq \sqrt{2} - \bar{q}_{\text{m}} < \sqrt{2}$ & Figure \ref {fig:KnownSolutionsEMdaFig} {\em (top)\/} & $\left(\bar{q}_{\text{m}}^2 + \bar{q}_{\text{e}}^2\right)/2$ & $\left(\bar{q}_{\text{m}}^2 - \bar{q}_{\text{e}}^2\right)^2/8$ \\
\end{tabular}

\begin{tablenotes}
 Notes: We denote by crosses and dashes spacetimes that are entirely ruled out and that are unaffected by the EHT measurements.
\end{tablenotes}
\end{ruledtabular}
\end{center}
\end{table*}
\subsection{Comparisons Between Metric Constraints}\label{sec:PPN}

In the discussion above, we used the inferred size of the black-hole shadow in \sgra\ in order to place constraints on parameters of metrics that deviate from Kerr. For the metrics that are solutions to particular modifications to General Relativity, these parameters (or ``charges'') correspond to particular properties of the theory or of the back hole itself. For the parametrized metrics, these parameters are phenomenological coefficients that are agnostic to any particular aspect of the underlying theory. Even though it might appear that these bounds are specific to the particular metric used, we will show here that they describe deviations from Kerr that are mathematically very similar to each other and nearly independent of the characteristics of the metric used.

First, as Figures~\ref{fig:numerical_param_constraints} and \ref{fig:KnownSolutionsBHFig} show, the constraints imposed by the measurement of the size of a black-hole shadow depend weakly on the spin of the black hole, for all the metrics explored. As discussed in Section~\ref{sec:parametric}, for metrics with zero spin, it can be shown analytically that the measurements lead to constraints only on the parameters that enter the $tt-$component of the metric in areal coordinates. The consequence of these two statements is that, for all spins, the primary constraints imposed by the measurement of a shadow size will be only on one of the metric components, largely independent of the other metric details.

\begin{figure}[t!]
\centering
\includegraphics[width=\columnwidth]{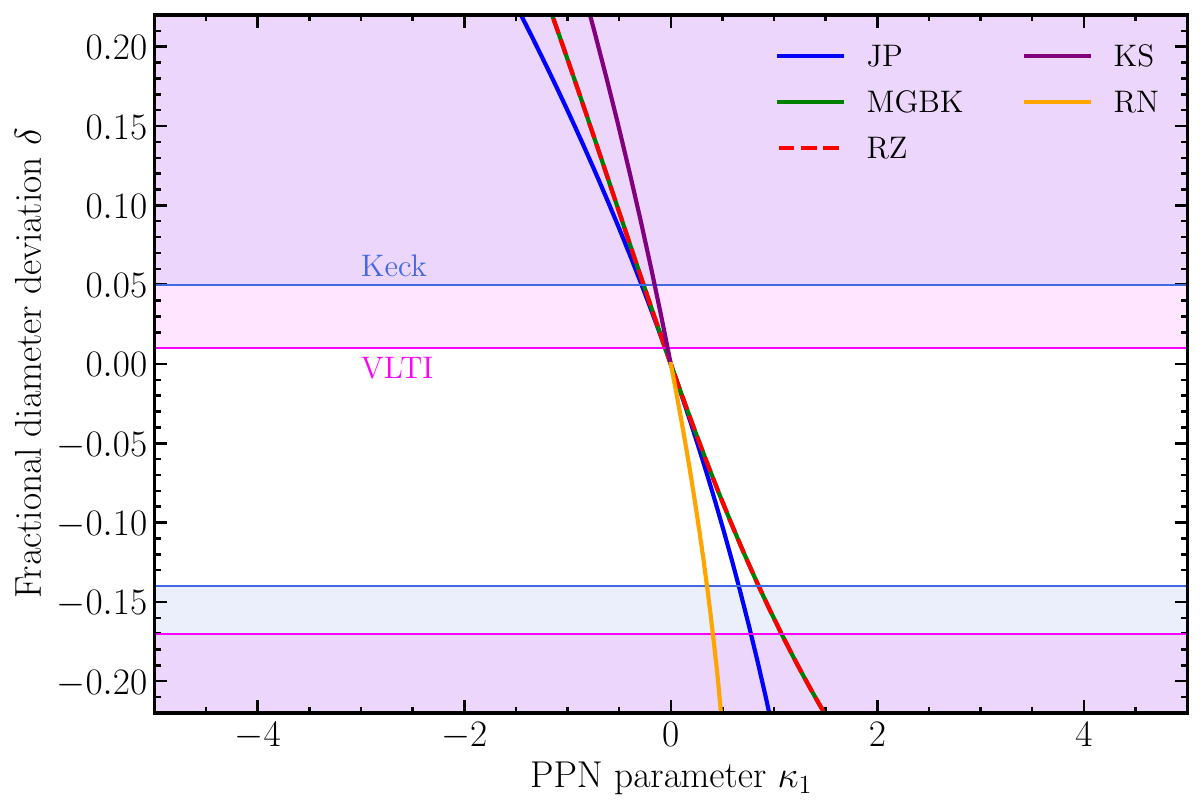}
\includegraphics[width=\columnwidth]{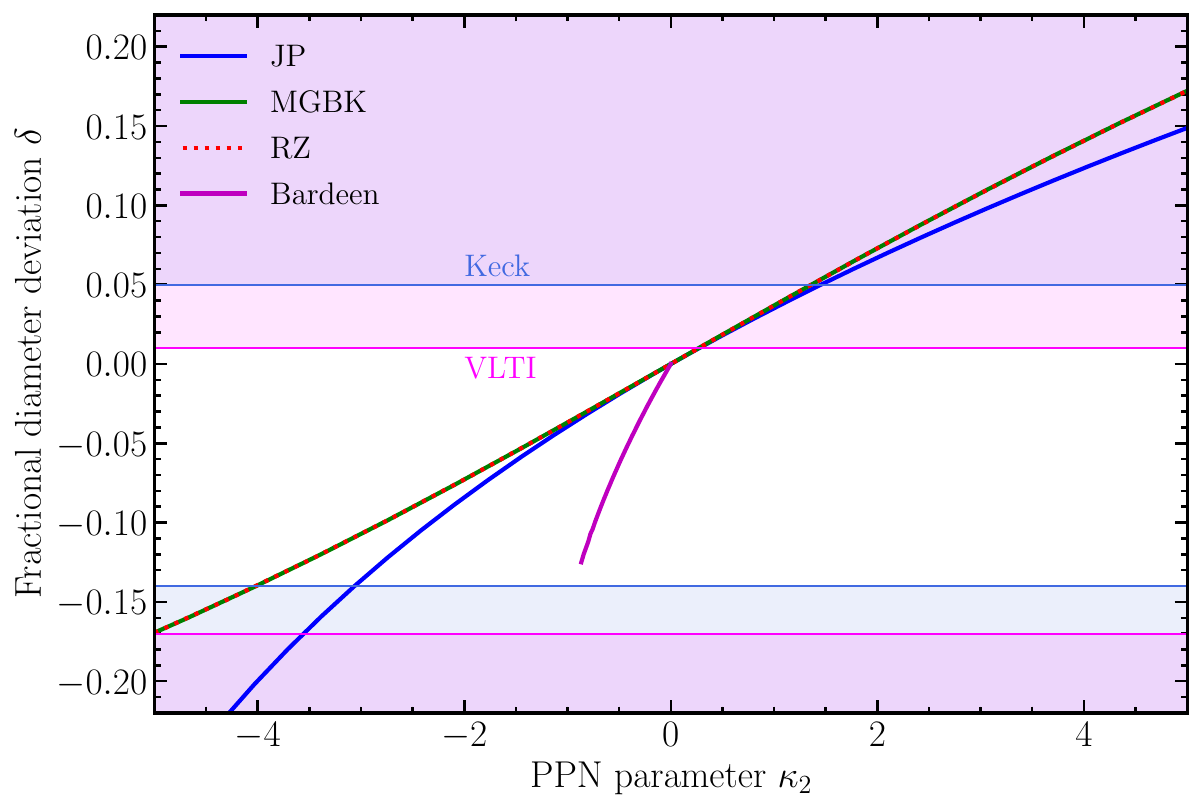}
\caption{Constraints on the post-Newtonian coefficients $\kappa_1$ (top panel) and $\kappa_2$ (bottom panel) of the \sgra\ metric imposed by the EHT images. We include three parametrized metrics as well as several metrics that are known solutions to particular modified gravity theories. For each curve, we only allow one of the first two post-Newtonian coefficient to vary and set the others to zero. The bounds on the post-Newtonian coefficients depend weakly on the specific properties of the metric used to obtain them.}
\label{fig:analytic_param_constraintsPPN}
\end{figure}

Translating directly the bounds on the parameters of one metric to those of another is non-trivial because of the usual coordinate and gauge ambiguities that are inherent to relativistic spacetimes. 
However, one avenue of making this comparison is by exploring the asymptotic behavior of these metrics towards radial infinity. In particular, we write the $tt-$component in areal coordinates in terms of the parametrized post-Newtonian (PPN) expansion
\begin{equation}
g_{tt} = -1 + \frac{2}{r_A} - 2\left(\frac{\kappa_1}{r_A^2}\right) + 2\left( \frac{\kappa_2}{r_A^3} \right) - 2\left( \frac{\kappa_3}{r_A^4} \right) + \mathcal{O}(r_A^{-5})
\end{equation}
and connect without ambiguity the post-Newtonian coefficients of each metric to each particular parameter. We then translate the bounds of the parameters to constraints on these post-Newtonian coefficients and compare the results obtained with different metrics.

We emphasize that we do not use these post-Newtonian expansions in order to calculate black-hole shadows, which would have been inappropriate given that the size of the shadow is comparable to the horizon radius. Instead, we calculate shadow sizes and place constraints on the particular parameters of each of the metrics that has been developed specifically for use in the strong-field regime. We only use the post-Newtonian coefficients as a mechanism to compare the asymptotic behavior of these metrics. Because each post-Newtonian coefficient at a given order $N$ is proportional to the derivative of the metric coefficient with respect to $1/r_A$ at order $N+1$, comparing post-Newtonian coefficients is equivalent to comparing the detailed functional forms of the metrics.

Tables~\ref{table:Solutions_Summary_parametrized_PPN} and \ref{table:Solutions_Summary} summarize the post-Newtonian coefficients at the various orders for the different metrics used in the previous sections (see, e.g., ~\citealt{Psaltis2021}). Using this correspondence between metric parameters and post-Newtonian coefficients, we show in Figure \ref{fig:analytic_param_constraintsPPN} the fractional deviation $\delta$ of the shadow size from the Schwarzschild prediction but plotted against the equivalent post-Newtonian coefficients for each metric, at the first and second order. 

As before, we show only two particular cross sections of the multi-dimensional parameter space for which the metric parameters were chosen such that only one of the first two post-Newtonian deviation coefficients has a non-zero value. As an example, for the JP metric we set $\alpha_{13}=2\alpha_{12}$ for the $\kappa_1$ plot in Figure~\ref{fig:analytic_param_constraintsPPN} and all deviation parameters other than $\alpha_{12}$ and $\alpha_{13}$ to zero. This forces the $\kappa_2$ term to be zero for this metric but does not set higher order terms to zero. Nevertheless, the influence on the shadow size of the higher order terms for this metric decrease quite rapidly (see also Equations 29-33 of \citealt{Psaltis2021}). For the $\kappa_2$ plot for the JP metric we set $\alpha_{12}=0$ to force $\kappa_{1}=0$ and allow only $\alpha_{13}$ to be non-zero. For the MGBK metric we set $\gamma_{4,2}=-\gamma_{1,2}/4$, and set all parameters other than $\gamma_{1,2}$ and $\gamma_{4,2}$ to zero for the $\kappa_{1}$ plot. For the $\kappa_2$ plot we set $\gamma_{4,2}=-\gamma_{1,2}/2$, and all other parameters to zero. For the RZ metric we set $r_0=2$, which in turn sets $\epsilon=0$ as discussed in Section~\ref{sec:parametric} and set $a_0=a_1$ for the $\kappa_1$ plot, and only $a_1$ to be non-zero for the $\kappa_2$ plot. For the metrics discussed in Section~\ref{sec:known} we include only metrics for which it is possible to allow only one of the first two PPN parameters to be non-zero. The full $tt-$component of the metrics are used for these calculations, not their expansions.

These figures demonstrate that the inferred size of the \sgra\ black hole shadow places bounds of order $\sim 1$ and $\sim 5$ on the first and second post-Newtonian coefficients of the underlying metric, with the specific values showing a weak dependence on the particular metric used to obtain these constraints. Constraints on higher order PN components would be factors of a few less stringent at each increasing order (see \citealt{Psaltis2021} for details).

\section{Comparisons To other tests of gravity}\label{sec:compare}

\begin{figure}[t]
\centering
\includegraphics[width=0.95\columnwidth]{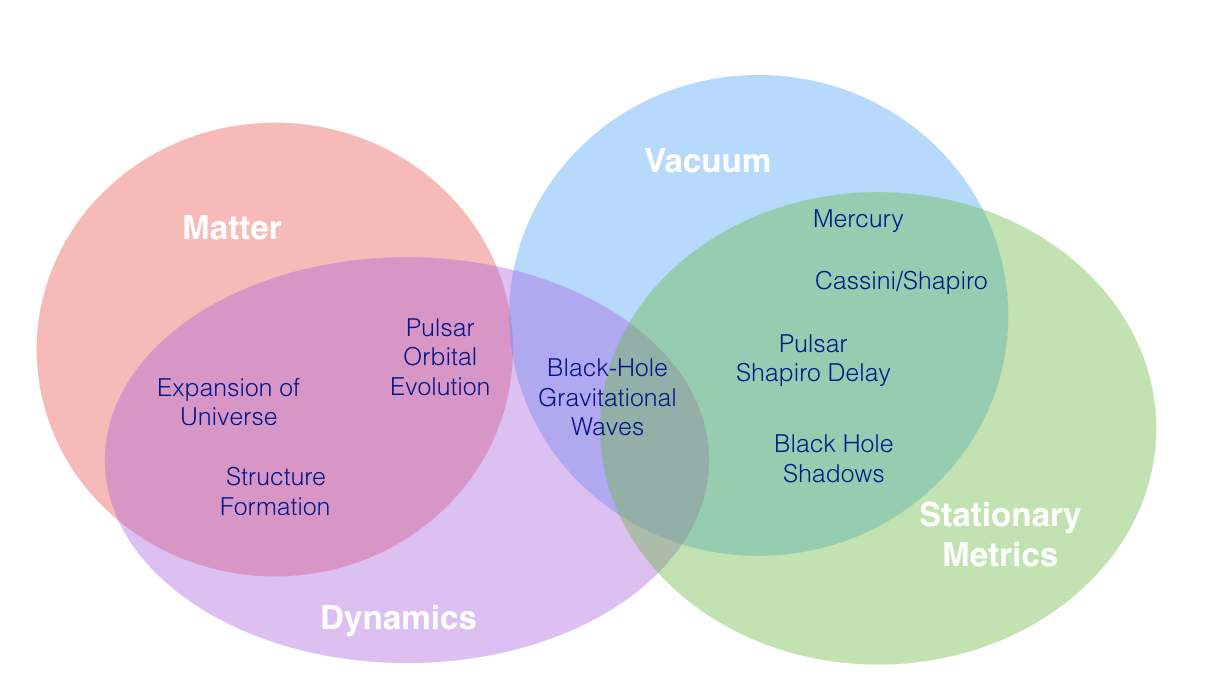}
\caption{An illustration of the different aspects of the theory of gravity probed by several examples of current tests in solar-system, compact-object, and cosmological settings. A group of tests explore {\em primarily\/} the dynamics and propagating modes of the theory, while others probe the stationary spacetimes of isolated objects. A different group of tests probe vacuum spacetimes, while others are sensitive to the coupling of matter to the gravitational fields. Any particular modification to General Relativity may alter one, several, or all of these aspects of the theory. Tests with horizon-scale images probe the characteristics of the stationary vacuum spacetimes of compact objects.}
\label{fig:tests}
\end{figure} 

\subsection{The Gravitational Field Probed by the Image of \sgra}

There exist a number of key qualitative differences between the aspects of the theory of gravity probed by various tests of General Relativity. For example, as Figure~\ref{fig:tests} illustrates, some of the tests are sensitive primarily to the dynamics and propagating modes of the gravitational fields, as is the case with pulsar timing, gravitational waves, and cosmology. \footnote{Note that these tests depend also on the properties of stationary spacetimes and can, therefore, provide information about them as well, as we will see below.} Other tests involve primarily measurements of photons in stationary spacetimes, as is the case of black-hole images and stellar orbits. Some tests involve orbits of massive particles, while others involve the propagation of photons in relativistic spacetimes. Moreover, tests performed with neutron stars and in cosmological settings also probe the coupling of matter to the gravitational field, whereas black-hole and most Solar-system tests are only sensitive to the properties of vacuum spacetimes.

\begin{figure}[t]
\centering
\includegraphics[width=0.95\columnwidth]{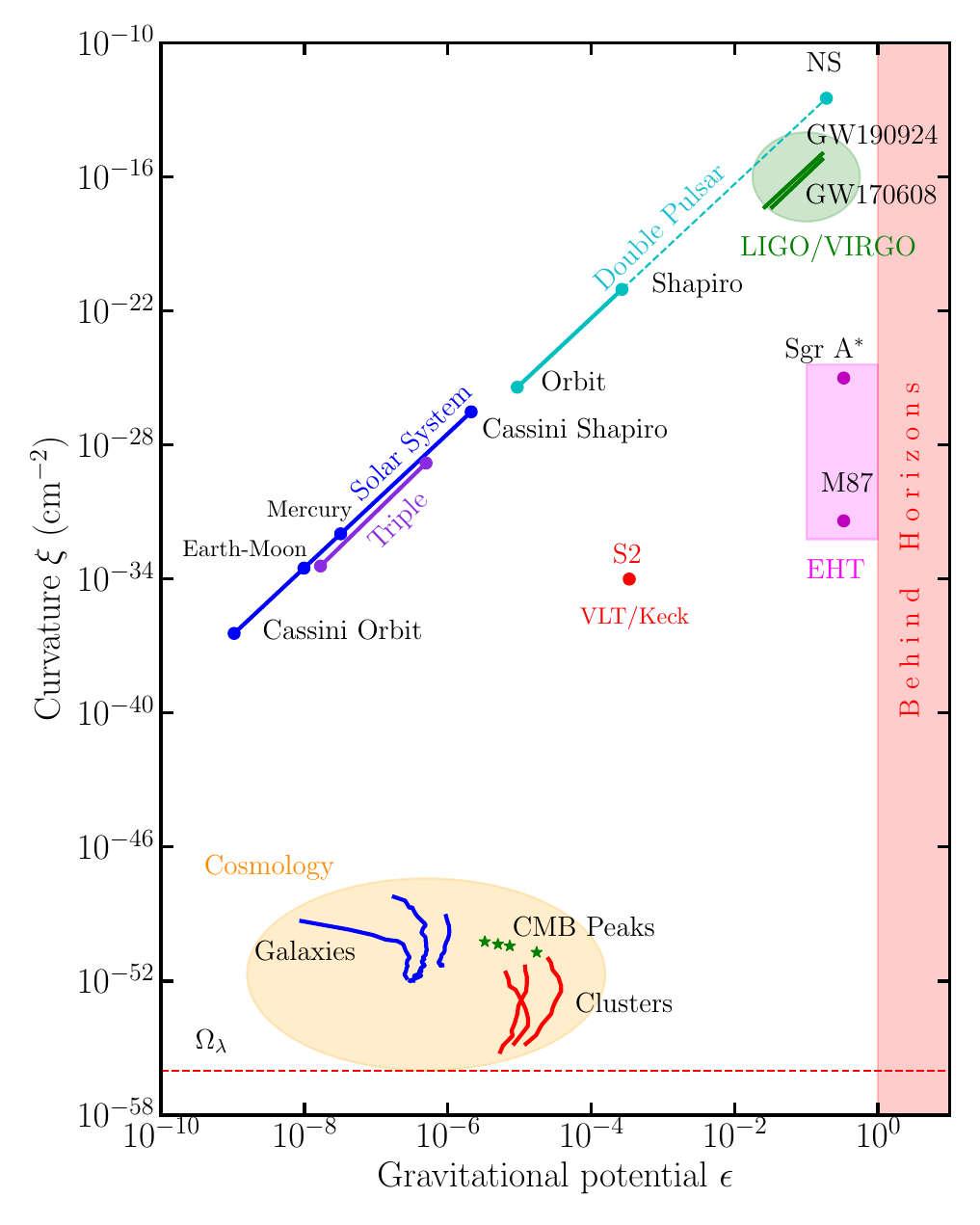}
\caption{A parameter space of tests of gravity with astrophysical and cosmological systems (after~\citealt{Baker2015}). For the Solar System and pulsar tests, the straight lines connect the range of gravitational fields that could affect, in principle, the outcome of each test, from the location of the outermost probe to the location of the central massive object; the dashed region of the cyan line indicates that the connection to the largest curvatures is theory-specific. The green lines connect the range of gravitational fields probed by two gravitational-wave tests with black-hole inspirals. Filled areas show the typical range of gravitational fields probed by (orange) cosmological, (green) gravitational-wave, and (magenta) black-hole imaging tests. Even though different tests explore, in principle, different aspects of the gravitational theory, as Figure~\ref{fig:tests} illustrates, they also probe vastly different scales. In particular, the horizon-scale images of \sgra\ that we report here probe a previously unexplored region of this parameter space of gravitational physics tests.}
\label{fig:paramspace}
\end{figure} 

Even within these qualitative distinctions, different tests probe vastly different regimes of gravitational potential and curvature, because of the large range of masses and length scales involved. This is illustrated in Figure~\ref{fig:paramspace}, following~\citet{Baker2015}. The horizontal axis in this figure shows the gravitational potential probed by each test; in the case of a test at distance $r$ from a Newtonian object of mass $M$, this dimensionless potential is equal to $\epsilon = GM/r c^2$. The vertical axis shows the spacetime curvature probed by each test, defined as the square root of the Kretschmann scalar; for a test in the Schwarzschild spacetime of an object, this is equal to $\xi=\sqrt{48}GM/r^3 c^2$ (see~\citealt{Baker2015}).

In order to highlight explicitly the fact that any test may probe a range of potentials and curvatures, we use straight lines to connect the smallest and largest potential and curvature that, in principle, may affect the outcome of each test. For example, in the case of the solar-System test with the Cassini spacecraft~\citep{Bertotti2003}, the Shapiro delay of radio signals was measured between the Earth and the spacecraft, when the latter was between Jupiter and Saturn and as these signals grazed the surface of the Sun; this test, therefore, probes the entire range of gravitational potentials and curvatures from the solar surface to the location of the Cassini spacecraft. We also use dashed lines to connect regions of the parameter space that may affect the outcome of a test, but in a theory-specific manner. For example, in double pulsar tests, the evolution of the orbital period caused by the emission of gravitational waves probes directly the potential and curvature at the orbital separation. However, in numerous modifications of the theory of gravity, enhanced rate of emission of gravitational waves becomes possible because of the coupling of neutron-star matter to the gravitational field at the highest potential and curvature (see, e.g.,~\citealt{Damour1993}). In other words, depending on the particular modification of gravity that is being tested, the test involving the evolution of the binary period may probe the entire range of field strengths covered by the solid and dashed cyan lines in Figure~\ref{fig:paramspace}.

As we will discuss below, these qualitative and quantitative differences between tests of gravity complicate our ability to cross compare and combine their results. However, these same differences also allow us to leverage the broad range of conditions that various tests probe in order to draw conclusions about the theory of gravity that could not have been reached by any test individually. For example, one of the key predictions of General Relativity is that the spacetime properties of black hole scale with their mass. This is a prediction that we can test by comparing the results of gravitational-wave tests that probe stellar-mass black holes to those of the imaging tests that probe supermassive black holes. At the same time, General Relativity predicts that, according to Birkhoff's theorem, the external spacetime of a slowly spinning object is independent of its internal structure and composition. We will test this prediction by comparing the results of black-hole tests to those that involve pulsars or the Sun.

\subsection{Comparing Gravitational Tests Across Scales}

Because of their qualitative differences, every type of test of General Relativity is performed with a unique theoretical framework that is optimal for the system under study. Solar-System tests use Parametrized Post-Newtonian (PPN) expansions~\citep{Will2014}, pulsar tests use post-Keplerian parametrizations and also a strong-field equivalent of the PPN-formulation~(see, e.g.,~\citealt{Wex2020}), shadow tests use parametric post-Kerr metrics~\citep{Johannsen2011,Johannsen2013b,Vigeland2011,Rezzolla2014,Konoplya2016},  gravitational-wave tests with inspirals use post-Newtonian~\citep{Khan2016}, effective-one-body~\citep{Buonanno1999,Buonanno2000}, parametrized post-Einstein (ppE) frameworks~\citep{Yunes2009}, etc. Unfortunately, this plurality of methods restricts our ability to combine and leverage the results of different tests since, in many cases, the parameters of each framework are not directly related to each other.

In principle, there are two ways we can combine tests across different scales and systems. In one approach, we can use a particular class of theories (e.g., scalar-tensor gravity) and compare the constraint of the parameters of that class from different tests. This is the most direct approach that requires typically no additional assumptions to be made. However, it is limited to the particular alternative to General Relativity that are described by the theory under study. In a second approach, as a practical solution, we can make simplifying assumptions (e.g., that the dynamics of the theory are the same as in General Relativity but the stationary spacetimes are not) and constrain phenomenological parameters of the metrics of the objects involved.

In order to make the latter approach independent of coordinate systems or gauges, and focusing here on tests of stationary metrics, we often convert the bounds on the parameters of a particular framework to constraints on the effective post-Newtonian parameters of the metrics of the objects involved. Within the particular assumptions inherent to each test, this approach is formally correct, even if one uses tests in the strong-field regime, as long as the framework used to obtain these constraints is itself applicable in that regime. Moreover, in doing so, there is no implicit assumption that the derived post-Newtonian parameters are universal constants. Indeed, in most modifications to General Relativity, the values of these parameters are specific to the situation under consideration, as they may depend on the strength of the gravitational field (curvature or potential) probed, the nature of the compact object (binary or not, with matter or pure vacuum, etc.), and the boundary conditions (the coupling of matter to the field at the center of the system, the asymptotic cosmological boundary conditions, the cosmic time of the test, etc.). This is the reason why it is important to measure potential deviations of such parameters in different astrophysical and cosmological settings that span a wide range of masses and physical conditions, as shown in Figure~\ref{fig:paramspace}.

\subsection{Tests with the S2 Orbit}
\label{sec:S2}

\sgra\ is unique in enabling us to probe the metric of the same black hole both at horizon scales, with the EHT images reported here, as well as at larger distances, with the orbits of S-stars. In performing the imaging test in Section \ref{sec:delta}, we have already used the measurement of the mass-to-distance ratio for the black hole that was obtained through monitoring the S-star orbits. In spacetime terms, this Keplerian mass is simply the coefficient of the asymptotic, Newtonian expansion of the metric. However, recent measurements of relativistic effects in the stellar orbits resulted in constraints on the metric properties beyond the Newtonian regime, which we explore here.

The motions of several S-stars have been monitored for almost three decades with adaptive optics instruments on VLT and Keck and their orbits have been well determined (see, e.g.,~\citealt{2008ApJ...689.1044G,2009ApJ...692.1075G}).  The detection of gravitational redshift~\citep{2018A&A...615L..15G,2019Sci...365..664D} and of the precession of the periapsis~\citep{2020A&A...636L...5G} in the S0-2 orbit has led to tests of the equivalence principle and of the Schwarzschild metric (Section~\ref{sec:priors}; see also~\citealt{2017PhRvL.118u1101H,Amorim2019}). Because in the gravitational test we report here with the \sgra\ images we explicitly assume the validity of the equivalence principle and only test the metric, we will focus on the connection of the imaging to the post-Newtonian tests of the metric using the precession of the S0-2 orbit.

In~\cite{2020A&A...636L...5G}, the measured rate of precession of the S0-2 orbit was quantified through a phenomenological parameter $f_{\rm SP}$, such that the precession per orbit at the first post-Newtonian order can be written as
\begin{equation}
\Delta \phi_\mathrm{1} = f_{\rm SP}  \frac{ 6\pi GM}{a(1-e^2)c^2}\;,
\end{equation}
where $a$ and $e$ are the orbital separation and eccentricity of the orbit, respectively. The best-fit value for $f_{\rm SP}$ was found to be consistent with the predictions of the Schwarzschild metric, i.e., $f_{\rm SP}=1.1\pm 0.19$.

In the PPN formalism, the phenomenological parameter $f_{\rm SP}$ is related to two of the first-order post-Newtonian parameters of the metric via~\citep{Will2014}
\begin{equation}
    f_{\rm SP}=\frac{1}{3}\left(2+2\gamma_{\rm S0-2}-\beta_{\rm S0-2}\right)\;,
    \label{eq:S2f}
\end{equation}
such  that
\begin{equation}
    2\gamma_{\rm S0-2}-\beta_{\rm S0-2}=1.3\pm 0.57\;.
    \label{eq:S2PPN}
\end{equation}
Here the subscripts explicitly denote the fact that these parameters are not universal constants but are specific to the metric of \sgra\ as measured at the location of the S0-2 orbit. In deriving this equation, we have also assumed that the mass of the S0-2 star is negligible with respect to the black-hole mass (see eq.~[\ref{eq:psrK}] below). 

We now assess the freedom these observations allow for possible deviations at higher post-Newtonian orders and hence the leverage of the strong-field imaging tests that we report here in constraining the metric of  \sgra.  We first write the precession per orbit at the second post-Newtonian orbit as~\citep{Will2018}
\begin{equation}
    \Delta\phi_{2}=
   - 6\pi\left[\frac{GM}{2a(1-e^2)c^2}\right]^2(10-e^2)\;.
\end{equation}
The ratio of the second to the first order post-Newtonian term for the S0-2 star is
\begin{equation}
    \frac{\Delta\phi_{2}}{\Delta\phi_{1}}=
    \frac{GM}{4a(1-e^2)c^2}(10-e^2)\simeq 7\times 10^{-4}\;.
\end{equation}
Because the first post-Newtonian term has been measured to an accuracy of $\sim 20$\% (see eq.~[\ref{eq:S2PPN}]), the second post-Newtonian term would have to be $\sim 0.2/7\times 10^{-4}\simeq 285$ times larger than the Schwarzschild prediction in order for it to cause deviations detectable with current instruments. We will consider this as a heuristic upper bound on possible deviations at the second post-Newtonian order imposed by the precession of the S0-2 orbit. 

Had the metric of \sgra\ deviated from Schwarzschild by, e.g., a factor of 250 at the second post-Newtonian order, this would have still been undetectable by the S0-2-precession test but would have led to a shadow size as large as $\sim 85 GM/c^2\simeq 425~\mu$as (see eq.~[33] of~\citealt{Psaltis2021}). This predicted size would have been at least a factor of 8 larger than the measured size we report in Section \ref{sec:delta} and, more importantly, would have been almost two orders of magnitude larger than any potential uncertainty introduced by systematics due to plasma physics (as captured by the $\alpha_1-1$ factor) or due to our measurement methods (as captured by the $\alpha_2-1$ factor). In other words, the horizon-scale images of \sgra\ provide substantial constraints to potential deviations of the black-hole spacetime from the GR predictions that could have evaded all prior bounds, beyond any astrophysical uncertainties.

\begin{figure}
\centering
\includegraphics[width=0.95\columnwidth]{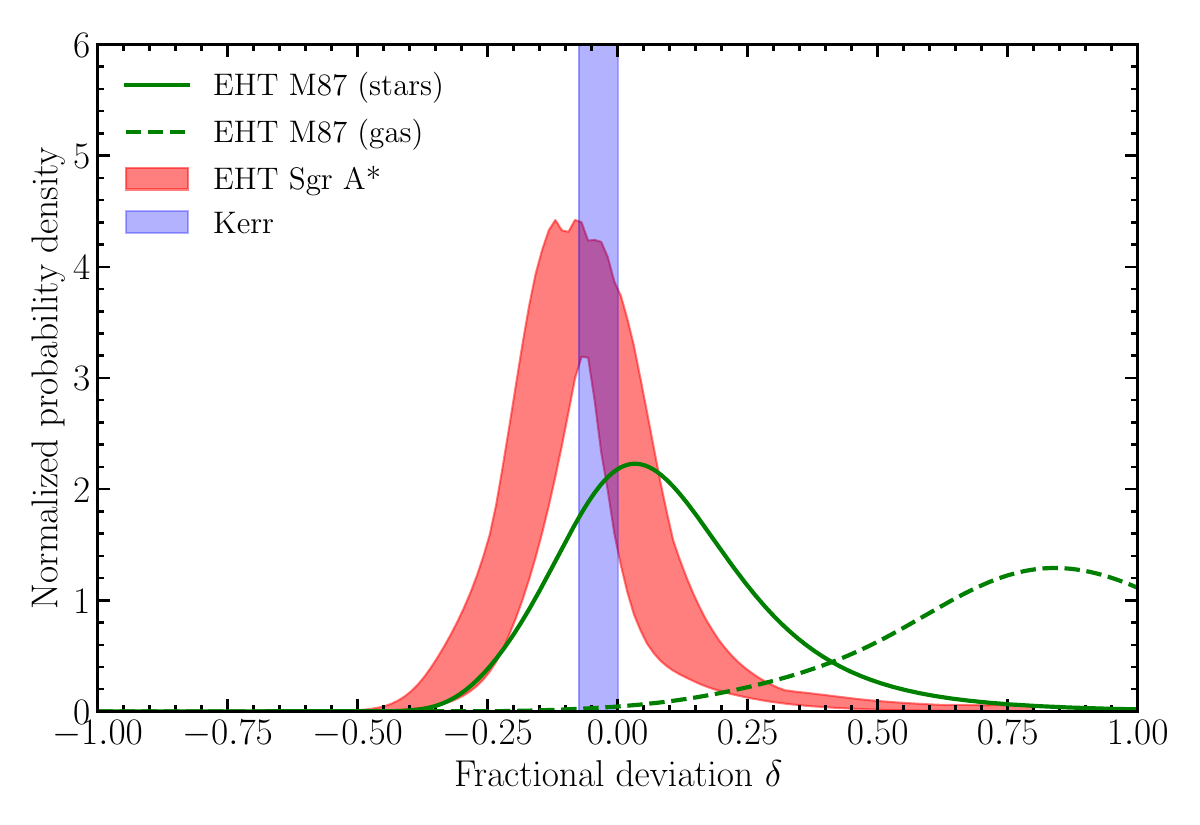}
\caption{Comparison of the posterior distributions for the fractional deviation $\delta$ from the Schwarzschild predictions, as inferred by the EHT measurement of the size of the black-hole shadows in \sgra\ and M87. The purple-shaded area shows the $\sim 8$\% range predicted for the Kerr metric, depending on the black-hole spin and observer inclination. The red-shaded area shows the small range of posteriors for \sgra, inferred with different imaging and calibration algorithms (see Fig.~\ref{fig:delta_post}). The solid and dashed lines show the posteriors for the M87 black hole, when the stellar- and gas-dynamics measurements of the mass-to-distance ratio have been used, respectively. The negligible uncertainties in the mass measurement of \sgra, which is the result of the detection of relativistic effects in the orbit of the S0-2 star, removes any ambiguity in the comparison with the Kerr predictions.}
\label{fig:Sgra_M87}
\end{figure} 

\subsection{M87 Imaging Tests}

The black hole at the center of the M87 galaxy has a mass that is approximately 1500 times larger than the one in \sgra. As a result, observations of horizon-scale images from the M87 black hole probe similar potentials but curvatures that are 6 orders of magnitude smaller than those of \sgra. In~\cite{VI_EHT2019_M87}, we used the 2017 EHT images of the M87 black hole to derive constraints on possible deviations of the inferred size of the black hole shadow from the Schwarzschild prediction and, in~\cite{Psaltis2020} and \cite{Kocherlakota2021}, we used these measurements to place constraints on possible deviations of metric parameters from Kerr. 

Contrary to the case of \sgra\ that we report here, there were two independent and distinct priors on the mass-to-distance ratio for the black hole in M87, based on either stellar-~\citep{Gebhardt2011} or gas-dynamic measurements~\citep{Walsh2013}. Adopting the former resulted in an upper bound on deviations from the Kerr predictions that was consistent with zero, within $\sim 17$\%. We opted to assign negligible prior likelihood to the latter prior, as it would have led us to conclude that there is significant tension between the Kerr predictions and the observations, and instead used the measurements as a null hypothesis test, i.e., concluded that the EHT images were not inconsistent with the Kerr predictions.

Figure~\ref{fig:Sgra_M87} compares the posteriors on the deviation parameter $\delta$ obtained here for \sgra\ to those reported earlier for the M87 black hole. In the case of the image in \sgra, we have a precise measurement of the mass-to-distance ratio for the black hole based on the detection of relativistic effects in the orbit of the S0-2 star, as discussed in Section \ref{sec:priors}. This removes any ambiguity in our calculation of the Kerr predictions. Moreover, the uncertainties in the mass-to-distance priors for \sgra\ are negligible compared to those in M87, even if we only adopt the stellar-dynamic measurement for the latter. This results in uncertainties on the bounds of the deviation parameter $\delta$ that are almost a factor of 2 smaller in \sgra\ compared to the M87 black hole. 

In both the \sgra\ and M87 cases, the inferred sizes of the black-hole shadows are consistent with the Kerr predictions, even though the black holes span 3 orders of magnitude in mass and 6 orders in curvature scale. This serves as a confirmation of the General Relativistic prediction that the spacetime properties of black holes scale with their mass and can be further reinforced by leveraging tests that involve stellar-mass black holes, as we discuss bellow.

\subsection{Gravitational Wave Tests}

Observations of gravitational waves from coalescing black-hole binaries with LIGO/Virgo provide strong-constraints on potential near-horizon modifications of the predictions of the theory of gravity for black holes~\citep{Abbott2016,Abbott2019,Abbott2021}. Because of the frequency range of these ground-based gravitational-wave detectors, the black-hole masses they are sensitive to are in the $10-100~M_\odot$ range. As a result, compared to the tests with the EHT black-hole images, existing gravitational wave observations probe similar potentials but curvatures that are different by 8-16 orders of magnitude.

A second important difference arises from the fact that, fundamentally, gravitational-wave observations measure the propagating gravitational modes of the theory, whereas black-hole images measure electromagnetic modes propagating on the black-hole spacetimes. It is possible that the number and polarization of the propagating modes of the fundamental theory of gravity are the same as those in General Relativity but the stationary metrics are not; in fact, it is possible that the fundamental theory of gravity is General Relativity but the stationary metrics of the supermassive compact objects in the centers of galaxies are not described by the Kerr metric (see, e.g., \citealt{Gair2008}). Alternatively, it is possible that the propagating modes of the theory are very different than those in General Relativity but the stationary spacetimes remain Kerr~\citep{Psaltis2008,Barausse2008}. In this way, the gravitational-wave and imaging observations of black holes provide complementary probes of potential modifications to General Relativity.

\begin{figure}
\centering
\includegraphics[width=0.95\columnwidth]{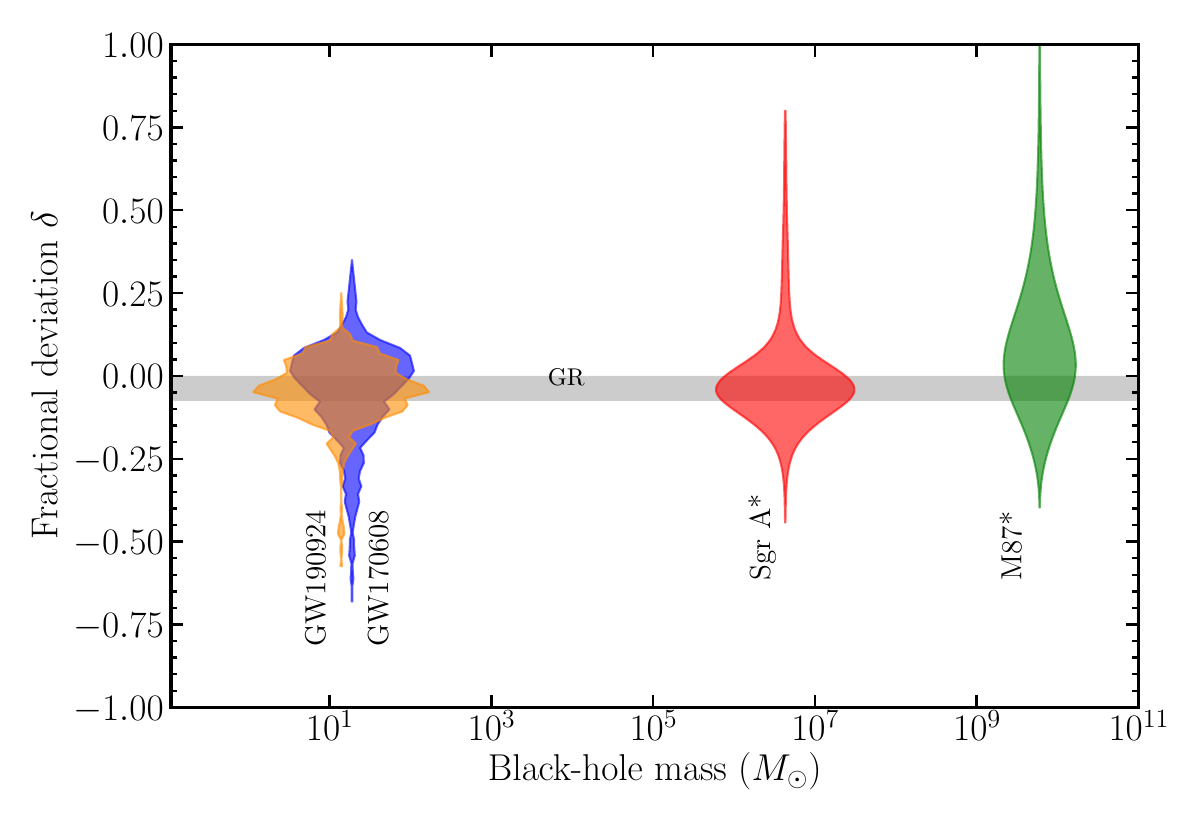}
\caption{Comparison of the posterior distributions for the fractional deviation $\delta$ from the Schwarzschild predictions, as inferred by the EHT measurement of the size of the black-hole shadow in \sgra\ (red curve for the fiducial priors) and M87 (green curve for the stellar-dynamics mass) as well as by the LIGO/Virgo measurements of the inspiral phases of GW170608 (blue) and GW190924 (orange). The posteriors corresponding to the latter two reflect the prediction on the fractional deviation $\delta$ in the shadow size one would have calculated based on  the constraints imposed by the gravitational-wave measurements, if the  coalescing, stellar-mass sources had the same form of metrics as those of the supermassive black holes observed with the EHT. The gray-shaded area shows the $\sim 8$\% range predicted for the Kerr metric, depending on the black-hole spin and observer inclination. Even though tests with gravitational waves and black-hole images span black-hole masses that are different by 8 orders of magnitude, they are all consistent with the GR predictions that all black holes are described by the same metric, independent of their mass.}
\label{fig:ligo}
\end{figure} 

Because of this fundamental difference, however, in order to compare directly gravitational-wave constraints to those of black-hole imaging, one needs to make specific assumptions. Our main goal in this section is to leverage the gravitational-wave tests in order to assess whether the black-hole metric properties scale with mass, as predicted by General Relativity. For this reason, we will focus on the inspiral phases of the observed gravitational waves, as these are ones that are mostly sensitive to modifications in the metrics of the coalescing black holes\footnote{There is a multitude of other tests of gravity that are possible with gravitational-wave observations, such as those that place bounds on the mass of the graviton~\citep{Abbott2016,Baker2017}. Albeit extremely important, these tests are not directly comparable to those we report here as imaging tests are sensitive only to the stationary black-hole metrics and not to other aspects of the theory} (see, also, \citealt{Volkel2020}). Moreover, following \citet{Psaltis2021}, we will assume here that the propagating modes of the theory are indistinguishable from those in General Relativity and assign any room for potential deviations to changes in the underlying metrics of the black holes. Unless the fundamental theory of gravity is finely tuned such that the modifications in the radiative sector exactly cancel those in the metrics, for the masses of the LIGO/Virgo black holes, our constraints will represent broad-brush upper limits on potential metric deviations.

Under the assumptions outlined above, the LIGO/Virgo measurements of the inspiral phases of coalescing black-hole binaries depend entirely on the $tt-$components of the metrics, as expressed in areal coordinates~\citep{Carson2020,Cardenas2020}. This is a consequence of the fact that the waveforms of the gravitational waves during the decay of quasi-circular orbits are determined by the binding energies of the orbits and their angular frequencies (see, e.g., eq.~[9]  of \citealt{Carson2020}), both of which are determined by the $tt-$components of the metric~\citep{Ryan1995}. This is the same component of the  metric that determines the size of the black-hole shadows measured with the EHT~\citep{Psaltis2020}. Remarkably, because of a coincidence related to the masses of the coalescing black holes, the degeneracies between the constraints from inspiral measurements on the various parameters of metrics that deviate from Kerr are nearly parallel to those of the constraints imposed from the EHT imaging observations~\citep{Psaltis2021}. In other words, this coincidence allows us to use the LIGO/Virgo constraints and make a prediction on the fractional deviation $\delta$ in the shadow-size one would have calculated, if the gravitational-wave sources had the same metrics as those of the supermassive black hole observed with the EHT.

Figure~\ref{fig:ligo} compares the results on the deviation parameter $\delta$ for the two most constraining gravitational-wave events, GW170608 and GW190924 \citep{Psaltis2021}, to those obtained in \S\ref{sec:delta} for \sgra, as well as those for the M87 black hole derived earlier~\citep{VI_EHT2019_M87}. As discussed above, all observations are consistent with the predictions of General Relativity, even though they utilize black holes with masses that are different by 8 orders of magnitude. This lends support not only to the Kerr nature of the black-hole spacetimes but also to the fact that the fundamental theory of gravity does not have a scale between those probed by stellar-mass and supermassive black holes. 

\subsection{Pulsar Timing Tests}
\label{sec:psr}

The potentials and curvatures probed by pulsar timing tests may depend on the underlying theory of gravity, as discussed above. In principle, in theories without a characteristic scale (such as screening) between the orbital separation of the binary and the size of the neutron star, pulsar tests probe the coupling of matter to the gravitational field, which takes place in the strong field regime of the neutron-star interior, i.e., at a potential of order unity and curvature of order $10^{-10}$ cm$^{-2}$. For such theories, the horizon-scale images of \sgra\ probe a similar potential but a curvature that is different by 15 orders of magnitude (see Fig.~\ref{fig:paramspace}).

In contrast, theories with a characteristic scale between the orbital separation of the binary and the size of the neutron star (or other similar effects as in~\citealt{Yagi2016}) are only probed at the potential of the periapsis distance for tests involving the orbital period derivative and orbital precession (for the double pulsar this is about $6\times 10^4 GM/c^2$) or at the distance of minimum approach for the Shapiro delay (for the double pulsar, this distance is about $1500 GM/c^2$). In this case, the horizon-scale images of \sgra probe similar curvatures but a potential that is larger by 5 orders of magnitude.

Following these two approaches, we are going to discuss the constraints imposed on the various deviation parameters in two complementary ways. First is in a theory-agnostic way, in terms of the effective post-Newtonian parameters of the metrics of the compact objects. Second is in terms of the scalar-tensor gravity theory of Damour \& Esposito-Farese~\citep{Damour1993}, with second-order couplings between the scalar field and the Einstein tensor defined by the parameters $\alpha_a\equiv \partial \ln m_a/\partial \phi_0$ and $\beta_a\equiv \partial \alpha_a/\partial \phi_0$, where the subscript $a$ corresponds to either the pulsar ``p'' or the companion ``c''. In such a theory, the effective 1PN parameters can be expressed in terms of the theory parameters $\alpha_a$ and $\beta_\alpha$ as
\begin{equation}
    \hat{\gamma}_{ab}=1-\frac{2\alpha_a\alpha_b}{1+\alpha_a\alpha_b}
\end{equation}
and
\begin{equation}
    \hat{\beta}^a_{bc}=1+\frac{\beta_a\alpha_b\alpha_c}{2(1+\alpha_a\alpha_b)(1+\alpha_a\alpha_c)}\;,
\end{equation}
where the ``hats'' and subscripts emphasize the fact that the strong-field equivalents to the
PPN parameters are not universal constants.

In principle, there are at least five  theory-independent post-Keplerian parameters in a binary system that can be measured from pulsar timing \citep{DamourTaylor1992}. Together with the normal Keplerian parameters, they comprise a set of inferred quantities that is larger than the free parameters in the system~(see \citealt{Wex2020} for a recent review).
The combination of any two post-Keplerian parameters is used to determine the masses of the two objects in the binary. Any additional measurement can then be used for testing
GR and a very broad class of alternative (boost-invariant) theories 
\citep{DamourTaylor1992,Will2014,Will2018,DeLaurentis2018}.
We will focus below on the constraints imposed by the measurement of Shapiro delay, of the precession of periapsis of the binary, and of the orbital period evolution caused by the emission of gravitational waves.

\noindent {\em Shapiro delay.---\/} Some of the main constraints on deviations from General Relativity come from the measurement of a Shapiro delay in binary pulsar systems. Like the imaging of black holes, such experiments provide rare opportunity to study the light-propagation near strongly self-gravitating
objects. Two parameters can be measured, the ``shape'' $s$ and the ``range'' $r$.
The Shapiro shape, $s$, can quite generally be identified with the sine of the orbital 
inclination (i.e., $s = \sin i$, e.g. \citealt{Kramer2021})
\footnote{There is an indirect dependence on the underlying metric via its relation to the masses of the orbiting objects, as measured within the particular gravity theory.} In comparison,
the Shapiro range $r$ relates to the companion mass, e.g., in General Relativity one finds $r= G m_c/c^3$.

For a theory agnostic constraint, one can relate $r$ to the 1PN parameter $\gamma$ via
\begin{equation}
    r=\frac{\hat{G}_{0,c} m_c(1+\hat{\gamma}_{0c})}{2c^3}
\end{equation}
where $m_c$ is the mass of the companion star and the subscript on the PN parameter makes explicit that this is the value for the coupling between a test particle and the pulsar companion,
i.e.~they are not universal constants but object-specific.
The subscript in the gravitational constant $G$ also denotes that this is the effective gravitational constant felt by a test particle in the field of the companion.

\begin{figure*}[h]
\centering
\includegraphics[width=0.95\textwidth]{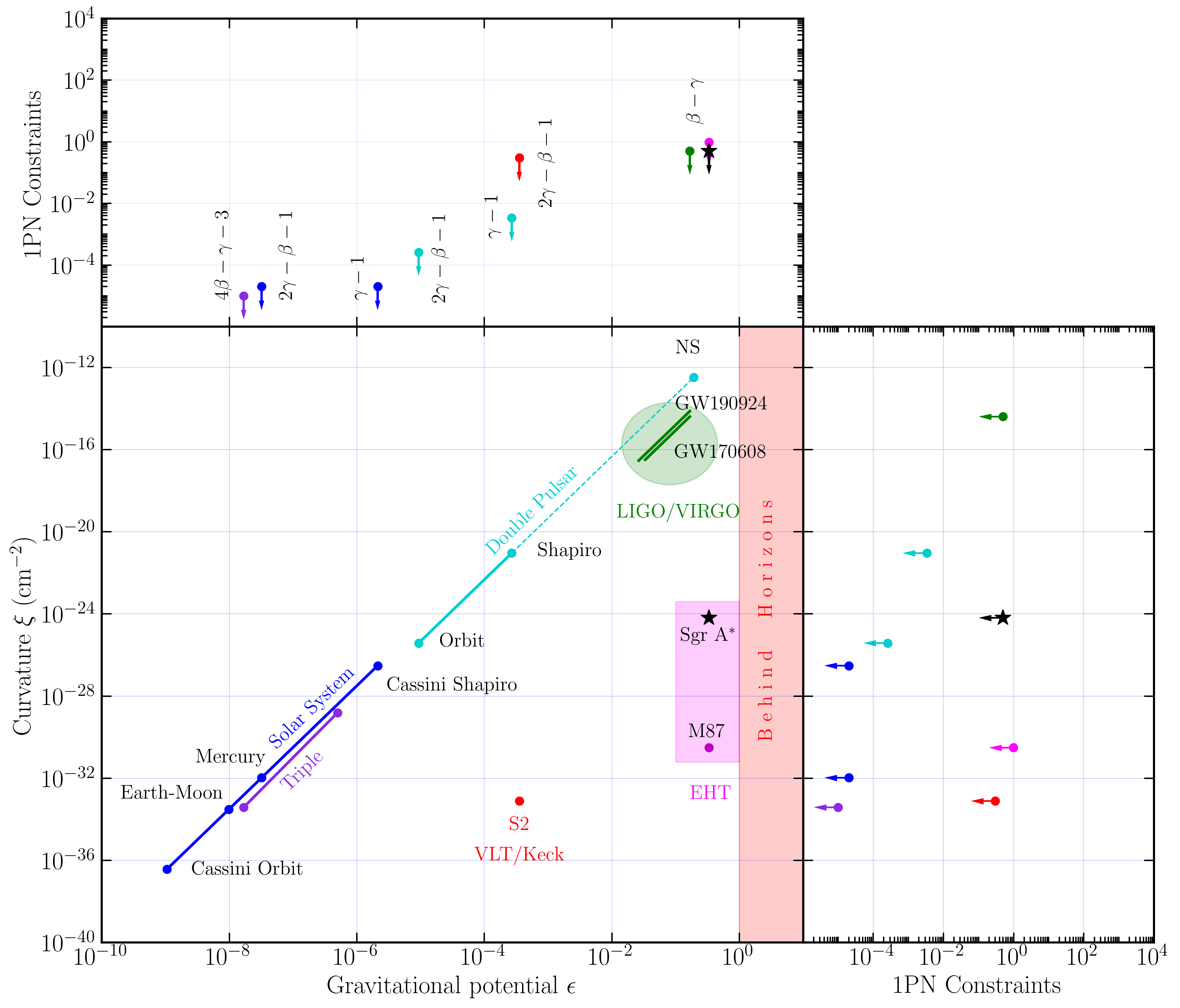}
\caption{Comparison of the current limits on various potential metric deviation parameters at the first post-Newtonian order obtained from different tests of gravity with astrophysical objects, as a function of the {\em (top panel)\/} gravitational potential and {\em (right panel)\/} curvature probed by each test. For visual clarity, each upper limit is calculated assuming that the deviations at all other post-Newtonian orders are negligible; this plot, therefore, represents only one cross section in the multi-dimensional parameter space of plaussible deviations. Every tests probes a different combination of post-Newtonian parameters, as shown on the upper panel. In most modifications to General Relativity, these post-Newtonian parameters are not universal constants but depend on the nature of the central object, its mass, composition, scale, etc. }

\label{fig:comparison_1pn}
\end{figure*} 

\noindent{\em Precession of periapsis.---\/} In GR, within the PPN framework, the rate of precession of the periapsis $\dot{\omega}$ depends on the combination $2+2\hat{\gamma}-\hat{\beta}$ (in the limit when the mass ratio is zero). Incorporating the effects due to the presence of two orbiting objects for which the strong-field coupling of matter to gravity cannot be neglected, $\dot\omega$ becomes proportional to
\begin{equation}
    k=\left(2+2\hat{\gamma}_{pc}-\frac{m_c}{M}\hat{\beta}_{cc}^p-\frac{m_p}{M}\hat{\beta}_{pp}^c\right)\frac{\left(\hat{G}_{pc}M n_b\right)^{2/3}}{c^2(1-e^2)}\;.
    \label{eq:psrK}
\end{equation}
Here $M\equiv m_c+m_p$ is the sum of the companion and pulsar masses, $m_c$ and $m_p$, respectively, and $e$ is the eccentricity of the orbit.

\noindent{\em Orbital period derivative.---\/}
The orbital period of a binary system may change due to a number of effects. In case that
only gravity plays a role, a decay in the orbital period results from  the emission of 
gravitational waves. The measurement of an orbital period derivative can then be used
to confront a given theory with its predictions. In General Relativity, to leading
order, the orbital period derivative is given by the quadrupole formula \citep{Peters1964}.
In contrast, many alternative theories of gravity violate the strong equivalence principle, resulting in
the emission of gravitational {\em dipolar} radiation. Observations of binary pulsars
can provide strict limits on the existence of dipolar radiation \citep{ShaoWex2016,Wex2020}.
Indeed, the Double Pulsar system provides currently the most precise test of
the General Relativistic quadrupolar description of gravitational waves, 
validating the prediction at a level of $1.3\times 10^{-4}$ (95\% conf.~level, \citealt{Kramer2021}).

Under the  assumption that the radiative sector is negligibly different from General Relativity, 
one can put, in a theory agnostic way, constraints on the post-Newtonian parameters 
of the waveform, obtaining at 1PN order
\begin{equation}
    \hat{\beta}-\hat{\gamma} = \Delta\phi_2 \frac{743+924\eta}{1344}\;,
\end{equation}
where $\Delta\phi_2$ is the 1PN order correction to the phase term of the gravitational waveform.

Following this framework, and emphasizing the mentioned assumptions, 
we can use the other precise measurements in the Double Pulsar
system for $r$ and $\dot\omega$ to derive constraints on the corresponding post-Newtonian parameters at the first and second orders. At the 68\% confidence limit, particular combinations of these parameters that give rise to the the Shapiro range, the periapsis advance and the orbital decay have been found to be consistent with the General Relativistic predictions at a precision of $3.4\times 10^{-3}$, $2.6\times 10^{-4}$, and $6.3\times 10^{-5}$, respectively \citep{Kramer2021}.

\noindent{\em Universality of free fall .---\/} One can use pulsars also to test
the strong equivalence principle (SEP), as first shown by \cite{ds91}. The discovery of a
pulsar in a triple system \citep{triple} allow a variation of the Damour-Sch\"afer experiment in very constraining way \citep{fwk12}. In the triple system an inner pulsar - white dwarf system is orbited by a second
white dwarf in an outer orbit. By tracking the orbital motion of the neutron star via pulsar timing,
one can study how the inner two objects with significantly different gravitational self-energy are falling in the
gravitational field of the third object. \cite{archibald} presented a limit on the strong-field
equivalent of the Nordtvedt parameter of $\eta < 3 \times 10^{-5}$. Following Damour \& Sch\"afer,
it relates to the strong-field SEP violation parameter as
\begin{equation}
     \Delta = \eta \epsilon_{\rm grav} + \eta’ \epsilon_{\rm grav}^2
\end{equation}
where $\eta$ and $\eta'$ are strong-field equivalents of the Nordtvedt parameter
and $\epsilon_{\rm grav} = E_{\rm grav} / mc^2$ is the normalized Newtonian gravitational binding energy. At the first post-Newtonian order, the Nordtvedt parameter is related to the post-Newtonian coefficients by $|\eta| = |4\beta - \gamma - 3|$. Ignoring higher terms, and using the results by \cite{archibald}, leads to $|4\beta - \gamma - 3| < 1.0 \times 10^{-5}$ (95\% C.L.). At the second post-Newtonian order, we make use of the Messenger limit obtained in the Solar System \citep{messenger} to constrain $\eta' < 1.2\times 10^{-3}$ (95\% C.L.).

\subsection{Leveraging Gravitational Tests Across Different Scales}

\begin{figure*}[h]
\centering
\includegraphics[width=0.95\textwidth]{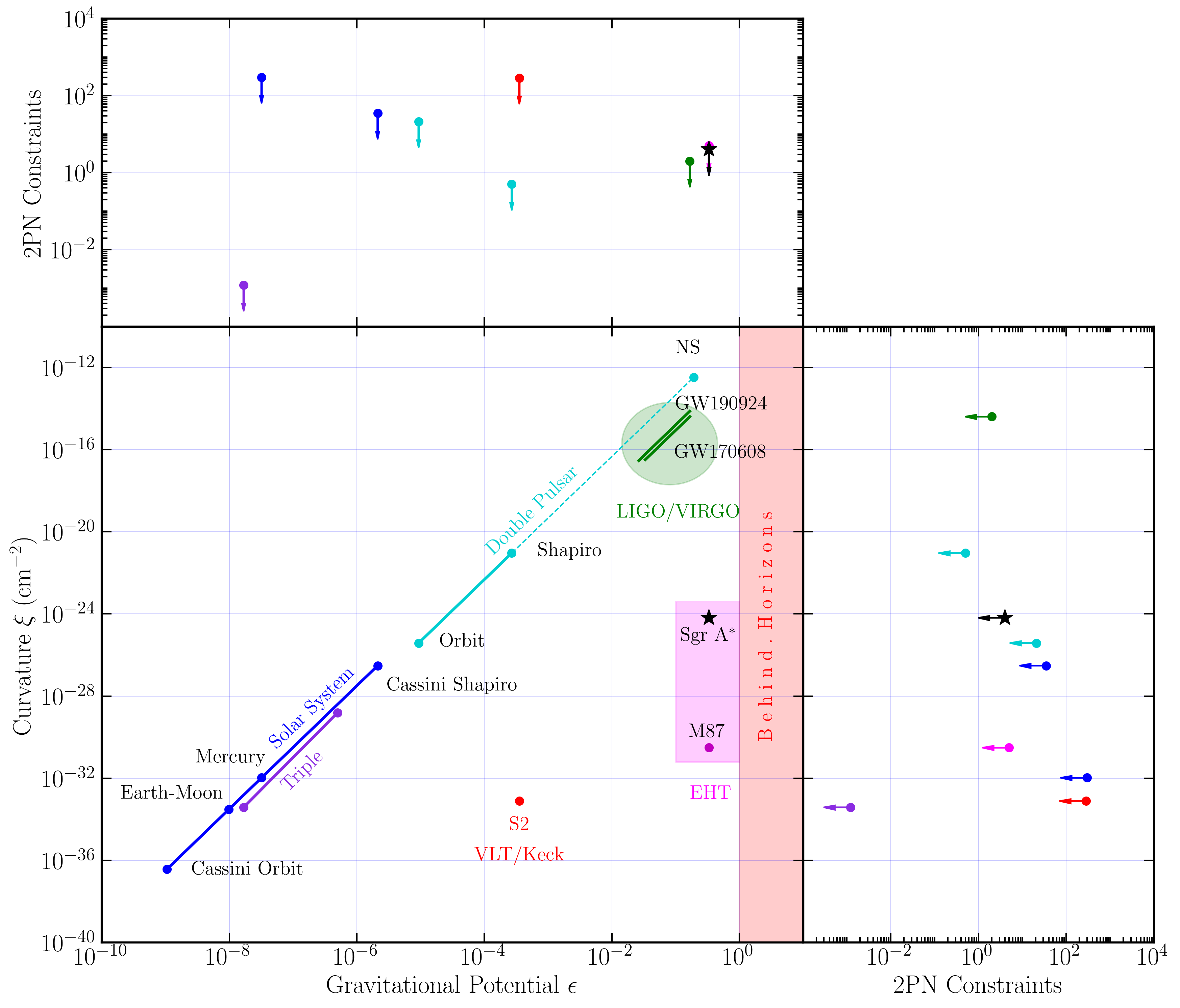}
\caption{Same as Fig.~\ref{fig:comparison_1pn} but for the constraints on the second post-Newtonian order, assuming that the deviations at all other post-Newtonian orders are negligible.}
\label{fig:comparison_2pn}
\end{figure*}

Figures~\ref{fig:comparison_1pn} and \ref{fig:comparison_2pn} compare the constraints on the various metric deviation parameters at the first and second post-Newtonian orders that are imposed by the gravity tests discussed above. For visual clarity reasons alone, the figures show only a particular cross section of the multi-dimensional parameter space of plausible deviations: in constructing this figure, we have assumed that all deviation parameters other than those plotted are negligible. Barring any fine tuning of the fundamental theory that would lead to fortuitous cancellations, the bounds plotted can be regarded as rough upper limits. It is important to emphasize here that since each test probes a different combination of the various post-Newtonian parameter, which are indicated on the figure for those at the first post-Newtonian order, it is mathematically impossible for any non-Kerr metric to evade simultaneously all constraints purely by fortuitous cancellations.

The bounds on the post-Newtonian parameters that are imposed by the imaging observations of \sgra and M87, the detection of periapsis precession in the S2 orbit, the gravitational-wave observations, and the double pulsar have been discussed in \S\ref{sec:PPN}, \cite{VI_EHT2019_M87}, \S\ref{sec:S2}, \cite{Psaltis2021}, and \S\ref{sec:psr}, respectively. The figures also incorporate the limits on the first- and second post-Newtonian parameters obtained by Solar System tests, which we briefly discuss below. 

The measurement of the solar Shapiro delay with the Cassini spacecraft constrained the $\gamma_\odot-1$ parameter at the first post-Newtonian order with an accuracy of $\sim 2\times 10^{-5}$~\citep{Bertotti2003}. At the second post-Newtonian order, the best constraint comes from the deflection of light measurements that are consistent with the Schwarzschild predictions to an accuracy of $\sim 10^{-4}$. The second-order post-Newtonian correction for light deflection at the solar surface is of order $\sim 3\times 10^{-6}$~\citep{Bodenner2003}. Therefore, any second-order post-Newtonian corrections to the solar metric cannot be larger than $\sim 10^{-4}/(3\times 10^{-6})\simeq 35$ times the Schwarzschild predictions.

The periastron precession of Mercury constraints the combination $2\gamma_\odot-\beta_\odot-1$ with an accuracy of $\sim 2\times 10^{-5}$. Given the strict limit on $\gamma_\odot$ for the Solar metric from Cassini, it can be translated into an upper bound on $\beta$ alone. Effects due to the second post-Newtonian order are subdominant compared to  relativistic effects involving other Solar-System bodies and are of order $\sim 7\times 10^{-8}$ smaller than the first-order effects~\citep{Will2018}. As a result, we can conclude that any second-order post-Newtonian corrections to the solar metric cannot be larger than $\sim 2\times 10^{-5}/(7\times 10^{-8})\simeq 300$ times the Schwarzschild predictions.

Examining the combined constraints shown in Figures~\ref{fig:comparison_1pn} and \ref{fig:comparison_2pn} allows to draw some general conclusions on modifications to the equilibrium metrics of massive, isolated objects from the predictions of General Relativity that can be accommodated within current observational bounds, barring any fortuitous cancellations: {\em (i)} They may appear at the first post-Newtonian order and attain up to order-unity magnitudes at the highest gravitational potentials of compact objects but only in theories with coupling to matter that evades the theory-specific bounds imposed by the double pulsar pulsar. {\em (ii)} They may appear primarily at the second or higher post-Newtonian orders but with magnitudes that are constrained only to within order-unity deviations from the general relativistic predictions. 

\section{Summary}\label{sec:conclusions}

The Galactic Center black hole, \sgra, is an ideal and natural laboratory for testing the strong-field predictions of General Relativity. Monitoring of the orbits of tens of stars within its radius of influence and the recent detection of two relativistic effects in one of them has led to a precise determination of the black-hole mass and distance from the Earth~\citep{2018A&A...615L..15G,2019Sci...365..664D,2020A&A...636L...5G}. Compared to any other black hole on the sky observed in the electromagnetic spectrum, these measurements lead to precise predictions for the magnitudes of gravitational effect, which can then be tested against other observations.

In this series of papers, we are reporting the first horizon-scale images of \sgra, obtained with the EHT at a wavelength of 1.3\,mm (\citetalias{PaperII}). The images are characterized by a bright ring of emission surrounded by a deep brightness depression (\citetalias{PaperIII}). This image structure is stable and remains present for at least the $\sim 32$hr span of observations on 2017 April 5-6, which corresponds to $\sim 60-500$ dynamical timescales at the radius of the innermost stable circular orbit, depending on black-hole spin. Using a variety of image-reconstruction and visibility-domain modeling tools, we measured the diameter of the ring-like structure to be equal to $50\pm XX~\mu$as (make consistent with \citetalias{PaperI}).

The structure, size, and persistent nature of the black-hole image leads us to identify the central brightness depression with the shadow that the black hole is expected to cast on the emission from the accreting plasma. Using the depth and size of this brightness depression as well as the observed broadband spectrum of the source, we rule out the possibility that \sgra\ has either a surface that fully absorbs and remits thermally the incoming energy flux or that it has a reflecting surface at 1.3\,mm with significant albedo.

Using an extensive suite of images and synthetic data based on time-dependent and semi-analytic plasma models in a variety of spacetimes, we calibrated the difference between the size of the observed emission ring and that of the shadow as well as potential systematic effects introduced by the sparse interferometric coverage of the EHT array and our analysis methods. We found the magnitudes of these effects to be of order $\sim 10$\%. This is subdominant compared to the significantly larger effect caused by changing the spacetime of \sgra, which can be as large as an order of magnitude while remaining consistent with the bounds imposed by the observation of the orbit of the S0-2 star in the spacetime of the same black hole. 

Because in the Kerr metric the size of the shadow depends primarily on the mass-to-distance ratio of the black hole, which is well determined by monitoring of stellar orbits, we can compare directly the predicted Kerr size to the observations, without any free parameters. We find that the Kerr predictions are consistent with observations at the $\sim 10$\% level. We then use this strong-field inference to place bounds of order unity to the parameters of metrics that deviate from Kerr.

The $\sim 4\times 10^6 M_\odot$ mass of \sgra\ places these tests of the Kerr metric in a region of the parameter space of gravitational objects that has never been probed before in the strong field regime. This mass is approximately 5 orders of magnitude larger than the masses of the objects probed by LIGO/Virgo via the detection of gravitational waves and a factor of 1500 smaller than the mass of the black hole in M87. Leveraging the fact that similar bounds on the strong-field predictions of General Relativity have been placed by gravitational-wave and imaging observations across these 8 orders of magnitude range in mass, we conclude that it is unlikely for the fundamental theory of gravity to possess a scale in this range. 

These conclusions are based predominantly on the identification of the central brightness depression with the shadow of the black hole and, at the $\sim 10$\% level, on the calibration of the relative size of the bright ring of emission, which we measure, to that of the shadow, which we infer. In the case of the tests involving the image of the M87 black hole that we reported earlier~\citep{VI_EHT2019_M87,Psaltis2020,Kocherlakota2021}, two issues related to that black hole left open the possibility that the observed brightness depression might not have been related to the black-hole shadow. First is the factor of $\sim 2$ difference in the prior measurements of the black-hole mass based on stellar and gas dynamics~\citep{Gebhardt2011,Walsh2013}. It could have been possible, in principle, for {\em (i)\/} the brightness depression to not be related to the black-hole shadow but be generated, for example, at a larger distance from the black hole and {\em (ii)\/} for the black-hole mass to be smaller than that inferred from stellar dynamics in such a way that, {\it by pure coincidence}, the size of the brightness depression is equal to the size inferred observationally. This is the approach taken, e.g., by~\cite{Gralla2019} (see also~\citealt{Gralla2019,Gralla2020}). Second is the fact that the time spread of the M87 observations was comparable to the dynamical timescale at its innermost stable circular orbit, allowing for the possibility that the image structure observed was transient and did not correspond to the persistent image of the bright ring surrounding the black-hole shadow. 

In the case of the \sgra\ images, neither of these considerations provide reasonable alternatives to our interpretation. Indeed, the mass of the black hole in \sgra\ is known to such a degree of precision that leaves very little room for uncertainties in the predicted diameter of the black-hole shadow. At the same time, the relatively small value of this mass, compared to the M87 black hole, allows us to observe the image over tens to hundreds of dynamical timescales and conclude that this structure is persistent and not transient. With these potential sources of uncertainty under control, we measured the size of the black-hole shadow in \sgra\ and found it to be in agreement with the Kerr prediction, as we did for the case of the M87 black hole. In order to argue that the observed brightness depression is not related to the black-hole shadow, we would have to assign this consistency not only to coincidence but also require that the same coincidence works for both M87 and \sgra. Given that the two black holes have masses that are different by a factor of $\sim 1500$, are accreting at widely different rates with one showing a prominent jet that is missing from the other, and are probably observed at different inclinations, we consider this alternative to be highly unlikely.

\acknowledgements
$\,$

The authors thank the anonymous referee for comments that improved the manuscript. The Event Horizon Telescope Collaboration thanks the following
organizations and programs: 
National Science Foundation (awards OISE-1743747, AST-1816420, AST-1716536, AST-1440254, AST-1935980);
the Black Hole Initiative, which is funded by grants from the John Templeton Foundation and the Gordon 
and Betty Moore Foundation (although the opinions expressed in this work are those of the author(s) 
and do not necessarily reflect the views of these Foundations);
NASA Hubble Fellowship grant
HST-HF2-51431.001-A awarded by the Space Telescope
Science Institute, which is operated by the Association
of Universities for Research in Astronomy, Inc.,
for NASA, under contract NAS5-26555;
the Academy
of Finland (projects 274477, 284495, 312496, 315721); the Agencia Nacional de Investigación 
y Desarrollo (ANID), Chile via NCN$19\_058$ (TITANs) and Fondecyt 1221421, the Alexander
von Humboldt Stiftung; an Alfred P. Sloan Research Fellowship;
Allegro, the European ALMA Regional Centre node in the Netherlands, the NL astronomy
research network NOVA and the astronomy institutes of the University of Amsterdam, Leiden University and Radboud University;
the Institute for Advanced Study;
the China Scholarship
Council;  Consejo
Nacional de Ciencia y Tecnolog\'{\i}a (CONACYT,
Mexico, projects  U0004-246083, U0004-259839, F0003-272050, M0037-279006, F0003-281692,
104497, 275201, 263356);
the Delaney Family via the Delaney Family John A.
Wheeler Chair at Perimeter Institute; Dirección General
de Asuntos del Personal Académico-—Universidad
Nacional Autónoma de México (DGAPA-—UNAM,
projects IN112417 and IN112820); the European Research Council Synergy
Grant ``BlackHoleCam: Imaging the Event Horizon
of Black Holes" (grant 610058); the Generalitat
Valenciana postdoctoral grant APOSTD/2018/177 and
GenT Program (project CIDEGENT/2018/021); MICINN Research Project PID2019-108995GB-C22;
the European Research Council for advanced grant `JETSET: Launching, propagation and 
emission of relativistic 
jets from binary mergers and across mass scales' (Grant No. 884631); 
the Istituto Nazionale di Fisica
Nucleare (INFN) sezione di Napoli, iniziative specifiche
TEONGRAV; the two Dutch National Supercomputers, Cartesius and Snellius  
(NWO Grant 2021.013);
the International Max Planck Research
School for Astronomy and Astrophysics at the
Universities of Bonn and Cologne; 
DFG research grant ``Jet physics on horizon scales and beyond'' (Grant No. FR 4069/2- 1);
Joint Princeton/Flatiron and Joint Columbia/Flatiron Postdoctoral Fellowships, 
research at the Flatiron Institute is supported by the Simons Foundation; 
the Japanese Government (Monbukagakusho:
MEXT) Scholarship; the Japan Society for
the Promotion of Science (JSPS) Grant-in-Aid for JSPS
Research Fellowship (JP17J08829); the Key Research
Program of Frontier Sciences, Chinese Academy of
Sciences (CAS, grants QYZDJ-SSW-SLH057, QYZDJSSW-SYS008, ZDBS-LY-SLH011); 
the Leverhulme Trust Early Career Research
Fellowship; the Max-Planck-Gesellschaft (MPG);
the Max Planck Partner Group of the MPG and the
CAS; the MEXT/JSPS KAKENHI (grants 18KK0090, JP21H01137,
JP18H03721, JP18K13594, 18K03709, JP19K14761, 18H01245, 25120007); the Malaysian Fundamental Research Grant Scheme (FRGS) FRGS/1/2019/STG02/UM/02/6; the MIT International Science
and Technology Initiatives (MISTI) Funds; 
the Ministry of Science and Technology (MOST) of Taiwan (103-2119-M-001-010-MY2, 105-2112-M-001-025-MY3, 105-2119-M-001-042, 106-2112-M-001-011, 106-2119-M-001-013, 106-2119-M-001-027, 106-2923-M-001-005, 107-2119-M-001-017, 107-2119-M-001-020, 107-2119-M-001-041, 107-2119-M-110-005, 107-2923-M-001-009, 108-2112-M-001-048, 108-2112-M-001-051, 108-2923-M-001-002, 109-2112-M-001-025, 109-2124-M-001-005, 109-2923-M-001-001, 110-2112-M-003-007-MY2, 110-2112-M-001-033, 110-2124-M-001-007, and 110-2923-M-001-001);
the Ministry of Education (MoE) of Taiwan Yushan Young Scholar Program;
the Physics Division, National Center for Theoretical Sciences of Taiwan;
the National Aeronautics and
Space Administration (NASA, Fermi Guest Investigator
grant 80NSSC20K1567, NASA Astrophysics Theory Program grant 80NSSC20K0527, NASA NuSTAR award 
80NSSC20K0645); 
the National
Institute of Natural Sciences (NINS) of Japan; the National
Key Research and Development Program of China
(grant 2016YFA0400704, 2017YFA0402703, 2016YFA0400702); the National
Science Foundation (NSF, grants AST-0096454,
AST-0352953, AST-0521233, AST-0705062, AST-0905844, AST-0922984, AST-1126433, AST-1140030,
DGE-1144085, AST-1207704, AST-1207730, AST-1207752, MRI-1228509, OPP-1248097, AST-1310896,  
AST-1555365, AST-1614868, AST-1615796, AST-1715061, AST-1716327,  AST-2034306); 
the Natural Science Foundation of China (grants 11650110427, 10625314, 11721303, 11725312, 11873028, 11933007, 11991052, 11991053, 12192220, 12192223);
NWO grant number OCENW.KLEIN.113; a 
fellowship of China Postdoctoral Science Foundation (2020M671266); the Natural
Sciences and Engineering Research Council of
Canada (NSERC, including a Discovery Grant and
the NSERC Alexander Graham Bell Canada Graduate
Scholarships-Doctoral Program); the National Youth
Thousand Talents Program of China; the National Research
Foundation of Korea (the Global PhD Fellowship
Grant: grants NRF-2015H1A2A1033752, the Korea Research Fellowship Program:
NRF-2015H1D3A1066561, Basic Research Support Grant 2019R1F1A1059721, 2022R1C1C1005255); the Dutch Organization
for Scientific Research (NWO) VICI award
(grant 639.043.513) and Spinoza Prize SPI 78-409; the YCAA Prize Postdoctoral Fellowship.
LM gratefully acknowledges support from an NSF Astronomy and Astrophysics Postdoctoral Fellowship under award no. AST-1903847. TK is supported by MEXT as "Program for Promoting Researches on the Supercomputer Fugaku" (Toward a unified view of the universe: from large scale structures to planets, JPMXP1020200109) and JICFuS. RPD and IN acknowledge funding by the South African Research Chairs Initiative, through the South African Radio Astronomy Observatory (SARAO, grant ID 77948),  which is a facility of the National Research Foundation (NRF), an agency of the Department of Science and Innovation (DSI) of South Africa.

We thank the Onsala Space Observatory
(OSO) national infrastructure, for the provisioning
of its facilities/observational support (OSO receives
funding through the Swedish Research Council under
grant 2017-00648);  the Perimeter Institute for Theoretical
Physics (research at Perimeter Institute is supported
by the Government of Canada through the Department
of Innovation, Science and Economic Development
and by the Province of Ontario through the
Ministry of Research, Innovation and Science); the Spanish Ministerio de Ciencia e Innovación (grants PGC2018-098915-B-C21, AYA2016-80889-P,
PID2019-108995GB-C21, PID2020-117404GB-C21); 
the University of Pretoria for financial aid in the provision of the new 
Cluster Server nodes and SuperMicro (USA) for a SEEDING GRANT approved towards these 
nodes in 2020;
the State
Agency for Research of the Spanish MCIU through
the ``Center of Excellence Severo Ochoa'' award for
the Instituto de Astrofísica de Andalucía (SEV-2017-
0709); the Toray Science Foundation; the Consejería de Economía, Conocimiento, 
Empresas y Universidad 
of the Junta de Andalucía (grant P18-FR-1769), the Consejo Superior de Investigaciones 
Científicas (grant 2019AEP112);
the M2FINDERS project which has received funding by the European Research Council (ERC) under 
the European Union’s Horizon 2020 Research and Innovation Programme (grant agreement No 101018682);
the US Department
of Energy (USDOE) through the Los Alamos National
Laboratory (operated by Triad National Security,
LLC, for the National Nuclear Security Administration
of the USDOE (Contract 89233218CNA000001);
 the European Union’s Horizon 2020
research and innovation programme under grant agreement
No 730562 RadioNet;
Shanghai Pilot Program for Basic Research, Chinese Academy of Science, 
Shanghai Branch (JCYJ-SHFY-2021-013);
ALMA North America Development
Fund; the Academia Sinica; Chandra DD7-18089X and TM6-
17006X; the GenT Program (Generalitat Valenciana)
Project CIDEGENT/2018/021. This work used the
Extreme Science and Engineering Discovery Environment
(XSEDE), supported by NSF grant ACI-1548562,
and CyVerse, supported by NSF grants DBI-0735191,
DBI-1265383, and DBI-1743442. XSEDE Stampede2 resource
at TACC was allocated through TG-AST170024
and TG-AST080026N. XSEDE JetStream resource at
PTI and TACC was allocated through AST170028.

The simulations were performed in part on the SuperMUC
cluster at the LRZ in Garching, on the
LOEWE cluster in CSC in Frankfurt, and on the
HazelHen cluster at the HLRS in Stuttgart. This
research was enabled in part by support provided
by Compute Ontario (http://computeontario.ca), Calcul
Quebec (http://www.calculquebec.ca) and Compute
Canada (http://www.computecanada.ca). 
CC acknowledges support from the Swedish Research Council (VR).

We thank
the staff at the participating observatories, correlation
centers, and institutions for their enthusiastic support.
This paper makes use of the following ALMA data:
ADS/JAO.ALMA\#2016.1.01154.V. ALMA is a partnership
of the European Southern Observatory (ESO;
Europe, representing its member states), NSF, and
National Institutes of Natural Sciences of Japan, together
with National Research Council (Canada), Ministry
of Science and Technology (MOST; Taiwan),
Academia Sinica Institute of Astronomy and Astrophysics
(ASIAA; Taiwan), and Korea Astronomy and
Space Science Institute (KASI; Republic of Korea), in
cooperation with the Republic of Chile. The Joint
ALMA Observatory is operated by ESO, Associated
Universities, Inc. (AUI)/NRAO, and the National Astronomical
Observatory of Japan (NAOJ). The NRAO
is a facility of the NSF operated under cooperative agreement
by AUI.
This research used resources of the Oak Ridge Leadership Computing Facility at the Oak Ridge National
Laboratory, which is supported by the Office of Science of the U.S. Department of Energy under Contract
No. DE-AC05-00OR22725. We also thank the Center for Computational Astrophysics, National Astronomical Observatory of Japan.

Support for this work was also provided by the NASA Hubble Fellowship 
grant HST-HF2-51431.001-A awarded 
by the Space Telescope Science Institute, which is operated by the Association of Universities for 
Research in Astronomy, Inc., for NASA, under contract NAS5-26555.
HO and GM were supported by Virtual Institute of Accretion (VIA) postdoctoral fellowships from the Netherlands Research School for Astronomy (NOVA).
APEX is a collaboration between the
Max-Planck-Institut f{\"u}r Radioastronomie (Germany),
ESO, and the Onsala Space Observatory (Sweden). The
SMA is a joint project between the SAO and ASIAA
and is funded by the Smithsonian Institution and the
Academia Sinica. The JCMT is operated by the East
Asian Observatory on behalf of the NAOJ, ASIAA, and
KASI, as well as the Ministry of Finance of China, Chinese
Academy of Sciences, and the National Key Research and Development
Program (No. 2017YFA0402700) of China
and Natural Science Foundation of China grant 11873028.
Additional
funding support for the JCMT is provided by the Science
and Technologies Facility Council (UK) and participating
universities in the UK and Canada. 
Simulations were performed in part on the SuperMUC cluster at the LRZ in Garching, 
on the 
LOEWE cluster in CSC in Frankfurt, on the HazelHen cluster at the HLRS in Stuttgart, 
and on the Pi2.0 and Siyuan Mark-I at Shanghai Jiao Tong University.
The computer resources of the Finnish IT Center for Science (CSC) and the Finnish Computing 
Competence Infrastructure (FCCI) project are acknowledged.
JO was supported by Basic Science Research Program through the National Research
Foundation of Korea(NRF) funded by the Ministry of Education(NRF-2021R1A6A3A01086420;
2022R1C1C1005255).
We thank Martin Shepherd for the addition of extra features in the Difmap software 
that were used for the CLEAN imaging results presented in this paper.
The computing cluster of Shanghai VLBI correlator supported by the Special Fund 
for Astronomy from the Ministry of Finance in China is acknowledged.
This work was supported by the Brain Pool Program through the National Research
Foundation 
of Korea (NRF) funded by the Ministry of Science and ICT (019H1D3A1A01102564).
This research is part of the Frontera computing project at the Texas Advanced 
Computing Center through the Frontera Large-Scale Community Partnerships allocation
AST20023. Frontera is made possible by National Science Foundation award OAC-1818253.
This research was carried out using resources provided by the Open Science Grid, 
which is supported by the National Science Foundation and the U.S. Department of 
Energy Office of Science.

The LMT is a project operated by the Instituto Nacional
de Astrófisica, Óptica, y Electrónica (Mexico) and the
University of Massachusetts at Amherst (USA). The
IRAM 30-m telescope on Pico Veleta, Spain is operated
by IRAM and supported by CNRS (Centre National de
la Recherche Scientifique, France), MPG (Max-Planck-Gesellschaft, Germany) 
and IGN (Instituto Geográfico
Nacional, Spain). The SMT is operated by the Arizona
Radio Observatory, a part of the Steward Observatory
of the University of Arizona, with financial support of
operations from the State of Arizona and financial support
for instrumentation development from the NSF.
Support for SPT participation in the EHT is provided by the National Science Foundation through award OPP-1852617 
to the University of Chicago. Partial support is also 
provided by the Kavli Institute of Cosmological Physics at the University of Chicago. The SPT hydrogen maser was 
provided on loan from the GLT, courtesy of ASIAA.
Support for this work was provided by NASA through the NASA Hubble Fellowship grant
\#HST--HF2--51494.001 awarded by the Space Telescope Science Institute, which is operated 
by the Association of Universities for Research in Astronomy, Inc., for NASA, 
under contract NAS5--26555.
Jongho Park acknowledges financial support through the EACOA Fellowship awarded by the East Asia Core
Observatories Association, which consists of the Academia Sinica Institute of Astronomy and
Astrophysics, the National Astronomical Observatory of Japan, Center for Astronomical Mega-Science,
Chinese Academy of Sciences, and the Korea Astronomy and Space Science Institute.

The EHTC has
received generous donations of FPGA chips from Xilinx
Inc., under the Xilinx University Program. The EHTC
has benefited from technology shared under open-source
license by the Collaboration for Astronomy Signal Processing
and Electronics Research (CASPER). The EHT
project is grateful to T4Science and Microsemi for their
assistance with Hydrogen Masers. This research has
made use of NASA’s Astrophysics Data System. We
gratefully acknowledge the support provided by the extended
staff of the ALMA, both from the inception of
the ALMA Phasing Project through the observational
campaigns of 2017 and 2018. We would like to thank
A. Deller and W. Brisken for EHT-specific support with
the use of DiFX. We acknowledge the significance that
Maunakea, where the SMA and JCMT EHT stations
are located, has for the indigenous Hawaiian people.


\facility{EHT, ALMA, APEX, IRAM:30m, JCMT, LMT, SMA, ARO:SMT, SPT}.
\bibliography{paperVI_SgrA,EHTCPapers}{}

\begin{thebibliography}{}
\expandafter\ifx\csname natexlab\endcsname\relax\def\natexlab#1{#1}\fi
\providecommand{\url}[1]{\href{#1}{#1}}
\providecommand{\dodoi}[1]{doi:~\href{http://doi.org/#1}{\nolinkurl{#1}}}
\providecommand{\doeprint}[1]{\href{http://ascl.net/#1}{\nolinkurl{http://ascl.net/#1}}}
\providecommand{\doarXiv}[1]{\href{https://arxiv.org/abs/#1}{\nolinkurl{https://arxiv.org/abs/#1}}}

\bibitem[{{Abbott} {et~al.}(2016){Abbott}, {Abbott}, {Abbott}, {Abernathy},
  {Acernese}, {Ackley}, {Adams}, {Adams}, {Addesso}, {Adhikari}, {Adya},
  {Affeldt}, {Agathos}, {Agatsuma}, {Aggarwal}, {Aguiar}, {Aiello}, {Ain},
  {Ajith}, {Allen}, {Allocca}, {Altin}, {Anderson}, {Anderson}, {Arai},
  {Araya}, {Arceneaux}, {Areeda}, {Arnaud}, {Arun}, {Ascenzi}, {Ashton}, {Ast},
  {Aston}, {Astone}, {Aufmuth}, {Aulbert}, {Babak}, {Bacon}, {Bader}, {Baker},
  {Baldaccini}, {Ballardin}, {Ballmer}, {Barayoga}, {Barclay}, {Barish},
  {Barker}, {Barone}, {Barr}, {Barsotti}, {Barsuglia}, {Barta}, {Bartlett},
  {Bartos}, {Bassiri}, {Basti}, {Batch}, {Baune}, {Bavigadda}, {Bazzan},
  {Behnke}, {Bejger}, {Bell}, {Bell}, {Berger}, {Bergman}, {Bergmann}, {Berry},
  {Bersanetti}, {Bertolini}, {Betzwieser}, {Bhagwat}, {Bhandare}, {Bilenko},
  {Billingsley}, {Birch}, {Birney}, {Birnholtz}, {Biscans}, {Bisht}, {Bitossi},
  {Biwer}, {Bizouard}, {Blackburn}, {Blair}, {Blair}, {Blair}, {Bloemen},
  {Bock}, {Bodiya}, {Boer}, {Bogaert}, {Bogan}, {Bohe}, {Bojtos}, {Bond},
  {Bondu}, {Bonnand}, {Boom}, {Bork}, {Boschi}, {Bose}, {Bouffanais}, {Bozzi},
  {Bradaschia}, {Brady}, {Braginsky}, {Branchesi}, {Brau}, {Briant}, {Brillet},
  {Brinkmann}, {Brisson}, {Brockill}, {Brooks}, {Brown}, {Brown}, {Brown},
  {Buchanan}, {Buikema}, {Bulik}, {Bulten}, {Buonanno}, {Buskulic}, {Buy},
  {Byer}, {Cadonati}, {Cagnoli}, {Cahillane}, {Calder{\'o}n Bustillo},
  {Callister}, {Calloni}, {Camp}, {Cannon}, {Cao}, {Capano}, {Capocasa},
  {Carbognani}, {Caride}, {Casanueva Diaz}, {Casentini}, {Caudill},
  {Cavagli{\`a}}, {Cavalier}, {Cavalieri}, {Cella}, {Cepeda}, {Cerboni
  Baiardi}, {Cerretani}, {Cesarini}, {Chakraborty}, {Chalermsongsak},
  {Chamberlin}, {Chan}, {Chao}, {Charlton}, {Chassande-Mottin}, {Chen}, {Chen},
  {Cheng}, {Chincarini}, {Chiummo}, {Cho}, {Cho}, {Chow}, {Christensen}, {Chu},
  {Chua}, {Chung}, {Ciani}, {Clara}, {Clark}, {Cleva}, {Coccia}, {Cohadon},
  {Colla}, {Collette}, {Cominsky}, {Constancio}, {Conte}, {Conti}, {Cook},
  {Corbitt}, {Cornish}, {Corsi}, {Cortese}, {Costa}, {Coughlin}, {Coughlin},
  {Coulon}, {Countryman}, {Couvares}, {Cowan}, {Coward}, {Cowart}, {Coyne},
  {Coyne}, {Craig}, {Creighton}, {Cripe}, {Crowder}, {Cumming}, {Cunningham},
  {Cuoco}, {Dal Canton}, {Danilishin}, {D'Antonio}, {Danzmann}, {Darman},
  {Dattilo}, {Dave}, {Daveloza}, {Davier}, {Davies}, {Daw}, {Day}, {DeBra},
  {Debreczeni}, {Degallaix}, {De Laurentis}, {Del{\'e}glise}, {Del Pozzo},
  {Denker}, {Dent}, {Dereli}, {Dergachev}, {De Rosa}, {DeRosa}, {DeSalvo},
  {Dhurandhar}, {D{\'\i}az}, {Di Fiore}, {Di Giovanni}, {Di Lieto}, {Di Pace},
  {Di Palma}, {Di Virgilio}, {Dojcinoski}, {Dolique}, {Donovan}, {Dooley},
  {Doravari}, {Douglas}, {Downes}, {Drago}, {Drever}, {Driggers}, {Du},
  {Ducrot}, {Dwyer}, {Edo}, {Edwards}, {Effler}, {Eggenstein}, {Ehrens},
  {Eichholz}, {Eikenberry}, {Engels}, {Essick}, {Etzel}, {Evans}, {Evans},
  {Everett}, {Factourovich}, {Fafone}, {Fair}, {Fairhurst}, {Fan}, {Fang},
  {Farinon}, {Farr}, {Farr}, {Favata}, {Fays}, {Fehrmann}, {Fejer}, {Ferrante},
  {Ferreira}, {Ferrini}, {Fidecaro}, {Fiori}, {Fiorucci}, {Fisher}, {Flaminio},
  {Fletcher}, {Fournier}, {Franco}, {Frasca}, {Frasconi}, {Frei}, {Freise},
  {Frey}, {Frey}, {Fricke}, {Fritschel}, {Frolov}, {Fulda}, {Fyffe}, {Gabbard},
  {Gair}, {Gammaitoni}, {Gaonkar}, {Garufi}, {Gatto}, {Gaur}, {Gehrels},
  {Gemme}, {Gendre}, {Genin}, {Gennai}, {George}, {Gergely}, {Germain},
  {Ghosh}, {Ghosh}, {Ghosh}, {Giaime}, {Giardina}, {Giazotto}, {Gill},
  {Glaefke}, {Goetz}, {Goetz}, {Gondan}, {Gonz{\'a}lez}, {Gonzalez Castro},
  {Gopakumar}, {Gordon}, {Gorodetsky}, {Gossan}, {Gosselin}, {Gouaty}, {Graef},
  {Graff}, {Granata}, {Grant}, {Gras}, {Gray}, {Greco}, {Green}, {Groot},
  {Grote}, {Grunewald}, {Guidi}, {Guo}, {Gupta}, {Gupta}, {Gushwa},
  {Gustafson}, {Gustafson}, {Hacker}, {Hall}, {Hall}, {Hammond}, {Haney},
  {Hanke}, {Hanks}, {Hanna}, {Hannam}, {Hanson}, {Hardwick}, {Harms}, {Harry},
  {Harry}, {Hart}, {Hartman}, {Haster}, {Haughian}, {Healy}, {Heidmann},
  {Heintze}, {Heitmann}, {Hello}, {Hemming}, {Hendry}, {Heng}, {Hennig},
  {Heptonstall}, {Heurs}, {Hild}, {Hoak}, {Hodge}, {Hofman}, {Hollitt}, {Holt},
  {Holz}, {Hopkins}, {Hosken}, {Hough}, {Houston}, {Howell}, {Hu}, {Huang},
  {Huerta}, {Huet}, {Hughey}, {Husa}, {Huttner}, {Huynh-Dinh}, {Idrisy},
  {Indik}, {Ingram}, {Inta}, {Isa}, {Isac}, {Isi}, {Islas}, {Isogai}, {Iyer},
  {Izumi}, {Jacqmin}, {Jang}, {Jani}, {Jaranowski}, {Jawahar},
  {Jim{\'e}nez-Forteza}, {Johnson}, {Johnson-McDaniel}, {Jones}, {Jones},
  {Jonker}, {Ju}, {Haris}, {Kalaghatgi}, {Kalogera}, {Kandhasamy}, {Kang},
  {Kanner}, {Karki}, {Kasprzack}, {Katsavounidis}, {Katzman}, {Kaufer}, {Kaur},
  {Kawabe}, {Kawazoe}, {K{\'e}f{\'e}lian}, {Kehl}, {Keitel}, {Kelley}, {Kells},
  {Kennedy}, {Key}, {Khalaidovski}, {Khalili}, {Khan}, {Khan}, {Khan},
  {Khazanov}, {Kijbunchoo}, {Kim}, {Kim}, {Kim}, {Kim}, {Kim}, {Kim}, {King},
  {King}, {Kinzel}, {Kissel}, {Kleybolte}, {Klimenko}, {Koehlenbeck},
  {Kokeyama}, {Koley}, {Kondrashov}, {Kontos}, {Korobko}, {Korth}, {Kowalska},
  {Kozak}, {Kringel}, {Krishnan}, {Kr{\'o}lak}, {Krueger}, {Kuehn}, {Kumar},
  {Kuo}, {Kutynia}, {Lackey}, {Landry}, {Lange}, {Lantz}, {Lasky}, {Lazzarini},
  {Lazzaro}, {Leaci}, {Leavey}, {Lebigot}, {Lee}, {Lee}, {Lee}, {Lee}, {Lenon},
  {Leonardi}, {Leong}, {Leroy}, {Letendre}, {Levin}, {Levine}, {Li}, {Libson},
  {Littenberg}, {Lockerbie}, {Logue}, {Lombardi}, {London}, {Lord},
  {Lorenzini}, {Loriette}, {Lormand}, {Losurdo}, {Lough}, {Lousto}, {Lovelace},
  {L{\"u}ck}, {Lundgren}, {Luo}, {Lynch}, {Ma}, {MacDonald}, {Machenschalk},
  {MacInnis}, {Macleod}, {Maga{\~n}a-Sandoval}, {Magee}, {Mageswaran},
  {Majorana}, {Maksimovic}, {Malvezzi}, {Man}, {Mandel}, {Mandic}, {Mangano},
  {Mansell}, {Manske}, {Mantovani}, {Marchesoni}, {Marion}, {M{\'a}rka},
  {M{\'a}rka}, {Markosyan}, {Maros}, {Martelli}, {Martellini}, {Martin},
  {Martin}, {Martynov}, {Marx}, {Mason}, {Masserot}, {Massinger}, {Masso-Reid},
  {Matichard}, {Matone}, {Mavalvala}, {Mazumder}, {Mazzolo}, {McCarthy},
  {McClelland}, {McCormick}, {McGuire}, {McIntyre}, {McIver}, {McManus},
  {McWilliams}, {Meacher}, {Meadors}, {Meidam}, {Melatos}, {Mendell},
  {Mendoza-Gandara}, {Mercer}, {Merilh}, {Merzougui}, {Meshkov}, {Messenger},
  {Messick}, {Meyers}, {Mezzani}, {Miao}, {Michel}, {Middleton}, {Mikhailov},
  {Milano}, {Miller}, {Millhouse}, {Minenkov}, {Ming}, {Mirshekari}, {Mishra},
  {Mitra}, {Mitrofanov}, {Mitselmakher}, {Mittleman}, {Moggi}, {Mohan},
  {Mohapatra}, {Montani}, {Moore}, {Moore}, {Moraru}, {Moreno}, {Morriss},
  {Mossavi}, {Mours}, {Mow-Lowry}, {Mueller}, {Mueller}, {Muir}, {Mukherjee},
  {Mukherjee}, {Mukherjee}, {Mukund}, {Mullavey}, {Munch}, {Murphy}, {Murray},
  {Mytidis}, {Nardecchia}, {Naticchioni}, {Nayak}, {Necula}, {Nedkova},
  {Nelemans}, {Neri}, {Neunzert}, {Newton}, {Nguyen}, {Nielsen}, {Nissanke},
  {Nitz}, {Nocera}, {Nolting}, {Normandin}, {Nuttall}, {Oberling}, {Ochsner},
  {O'Dell}, {Oelker}, {Ogin}, {Oh}, {Oh}, {Ohme}, {Oliver}, {Oppermann},
  {Oram}, {O'Reilly}, {O'Shaughnessy}, {Ottaway}, {Ottens}, {Overmier}, {Owen},
  {Pai}, {Pai}, {Palamos}, {Palashov}, {Palomba}, {Pal-Singh}, {Pan}, {Pan},
  {Pankow}, {Pannarale}, {Pant}, {Paoletti}, {Paoli}, {Papa}, {Paris},
  {Parker}, {Pascucci}, {Pasqualetti}, {Passaquieti}, {Passuello},
  {Patricelli}, {Patrick}, {Pearlstone}, {Pedraza}, {Pedurand}, {Pekowsky},
  {Pele}, {Penn}, {Perreca}, {Pfeiffer}, {Phelps}, {Piccinni}, {Pichot},
  {Piergiovanni}, {Pierro}, {Pillant}, {Pinard}, {Pinto}, {Pitkin}, {Poggiani},
  {Popolizio}, {Post}, {Powell}, {Prasad}, {Predoi}, {Premachandra},
  {Prestegard}, {Price}, {Prijatelj}, {Principe}, {Privitera}, {Prix}, {Prodi},
  {Prokhorov}, {Puncken}, {Punturo}, {Puppo}, {P{\"u}rrer}, {Qi}, {Qin},
  {Quetschke}, {Quintero}, {Quitzow-James}, {Raab}, {Rabeling}, {Radkins},
  {Raffai}, {Raja}, {Rakhmanov}, {Rapagnani}, {Raymond}, {Razzano}, {Re},
  {Read}, {Reed}, {Regimbau}, {Rei}, {Reid}, {Reitze}, {Rew}, {Reyes}, {Ricci},
  {Riles}, {Robertson}, {Robie}, {Robinet}, {Rocchi}, {Rolland}, {Rollins},
  {Roma}, {Romano}, {Romanov}, {Romie}, {Rosi{\'n}ska}, {Rowan}, {R{\"u}diger},
  {Ruggi}, {Ryan}, {Sachdev}, {Sadecki}, {Sadeghian}, {Salconi}, {Saleem},
  {Salemi}, {Samajdar}, {Sammut}, {Sanchez}, {Sandberg}, {Sandeen}, {Sanders},
  {Sassolas}, {Sathyaprakash}, {Saulson}, {Sauter}, {Savage}, {Sawadsky},
  {Schale}, {Schilling}, {Schmidt}, {Schmidt}, {Schnabel}, {Schofield},
  {Sch{\"o}nbeck}, {Schreiber}, {Schuette}, {Schutz}, {Scott}, {Scott},
  {Sellers}, {Sengupta}, {Sentenac}, {Sequino}, {Sergeev}, {Serna},
  {Setyawati}, {Sevigny}, {Shaddock}, {Shah}, {Shahriar}, {Shaltev}, {Shao},
  {Shapiro}, {Shawhan}, {Sheperd}, {Shoemaker}, {Shoemaker}, {Siellez},
  {Siemens}, {Sigg}, {Silva}, {Simakov}, {Singer}, {Singer}, {Singh}, {Singh},
  {Singhal}, {Sintes}, {Slagmolen}, {Smith}, {Smith}, {Smith}, {Son}, {Sorazu},
  {Sorrentino}, {Souradeep}, {Srivastava}, {Staley}, {Steinke}, {Steinlechner},
  {Steinlechner}, {Steinmeyer}, {Stephens}, {Stone}, {Strain}, {Straniero},
  {Stratta}, {Strauss}, {Strigin}, {Sturani}, {Stuver}, {Summerscales}, {Sun},
  {Sutton}, {Swinkels}, {Szczepa{\'n}czyk}, {Tacca}, {Talukder}, {Tanner},
  {T{\'a}pai}, {Tarabrin}, {Taracchini}, {Taylor}, {Theeg},
  {Thirugnanasambandam}, {Thomas}, {Thomas}, {Thomas}, {Thorne}, {Thorne},
  {Thrane}, {Tiwari}, {Tiwari}, {Tokmakov}, {Tomlinson}, {Tonelli}, {Torres},
  {Torrie}, {T{\"o}yr{\"a}}, {Travasso}, {Traylor}, {Trifir{\`o}}, {Tringali},
  {Trozzo}, {Tse}, {Turconi}, {Tuyenbayev}, {Ugolini}, {Unnikrishnan}, {Urban},
  {Usman}, {Vahlbruch}, {Vajente}, {Valdes}, {Vallisneri}, {van Bakel}, {van
  Beuzekom}, {van den Brand}, {Van Den Broeck}, {Vander-Hyde}, {van der
  Schaaf}, {van Heijningen}, {van Veggel}, {Vardaro}, {Vass}, {Vas{\'u}th},
  {Vaulin}, {Vecchio}, {Vedovato}, {Veitch}, {Veitch}, {Venkateswara},
  {Verkindt}, {Vetrano}, {Vicer{\'e}}, {Vinciguerra}, {Vine}, {Vinet},
  {Vitale}, {Vo}, {Vocca}, {Vorvick}, {Voss}, {Vousden}, {Vyatchanin}, {Wade},
  {Wade}, {Wade}, {Walker}, {Wallace}, {Walsh}, {Wang}, {Wang}, {Wang}, {Wang},
  {Wang}, {Ward}, {Warner}, {Was}, {Weaver}, {Wei}, {Weinert}, {Weinstein},
  {Weiss}, {Welborn}, {Wen}, {We{\ss}els}, {Westphal}, {Wette}, {Whelan},
  {White}, {Whiting}, {Williams}, {Williams}, {Williamson}, {Willis}, {Willke},
  {Wimmer}, {Winkler}, {Wipf}, {Wittel}, {Woan}, {Worden}, {Wright}, {Wu},
  {Yablon}, {Yam}, {Yamamoto}, {Yancey}, {Yap}, {Yu}, {Yvert}, {Zadro{\.Z}ny},
  {Zangrando}, {Zanolin}, {Zendri}, {Zevin}, {Zhang}, {Zhang}, {Zhang},
  {Zhang}, {Zhao}, {Zhou}, {Zhou}, {Zhu}, {Zucker}, {Zuraw}, {Zweizig},
  {Boyle}, {Campanelli}, {Hemberger}, {Kidder}, {Ossokine}, {Scheel},
  {Szilagyi}, {Teukolsky}, {Zlochower}, {LIGO Scientific}, \& {Virgo
  Collaborations}}]{Abbott2016}
{Abbott}, B.~P., {Abbott}, R., {Abbott}, T.~D., {et~al.} 2016, \prl, 116,
  221101, \dodoi{10.1103/PhysRevLett.116.221101}

\bibitem[{{Abbott} {et~al.}(2019){Abbott}, {Abbott}, {Abbott}, {Abraham},
  {Acernese}, {Ackley}, {Adams}, {Adhikari}, {Adya}, {Affeldt}, {Agathos},
  {Agatsuma}, {Aggarwal}, {Aguiar}, {Aiello}, {Ain}, {Ajith}, {Allen},
  {Allocca}, {Aloy}, {Altin}, {Amato}, {Ananyeva}, {Anderson}, {Anderson},
  {Angelova}, {Antier}, {Appert}, {Arai}, {Araya}, {Areeda}, {Ar{\`e}ne},
  {Arnaud}, {Arun}, {Ascenzi}, {Ashton}, {Aston}, {Astone}, {Aubin}, {Aufmuth},
  {AultONeal}, {Austin}, {Avendano}, {Avila-Alvarez}, {Babak}, {Bacon},
  {Badaracco}, {Bader}, {Bae}, {Baker}, {Baldaccini}, {Ballardin}, {Ballmer},
  {Banagiri}, {Barayoga}, {Barclay}, {Barish}, {Barker}, {Barkett}, {Barnum},
  {Barone}, {Barr}, {Barsotti}, {Barsuglia}, {Barta}, {Bartlett}, {Bartos},
  {Bassiri}, {Basti}, {Bawaj}, {Bayley}, {Bazzan}, {B{\'e}csy}, {Bejger},
  {Belahcene}, {Bell}, {Beniwal}, {Berger}, {Bergmann}, {Bernuzzi}, {Bero},
  {Berry}, {Bersanetti}, {Bertolini}, {Betzwieser}, {Bhandare}, {Bidler},
  {Bilenko}, {Bilgili}, {Billingsley}, {Birch}, {Birney}, {Birnholtz},
  {Biscans}, {Biscoveanu}, {Bisht}, {Bitossi}, {Bizouard}, {Blackburn},
  {Blair}, {Blair}, {Blair}, {Bloemen}, {Bode}, {Boer}, {Boetzel}, {Bogaert},
  {Bondu}, {Bonilla}, {Bonnand}, {Booker}, {Boom}, {Booth}, {Bork}, {Boschi},
  {Bose}, {Bossie}, {Bossilkov}, {Bosveld}, {Bouffanais}, {Bozzi},
  {Bradaschia}, {Brady}, {Bramley}, {Branchesi}, {Brau}, {Breschi}, {Briant},
  {Briggs}, {Brighenti}, {Brillet}, {Brinkmann}, {Brisson}, {Brito},
  {Brockill}, {Brooks}, {Brown}, {Brunett}, {Buikema}, {Bulik}, {Bulten},
  {Buonanno}, {Buskulic}, {Rosell}, {Buy}, {Byer}, {Cabero}, {Cadonati},
  {Cagnoli}, {Cahillane}, {Bustillo}, {Callister}, {Calloni}, {Camp},
  {Campbell}, {Canepa}, {Cannon}, {Cao}, {Cao}, {Capano}, {Capocasa},
  {Carbognani}, {Caride}, {Carney}, {Carullo}, {Diaz}, {Casentini}, {Caudill},
  {Cavagli{\`a}}, {Cavalier}, {Cavalieri}, {Cella}, {Cerd{\'a}-Dur{\'a}n},
  {Cerretani}, {Cesarini}, {Chaibi}, {Chakravarti}, {Chamberlin}, {Chan},
  {Chao}, {Charlton}, {Chase}, {Chassande-Mottin}, {Chatterjee}, {Chaturvedi},
  {Chatziioannou}, {Cheeseboro}, {Chen}, {Chen}, {Chen}, {Cheng}, {Cheong},
  {Chia}, {Chincarini}, {Chiummo}, {Cho}, {Cho}, {Cho}, {Christensen}, {Chu},
  {Chua}, {Chung}, {Chung}, {Ciani}, {Ciobanu}, {Ciolfi}, {Cipriano}, {Cirone},
  {Clara}, {Clark}, {Clearwater}, {Cleva}, {Cocchieri}, {Coccia}, {Cohadon},
  {Cohen}, {Colgan}, {Colleoni}, {Collette}, {Collins}, {Cominsky},
  {Constancio}, {Conti}, {Cooper}, {Corban}, {Corbitt}, {Cordero-Carri{\'o}n},
  {Corley}, {Cornish}, {Corsi}, {Cortese}, {Costa}, {Cotesta}, {Coughlin},
  {Coughlin}, {Coulon}, {Countryman}, {Couvares}, {Covas}, {Cowan}, {Coward},
  {Cowart}, {Coyne}, {Coyne}, {Creighton}, {Creighton}, {Cripe}, {Croquette},
  {Crowder}, {Cullen}, {Cumming}, {Cunningham}, {Cuoco}, {Canton}, {D{\'a}lya},
  {Danilishin}, {D'Antonio}, {Danzmann}, {Dasgupta}, {Costa}, {Datrier},
  {Dattilo}, {Dave}, {Davier}, {Davis}, {Daw}, {DeBra}, {Deenadayalan},
  {Degallaix}, {De Laurentis}, {Del{\'e}glise}, {Del Pozzo}, {DeMarchi},
  {Demos}, {Dent}, {De Pietri}, {Derby}, {De Rosa}, {De Rossi}, {DeSalvo}, {de
  Varona}, {Dhurandhar}, {D{\'\i}az}, {Dietrich}, {Di Fiore}, {Di Giovanni},
  {Di Girolamo}, {Di Lieto}, {Ding}, {Di Pace}, {Di Palma}, {Di Renzo},
  {Dmitriev}, {Doctor}, {Donovan}, {Dooley}, {Doravari}, {Dorrington},
  {Downes}, {Drago}, {Driggers}, {Du}, {Ducoin}, {Dupej}, {Dwyer}, {Easter},
  {Edo}, {Edwards}, {Effler}, {Ehrens}, {Eichholz}, {Eikenberry}, {Eisenmann},
  {Eisenstein}, {Essick}, {Estelles}, {Estevez}, {Etienne}, {Etzel}, {Evans},
  {Evans}, {Fafone}, {Fair}, {Fairhurst}, {Fan}, {Farinon}, {Farr}, {Farr},
  {Fauchon-Jones}, {Favata}, {Fays}, {Fazio}, {Fee}, {Feicht}, {Fejer}, {Feng},
  {Fernandez-Galiana}, {Ferrante}, {Ferreira}, {Ferreira}, {Ferrini},
  {Fidecaro}, {Fiori}, {Fiorucci}, {Fishbach}, {Fisher}, {Fishner},
  {Fitz-Axen}, {Flaminio}, {Fletcher}, {Flynn}, {Fong}, {Font}, {Forsyth},
  {Fournier}, {Frasca}, {Frasconi}, {Frei}, {Freise}, {Frey}, {Frey},
  {Fritschel}, {Frolov}, {Fulda}, {Fyffe}, {Gabbard}, {Gadre}, {Gaebel},
  {Gair}, {Gammaitoni}, {Ganija}, {Gaonkar}, {Garcia},
  {Garc{\'\i}a-Quir{\'o}s}, {Garufi}, {Gateley}, {Gaudio}, {Gaur}, {Gayathri},
  {Gemme}, {Genin}, {Gennai}, {George}, {George}, {Gergely}, {Germain},
  {Ghonge}, {Ghosh}, {Ghosh}, {Ghosh}, {Giacomazzo}, {Giaime}, {Giardina},
  {Giazotto}, {Gill}, {Giordano}, {Glover}, {Godwin}, {Goetz}, {Goetz},
  {Goncharov}, {Gonz{\'a}lez}, {Castro}, {Gopakumar}, {Gorodetsky}, {Gossan},
  {Gosselin}, {Gouaty}, {Grado}, {Graef}, {Granata}, {Grant}, {Gras},
  {Grassia}, {Gray}, {Gray}, {Greco}, {Green}, {Green}, {Gretarsson}, {Groot},
  {Grote}, {Grunewald}, {Gruning}, {Guidi}, {Gulati}, {Guo}, {Gupta}, {Gupta},
  {Gustafson}, {Gustafson}, {Haegel}, {Halim}, {Hall}, {Hall}, {Hamilton},
  {Hammond}, {Haney}, {Hanke}, {Hanks}, {Hanna}, {Hannam}, {Hannuksela},
  {Hanson}, {Hardwick}, {Haris}, {Harms}, {Harry}, {Harry}, {Haster},
  {Haughian}, {Hayes}, {Healy}, {Heidmann}, {Heintze}, {Heitmann}, {Hello},
  {Hemming}, {Hendry}, {Heng}, {Hennig}, {Heptonstall}, {Vivanco}, {Heurs},
  {Hild}, {Hinderer}, {Hoak}, {Hochheim}, {Hofman}, {Holgado}, {Holland},
  {Holt}, {Holz}, {Hopkins}, {Horst}, {Hough}, {Howell}, {Hoy}, {Hreibi},
  {Huerta}, {Huet}, {Hughey}, {Hulko}, {Husa}, {Huttner}, {Huynh-Dinh},
  {Idzkowski}, {Iess}, {Ingram}, {Inta}, {Intini}, {Irwin}, {Isa}, {Isac},
  {Isi}, {Iyer}, {Izumi}, {Jacqmin}, {Jadhav}, {Jani}, {Janthalur},
  {Jaranowski}, {Jenkins}, {Jiang}, {Johnson}, {Johnson-McDaniel}, {Jones},
  {Jones}, {Jones}, {Jonker}, {Ju}, {Junker}, {Kalaghatgi}, {Kalogera},
  {Kamai}, {Kandhasamy}, {Kang}, {Kanner}, {Kapadia}, {Karki}, {Karvinen},
  {Kashyap}, {Kasprzack}, {Katsanevas}, {Katsavounidis}, {Katzman}, {Kaufer},
  {Kawabe}, {Keerthana}, {K{\'e}f{\'e}lian}, {Keitel}, {Kennedy}, {Key},
  {Khalili}, {Khan}, {Khan}, {Khan}, {Khan}, {Khazanov}, {Khursheed},
  {Kijbunchoo}, {Kim}, {Kim}, {Kim}, {Kim}, {Kim}, {Kim}, {Kimball}, {King},
  {King}, {Kinley-Hanlon}, {Kirchhoff}, {Kissel}, {Kleybolte}, {Klika},
  {Klimenko}, {Knowles}, {Koch}, {Koehlenbeck}, {Koekoek}, {Koley},
  {Kondrashov}, {Kontos}, {Koper}, {Korobko}, {Korth}, {Kowalska}, {Kozak},
  {Kringel}, {Krishnendu}, {Kr{\'o}lak}, {Kuehn}, {Kumar}, {Kumar}, {Kumar},
  {Kumar}, {Kuo}, {Kutynia}, {Kwang}, {Lackey}, {Lai}, {Lam}, {Landry}, {Lane},
  {Lang}, {Lange}, {Lantz}, {Lanza}, {Lartaux-Vollard}, {Lasky}, {Laxen},
  {Lazzarini}, {Lazzaro}, {Leaci}, {Leavey}, {Lecoeuche}, {Lee}, {Lee}, {Lee},
  {Lee}, {Lee}, {Lee}, {Lehmann}, {Lenon}, {Leroy}, {Letendre}, {Levin}, {Li},
  {Li}, {Li}, {Li}, {Lin}, {Linde}, {Linker}, {Littenberg}, {Liu}, {Liu}, {Lo},
  {Lockerbie}, {London}, {Longo}, {Lorenzini}, {Loriette}, {Lormand},
  {Losurdo}, {Lough}, {Lousto}, {Lovelace}, {Lower}, {L{\"u}ck}, {Lumaca},
  {Lundgren}, {Lynch}, {Ma}, {Macas}, {Macfoy}, {MacInnis}, {Macleod},
  {Macquet}, {Maga{\~n}a-Sandoval}, {Zertuche}, {Magee}, {Majorana},
  {Maksimovic}, {Malik}, {Man}, {Mandic}, {Mangano}, {Mansell}, {Manske},
  {Mantovani}, {Marchesoni}, {Marion}, {M{\'a}rka}, {M{\'a}rka}, {Markakis},
  {Markosyan}, {Markowitz}, {Maros}, {Marquina}, {Marsat}, {Martelli},
  {Martin}, {Martin}, {Martynov}, {Mason}, {Massera}, {Masserot}, {Massinger},
  {Masso-Reid}, {Mastrogiovanni}, {Matas}, {Matichard}, {Matone}, {Mavalvala},
  {Mazumder}, {McCann}, {McCarthy}, {McClelland}, {McCormick}, {McCuller},
  {McGuire}, {McIver}, {McManus}, {McRae}, {McWilliams}, {Meacher}, {Meadors},
  {Mehmet}, {Mehta}, {Meidam}, {Melatos}, {Mendell}, {Mercer}, {Mereni},
  {Merilh}, {Merzougui}, {Meshkov}, {Messenger}, {Messick}, {Metzdorff},
  {Meyers}, {Miao}, {Michel}, {Middleton}, {Mikhailov}, {Milano}, {Miller},
  {Miller}, {Millhouse}, {Mills}, {Milovich-Goff}, {Minazzoli}, {Minenkov},
  {Mishkin}, {Mishra}, {Mistry}, {Mitra}, {Mitrofanov}, {Mitselmakher},
  {Mittleman}, {Mo}, {Moffa}, {Mogushi}, {Mohapatra}, {Montani}, {Moore},
  {Moraru}, {Moreno}, {Morisaki}, {Mours}, {Mow-Lowry}, {Mukherjee},
  {Mukherjee}, {Mukherjee}, {Mukund}, {Mullavey}, {Munch}, {Mu{\~n}iz},
  {Muratore}, {Murray}, {Nagar}, {Nardecchia}, {Naticchioni}, {Nayak},
  {Neilson}, {Nelemans}, {Nelson}, {Nery}, {Neunzert}, {Ng}, {Ng}, {Nguyen},
  {Nichols}, {Nielsen}, {Nissanke}, {Nitz}, {Nocera}, {North}, {Nuttall},
  {Obergaulinger}, {Oberling}, {O'Brien}, {O'Dea}, {Ogin}, {Oh}, {Oh}, {Ohme},
  {Ohta}, {Okada}, {Oliver}, {Oppermann}, {Oram}, {O'Reilly}, {Ormiston},
  {Ortega}, {O'Shaughnessy}, {Ossokine}, {Ottaway}, {Overmier}, {Owen}, {Pace},
  {Pagano}, {Page}, {Pai}, {Pai}, {Palamos}, {Palashov}, {Palomba},
  {Pal-Singh}, {Pan}, {Pang}, {Pang}, {Pankow}, {Pannarale}, {Pant},
  {Paoletti}, {Paoli}, {Parida}, {Parker}, {Pascucci}, {Pasqualetti},
  {Passaquieti}, {Passuello}, {Patil}, {Patricelli}, {Pearlstone}, {Pedersen},
  {Pedraza}, {Pedurand}, {Pele}, {Penn}, {Perez}, {Perreca}, {Pfeiffer},
  {Phelps}, {Phukon}, {Piccinni}, {Pichot}, {Piergiovanni}, {Pillant},
  {Pinard}, {Pirello}, {Pitkin}, {Poggiani}, {Pong}, {Ponrathnam}, {Popolizio},
  {Porter}, {Powell}, {Prajapati}, {Prasad}, {Prasai}, {Prasanna}, {Pratten},
  {Prestegard}, {Privitera}, {Prodi}, {Prokhorov}, {Puncken}, {Punturo},
  {Puppo}, {P{\"u}rrer}, {Qi}, {Quetschke}, {Quinonez}, {Quintero},
  {Quitzow-James}, {Raab}, {Radkins}, {Radulescu}, {Raffai}, {Raja}, {Rajan},
  {Rajbhandari}, {Rakhmanov}, {Ramirez}, {Ramos-Buades}, {Rana}, {Rao},
  {Rapagnani}, {Raymond}, {Razzano}, {Read}, {Regimbau}, {Rei}, {Reid},
  {Reitze}, {Ren}, {Ricci}, {Richardson}, {Richardson}, {Ricker}, {Riles},
  {Rizzo}, {Robertson}, {Robie}, {Robinet}, {Rocchi}, {Rolland}, {Rollins},
  {Roma}, {Romanelli}, {Romano}, {Romel}, {Romie}, {Rose}, {Rosi{\'n}ska},
  {Rosofsky}, {Ross}, {Rowan}, {R{\"u}diger}, {Ruggi}, {Rutins}, {Ryan},
  {Sachdev}, {Sadecki}, {Sakellariadou}, {Salconi}, {Saleem}, {Samajdar},
  {Sammut}, {Sanchez}, {Sanchez}, {Sanchis-Gual}, {Sandberg}, {Sanders},
  {Santiago}, {Sarin}, {Sassolas}, {Sathyaprakash}, {Saulson}, {Sauter},
  {Savage}, {Schale}, {Scheel}, {Scheuer}, {Schmidt}, {Schnabel}, {Schofield},
  {Sch{\"o}nbeck}, {Schreiber}, {Schulte}, {Schutz}, {Schwalbe}, {Scott},
  {Scott}, {Seidel}, {Sellers}, {Sengupta}, {Sennett}, {Sentenac}, {Sequino},
  {Sergeev}, {Setyawati}, {Shaddock}, {Shaffer}, {Shahriar}, {Shaner}, {Shao},
  {Sharma}, {Shawhan}, {Shen}, {Shink}, {Shoemaker}, {Shoemaker},
  {ShyamSundar}, {Siellez}, {Sieniawska}, {Sigg}, {Silva}, {Singer}, {Singh},
  {Singhal}, {Sintes}, {Sitmukhambetov}, {Skliris}, {Slagmolen},
  {Slaven-Blair}, {Smith}, {Smith}, {Somala}, {Son}, {Sorazu}, {Sorrentino},
  {Souradeep}, {Sowell}, {Spencer}, {Srivastava}, {Srivastava}, {Staats},
  {Stachie}, {Standke}, {Steer}, {Steinke}, {Steinlechner}, {Steinlechner},
  {Steinmeyer}, {Stevenson}, {Stocks}, {Stone}, {Stops}, {Strain}, {Stratta},
  {Strigin}, {Strunk}, {Sturani}, {Stuver}, {Sudhir}, {Summerscales}, {Sun},
  {Sunil}, {Suresh}, {Sutton}, {Swinkels}, {Szczepa{\'n}czyk}, {Tacca}, {Tait},
  {Talbot}, {Talukder}, {Tanner}, {T{\'a}pai}, {Taracchini}, {Tasson},
  {Taylor}, {Thies}, {Thomas}, {Thomas}, {Thondapu}, {Thorne}, {Thrane},
  {Tiwari}, {Tiwari}, {Tiwari}, {Toland}, {Tonelli}, {Tornasi},
  {Torres-Forn{\'e}}, {Torrie}, {T{\"o}yr{\"a}}, {Travasso}, {Traylor},
  {Tringali}, {Trovato}, {Trozzo}, {Trudeau}, {Tsang}, {Tse}, {Tso}, {Tsukada},
  {Tsuna}, {Tuyenbayev}, {Ueno}, {Ugolini}, {Unnikrishnan}, {Urban}, {Usman},
  {Vahlbruch}, {Vajente}, {Valdes}, {van Bakel}, {van Beuzekom}, {van den
  Brand}, {Van Den Broeck}, {Vander-Hyde}, {van Heijningen}, {van der Schaaf},
  {van Veggel}, {Vardaro}, {Varma}, {Vass}, {Vas{\'u}th}, {Vecchio},
  {Vedovato}, {Veitch}, {Veitch}, {Venkateswara}, {Venugopalan}, {Verkindt},
  {Vetrano}, {Vicer{\'e}}, {Viets}, {Vine}, {Vinet}, {Vitale}, {Vo}, {Vocca},
  {Vorvick}, {Vyatchanin}, {Wade}, {Wade}, {Wade}, {Wald}, {Walet}, {Walker},
  {Wallace}, {Walsh}, {Wang}, {Wang}, {Wang}, {Wang}, {Wang}, {Ward}, {Warden},
  {Warner}, {Was}, {Watchi}, {Weaver}, {Wei}, {Weinert}, {Weinstein}, {Weiss},
  {Wellmann}, {Wen}, {Wessel}, {We{\ss}els}, {Westhouse}, {Wette}, {Whelan},
  {Whiting}, {Whittle}, {Wilken}, {Williams}, {Williamson}, {Willis}, {Willke},
  {Wimmer}, {Winkler}, {Wipf}, {Wittel}, {Woan}, {Woehler}, {Wofford},
  {Worden}, {Wright}, {Wu}, {Wysocki}, {Xiao}, {Yamamoto}, {Yancey}, {Yang},
  {Yap}, {Yazback}, {Yeeles}, {Yu}, {Yu}, {Yuen}, {Yvert}, {Zadro{\.Z}ny},
  {Zanolin}, {Zelenova}, {Zendri}, {Zevin}, {Zhang}, {Zhang}, {Zhang}, {Zhao},
  {Zhou}, {Zhou}, {Zhu}, {Zimmerman}, {Zucker}, {Zweizig}, {LIGO Scientific
  Collaboration}, \& {Virgo Collaboration}}]{Abbott2019}
---. 2019, \prd, 100, 104036, \dodoi{10.1103/PhysRevD.100.104036}

\bibitem[{{Abbott} {et~al.}(2021){Abbott}, {Abbott}, {Abraham}, {Acernese},
  {Ackley}, {Adams}, {Adams}, {Adhikari}, {Adya}, {Affeldt}, {Agathos},
  {Agatsuma}, {Aggarwal}, {Aguiar}, {Aiello}, {Ain}, {Ajith}, {Akcay}, {Allen},
  {Allocca}, {Altin}, {Amato}, {Anand}, {Ananyeva}, {Anderson}, {Anderson},
  {Angelova}, {Ansoldi}, {Antelis}, {Antier}, {Appert}, {Arai}, {Araya},
  {Areeda}, {Ar{\`e}ne}, {Arnaud}, {Aronson}, {Arun}, {Asali}, {Ascenzi},
  {Ashton}, {Aston}, {Astone}, {Aubin}, {Aufmuth}, {AultONeal}, {Austin},
  {Avendano}, {Babak}, {Badaracco}, {Bader}, {Bae}, {Baer}, {Bagnasco},
  {Baird}, {Ball}, {Ballardin}, {Ballmer}, {Bals}, {Balsamo}, {Baltus},
  {Banagiri}, {Bankar}, {Bankar}, {Barayoga}, {Barbieri}, {Barish}, {Barker},
  {Barneo}, {Barnum}, {Barone}, {Barr}, {Barsotti}, {Barsuglia}, {Barta},
  {Bartlett}, {Bartos}, {Bassiri}, {Basti}, {Bawaj}, {Bayley}, {Bazzan},
  {Becher}, {B{\'e}csy}, {Bedakihale}, {Bejger}, {Belahcene}, {Beniwal},
  {Benjamin}, {Benkel}, {Bennett}, {Bentley}, {Bergamin}, {Berger}, {Bergmann},
  {Bernuzzi}, {Berry}, {Bersanetti}, {Bertolini}, {Betzwieser}, {Bhandare},
  {Bhandari}, {Bhattacharjee}, {Bidler}, {Bilenko}, {Billingsley}, {Birney},
  {Birnholtz}, {Biscans}, {Bischi}, {Biscoveanu}, {Bisht}, {Bitossi},
  {Bizouard}, {Blackburn}, {Blackman}, {Blair}, {Blair}, {Blair}, {Blanch},
  {Bobba}, {Bode}, {Boer}, {Boetzel}, {Bogaert}, {Boldrini}, {Bondu},
  {Bonilla}, {Bonnand}, {Booker}, {Boom}, {Borhanian}, {Bork}, {Boschi},
  {Bose}, {Bose}, {Bossilkov}, {Boudart}, {Bouffanais}, {Bozzi}, {Bradaschia},
  {Brady}, {Bramley}, {Branchesi}, {Brau}, {Breschi}, {Briant}, {Briggs},
  {Brighenti}, {Brillet}, {Brinkmann}, {Brito}, {Brockill}, {Brooks}, {Brooks},
  {Brown}, {Brunett}, {Bruno}, {Bruntz}, {Buikema}, {Bulik}, {Bulten},
  {Buonanno}, {Buskulic}, {Byer}, {Cabero}, {Cadonati}, {Caesar}, {Cagnoli},
  {Cahillane}, {Bustillo}, {Callaghan}, {Callister}, {Calloni}, {Camp},
  {Canepa}, {Cannon}, {Cao}, {Cao}, {Carapella}, {Carbognani}, {Carney},
  {Carpinelli}, {Carullo}, {Carver}, {Diaz}, {Casentini}, {Caudill},
  {Cavagli{\`a}}, {Cavalier}, {Cavalieri}, {Cella}, {Cerd{\'a}-Dur{\'a}n},
  {Cesarini}, {Chaibi}, {Chakravarti}, {Chan}, {Chan}, {Chandra}, {Chanial},
  {Chao}, {Charlton}, {Chase}, {Chassande-Mottin}, {Chatterjee}, {Chaturvedi},
  {Chatziioannou}, {Chen}, {Chen}, {Chen}, {Chen}, {Cheng}, {Cheong}, {Chia},
  {Chiadini}, {Chierici}, {Chincarini}, {Chiummo}, {Cho}, {Cho}, {Cho},
  {Choate}, {Christensen}, {Chu}, {Chua}, {Chung}, {Chung}, {Ciani},
  {Ciecielag}, {Cie{\'s}lar}, {Cifaldi}, {Ciobanu}, {Ciolfi}, {Cipriano},
  {Cirone}, {Clara}, {Clark}, {Clark}, {Clarke}, {Clearwater}, {Clesse},
  {Cleva}, {Coccia}, {Cohadon}, {Cohen}, {Colleoni}, {Collette}, {Collins},
  {Colpi}, {Constancio}, {Conti}, {Cooper}, {Corban}, {Corbitt},
  {Cordero-Carri{\'o}n}, {Corezzi}, {Corley}, {Cornish}, {Corre}, {Corsi},
  {Cortese}, {Costa}, {Cotesta}, {Coughlin}, {Coughlin}, {Coulon},
  {Countryman}, {Couvares}, {Covas}, {Coward}, {Cowart}, {Coyne}, {Coyne},
  {Creighton}, {Creighton}, {Croquette}, {Crowder}, {Cudell}, {Cullen},
  {Cumming}, {Cummings}, {Cunningham}, {Cuoco}, {Curylo}, {Canton},
  {D{\'a}lya}, {Dana}, {DaneshgaranBajastani}, {D'Angelo}, {Danilishin},
  {D'Antonio}, {Danzmann}, {Darsow-Fromm}, {Dasgupta}, {Datrier}, {Dattilo},
  {Dave}, {Davier}, {Davies}, {Davis}, {Daw}, {Dean}, {DeBra}, {Deenadayalan},
  {Degallaix}, {De Laurentis}, {Del{\'e}glise}, {Del Favero}, {De Lillo}, {De
  Lillo}, {Del Pozzo}, {DeMarchi}, {De Matteis}, {D'Emilio}, {Demos}, {Denker},
  {Dent}, {Depasse}, {De Pietri}, {De Rosa}, {De Rossi}, {DeSalvo}, {de
  Varona}, {Dhani}, {Dhurandhar}, {D{\'\i}az}, {Diaz-Ortiz}, {Didio},
  {Dietrich}, {Di Fiore}, {DiFronzo}, {Di Giorgio}, {Di Giovanni}, {Di
  Giovanni}, {Di Girolamo}, {Di Lieto}, {Ding}, {Di Pace}, {Di Palma}, {Di
  Renzo}, {Divakarla}, {Dmitriev}, {Doctor}, {D'Onofrio}, {Donovan}, {Dooley},
  {Doravari}, {Dorrington}, {Downes}, {Drago}, {Driggers}, {Du}, {Ducoin},
  {Dudi}, {Dupej}, {Durante}, {D'Urso}, {Duverne}, {Dwyer}, {Easter},
  {Eddolls}, {Edelman}, {Edo}, {Edy}, {Effler}, {Eichholz}, {Eikenberry},
  {Eisenmann}, {Eisenstein}, {Ejlli}, {Errico}, {Essick}, {Estell{\'e}s},
  {Estevez}, {Etienne}, {Etzel}, {Evans}, {Evans}, {Ewing}, {Fafone}, {Fair},
  {Fairhurst}, {Fan}, {Farah}, {Farinon}, {Farr}, {Farr}, {Fauchon-Jones},
  {Favata}, {Fays}, {Fazio}, {Feicht}, {Fejer}, {Feng}, {Fenyvesi}, {Ferguson},
  {Fernandez-Galiana}, {Ferrante}, {Ferreira}, {Fidecaro}, {Figura}, {Fiori},
  {Fiorucci}, {Fishbach}, {Fisher}, {Fishner}, {Fittipaldi}, {Fitz-Axen},
  {Fiumara}, {Flaminio}, {Floden}, {Flynn}, {Fong}, {Font}, {Forsyth},
  {Fournier}, {Frasca}, {Frasconi}, {Frei}, {Freise}, {Frey}, {Frey},
  {Fritschel}, {Frolov}, {Fronz{\'e}}, {Fulda}, {Fyffe}, {Gabbard}, {Gadre},
  {Gaebel}, {Gair}, {Gais}, {Galaudage}, {Gamba}, {Ganapathy}, {Ganguly},
  {Gaonkar}, {Garaventa}, {Garc{\'\i}a-Quir{\'o}s}, {Garufi}, {Gateley},
  {Gaudio}, {Gayathri}, {Gemme}, {Gennai}, {George}, {George}, {George},
  {Gergely}, {Ghonge}, {Ghosh}, {Ghosh}, {Ghosh}, {Giacomazzo}, {Giacoppo},
  {Giaime}, {Giardina}, {Gibson}, {Gier}, {Gill}, {Giri}, {Glanzer}, {Gleckl},
  {Godwin}, {Goetz}, {Goetz}, {Gohlke}, {Goncharov}, {Gonz{\'a}lez},
  {Gopakumar}, {Gossan}, {Gosselin}, {Gouaty}, {Grace}, {Grado}, {Granata},
  {Granata}, {Grant}, {Gras}, {Grassia}, {Gray}, {Gray}, {Greco}, {Green},
  {Green}, {Gretarsson}, {Griggs}, {Grignani}, {Grimaldi}, {Grimes}, {Grimm},
  {Grote}, {Grunewald}, {Gruning}, {Guerrero}, {Guidi}, {Guimaraes},
  {Guix{\'e}}, {Gulati}, {Guo}, {Gupta}, {Gupta}, {Gupta}, {Gustafson},
  {Gustafson}, {Guzman}, {Haegel}, {Halim}, {Hall}, {Hamilton}, {Hammond},
  {Haney}, {Hanke}, {Hanks}, {Hanna}, {Hannam}, {Hannuksela}, {Hannuksela},
  {Hansen}, {Hansen}, {Hanson}, {Harder}, {Hardwick}, {Haris}, {Harms},
  {Harry}, {Harry}, {Hartwig}, {Hasskew}, {Haster}, {Haughian}, {Hayes},
  {Healy}, {Heidmann}, {Heintze}, {Heinze}, {Heinzel}, {Heitmann}, {Hellman},
  {Hello}, {Helmling-Cornell}, {Hemming}, {Hendry}, {Heng}, {Hennes}, {Hennig},
  {Hennig}, {Vivanco}, {Heurs}, {Hild}, {Hill}, {Hines}, {Hochheim}, {Hofgard},
  {Hofman}, {Hohmann}, {Holgado}, {Holland}, {Hollows}, {Holmes}, {Holt},
  {Holz}, {Hopkins}, {Horst}, {Hough}, {Howell}, {Hoy}, {Hoyland}, {Huang},
  {H{\"u}bner}, {Huddart}, {Huerta}, {Hughey}, {Hui}, {Husa}, {Huttner},
  {Hutzler}, {Huxford}, {Huynh-Dinh}, {Idzkowski}, {Iess}, {Imperato},
  {Inchauspe}, {Ingram}, {Intini}, {Isi}, {Iyer}, {JaberianHamedan}, {Jacqmin},
  {Jadhav}, {Jadhav}, {James}, {Jani}, {Janssens}, {Janthalur}, {Jaranowski},
  {Jariwala}, {Jaume}, {Jenkins}, {Jeunon}, {Jiang}, {Johns},
  {Johnson-McDaniel}, {Jones}, {Jones}, {Jones}, {Jones}, {Jones}, {Jonker},
  {Ju}, {Junker}, {Kalaghatgi}, {Kalogera}, {Kamai}, {Kandhasamy}, {Kang},
  {Kanner}, {Kapadia}, {Kapasi}, {Karathanasis}, {Karki}, {Kashyap},
  {Kasprzack}, {Kastaun}, {Katsanevas}, {Katsavounidis}, {Katzman}, {Kawabe},
  {K{\'e}f{\'e}lian}, {Keitel}, {Key}, {Khadka}, {Khalili}, {Khan}, {Khan},
  {Khazanov}, {Khetan}, {Khursheed}, {Kijbunchoo}, {Kim}, {Kim}, {Kim}, {Kim},
  {Kim}, {Kim}, {Kimball}, {King}, {Kinley-Hanlon}, {Kirchhoff}, {Kissel},
  {Kleybolte}, {Klimenko}, {Knowles}, {Knyazev}, {Koch}, {Koehlenbeck},
  {Koekoek}, {Koley}, {Kolstein}, {Komori}, {Kondrashov}, {Kontos}, {Koper},
  {Korobko}, {Korth}, {Kovalam}, {Kozak}, {Kr{\"a}mer}, {Kringel},
  {Krishnendu}, {Kr{\'o}lak}, {Kuehn}, {Kumar}, {Kumar}, {Kumar}, {Kumar},
  {Kuns}, {Kwang}, {Lackey}, {Laghi}, {Lalande}, {Lam}, {Lamberts}, {Landry},
  {Lane}, {Lang}, {Lange}, {Lantz}, {Lanza}, {La Rosa}, {Lartaux-Vollard},
  {Lasky}, {Laxen}, {Lazzarini}, {Lazzaro}, {Leaci}, {Leavey}, {Lecoeuche},
  {Lee}, {Lee}, {Lee}, {Lee}, {Lehmann}, {Leon}, {Leroy}, {Letendre}, {Levin},
  {Li}, {Li}, {Li}, {Li}, {Li}, {Linde}, {Linker}, {Linley}, {Littenberg},
  {Liu}, {Liu}, {Llorens-Monteagudo}, {Lo}, {Lockwood}, {London}, {Longo},
  {Lorenzini}, {Loriette}, {Lormand}, {Losurdo}, {Lough}, {Lousto}, {Lovelace},
  {L{\"u}ck}, {Lumaca}, {Lundgren}, {Ma}, {Macas}, {MacInnis}, {Macleod},
  {MacMillan}, {Macquet}, {Hernandez}, {Maga{\~n}a-Sandoval}, {Magazz{\`u}},
  {Magee}, {Majorana}, {Maksimovic}, {Maliakal}, {Malik}, {Man}, {Mandic},
  {Mangano}, {Mansell}, {Manske}, {Mantovani}, {Mapelli}, {Marchesoni},
  {Marion}, {M{\'a}rka}, {M{\'a}rka}, {Markakis}, {Markosyan}, {Markowitz},
  {Maros}, {Marquina}, {Marsat}, {Martelli}, {Martin}, {Martin}, {Martinez},
  {Martinez}, {Martynov}, {Masalehdan}, {Mason}, {Massera}, {Masserot},
  {Massinger}, {Masso-Reid}, {Mastrogiovanni}, {Matas}, {Mateu-Lucena},
  {Matichard}, {Matiushechkina}, {Mavalvala}, {Maynard}, {McCann}, {McCarthy},
  {McClelland}, {McCormick}, {McCuller}, {McGuire}, {McIsaac}, {McIver},
  {McManus}, {McRae}, {McWilliams}, {Meacher}, {Meadors}, {Mehmet}, {Mehta},
  {Melatos}, {Melchor}, {Mendell}, {Menendez-Vazquez}, {Mercer}, {Mereni},
  {Merfeld}, {Merilh}, {Merritt}, {Merzougui}, {Meshkov}, {Messenger},
  {Messick}, {Metzdorff}, {Meyers}, {Meylahn}, {Mhaske}, {Miani}, {Miao},
  {Michaloliakos}, {Michel}, {Middleton}, {Milano}, {Miller}, {Millhouse},
  {Mills}, {Milotti}, {Milovich-Goff}, {Minazzoli}, {Minenkov}, {Mir},
  {Mishkin}, {Mishra}, {Mistry}, {Mitra}, {Mitrofanov}, {Mitselmakher},
  {Mittleman}, {Mo}, {Mogushi}, {Mohapatra}, {Mohite}, {Molina}, {Molina-Ruiz},
  {Mondin}, {Montani}, {Moore}, {Moraru}, {Morawski}, {Moreno}, {Morisaki},
  {Mours}, {Mow-Lowry}, {Mozzon}, {Muciaccia}, {Mukherjee}, {Mukherjee},
  {Mukherjee}, {Mukherjee}, {Mukund}, {Mullavey}, {Munch}, {Mu{\~n}iz},
  {Murray}, {Nadji}, {Nagar}, {Nardecchia}, {Naticchioni}, {Nayak}, {Neil},
  {Neilson}, {Nelemans}, {Nelson}, {Nery}, {Neunzert}, {Ng}, {Ng}, {Nguyen},
  {Nguyen}, {Nguyen}, {Nichols}, {Nissanke}, {Nocera}, {Noh}, {North},
  {Nothard}, {Nuttall}, {Oberling}, {O'Brien}, {O'Dell}, {Oganesyan}, {Ogin},
  {Oh}, {Oh}, {Ohme}, {Ohta}, {Okada}, {Olivetto}, {Oppermann}, {Oram},
  {O'Reilly}, {Ormiston}, {Ormsby}, {Ortega}, {O'Shaughnessy}, {Ossokine},
  {Osthelder}, {Ottaway}, {Overmier}, {Owen}, {Pace}, {Pagano}, {Page},
  {Pagliaroli}, {Pai}, {Pai}, {Palamos}, {Palashov}, {Palomba}, {Pan}, {Panda},
  {Pang}, {Pankow}, {Pannarale}, {Pant}, {Paoletti}, {Paoli}, {Paolone},
  {Parker}, {Pascucci}, {Pasqualetti}, {Passaquieti}, {Passuello}, {Patel},
  {Patricelli}, {Payne}, {Pechsiri}, {Pedraza}, {Pegoraro}, {Pele}, {Penn},
  {Perego}, {Perez}, {P{\'e}rigois}, {Perreca}, {Perri{\`e}s}, {Petermann},
  {Petterson}, {Pfeiffer}, {Pham}, {Phukon}, {Piccinni}, {Pichot},
  {Piendibene}, {Piergiovanni}, {Pierini}, {Pierro}, {Pillant}, {Pilo},
  {Pinard}, {Pinto}, {Piotrzkowski}, {Pirello}, {Pitkin}, {Placidi},
  {Plastino}, {Pluchar}, {Poggiani}, {Polini}, {Pong}, {Ponrathnam},
  {Popolizio}, {Porter}, {Poverman}, {Powell}, {Pracchia}, {Prajapati},
  {Prasai}, {Prasanna}, {Pratten}, {Prestegard}, {Principe}, {Prodi},
  {Prokhorov}, {Prosposito}, {Puecher}, {Punturo}, {Puosi}, {Puppo},
  {P{\"u}rrer}, {Qi}, {Quetschke}, {Quinonez}, {Quitzow-James}, {Raab},
  {Raaijmakers}, {Radkins}, {Radulesco}, {Raffai}, {Rafferty}, {Rail}, {Raja},
  {Rajan}, {Rajbhandari}, {Rakhmanov}, {Ramirez}, {Ramirez}, {Ramos-Buades},
  {Rana}, {Rao}, {Rapagnani}, {Rapol}, {Ratto}, {Raymond}, {Razzano}, {Read},
  {Regimbau}, {Rei}, {Reid}, {Reitze}, {Rettegno}, {Ricci}, {Richardson},
  {Richardson}, {Richardson}, {Ricker}, {Riemenschneider}, {Riles}, {Rizzo},
  {Robertson}, {Robinet}, {Rocchi}, {Rocha}, {Rodriguez}, {Rodriguez-Soto},
  {Rolland}, {Rollins}, {Roma}, {Romanelli}, {Romano}, {Romel}, {Romero},
  {Romero-Shaw}, {Romie}, {Ronchini}, {Rose}, {Rose}, {Rose}, {Rosi{\'n}ska},
  {Rosofsky}, {Ross}, {Rowan}, {Rowlinson}, {Roy}, {Roy}, {Ruggi}, {Ryan},
  {Sachdev}, {Sadecki}, {Sakellariadou}, {Salafia}, {Salconi}, {Saleem},
  {Samajdar}, {Sanchez}, {Sanchez}, {Sanchez}, {Sanchis-Gual}, {Sanders},
  {Santiago}, {Santos}, {Saravanan}, {Sarin}, {Sassolas}, {Sathyaprakash},
  {Sauter}, {Savage}, {Savant}, {Sawant}, {Sayah}, {Schaetzl}, {Schale},
  {Scheel}, {Scheuer}, {Schindler-Tyka}, {Schmidt}, {Schnabel}, {Schofield},
  {Sch{\"o}nbeck}, {Schreiber}, {Schulte}, {Schutz}, {Schwarm}, {Schwartz},
  {Scott}, {Scott}, {Seglar-Arroyo}, {Seidel}, {Sellers}, {Sengupta},
  {Sennett}, {Sentenac}, {Sequino}, {Sergeev}, {Setyawati}, {Shaffer},
  {Shahriar}, {Sharifi}, {Sharma}, {Sharma}, {Shawhan}, {Shen}, {Shikauchi},
  {Shink}, {Shoemaker}, {Shoemaker}, {Shukla}, {ShyamSundar}, {Sieniawska},
  {Sigg}, {Singer}, {Singh}, {Singh}, {Singha}, {Singhal}, {Sintes}, {Sipala},
  {Skliris}, {Slagmolen}, {Slaven-Blair}, {Smetana}, {Smith}, {Smith},
  {Somala}, {Son}, {Soni}, {Sorazu}, {Sordini}, {Sorrentino}, {Sorrentino},
  {Soulard}, {Souradeep}, {Sowell}, {Spencer}, {Spera}, {Srivastava},
  {Srivastava}, {Staats}, {Stachie}, {Steer}, {Steinhoff}, {Steinke},
  {Steinlechner}, {Steinlechner}, {Steinmeyer}, {Stolle-McAllister}, {Stops},
  {Stover}, {Strain}, {Stratta}, {Strunk}, {Sturani}, {Stuver}, {S{\"u}dbeck},
  {Sudhagar}, {Sudhir}, {Suh}, {Summerscales}, {Sun}, {Sun}, {Sunil}, {Sur},
  {Suresh}, {Sutton}, {Swinkels}, {Szczepa{\'n}czyk}, {Tacca}, {Tait},
  {Talbot}, {Tanasijczuk}, {Tanner}, {Tao}, {Tapia}, {Martin}, {Tasson},
  {Taylor}, {Tenorio}, {Terkowski}, {Thirugnanasambandam}, {Thomas}, {Thomas},
  {Thomas}, {Thompson}, {Thondapu}, {Thorne}, {Thrane}, {Tiwari}, {Tiwari},
  {Tiwari}, {Toland}, {Tolley}, {Tonelli}, {Tornasi}, {Torres-Forn{\'e}},
  {Torrie}, {e Melo}, {T{\"o}yr{\"a}}, {Tran}, {Trapananti}, {Travasso},
  {Traylor}, {Tringali}, {Tripathee}, {Trovato}, {Trudeau}, {Tsai}, {Tsang},
  {Tse}, {Tso}, {Tsukada}, {Tsuna}, {Tsutsui}, {Turconi}, {Ubhi}, {Udall},
  {Ueno}, {Ugolini}, {Unnikrishnan}, {Urban}, {Usman}, {Utina}, {Vahlbruch},
  {Vajente}, {Vajpeyi}, {Valdes}, {Valentini}, {Valsan}, {van Bakel}, {van
  Beuzekom}, {van den Brand}, {Van Den Broeck}, {Vander-Hyde}, {van der
  Schaaf}, {van Heijningen}, {Vardaro}, {Vargas}, {Varma}, {Vass},
  {Vas{\'u}th}, {Vecchio}, {Vedovato}, {Veitch}, {Veitch}, {Venkateswara},
  {Venneberg}, {Venugopalan}, {Verkindt}, {Verma}, {Veske}, {Vetrano},
  {Vicer{\'e}}, {Viets}, {Vijaykumar}, {Villa-Ortega}, {Vinet}, {Vitale}, {Vo},
  {Vocca}, {Vorvick}, {Vyatchanin}, {Wade}, {Wade}, {Wade}, {Wald}, {Walet},
  {Walker}, {Wallace}, {Wallace}, {Walsh}, {Wang}, {Wang}, {Wang}, {Wang},
  {Ward}, {Warner}, {Was}, {Washington}, {Watchi}, {Weaver}, {Wei}, {Weinert},
  {Weinstein}, {Weiss}, {Wellmann}, {Wen}, {We{\ss}els}, {Westhouse}, {Wette},
  {Whelan}, {White}, {White}, {Whiting}, {Whittle}, {Wilken}, {Williams},
  {Williams}, {Williamson}, {Willis}, {Willke}, {Wilson}, {Wimmer}, {Winkler},
  {Wipf}, {Woan}, {Woehler}, {Wofford}, {Wong}, {Wrangel}, {Wright}, {Wu},
  {Wysocki}, {Xiao}, {Yamamoto}, {Yang}, {Yang}, {Yang}, {Yap}, {Yeeles},
  {Yoon}, {Yu}, {Yu}, {Yuen}, {Zadro{\.Z}ny}, {Zanolin}, {Zelenova}, {Zendri},
  {Zevin}, {Zhang}, {Zhang}, {Zhang}, {Zhang}, {Zhao}, {Zhao}, {Zhou}, {Zhou},
  {Zhu}, {Zimmerman}, {Zucker}, {Zweizig}, {LIGO Scientific Collaboration}, \&
  {Virgo Collaboration}}]{Abbott2021}
{Abbott}, R., {Abbott}, T.~D., {Abraham}, S., {et~al.} 2021, \prd, 103, 122002,
  \dodoi{10.1103/PhysRevD.103.122002}

\bibitem[{{Abdujabbarov} {et~al.}(2016){Abdujabbarov}, {Amir}, {Ahmedov}, \&
  {Ghosh}}]{Abdujabbarov+2016}
{Abdujabbarov}, A., {Amir}, M., {Ahmedov}, B., \& {Ghosh}, S.~G. 2016, \prd,
  93, 104004, \dodoi{10.1103/PhysRevD.93.104004}

\bibitem[{{Abedi} {et~al.}(2017){Abedi}, {Dykaar}, \& {Afshordi}}]{Abedi+2017}
{Abedi}, J., {Dykaar}, H., \& {Afshordi}, N. 2017, \prd, 96, 082004,
  \dodoi{10.1103/PhysRevD.96.082004}

\bibitem[{{Abramowicz} {et~al.}(2002){Abramowicz}, {Klu{\'z}niak}, \&
  {Lasota}}]{Abramowicz+2002}
{Abramowicz}, M.~A., {Klu{\'z}niak}, W., \& {Lasota}, J.~P. 2002, \aap, 396,
  L31, \dodoi{10.1051/0004-6361:20021645}

\bibitem[{Amorim {et~al.}(2019)Amorim, Baub\"ock, Berger, Brandner, Cl\'enet,
  Coud\'e~du Foresto, de~Zeeuw, Dexter, Duvert, Ebert, Eckart, Eisenhauer,
  F\"orster~Schreiber, Garcia, Gao, Gendron, Genzel, Gillessen, Habibi,
  Haubois, Henning, Hippler, Horrobin, Hubert, Jim\'enez~Rosales, Jocou,
  Kervella, Lacour, Lapeyr\`ere, Le~Bouquin, L\'ena, Ott, Paumard, Perraut,
  Perrin, Pfuhl, Rabien, Rodr\'{\i}guez-Coira, Rousset, Scheithauer, Sternberg,
  Straub, Straubmeier, Sturm, Tacconi, Vincent, von Fellenberg, Waisberg,
  Widmann, Wieprecht, Wiezorrek, \& Yazici}]{PhysRevLett.122.101102}
Amorim, A., Baub\"ock, M., Berger, J.~P., {et~al.} 2019, Phys. Rev. Lett., 122,
  101102, \dodoi{10.1103/PhysRevLett.122.101102}

\bibitem[{{Amorim} {et~al.}(2019){Amorim}, {Baub{\"o}ck}, {Berger}, {Brandner},
  {Cl{\'e}net}, {Coud{\'e} Du Foresto}, {de Zeeuw}, {Dexter}, {Duvert},
  {Ebert}, {Eckart}, {Eisenhauer}, {F{\"o}rster Schreiber}, {Garcia}, {Gao},
  {Gendron}, {Genzel}, {Gillessen}, {Habibi}, {Haubois}, {Henning}, {Hippler},
  {Horrobin}, {Hubert}, {Jim{\'e}nez Rosales}, {Jocou}, {Kervella}, {Lacour},
  {Lapeyr{\`e}re}, {Le Bouquin}, {L{\'e}na}, {Ott}, {Paumard}, {Perraut},
  {Perrin}, {Pfuhl}, {Rabien}, {Rodr{\'\i}guez-Coira}, {Rousset},
  {Scheithauer}, {Sternberg}, {Straub}, {Straubmeier}, {Sturm}, {Tacconi},
  {Vincent}, {von Fellenberg}, {Waisberg}, {Widmann}, {Wieprecht}, {Wiezorrek},
  {Yazici}, \& {Gravity Collaboration}}]{Amorim2019}
{Amorim}, A., {Baub{\"o}ck}, M., {Berger}, J.~P., {et~al.} 2019, \prl, 122,
  101102, \dodoi{10.1103/PhysRevLett.122.101102}

\bibitem[{{An} {et~al.}(2005){An}, {Goss}, {Zhao}, {Hong}, {Roy}, {Rao}, \&
  {Shen}}]{2005ApJ...634L..49A}
{An}, T., {Goss}, W.~M., {Zhao}, J.-H., {et~al.} 2005, \apjl, 634, L49,
  \dodoi{10.1086/498687}

\bibitem[{{Archibald} {et~al.}(2018){Archibald}, {Gusinskaia}, {Hessels},
  {Deller}, {Kaplan}, {Lorimer}, {Lynch}, {Ransom}, \& {Stairs}}]{archibald}
{Archibald}, A.~M., {Gusinskaia}, N.~V., {Hessels}, J. W.~T., {et~al.} 2018,
  \nat, 559, 73, \dodoi{10.1038/s41586-018-0265-1}

\bibitem[{{Ay{\'o}n-Beato} \& {Garc{\'\i}a}(1998)}]{Ayon-Beato+1998}
{Ay{\'o}n-Beato}, E., \& {Garc{\'\i}a}, A. 1998, \prl, 80, 5056,
  \dodoi{10.1103/PhysRevLett.80.5056}

\bibitem[{{Ay{\'o}n-Beato} \& {Garc{\'\i}a}(2000)}]{Ayon-Beato+2000}
---. 2000, Physics Letters B, 493, 149, \dodoi{10.1016/S0370-2693(00)01125-4}

\bibitem[{{Baganoff} {et~al.}(2001){Baganoff}, {Bautz}, {Brandt}, {Chartas},
  {Feigelson}, {Garmire}, {Maeda}, {Morris}, {Ricker}, {Townsley}, \&
  {Walter}}]{2001Natur.413...45B}
{Baganoff}, F.~K., {Bautz}, M.~W., {Brandt}, W.~N., {et~al.} 2001, \nat, 413,
  45, \dodoi{10.1038/35092510}

\bibitem[{{Baker} {et~al.}(2017){Baker}, {Bellini}, {Ferreira}, {Lagos},
  {Noller}, \& {Sawicki}}]{Baker2017}
{Baker}, T., {Bellini}, E., {Ferreira}, P.~G., {et~al.} 2017, \prl, 119,
  251301, \dodoi{10.1103/PhysRevLett.119.251301}

\bibitem[{{Baker} {et~al.}(2015){Baker}, {Psaltis}, \& {Skordis}}]{Baker2015}
{Baker}, T., {Psaltis}, D., \& {Skordis}, C. 2015, \apj, 802, 63,
  \dodoi{10.1088/0004-637X/802/1/63}

\bibitem[{{Ball} {et~al.}(2021){Ball}, {{\"O}zel}, {Christian}, {Chan}, \&
  {Psaltis}}]{Ball+2021}
{Ball}, D., {{\"O}zel}, F., {Christian}, P., {Chan}, C.-K., \& {Psaltis}, D.
  2021, \apj, 917, 8, \dodoi{10.3847/1538-4357/abf8ae}

\bibitem[{{Ball} {et~al.}(2016){Ball}, {{\"O}zel}, {Psaltis}, \&
  {Chan}}]{Ball+2016}
{Ball}, D., {{\"O}zel}, F., {Psaltis}, D., \& {Chan}, C.-k. 2016, \apj, 826,
  77, \dodoi{10.3847/0004-637X/826/1/77}

\bibitem[{{Bambi}(2013)}]{Bambi2013}
{Bambi}, C. 2013, \prd, 87, 084039, \dodoi{10.1103/PhysRevD.87.084039}

\bibitem[{{Barausse} \& {Sotiriou}(2008)}]{Barausse2008}
{Barausse}, E., \& {Sotiriou}, T.~P. 2008, \prl, 101, 099001,
  \dodoi{10.1103/PhysRevLett.101.099001}

\bibitem[{Bardeen(1968)}]{Bardeen1968}
Bardeen, J. 1968, in Proceedings of GR5, Tbilisi, USSR, 174

\bibitem[{{Bardeen}(1973)}]{Bardeen1973}
{Bardeen}, J.~M. 1973, in Black Holes (Les Astres Occlus), 215--239

\bibitem[{{Bertotti} {et~al.}(2003){Bertotti}, {Iess}, \&
  {Tortora}}]{Bertotti2003}
{Bertotti}, B., {Iess}, L., \& {Tortora}, P. 2003, \nat, 425, 374,
  \dodoi{10.1038/nature01997}

\bibitem[{{Bodenner} \& {Will}(2003)}]{Bodenner2003}
{Bodenner}, J., \& {Will}, C.~M. 2003, American Journal of Physics, 71, 770,
  \dodoi{10.1119/1.1570416}

\bibitem[{{Boehle} {et~al.}(2016){Boehle}, {Ghez}, {Sch{\"o}del}, {Meyer},
  {Yelda}, {Albers}, {Martinez}, {Becklin}, {Do}, {Lu}, {Matthews}, {Morris},
  {Sitarski}, \& {Witzel}}]{2016ApJ...830...17B}
{Boehle}, A., {Ghez}, A.~M., {Sch{\"o}del}, R., {et~al.} 2016, \apj, 830, 17,
  \dodoi{10.3847/0004-637X/830/1/17}

\bibitem[{{Bower} {et~al.}(2015){Bower}, {Markoff}, {Dexter}, {Gurwell},
  {Moran}, {Brunthaler}, {Falcke}, {Fragile}, {Maitra}, {Marrone}, {Peck},
  {Rushton}, \& {Wright}}]{2015ApJ...802...69B}
{Bower}, G.~C., {Markoff}, S., {Dexter}, J., {et~al.} 2015, \apj, 802, 69,
  \dodoi{10.1088/0004-637X/802/1/69}

\bibitem[{{Bower} {et~al.}(2019){Bower}, {Dexter}, {Asada}, {Brinkerink},
  {Falcke}, {Ho}, {Inoue}, {Markoff}, {Marrone}, {Matsushita}, {Moscibrodzka},
  {Nakamura}, {Peck}, \& {Rao}}]{2019ApJ...881L...2B}
{Bower}, G.~C., {Dexter}, J., {Asada}, K., {et~al.} 2019, \apjl, 881, L2,
  \dodoi{10.3847/2041-8213/ab3397}

\bibitem[{{Brinkerink} {et~al.}(2021){Brinkerink}, {Falcke}, {Brunthaler}, \&
  {Law}}]{Brinkerink+2021}
{Brinkerink}, C., {Falcke}, H., {Brunthaler}, A., \& {Law}, C. 2021, arXiv
  e-prints, arXiv:2107.13402.
\newblock \doarXiv{2107.13402}

\bibitem[{{Brinkerink} {et~al.}(2015){Brinkerink}, {Falcke}, {Law}, {Barkats},
  {Bower}, {Brunthaler}, {Gammie}, {Impellizzeri}, {Markoff}, {Menten},
  {Moscibrodzka}, {Peck}, {Rushton}, {Schaaf}, \&
  {Wright}}]{2015A&A...576A..41B}
{Brinkerink}, C.~D., {Falcke}, H., {Law}, C.~J., {et~al.} 2015, \aap, 576, A41,
  \dodoi{10.1051/0004-6361/201424783}

\bibitem[{{Broderick} {et~al.}(2009){Broderick}, {Loeb}, \&
  {Narayan}}]{Broderick+2009}
{Broderick}, A.~E., {Loeb}, A., \& {Narayan}, R. 2009, \apj, 701, 1357,
  \dodoi{10.1088/0004-637X/701/2/1357}

\bibitem[{{Broderick} \& {Narayan}(2006)}]{Broderick_Narayan_2006}
{Broderick}, A.~E., \& {Narayan}, R. 2006, \apjl, 638, L21,
  \dodoi{10.1086/500930}

\bibitem[{{Broderick} \& {Narayan}(2007)}]{Broderick_Narayan_2007}
---. 2007, Classical and Quantum Gravity, 24, 659,
  \dodoi{10.1088/0264-9381/24/3/009}

\bibitem[{{Broderick} {et~al.}(2015){Broderick}, {Narayan}, {Kormendy},
  {Perlman}, {Rieke}, \& {Doeleman}}]{Broderick+2015}
{Broderick}, A.~E., {Narayan}, R., {Kormendy}, J., {et~al.} 2015, \apj, 805,
  179, \dodoi{10.1088/0004-637X/805/2/179}

\bibitem[{{Bronzwaer} \& {Falcke}(2021)}]{Bronzwaer2021}
{Bronzwaer}, T., \& {Falcke}, H. 2021, \apj, 920, 155,
  \dodoi{10.3847/1538-4357/ac1738}

\bibitem[{{Buonanno} \& {Damour}(1999)}]{Buonanno1999}
{Buonanno}, A., \& {Damour}, T. 1999, \prd, 59, 084006,
  \dodoi{10.1103/PhysRevD.59.084006}

\bibitem[{{Buonanno} \& {Damour}(2000)}]{Buonanno2000}
---. 2000, \prd, 62, 064015, \dodoi{10.1103/PhysRevD.62.064015}

\bibitem[{{Carballo-Rubio} {et~al.}(2018){Carballo-Rubio}, {Kumar}, \&
  {Lu}}]{Carballo-Rubio+2018}
{Carballo-Rubio}, R., {Kumar}, P., \& {Lu}, W. 2018, \prd, 97, 123012,
  \dodoi{10.1103/PhysRevD.97.123012}

\bibitem[{{C{\'a}rdenas-Avenda{\~n}o}
  {et~al.}(2020){C{\'a}rdenas-Avenda{\~n}o}, {Nampalliwar}, \&
  {Yunes}}]{Cardenas2020}
{C{\'a}rdenas-Avenda{\~n}o}, A., {Nampalliwar}, S., \& {Yunes}, N. 2020,
  Classical and Quantum Gravity, 37, 135008, \dodoi{10.1088/1361-6382/ab8f64}

\bibitem[{{Cardoso} \& {Pani}(2019)}]{Cardoso2019}
{Cardoso}, V., \& {Pani}, P. 2019, Living Reviews in Relativity, 22, 4,
  \dodoi{10.1007/s41114-019-0020-4}

\bibitem[{{Carson} \& {Yagi}(2020)}]{Carson2020}
{Carson}, Z., \& {Yagi}, K. 2020, \prd, 101, 084050,
  \dodoi{10.1103/PhysRevD.101.084050}

\bibitem[{{Carter}(1968)}]{Carter1968}
{Carter}, B. 1968, Physical Review, 174, 1559, \dodoi{10.1103/PhysRev.174.1559}

\bibitem[{{Carter}(1971)}]{Carter1971}
---. 1971, \prl, 26, 331, \dodoi{10.1103/PhysRevLett.26.331}

\bibitem[{{Chael} {et~al.}(2021){Chael}, {Johnson}, \& {Lupsasca}}]{Chael+2021}
{Chael}, A., {Johnson}, M.~D., \& {Lupsasca}, A. 2021, \apj, 918, 6,
  \dodoi{10.3847/1538-4357/ac09ee}

\bibitem[{{Chael} {et~al.}(2018){Chael}, {Rowan}, {Narayan}, {Johnson}, \&
  {Sironi}}]{Chael+2018}
{Chael}, A., {Rowan}, M., {Narayan}, R., {Johnson}, M., \& {Sironi}, L. 2018,
  \mnras, 478, 5209, \dodoi{10.1093/mnras/sty1261}

\bibitem[{{Chan} {et~al.}(2013){Chan}, {Psaltis}, \& {{\"O}zel}}]{Chan2013}
{Chan}, C.-k., {Psaltis}, D., \& {{\"O}zel}, F. 2013, \apj, 777, 13,
  \dodoi{10.1088/0004-637X/777/1/13}

\bibitem[{{Chapline}(2003)}]{Chapline_2003}
{Chapline}, G. 2003, International Journal of Modern Physics A, 18, 3587,
  \dodoi{10.1142/S0217751X03016380}

\bibitem[{{Chatterjee} {et~al.}(2021){Chatterjee}, {Markoff}, {Neilsen},
  {Younsi}, {Witzel}, {Tchekhovskoy}, {Yoon}, {Ingram}, {van der Klis},
  {Boyce}, {Do}, {Haggard}, \& {Nowak}}]{Chatterjee+2021}
{Chatterjee}, K., {Markoff}, S., {Neilsen}, J., {et~al.} 2021, \mnras, 507,
  5281, \dodoi{10.1093/mnras/stab2466}

\bibitem[{{Chirenti} \& {Rezzolla}(2007)}]{Chirenti+2007}
{Chirenti}, C. B.~M.~H., \& {Rezzolla}, L. 2007, Classical and Quantum Gravity,
  24, 4191, \dodoi{10.1088/0264-9381/24/16/013}

\bibitem[{{Chu} {et~al.}(2018){Chu}, {Do}, {Hees}, {Ghez}, {Naoz}, {Witzel},
  {Sakai}, {Chappell}, {Gautam}, {Lu}, \& {Matthews}}]{2018ApJ...854...12C}
{Chu}, D.~S., {Do}, T., {Hees}, A., {et~al.} 2018, \apj, 854, 12,
  \dodoi{10.3847/1538-4357/aaa3eb}

\bibitem[{{Clavel} {et~al.}(2013){Clavel}, {Terrier}, {Goldwurm}, {Morris},
  {Ponti}, {Soldi}, \& {Trap}}]{Clavel+2013}
{Clavel}, M., {Terrier}, R., {Goldwurm}, A., {et~al.} 2013, \aap, 558, A32,
  \dodoi{10.1051/0004-6361/201321667}

\bibitem[{{Cotera} {et~al.}(1999){Cotera}, {Morris}, {Ghez}, {Becklin},
  {Tanner}, {Werner}, \& {Stolovy}}]{1999ASPC..186..240C}
{Cotera}, A., {Morris}, M., {Ghez}, A.~M., {et~al.} 1999, in Astronomical
  Society of the Pacific Conference Series, Vol. 186, The Central Parsecs of
  the Galaxy, ed. H.~{Falcke}, A.~{Cotera}, W.~J. {Duschl}, F.~{Melia}, \&
  M.~J. {Rieke}, 240

\bibitem[{{Cunha} \& {Herdeiro}(2018)}]{Cunha2018}
{Cunha}, P. V.~P., \& {Herdeiro}, C. A.~R. 2018, General Relativity and
  Gravitation, 50, 42, \dodoi{10.1007/s10714-018-2361-9}

\bibitem[{Curiel(2017)}]{Curiel2014}
Curiel, E. 2017, in Towards a Theory of Spacetime Theories, ed. D.~Lehmkuhl,
  G.~Schiemann, \& E.~Scholz (New York, NY: Springer New York), 43--104,
  \dodoi{10.1007/978-1-4939-3210-8_3}

\bibitem[{{Damour} \& {Esposito-Farese}(1993)}]{Damour1993}
{Damour}, T., \& {Esposito-Farese}, G. 1993, \prl, 70, 2220,
  \dodoi{10.1103/PhysRevLett.70.2220}

\bibitem[{{Damour} \& {Schaefer}(1991)}]{ds91}
{Damour}, T., \& {Schaefer}, G. 1991, \prl, 66, 2549,
  \dodoi{10.1103/PhysRevLett.66.2549}

\bibitem[{{Damour} \& {Taylor}(1992)}]{DamourTaylor1992}
{Damour}, T., \& {Taylor}, J.~H. 1992, \prd, 45, 1840,
  \dodoi{10.1103/PhysRevD.45.1840}

\bibitem[{{Danielsson} {et~al.}(2021){Danielsson}, {Lehner}, \&
  {Pretorius}}]{Danielsson2021}
{Danielsson}, U., {Lehner}, L., \& {Pretorius}, F. 2021, \prd, 104, 124011,
  \dodoi{10.1103/PhysRevD.104.124011}

\bibitem[{{Davelaar} {et~al.}(2018){Davelaar}, {Mo{\'s}cibrodzka}, {Bronzwaer},
  \& {Falcke}}]{Davelaar2018}
{Davelaar}, J., {Mo{\'s}cibrodzka}, M., {Bronzwaer}, T., \& {Falcke}, H. 2018,
  \aap, 612, A34, \dodoi{10.1051/0004-6361/201732025}

\bibitem[{{De Laurentis} {et~al.}(2018){De Laurentis}, {Younsi}, {Porth},
  {Mizuno}, \& {Rezzolla}}]{DeLaurentis2018}
{De Laurentis}, M., {Younsi}, Z., {Porth}, O., {Mizuno}, Y., \& {Rezzolla}, L.
  2018, \prd, 97, 104024, \dodoi{10.1103/PhysRevD.97.104024}

\bibitem[{{De Martino} {et~al.}(2021){De Martino}, {Della Monica}, \& {De
  Laurentis}}]{DeMartinoI}
{De Martino}, I., {Della Monica}, R., \& {De Laurentis}, M. 2021, \prd, 104,
  L101502, \dodoi{10.1103/PhysRevD.104.L101502}

\bibitem[{Della~Monica {et~al.}(2021)Della~Monica, De~Martino, \&
  De~Laurentis}]{DellaMonica}
Della~Monica, R., De~Martino, I., \& De~Laurentis, M. 2021, Monthly Notices of
  the Royal Astronomical Society, \dodoi{10.1093/mnras/stab3727}

\bibitem[{{Dexter} {et~al.}(2014){Dexter}, {Kelly}, {Bower}, {Marrone},
  {Stone}, \& {Plambeck}}]{Dexter2014}
{Dexter}, J., {Kelly}, B., {Bower}, G.~C., {et~al.} 2014, \mnras, 442, 2797,
  \dodoi{10.1093/mnras/stu1039}

\bibitem[{{Dexter} {et~al.}(2020){Dexter}, {Tchekhovskoy},
  {Jim{\'e}nez-Rosales}, {Ressler}, {Baub{\"o}ck}, {Dallilar}, {de Zeeuw},
  {Eisenhauer}, {von Fellenberg}, {Gao}, {Genzel}, {Gillessen}, {Habibi},
  {Ott}, {Stadler}, {Straub}, \& {Widmann}}]{Dexter+2020}
{Dexter}, J., {Tchekhovskoy}, A., {Jim{\'e}nez-Rosales}, A., {et~al.} 2020,
  \mnras, 497, 4999, \dodoi{10.1093/mnras/staa2288}

\bibitem[{{Dey} {et~al.}(2019){Dey}, {Kocherlakota}, \& {Joshi}}]{Dey+2019}
{Dey}, D., {Kocherlakota}, P., \& {Joshi}, P.~S. 2019, arXiv e-prints,
  arXiv:1907.12792.
\newblock \doarXiv{1907.12792}

\bibitem[{{Dicke}(2019)}]{Dicke1964}
{Dicke}, R.~H. 2019, General Relativity and Gravitation, 51, 57,
  \dodoi{10.1007/s10714-019-2509-2}

\bibitem[{{Do} {et~al.}(2009){Do}, {Ghez}, {Morris}, {Yelda}, {Meyer}, {Lu},
  {Hornstein}, \& {Matthews}}]{2009ApJ...691.1021D}
{Do}, T., {Ghez}, A.~M., {Morris}, M.~R., {et~al.} 2009, \apj, 691, 1021,
  \dodoi{10.1088/0004-637X/691/2/1021}

\bibitem[{{Do} {et~al.}(2013){Do}, {Martinez}, {Yelda}, {Ghez}, {Bullock},
  {Kaplinghat}, {Lu}, {Peter}, \& {Phifer}}]{2013ApJ...779L...6D}
{Do}, T., {Martinez}, G.~D., {Yelda}, S., {et~al.} 2013, \apjl, 779, L6,
  \dodoi{10.1088/2041-8205/779/1/L6}

\bibitem[{{Do} {et~al.}(2019){Do}, {Hees}, {Ghez}, {Martinez}, {Chu}, {Jia},
  {Sakai}, {Lu}, {Gautam}, {O'Neil}, {Becklin}, {Morris}, {Matthews},
  {Nishiyama}, {Campbell}, {Chappell}, {Chen}, {Ciurlo}, {Dehghanfar},
  {Gallego-Cano}, {Kerzendorf}, {Lyke}, {Naoz}, {Saida}, {Sch{\"o}del},
  {Takahashi}, {Takamori}, {Witzel}, \& {Wizinowich}}]{2019Sci...365..664D}
{Do}, T., {Hees}, A., {Ghez}, A., {et~al.} 2019, Science, 365, 664,
  \dodoi{10.1126/science.aav8137}

\bibitem[{{Dodds-Eden} {et~al.}(2009){Dodds-Eden}, {Porquet}, {Trap},
  {Quataert}, {Haubois}, {Gillessen}, {Grosso}, {Pantin}, {Falcke}, {Rouan},
  {Genzel}, {Hasinger}, {Goldwurm}, {Yusef-Zadeh}, {Clenet}, {Trippe},
  {Lagage}, {Bartko}, {Eisenhauer}, {Ott}, {Paumard}, {Perrin}, {Yuan},
  {Fritz}, \& {Mascetti}}]{2009ApJ...698..676D}
{Dodds-Eden}, K., {Porquet}, D., {Trap}, G., {et~al.} 2009, \apj, 698, 676,
  \dodoi{10.1088/0004-637X/698/1/676}

\bibitem[{{Dodds-Eden} {et~al.}(2011){Dodds-Eden}, {Gillessen}, {Fritz},
  {Eisenhauer}, {Trippe}, {Genzel}, {Ott}, {Bartko}, {Pfuhl}, {Bower},
  {Goldwurm}, {Porquet}, {Trap}, \& {Yusef-Zadeh}}]{2011ApJ...728...37D}
{Dodds-Eden}, K., {Gillessen}, S., {Fritz}, T.~K., {et~al.} 2011, \apj, 728,
  37, \dodoi{10.1088/0004-637X/728/1/37}

\bibitem[{{Doeleman} {et~al.}(2019){Doeleman}, {Blackburn}, {Dexter}, {Gomez},
  {Johnson}, {Palumbo}, {Weintroub}, {Farah}, {Fish}, {Loinard}, {Lonsdale},
  {Narayanan}, {Patel}, {Pesce}, {Raymond}, {Tilanus}, {Wielgus}, {Akiyama},
  {Bower}, {Broderick}, {Deane}, {Fromm}, {Gammie}, {Gold}, {Janssen},
  {Kawashima}, {Krichbaum}, {Marrone}, {Matthews}, {Mizuno}, {Rezzolla},
  {Roelofs}, {Ros}, {Savolainen}, {Yuan}, {Zhao}, {Blackburn}, {Doeleman},
  {Dexter}, {Gomez}, {Johnson}, {Palumbo}, {Weintroub}, {Farah}, {Fish},
  {Loinard}, {Lonsdale}, {Narayanan}, {Patel}, {Pesce}, {Raymond}, {Tilanus},
  {Wielgus}, {Akiyama}, {Bower}, {Broderick}, {Deane}, {Fromm}, {Gammie},
  {Gold}, {Janssen}, {Kawashima}, {Krichbaum}, {Marrone}, {Matthews}, {Mizuno},
  {Rezzolla}, {Roelofs}, {Ros}, {Savolainen}, {Yuan}, \&
  {Zhao}}]{Doeleman+2019}
{Doeleman}, S., {Blackburn}, L., {Dexter}, J., {et~al.} 2019, in Bulletin of
  the American Astronomical Society, Vol.~51, 256.
\newblock \doarXiv{1909.01411}

\bibitem[{{Done} \& {Gierli{\'n}ski}(2003)}]{Done_Gierlinski_2003}
{Done}, C., \& {Gierli{\'n}ski}, M. 2003, \mnras, 342, 1041,
  \dodoi{10.1046/j.1365-8711.2003.06614.x}

\bibitem[{{Eckart} {et~al.}(2002){Eckart}, {Genzel}, {Ott}, \&
  {Sch{\"o}del}}]{2002MNRAS.331..917E}
{Eckart}, A., {Genzel}, R., {Ott}, T., \& {Sch{\"o}del}, R. 2002, \mnras, 331,
  917, \dodoi{10.1046/j.1365-8711.2002.05237.x}

\bibitem[{{Eckart} {et~al.}(1999){Eckart}, {Ott}, \&
  {Genzel}}]{1999A&A...352L..22E}
{Eckart}, A., {Ott}, T., \& {Genzel}, R. 1999, \aap, 352, L22.
\newblock \doarXiv{astro-ph/9911011}

\bibitem[{{Eckart} {et~al.}(2006){Eckart}, {Sch{\"o}del}, {Meyer}, {Trippe},
  {Ott}, \& {Genzel}}]{Eckart+2006}
{Eckart}, A., {Sch{\"o}del}, R., {Meyer}, L., {et~al.} 2006, \aap, 455, 1,
  \dodoi{10.1051/0004-6361:20064948}

\bibitem[{{Eckart} {et~al.}(2004{\natexlab{a}}){Eckart}, {Baganoff}, {Morris},
  {Bautz}, {Brandt}, {Garmire}, {Genzel}, {Ott}, {Ricker}, {Straubmeier},
  {Viehmann}, {Sch{\"o}del}, {Bower}, \& {Goldston}}]{2004A&A...427....1E}
{Eckart}, A., {Baganoff}, F.~K., {Morris}, M., {et~al.} 2004{\natexlab{a}},
  \aap, 427, 1, \dodoi{10.1051/0004-6361:20040495}

\bibitem[{{Eckart} {et~al.}(2004{\natexlab{b}}){Eckart}, {Baganoff}, {Morris},
  {Bautz}, {Brandt}, {Garmire}, {Genzel}, {Ott}, {Ricker}, {Straubmeier},
  {Viehmann}, {Sch{\"o}del}, {Bower}, \& {Goldston}}]{Eckart+2004}
---. 2004{\natexlab{b}}, \aap, 427, 1, \dodoi{10.1051/0004-6361:20040495}

\bibitem[{{Eisenhauer} {et~al.}(2005){Eisenhauer}, {Genzel}, {Alexander},
  {Abuter}, {Paumard}, {Ott}, {Gilbert}, {Gillessen}, {Horrobin}, {Trippe},
  {Bonnet}, {Dumas}, {Hubin}, {Kaufer}, {Kissler-Patig}, {Monnet},
  {Str{\"o}bele}, {Szeifert}, {Eckart}, {Sch{\"o}del}, \&
  {Zucker}}]{2005ApJ...628..246E}
{Eisenhauer}, F., {Genzel}, R., {Alexander}, T., {et~al.} 2005, \apj, 628, 246,
  \dodoi{10.1086/430667}

\bibitem[{{Event Horizon Telescope Collaboration}
  {et~al.}(2019{\natexlab{a}}){Event Horizon Telescope Collaboration},
  {Akiyama}, {Alberdi}, {Alef}, {Asada}, {Azulay}, {Baczko}, {Ball},
  {Balokovi{\'c}}, {Barrett}, \& et~al.}]{IV_EHT2019_M87}
{Event Horizon Telescope Collaboration}, {Akiyama}, K., {Alberdi}, A., {et~al.}
  2019{\natexlab{a}}, \apjl, 875, L4, \dodoi{10.3847/2041-8213/ab0e85}

\bibitem[{{Event Horizon Telescope Collaboration}
  {et~al.}(2019{\natexlab{b}}){Event Horizon Telescope Collaboration},
  {Akiyama}, {Alberdi}, {Alef}, {Asada}, {Azulay}, {Baczko}, {Ball},
  {Balokovi{\'c}}, {Barrett}, \& et~al.}]{V_EHT2019_M87}
---. 2019{\natexlab{b}}, \apjl, 875, L5, \dodoi{10.3847/2041-8213/ab0f43}

\bibitem[{{Event Horizon Telescope Collaboration}
  {et~al.}(2019{\natexlab{c}}){Event Horizon Telescope Collaboration},
  {Akiyama}, {Alberdi}, {Alef}, {Asada}, {Azulay}, {Baczko}, {Ball},
  {Balokovi{\'c}}, {Barrett}, \& et~al.}]{VI_EHT2019_M87}
---. 2019{\natexlab{c}}, \apjl, 875, L6, \dodoi{10.3847/2041-8213/ab1141}

\bibitem[{{Event Horizon Telescope Collaboration}
  {et~al.}(2022{\natexlab{a}}){Event Horizon Telescope Collaboration},
  {Akiyama}, {Alberdi}, {Alef}, {Asada}, {Azulay}, {Baczko}, {Ball},
  {Balokovi{\'c}}, {Barrett}, \& et~al.}]{PaperII}
---. 2022{\natexlab{a}}, \apjl, 875, 2

\bibitem[{{Event Horizon Telescope Collaboration}
  {et~al.}(2022{\natexlab{b}}){Event Horizon Telescope Collaboration},
  {Akiyama}, {Alberdi}, {Alef}, {Asada}, {Azulay}, {Baczko}, {Ball},
  {Balokovi{\'c}}, {Barrett}, \& et~al.}]{PaperIII}
---. 2022{\natexlab{b}}, \apjl, 875, 3

\bibitem[{{Event Horizon Telescope Collaboration}
  {et~al.}(2022{\natexlab{c}}){Event Horizon Telescope Collaboration},
  {Akiyama}, {Alberdi}, {Alef}, {Asada}, {Azulay}, {Baczko}, {Ball},
  {Balokovi{\'c}}, {Barrett}, \& et~al.}]{PaperIV}
---. 2022{\natexlab{c}}, \apjl, 875, 4

\bibitem[{{Event Horizon Telescope Collaboration}
  {et~al.}(2022{\natexlab{d}}){Event Horizon Telescope Collaboration},
  {Akiyama}, {Alberdi}, {Alef}, {Asada}, {Azulay}, {Baczko}, {Ball},
  {Balokovi{\'c}}, {Barrett}, \& et~al.}]{PaperV}
---. 2022{\natexlab{d}}, \apjl, 875, 5

\bibitem[{{Event Horizon Telescope Collaboration}
  {et~al.}(2022{\natexlab{e}}){Event Horizon Telescope Collaboration},
  {Akiyama}, {Alberdi}, {Alef}, {Asada}, {Azulay}, {Baczko}, {Ball},
  {Balokovi{\'c}}, {Barrett}, \& et~al.}]{PaperI}
---. 2022{\natexlab{e}}, \apjl, 875, 1

\bibitem[{{Falcke}(1999)}]{1999ASPC..186..113F}
{Falcke}, H. 1999, in Astronomical Society of the Pacific Conference Series,
  Vol. 186, The Central Parsecs of the Galaxy, ed. H.~{Falcke}, A.~{Cotera},
  W.~J. {Duschl}, F.~{Melia}, \& M.~J. {Rieke}, 113.
\newblock \doarXiv{astro-ph/9909441}

\bibitem[{{Falcke} {et~al.}(1998){Falcke}, {Goss}, {Matsuo}, {Teuben}, {Zhao},
  \& {Zylka}}]{1998ApJ...499..731F}
{Falcke}, H., {Goss}, W.~M., {Matsuo}, H., {et~al.} 1998, \apj, 499, 731,
  \dodoi{10.1086/305687}

\bibitem[{{Falcke} {et~al.}(1993){Falcke}, {Mannheim}, \&
  {Biermann}}]{Falcke1993}
{Falcke}, H., {Mannheim}, K., \& {Biermann}, P.~L. 1993, \aap, 278, L1.
\newblock \doarXiv{astro-ph/9308031}

\bibitem[{{Falcke} \& {Markoff}(2000)}]{Falcke_Markoff2000}
{Falcke}, H., \& {Markoff}, S. 2000, \aap, 362, 113.
\newblock \doarXiv{astro-ph/0102186}

\bibitem[{{Falcke} \& {Markoff}(2013)}]{Falcke2013}
{Falcke}, H., \& {Markoff}, S.~B. 2013, Classical and Quantum Gravity, 30,
  244003, \dodoi{10.1088/0264-9381/30/24/244003}

\bibitem[{{Falcke} {et~al.}(2000){Falcke}, {Melia}, \& {Agol}}]{Falcke2000}
{Falcke}, H., {Melia}, F., \& {Agol}, E. 2000, \apjl, 528, L13,
  \dodoi{10.1086/312423}

\bibitem[{{Fazio} {et~al.}(2018){Fazio}, {Hora}, {Witzel}, {Willner}, {Ashby},
  {Baganoff}, {Becklin}, {Carey}, {Haggard}, {Gammie}, {Ghez}, {Gurwell},
  {Ingalls}, {Marrone}, {Morris}, \& {Smith}}]{2018ApJ...864...58F}
{Fazio}, G.~G., {Hora}, J.~L., {Witzel}, G., {et~al.} 2018, \apj, 864, 58,
  \dodoi{10.3847/1538-4357/aad4a2}

\bibitem[{{Ferreira}(2019)}]{Ferreira2019}
{Ferreira}, P.~G. 2019, \araa, 57, 335,
  \dodoi{10.1146/annurev-astro-091918-104423}

\bibitem[{{Fish} {et~al.}(2020){Fish}, {Shea}, \& {Akiyama}}]{Fish+2020}
{Fish}, V.~L., {Shea}, M., \& {Akiyama}, K. 2020, Advances in Space Research,
  65, 821, \dodoi{10.1016/j.asr.2019.03.029}

\bibitem[{{Freire} {et~al.}(2012){Freire}, {Kramer}, \& {Wex}}]{fwk12}
{Freire}, P. C.~C., {Kramer}, M., \& {Wex}, N. 2012, Classical and Quantum
  Gravity, 29, 184007, \dodoi{10.1088/0264-9381/29/18/184007}

\bibitem[{{Frolov}(2016)}]{Frolov2016}
{Frolov}, V.~P. 2016, \prd, 94, 104056, \dodoi{10.1103/PhysRevD.94.104056}

\bibitem[{{Fromm} {et~al.}(2021){Fromm}, {Mizuno}, {Younsi}, {Olivares},
  {Porth}, {De Laurentis}, {Falcke}, {Kramer}, \& {Rezzolla}}]{Fromm2021}
{Fromm}, C.~M., {Mizuno}, Y., {Younsi}, Z., {et~al.} 2021, \aap, 649, A116,
  \dodoi{10.1051/0004-6361/201937335}

\bibitem[{{Gair} \& {Yunes}(2011)}]{Gair2011}
{Gair}, J., \& {Yunes}, N. 2011, \prd, 84, 064016,
  \dodoi{10.1103/PhysRevD.84.064016}

\bibitem[{{Gair} {et~al.}(2008){Gair}, {Li}, \& {Mandel}}]{Gair2008}
{Gair}, J.~R., {Li}, C., \& {Mandel}, I. 2008, \prd, 77, 024035,
  \dodoi{10.1103/PhysRevD.77.024035}

\bibitem[{{Gan} {et~al.}(2021){Gan}, {Wang}, {Wu}, \& {Yang}}]{Gan+2021}
{Gan}, Q., {Wang}, P., {Wu}, H., \& {Yang}, H. 2021, \prd, 104, 024003,
  \dodoi{10.1103/PhysRevD.104.024003}

\bibitem[{{Garc{\'{\i}}a} {et~al.}(1995){Garc{\'{\i}}a}, {Galtsov}, \&
  {Kechkin}}]{Garcia+1995}
{Garc{\'{\i}}a}, A., {Galtsov}, D., \& {Kechkin}, O. 1995, Phys. Rev. Lett.,
  74, 1276, \dodoi{10.1103/PhysRevLett.74.1276}

\bibitem[{{Garcia} {et~al.}(2001){Garcia}, {McClintock}, {Narayan}, {Callanan},
  {Barret}, \& {Murray}}]{Garcia+2001}
{Garcia}, M.~R., {McClintock}, J.~E., {Narayan}, R., {et~al.} 2001, \apjl, 553,
  L47, \dodoi{10.1086/320494}

\bibitem[{{Garfinkle} {et~al.}(1991){Garfinkle}, {Horowitz}, \&
  {Strominger}}]{Garfinkle+1991}
{Garfinkle}, D., {Horowitz}, G.~T., \& {Strominger}, A. 1991, Phys. Rev. D, 43,
  3140, \dodoi{10.1103/PhysRevD.43.3140}

\bibitem[{{Gebhardt} {et~al.}(2011){Gebhardt}, {Adams}, {Richstone}, {Lauer},
  {Faber}, {G{\"u}ltekin}, {Murphy}, \& {Tremaine}}]{Gebhardt2011}
{Gebhardt}, K., {Adams}, J., {Richstone}, D., {et~al.} 2011, \apj, 729, 119,
  \dodoi{10.1088/0004-637X/729/2/119}

\bibitem[{{Genova} {et~al.}(2018){Genova}, {Mazarico}, {Goossens}, {Lemoine},
  {Neumann}, {Smith}, \& {Zuber}}]{messenger}
{Genova}, A., {Mazarico}, E., {Goossens}, S., {et~al.} 2018, Nature
  Communications, 9, 289, \dodoi{10.1038/s41467-017-02558-1}

\bibitem[{{Genzel} {et~al.}(1997){Genzel}, {Eckart}, {Ott}, \&
  {Eisenhauer}}]{1997MNRAS.291..219G}
{Genzel}, R., {Eckart}, A., {Ott}, T., \& {Eisenhauer}, F. 1997, \mnras, 291,
  219, \dodoi{10.1093/mnras/291.1.219}

\bibitem[{{Genzel} {et~al.}(2010){Genzel}, {Eisenhauer}, \&
  {Gillessen}}]{2010RvMP...82.3121G}
{Genzel}, R., {Eisenhauer}, F., \& {Gillessen}, S. 2010, Reviews of Modern
  Physics, 82, 3121, \dodoi{10.1103/RevModPhys.82.3121}

\bibitem[{{Genzel} {et~al.}(2000){Genzel}, {Pichon}, {Eckart}, {Gerhard}, \&
  {Ott}}]{2000MNRAS.317..348G}
{Genzel}, R., {Pichon}, C., {Eckart}, A., {Gerhard}, O.~E., \& {Ott}, T. 2000,
  \mnras, 317, 348, \dodoi{10.1046/j.1365-8711.2000.03582.x}

\bibitem[{{Genzel} {et~al.}(2003{\natexlab{a}}){Genzel}, {Sch{\"o}del}, {Ott},
  {Eckart}, {Alexander}, {Lacombe}, {Rouan}, \&
  {Aschenbach}}]{2003Natur.425..934G}
{Genzel}, R., {Sch{\"o}del}, R., {Ott}, T., {et~al.} 2003{\natexlab{a}}, \nat,
  425, 934

\bibitem[{{Genzel} {et~al.}(2003{\natexlab{b}}){Genzel}, {Sch{\"o}del}, {Ott},
  {Eisenhauer}, {Hofmann}, {Lehnert}, {Eckart}, {Alexander}, {Sternberg},
  {Lenzen}, {Cl{\'e}net}, {Lacombe}, {Rouan}, {Renzini}, \&
  {Tacconi-Garman}}]{2003ApJ...594..812G}
---. 2003{\natexlab{b}}, \apj, 594, 812, \dodoi{10.1086/377127}

\bibitem[{{Gezari} {et~al.}(2002){Gezari}, {Ghez}, {Becklin}, {Larkin},
  {McLean}, \& {Morris}}]{2002ApJ...576..790G}
{Gezari}, S., {Ghez}, A.~M., {Becklin}, E.~E., {et~al.} 2002, \apj, 576, 790,
  \dodoi{10.1086/341807}

\bibitem[{{Ghez} {et~al.}(1998){Ghez}, {Klein}, {Morris}, \&
  {Becklin}}]{1998ApJ...509..678G}
{Ghez}, A.~M., {Klein}, B.~L., {Morris}, M., \& {Becklin}, E.~E. 1998, \apj,
  509, 678, \dodoi{10.1086/306528}

\bibitem[{{Ghez} {et~al.}(2000){Ghez}, {Morris}, {Becklin}, {Tanner}, \&
  {Kremenek}}]{2000Natur.407..349G}
{Ghez}, A.~M., {Morris}, M., {Becklin}, E.~E., {Tanner}, A., \& {Kremenek}, T.
  2000, \nat, 407, 349, \dodoi{10.1038/35030032}

\bibitem[{{Ghez} {et~al.}(2005{\natexlab{a}}){Ghez}, {Salim}, {Hornstein},
  {Tanner}, {Lu}, {Morris}, {Becklin}, \& {Duch{\^e}ne}}]{2005ApJ...620..744G}
{Ghez}, A.~M., {Salim}, S., {Hornstein}, S.~D., {et~al.} 2005{\natexlab{a}},
  \apj, 620, 744, \dodoi{10.1086/427175}

\bibitem[{{Ghez} {et~al.}(2005{\natexlab{b}}){Ghez}, {Hornstein}, {Lu},
  {Bouchez}, {Le Mignant}, {van Dam}, {Wizinowich}, {Matthews}, {Morris},
  {Becklin}, {Campbell}, {Chin}, {Hartman}, {Johansson}, {Lafon}, {Stomski}, \&
  {Summers}}]{2005ApJ...635.1087G}
{Ghez}, A.~M., {Hornstein}, S.~D., {Lu}, J.~R., {et~al.} 2005{\natexlab{b}},
  \apj, 635, 1087, \dodoi{10.1086/497576}

\bibitem[{{Ghez} {et~al.}(2008){Ghez}, {Salim}, {Weinberg}, {Lu}, {Do}, {Dunn},
  {Matthews}, {Morris}, {Yelda}, {Becklin}, {Kremenek}, {Milosavljevic}, \&
  {Naiman}}]{2008ApJ...689.1044G}
{Ghez}, A.~M., {Salim}, S., {Weinberg}, N.~N., {et~al.} 2008, \apj, 689, 1044,
  \dodoi{10.1086/592738}

\bibitem[{{Gibbons} \& {Maeda}(1988)}]{Gibbons+1988}
{Gibbons}, G.~W., \& {Maeda}, K.-I. 1988, Nuclear Physics B, 298, 741,
  \dodoi{10.1016/0550-3213(88)90006-5}

\bibitem[{{Gillessen} {et~al.}(2009{\natexlab{a}}){Gillessen}, {Eisenhauer},
  {Fritz}, {Bartko}, {Dodds-Eden}, {Pfuhl}, {Ott}, \&
  {Genzel}}]{2009ApJ...707L.114G}
{Gillessen}, S., {Eisenhauer}, F., {Fritz}, T.~K., {et~al.} 2009{\natexlab{a}},
  \apjl, 707, L114, \dodoi{10.1088/0004-637X/707/2/L114}

\bibitem[{{Gillessen} {et~al.}(2009{\natexlab{b}}){Gillessen}, {Eisenhauer},
  {Trippe}, {Alexander}, {Genzel}, {Martins}, \& {Ott}}]{2009ApJ...692.1075G}
{Gillessen}, S., {Eisenhauer}, F., {Trippe}, S., {et~al.} 2009{\natexlab{b}},
  \apj, 692, 1075, \dodoi{10.1088/0004-637X/692/2/1075}

\bibitem[{{Gillessen} {et~al.}(2017){Gillessen}, {Plewa}, {Eisenhauer}, {Sari},
  {Waisberg}, {Habibi}, {Pfuhl}, {George}, {Dexter}, {von Fellenberg}, {Ott},
  \& {Genzel}}]{2017ApJ...837...30G}
{Gillessen}, S., {Plewa}, P.~M., {Eisenhauer}, F., {et~al.} 2017, \apj, 837,
  30, \dodoi{10.3847/1538-4357/aa5c41}

\bibitem[{{Goddi} {et~al.}(2017){Goddi}, {Falcke}, {Kramer}, {Rezzolla},
  {Brinkerink}, {Bronzwaer}, {Davelaar}, {Deane}, {de Laurentis}, {Desvignes},
  {Eatough}, {Eisenhauer}, {Fraga-Encinas}, {Fromm}, {Gillessen}, {Grenzebach},
  {Issaoun}, {Jan{\ss}en}, {Konoplya}, {Krichbaum}, {Laing}, {Liu}, {Lu},
  {Mizuno}, {Moscibrodzka}, {M{\"u}ller}, {Olivares}, {Pfuhl}, {Porth},
  {Roelofs}, {Ros}, {Schuster}, {Tilanus}, {Torne}, {van Bemmel}, {van
  Langevelde}, {Wex}, {Younsi}, \& {Zhidenko}}]{Goddi2017}
{Goddi}, C., {Falcke}, H., {Kramer}, M., {et~al.} 2017, International Journal
  of Modern Physics D, 26, 1730001, \dodoi{10.1142/S0218271817300014}

\bibitem[{{Goddi} {et~al.}(2021){Goddi}, {Mart{\'\i}-Vidal}, {Messias},
  {Bower}, {Broderick}, {Dexter}, {Marrone}, {Moscibrodzka}, {Nagai}, {Algaba},
  {Asada}, {Crew}, {G{\'o}mez}, {Impellizzeri}, {Janssen}, {Kadler},
  {Krichbaum}, {Lico}, {Matthews}, {Nathanail}, {Ricarte}, {Ros}, {Younsi},
  {Akiyama}, {Alberdi}, {Alef}, {Anantua}, {Azulay}, {Baczko}, {Ball},
  {Balokovi{\'c}}, {Barrett}, {Benson}, {Bintley}, {Blackburn}, {Blundell},
  {Boland}, {Bouman}, {Boyce}, {Bremer}, {Brinkerink}, {Brissenden}, {Britzen},
  {Broguiere}, {Bronzwaer}, {Byun}, {Carlstrom}, {Chael}, {Chan}, {Chatterjee},
  {Chatterjee}, {Chen}, {Chen}, {Chesler}, {Cho}, {Christian}, {Conway},
  {Cordes}, {Crawford}, {Cruz-Osorio}, {Cui}, {Davelaar}, {De Laurentis},
  {Deane}, {Dempsey}, {Desvignes}, {Doeleman}, {Eatough}, {Falcke}, {Farah},
  {Fish}, {Fomalont}, {Ford}, {Fraga-Encinas}, {Freeman}, {Friberg}, {Fromm},
  {Fuentes}, {Galison}, {Gammie}, {Garc{\'\i}a}, {Gentaz}, {Georgiev}, {Gold},
  {G{\'o}mez-Ruiz}, {Gu}, {Gurwell}, {Hada}, {Haggard}, {Hecht}, {Hesper},
  {Ho}, {Ho}, {Honma}, {Huang}, {Huang}, {Hughes}, {Inoue}, {Issaoun}, {James},
  {Jannuzi}, {Jeter}, {Jiang}, {Jimenez-Rosales}, {Johnson}, {Jorstad}, {Jung},
  {Karami}, {Karuppusamy}, {Kawashima}, {Keating}, {Kettenis}, {Kim}, {Kim},
  {Kim}, {Kim}, {Kino}, {Koay}, {Kofuji}, {Koch}, {Koyama}, {Kramer}, {Kramer},
  {Kuo}, {Lauer}, {Lee}, {Levis}, {Li}, {Li}, {Lindqvist}, {Lindahl}, {Liu},
  {Liu}, {Liuzzo}, {Lo}, {Lobanov}, {Loinard}, {Lonsdale}, {Lu}, {MacDonald},
  {Mao}, {Marchili}, {Markoff}, {Marscher}, {Matsushita}, {Medeiros}, {Menten},
  {Mizuno}, {Mizuno}, {Moran}, {Moriyama}, {M{\"u}ller}, {Musoke},
  {Mej{\'\i}as}, {Nagar}, {Nakamura}, {Narayan}, {Narayanan}, {Natarajan},
  {Neilsen}, {Neri}, {Ni}, {Noutsos}, {Nowak}, {Okino}, {Olivares},
  {Ortiz-Le{\'o}n}, {Oyama}, {{\"O}zel}, {Palumbo}, {Park}, {Patel}, {Pen},
  {Pesce}, {Pi{\'e}tu}, {Plambeck}, {PopStefanija}, {Porth}, {P{\"o}tzl},
  {Prather}, {Preciado-L{\'o}pez}, {Psaltis}, {Pu}, {Ramakrishnan}, {Rao},
  {Rawlings}, {Raymond}, {Rezzolla}, {Ripperda}, {Roelofs}, {Rogers}, {Rose},
  {Roshanineshat}, {Rottmann}, {Roy}, {Ruszczyk}, {Rygl}, {S{\'a}nchez},
  {S{\'a}nchez-Arguelles}, {Sasada}, {Savolainen}, {Schloerb}, {Schuster},
  {Shao}, {Shen}, {Small}, {Sohn}, {SooHoo}, {Sun}, {Tazaki}, {Tetarenko},
  {Tiede}, {Tilanus}, {Titus}, {Toma}, {Torne}, {Trent}, {Traianou}, {Trippe},
  {van Bemmel}, {van Langevelde}, {van Rossum}, {Wagner}, {Ward-Thompson},
  {Wardle}, {Weintroub}, {Wex}, {Wharton}, {Wielgus}, {Wong}, {Wu}, {Yoon},
  {Young}, {Young}, {Yuan}, {Yuan}, {Zensus}, {Zhao}, {Zhao}, {Bruni},
  {Gopakumar}, {Hern{\'a}ndez-G{\'o}mez}, {Herrero-Illana}, {Ingram},
  {Komossa}, {Kovalev}, {Muders}, {Perucho}, {R{\"o}sch}, \&
  {Valtonen}}]{Goddi2021}
{Goddi}, C., {Mart{\'\i}-Vidal}, I., {Messias}, H., {et~al.} 2021, \apjl, 910,
  L14, \dodoi{10.3847/2041-8213/abee6a}

\bibitem[{{Gralla} {et~al.}(2019){Gralla}, {Holz}, \& {Wald}}]{Gralla2019}
{Gralla}, S.~E., {Holz}, D.~E., \& {Wald}, R.~M. 2019, \prd, 100, 024018,
  \dodoi{10.1103/PhysRevD.100.024018}

\bibitem[{{Gralla} {et~al.}(2020){Gralla}, {Lupsasca}, \&
  {Marrone}}]{Gralla2020}
{Gralla}, S.~E., {Lupsasca}, A., \& {Marrone}, D.~P. 2020, \prd, 102, 124004,
  \dodoi{10.1103/PhysRevD.102.124004}

\bibitem[{{Gravity Collaboration} {et~al.}(2018{\natexlab{a}}){Gravity
  Collaboration}, {Abuter}, {Amorim}, {Anugu}, {Baub{\"o}ck}, {Benisty},
  {Berger}, {Blind}, {Bonnet}, {Brandner}, {Buron}, {Collin}, {Chapron},
  {Cl{\'e}net}, {Coud{\'e} Du Foresto}, {de Zeeuw}, {Deen},
  {Delplancke-Str{\"o}bele}, {Dembet}, {Dexter}, {Duvert}, {Eckart},
  {Eisenhauer}, {Finger}, {F{\"o}rster Schreiber}, {F{\'e}dou}, {Garcia},
  {Garcia Lopez}, {Gao}, {Gendron}, {Genzel}, {Gillessen}, {Gordo}, {Habibi},
  {Haubois}, {Haug}, {Hau{\ss}mann}, {Henning}, {Hippler}, {Horrobin},
  {Hubert}, {Hubin}, {Jimenez Rosales}, {Jochum}, {Jocou}, {Kaufer}, {Kellner},
  {Kendrew}, {Kervella}, {Kok}, {Kulas}, {Lacour}, {Lapeyr{\`e}re}, {Lazareff},
  {Le Bouquin}, {L{\'e}na}, {Lippa}, {Lenzen}, {M{\'e}rand}, {M{\"u}ler},
  {Neumann}, {Ott}, {Palanca}, {Paumard}, {Pasquini}, {Perraut}, {Perrin},
  {Pfuhl}, {Plewa}, {Rabien}, {Ram{\'\i}rez}, {Ramos}, {Rau},
  {Rodr{\'\i}guez-Coira}, {Rohloff}, {Rousset}, {Sanchez-Bermudez},
  {Scheithauer}, {Sch{\"o}ller}, {Schuler}, {Spyromilio}, {Straub},
  {Straubmeier}, {Sturm}, {Tacconi}, {Tristram}, {Vincent}, {von Fellenberg},
  {Wank}, {Waisberg}, {Widmann}, {Wieprecht}, {Wiest}, {Wiezorrek}, {Woillez},
  {Yazici}, {Ziegler}, \& {Zins}}]{2018A&A...615L..15G}
{Gravity Collaboration}, {Abuter}, R., {Amorim}, A., {et~al.}
  2018{\natexlab{a}}, \aap, 615, L15, \dodoi{10.1051/0004-6361/201833718}

\bibitem[{{Gravity Collaboration} {et~al.}(2018{\natexlab{b}}){Gravity
  Collaboration}, {Abuter}, {Amorim}, {Baub{\"o}ck}, {Berger}, {Bonnet},
  {Brandner}, {Cl{\'e}net}, {Coud{\'e} Du Foresto}, {de Zeeuw}, {Deen},
  {Dexter}, {Duvert}, {Eckart}, {Eisenhauer}, {F{\"o}rster Schreiber},
  {Garcia}, {Gao}, {Gendron}, {Genzel}, {Gillessen}, {Guajardo}, {Habibi},
  {Haubois}, {Henning}, {Hippler}, {Horrobin}, {Huber}, {Jim{\'e}nez-Rosales},
  {Jocou}, {Kervella}, {Lacour}, {Lapeyr{\`e}re}, {Lazareff}, {Le Bouquin},
  {L{\'e}na}, {Lippa}, {Ott}, {Panduro}, {Paumard}, {Perraut}, {Perrin},
  {Pfuhl}, {Plewa}, {Rabien}, {Rodr{\'\i}guez-Coira}, {Rousset}, {Sternberg},
  {Straub}, {Straubmeier}, {Sturm}, {Tacconi}, {Vincent}, {von Fellenberg},
  {Waisberg}, {Widmann}, {Wieprecht}, {Wiezorrek}, {Woillez}, \&
  {Yazici}}]{2018A&A...618L..10G}
---. 2018{\natexlab{b}}, \aap, 618, L10, \dodoi{10.1051/0004-6361/201834294}

\bibitem[{{Gravity Collaboration} {et~al.}(2019){Gravity Collaboration},
  {Abuter}, {Amorim}, {Baub{\"o}ck}, {Berger}, {Bonnet}, {Brandner},
  {Cl{\'e}net}, {Coud{\'e} Du Foresto}, {de Zeeuw}, {Dexter}, {Duvert},
  {Eckart}, {Eisenhauer}, {F{\"o}rster Schreiber}, {Garcia}, {Gao}, {Gendron},
  {Genzel}, {Gerhard}, {Gillessen}, {Habibi}, {Haubois}, {Henning}, {Hippler},
  {Horrobin}, {Jim{\'e}nez-Rosales}, {Jocou}, {Kervella}, {Lacour},
  {Lapeyr{\`e}re}, {Le Bouquin}, {L{\'e}na}, {Ott}, {Paumard}, {Perraut},
  {Perrin}, {Pfuhl}, {Rabien}, {Rodriguez Coira}, {Rousset}, {Scheithauer},
  {Sternberg}, {Straub}, {Straubmeier}, {Sturm}, {Tacconi}, {Vincent}, {von
  Fellenberg}, {Waisberg}, {Widmann}, {Wieprecht}, {Wiezorrek}, {Woillez}, \&
  {Yazici}}]{2019A&A...625L..10G}
---. 2019, \aap, 625, L10, \dodoi{10.1051/0004-6361/201935656}

\bibitem[{{Gravity Collaboration} {et~al.}(2020{\natexlab{a}}){Gravity
  Collaboration}, {Abuter}, {Amorim}, {Baub{\"o}ck}, {Berger}, {Bonnet},
  {Brandner}, {Cardoso}, {Cl{\'e}net}, {de Zeeuw}, {Dexter}, {Eckart},
  {Eisenhauer}, {F{\"o}rster Schreiber}, {Garcia}, {Gao}, {Gendron}, {Genzel},
  {Gillessen}, {Habibi}, {Haubois}, {Henning}, {Hippler}, {Horrobin},
  {Jim{\'e}nez-Rosales}, {Jochum}, {Jocou}, {Kaufer}, {Kervella}, {Lacour},
  {Lapeyr{\`e}re}, {Le Bouquin}, {L{\'e}na}, {Nowak}, {Ott}, {Paumard},
  {Perraut}, {Perrin}, {Pfuhl}, {Rodr{\'\i}guez-Coira}, {Shangguan},
  {Scheithauer}, {Stadler}, {Straub}, {Straubmeier}, {Sturm}, {Tacconi},
  {Vincent}, {von Fellenberg}, {Waisberg}, {Widmann}, {Wieprecht}, {Wiezorrek},
  {Woillez}, {Yazici}, \& {Zins}}]{2020A&A...636L...5G}
---. 2020{\natexlab{a}}, \aap, 636, L5, \dodoi{10.1051/0004-6361/202037813}

\bibitem[{{Gravity Collaboration} {et~al.}(2020{\natexlab{b}}){Gravity
  Collaboration}, {Abuter}, {Amorim}, {Baub{\"o}ck}, {Berger}, {Bonnet}, {Brand
  ner}, {Cardoso}, {Cl{\'e}net}, {de Zeeuw}, {Dallilar}, {Dexter}, {Eckart},
  {Eisenhauer}, {F{\"o}rster Schreiber}, {Garcia}, {Gao}, {Gendron}, {Genzel},
  {Gillessen}, {Habibi}, {Haubois}, {Henning}, {Hippler}, {Horrobin},
  {Jim{\'e}nez-Rosales}, {Jochum}, {Jocou}, {Kaufer}, {Kervella}, {Lacour},
  {Lapeyr{\`e}re}, {Le Bouquin}, {L{\'e}na}, {Nowak}, {Ott}, {Paumard},
  {Perraut}, {Perrin}, {Pfuhl}, {Ponti}, {Rodriguez Coira}, {Shangguan},
  {Scheithauer}, {Stadler}, {Straub}, {Straubmeier}, {Sturm}, {Tacconi},
  {Vincent}, {von Fellenberg}, {Waisberg}, {Widmann}, {Wieprecht}, {Wiezorrek},
  {Woillez}, {Yazici}, \& {Zins}}]{2020A&A...638A...2G}
---. 2020{\natexlab{b}}, \aap, 638, A2, \dodoi{10.1051/0004-6361/202037717}

\bibitem[{{Gravity Collaboration} {et~al.}(2020{\natexlab{c}}){Gravity
  Collaboration}, {Jim{\'e}nez-Rosales}, {Dexter}, {Widmann}, {Baub{\"o}ck},
  {Abuter}, {Amorim}, {Berger}, {Bonnet}, {Brandner}, {Cl{\'e}net}, {de Zeeuw},
  {Eckart}, {Eisenhauer}, {F{\"o}rster Schreiber}, {Garcia}, {Gao}, {Gendron},
  {Genzel}, {Gillessen}, {Habibi}, {Haubois}, {Hei{\ss}el}, {Henning},
  {Hippler}, {Horrobin}, {Jochum}, {Jocou}, {Kaufer}, {Kervella}, {Lacour},
  {Lapeyr{\`e}re}, {Le Bouquin}, {L{\'e}na}, {Nowak}, {Ott}, {Paumard},
  {Perraut}, {Perrin}, {Pfuhl}, {Rodr{\'\i}guez-Coira}, {Shangguan},
  {Scheithauer}, {Stadler}, {Straub}, {Straubmeier}, {Sturm}, {Tacconi},
  {Vincent}, {von Fellenberg}, {Waisberg}, {Wieprecht}, {Wiezorrek}, {Woillez},
  {Yazici}, \& {Zins}}]{GRAVITY+2020}
{Gravity Collaboration}, {Jim{\'e}nez-Rosales}, A., {Dexter}, J., {et~al.}
  2020{\natexlab{c}}, \aap, 643, A56, \dodoi{10.1051/0004-6361/202038283}

\bibitem[{{Gravity Collaboration} {et~al.}(2021{\natexlab{a}}){Gravity
  Collaboration}, {Abuter}, {Amorim}, {Baub{\"o}ck}, {Berger}, {Bonnet},
  {Brandner}, {Cl{\'e}net}, {Dallilar}, {Davies}, {de Zeeuw}, {Dexter},
  {Drescher}, {Eisenhauer}, {F{\"o}rster Schreiber}, {Garcia}, {Gao},
  {Gendron}, {Genzel}, {Gillessen}, {Habibi}, {Haubois}, {Hei{\ss}el},
  {Henning}, {Hippler}, {Horrobin}, {Jim{\'e}nez-Rosales}, {Jochum}, {Jocou},
  {Kaufer}, {Kervella}, {Lacour}, {Lapeyr{\`e}re}, {Le Bouquin}, {L{\'e}na},
  {Lutz}, {Nowak}, {Ott}, {Paumard}, {Perraut}, {Perrin}, {Pfuhl}, {Rabien},
  {Rodr{\'\i}guez-Coira}, {Shangguan}, {Shimizu}, {Scheithauer}, {Stadler},
  {Straub}, {Straubmeier}, {Sturm}, {Tacconi}, {Vincent}, {von Fellenberg},
  {Waisberg}, {Widmann}, {Wieprecht}, {Wiezorrek}, {Woillez}, {Yazici}, \&
  {Zins}}]{2021A&A...645A.127G}
{Gravity Collaboration}, {Abuter}, R., {Amorim}, A., {et~al.}
  2021{\natexlab{a}}, \aap, 645, A127, \dodoi{10.1051/0004-6361/202039544}

\bibitem[{{Gravity Collaboration} {et~al.}(2021{\natexlab{b}}){Gravity
  Collaboration}, {Abuter}, {Amorim}, {Baub{\"o}ck}, {Berger}, {Bonnet},
  {Brandner}, {Cl{\'e}net}, {Davies}, {de Zeeuw}, {Dexter}, {Dallilar},
  {Drescher}, {Eckart}, {Eisenhauer}, {F{\"o}rster Schreiber}, {Garcia}, {Gao},
  {Gendron}, {Genzel}, {Gillessen}, {Habibi}, {Haubois}, {Hei{\ss}el},
  {Henning}, {Hippler}, {Horrobin}, {Jim{\'e}nez-Rosales}, {Jochum}, {Jocou},
  {Kaufer}, {Kervella}, {Lacour}, {Lapeyr{\`e}re}, {Le Bouquin}, {L{\'e}na},
  {Lutz}, {Nowak}, {Ott}, {Paumard}, {Perraut}, {Perrin}, {Pfuhl}, {Rabien},
  {Rodr{\'\i}guez-Coira}, {Shangguan}, {Shimizu}, {Scheithauer}, {Stadler},
  {Straub}, {Straubmeier}, {Sturm}, {Tacconi}, {Vincent}, {von Fellenberg},
  {Waisberg}, {Widmann}, {Wieprecht}, {Wiezorrek}, {Woillez}, {Yazici},
  {Young}, \& {Zins}}]{2021A&A...647A..59G}
---. 2021{\natexlab{b}}, \aap, 647, A59, \dodoi{10.1051/0004-6361/202040208}

\bibitem[{{GRAVITY Collaboration} {et~al.}(2021){GRAVITY Collaboration},
  {Abuter}, {Amorim}, {Baub{\"o}ck}, {Baganoff}, {Berger}, {Boyce}, {Bonnet},
  {Brandner}, {Cl{\'e}net}, {Davies}, {de Zeeuw}, {Dexter}, {Dallilar},
  {Drescher}, {Eckart}, {Eisenhauer}, {Fazio}, {F{\"o}rster Schreiber},
  {Foster}, {Gammie}, {Garcia}, {Gao}, {Gendron}, {Genzel}, {Ghisellini},
  {Gillessen}, {Gurwell}, {Habibi}, {Haggard}, {Hailey}, {Harrison}, {Haubois},
  {Hei{\ss}el}, {Henning}, {Hippler}, {Hora}, {Horrobin},
  {Jim{\'e}nez-Rosales}, {Jochum}, {Jocou}, {Kaufer}, {Kervella}, {Lacour},
  {Lapeyr{\`e}re}, {Le Bouquin}, {L{\'e}na}, {Lowrance}, {Lutz}, {Markoff},
  {Mori}, {Morris}, {Neilsen}, {Nowak}, {Ott}, {Paumard}, {Perraut}, {Perrin},
  {Ponti}, {Pfuhl}, {Rabien}, {Rodr{\'\i}guez-Coira}, {Shangguan}, {Shimizu},
  {Scheithauer}, {Smith}, {Stadler}, {Stern}, {Straub}, {Straubmeier}, {Sturm},
  {Tacconi}, {Vincent}, {von Fellenberg}, {Waisberg}, {Widmann}, {Wieprecht},
  {Wiezorrek}, {Willner}, {Witzel}, {Woillez}, {Yazici}, {Young}, {Zhang}, \&
  {Zins}}]{2021A&A...654A..22G}
{GRAVITY Collaboration}, {Abuter}, R., {Amorim}, A., {et~al.} 2021, \aap, 654,
  A22, \dodoi{10.1051/0004-6361/202140981}

\bibitem[{{GRAVITY Collaboration} {et~al.}(2022){GRAVITY Collaboration},
  {Abuter, R.}, {Aimar, N.}, {Amorim, A.}, {Ball, J.}, {Baub\"ock, M.},
  {Berger, J. P.}, {Bonnet, H.}, {Bourdarot, G.}, {Brandner, W.}, {Cardoso,
  V.}, {Cl\'enet, Y.}, {Dallilar, Y.}, {Davies, R.}, {de Zeeuw, P. T.},
  {Dexter, J.}, {Drescher, A.}, {Eisenhauer, F.}, {F\"orster Schreiber, N. M.},
  {Foschi, A.}, {Garcia, P.}, {Gao, F.}, {Gendron, E.}, {Genzel, R.},
  {Gillessen, S.}, {Habibi, M.}, {Haubois, X.}, {Hei\ss{}el, G.}, {Henning,
  T.}, {Hippler, S.}, {Horrobin, M.}, {Jochum, L.}, {Jocou, L.}, {Kaufer, A.},
  {Kervella, P.}, {Lacour, S.}, {Lapeyr\`ere, V.}, {Le Bouquin, J.-B.},
  {L\'ena, P.}, {Lutz, D.}, {Ott, T.}, {Paumard, T.}, {Perraut, K.}, {Perrin,
  G.}, {Pfuhl, O.}, {Rabien, S.}, {Shangguan, J.}, {Shimizu, T.}, {Scheithauer,
  S.}, {Stadler, J.}, {Stephens, A.W.}, {Straub, O.}, {Straubmeier, C.},
  {Sturm, E.}, {Tacconi, L. J.}, {Tristram, K. R. W.}, {Vincent, F.}, {von
  Fellenberg, S.}, {Widmann, F.}, {Wieprecht, E.}, {Wiezorrek, E.}, {Woillez,
  J.}, {Yazici, S.}, \& {Young, A.}}]{2021arXiv211207478G}
{GRAVITY Collaboration}, {Abuter, R.}, {Aimar, N.}, {et~al.} 2022, A\&A, 657,
  L12, \dodoi{10.1051/0004-6361/202142465}

\bibitem[{{Gurvits} {et~al.}(2021){Gurvits}, {Paragi}, {Casasola}, {Conway},
  {Davelaar}, {Falcke}, {Fender}, {Frey}, {Fromm}, {Mir{\'o}}, {Garrett},
  {Giroletti}, {Goddi}, {G{\'o}mez}, {van der Gucht}, {Guirado}, {Haiman},
  {Helmich}, {Humphreys}, {Impellizzeri}, {Kramer}, {Lindqvist}, {Linz},
  {Liuzzo}, {Lobanov}, {Mizuno}, {Rezzolla}, {Roelofs}, {Ros}, {Rygl},
  {Savolainen}, {Schuster}, {Venturi}, {Wiedner}, \& {Zensus}}]{Gurvits2021}
{Gurvits}, L.~I., {Paragi}, Z., {Casasola}, V., {et~al.} 2021, Experimental
  Astronomy, 51, 559, \dodoi{10.1007/s10686-021-09714-y}

\bibitem[{{Haggard} {et~al.}(2019){Haggard}, {Nynka}, {Mon}, {de la Cruz
  Hernandez}, {Nowak}, {Heinke}, {Neilsen}, {Dexter}, {Fragile}, {Baganoff},
  {Bower}, {Corrales}, {Coti Zelati}, {Degenaar}, {Markoff}, {Morris}, {Ponti},
  {Rea}, {Wilms}, \& {Yusef-Zadeh}}]{2019ApJ...886...96H}
{Haggard}, D., {Nynka}, M., {Mon}, B., {et~al.} 2019, \apj, 886, 96,
  \dodoi{10.3847/1538-4357/ab4a7f}

\bibitem[{{Harko} {et~al.}(2009){Harko}, {Kov{\'a}cs}, \& {Lobo}}]{Harko+2009}
{Harko}, T., {Kov{\'a}cs}, Z., \& {Lobo}, F. S.~N. 2009, \prd, 79, 064001,
  \dodoi{10.1103/PhysRevD.79.064001}

\bibitem[{Harlow(2016)}]{Harlow2014}
Harlow, D. 2016, Rev. Mod. Phys., 88, 015002,
  \dodoi{10.1103/RevModPhys.88.015002}

\bibitem[{{Hawking}(1972)}]{Hawking1972}
{Hawking}, S.~W. 1972, Communications in Mathematical Physics, 25, 152,
  \dodoi{10.1007/BF01877517}

\bibitem[{Hawking \& Ellis(1973)}]{Hawking+1973}
Hawking, S.~W., \& Ellis, G. F.~R. 1973, The large scale structure of spacetime
  (Cambridge, England: Cambridge University Press)

\bibitem[{{Hayward}(2006)}]{Hayward2006}
{Hayward}, S.~A. 2006, Phys. Rev. Lett., 96, 031103,
  \dodoi{10.1103/PhysRevLett.96.031103}

\bibitem[{{Hees} {et~al.}(2019){Hees}, {Dehghanfar}, {Do}, {Ghez}, {Martinez},
  {Campbell}, \& {Lu}}]{2019ApJ...880...87H}
{Hees}, A., {Dehghanfar}, A., {Do}, T., {et~al.} 2019, \apj, 880, 87,
  \dodoi{10.3847/1538-4357/ab2ae0}

\bibitem[{{Hees} {et~al.}(2017){Hees}, {Do}, {Ghez}, {Martinez}, {Naoz},
  {Becklin}, {Boehle}, {Chappell}, {Chu}, {Dehghanfar}, {Kosmo}, {Lu},
  {Matthews}, {Morris}, {Sakai}, {Sch{\"o}del}, \&
  {Witzel}}]{2017PhRvL.118u1101H}
{Hees}, A., {Do}, T., {Ghez}, A.~M., {et~al.} 2017, \prl, 118, 211101,
  \dodoi{10.1103/PhysRevLett.118.211101}

\bibitem[{{Hees} {et~al.}(2020){Hees}, {Do}, {Roberts}, {Ghez}, {Nishiyama},
  {Bentley}, {Gautam}, {Jia}, {Kara}, {Lu}, {Saida}, {Sakai}, {Takahashi}, \&
  {Takamori}}]{2020PhRvL.124h1101H}
{Hees}, A., {Do}, T., {Roberts}, B.~M., {et~al.} 2020, \prl, 124, 081101,
  \dodoi{10.1103/PhysRevLett.124.081101}

\bibitem[{{Held} {et~al.}(2019){Held}, {Gold}, \& {Eichhorn}}]{Held+2019}
{Held}, A., {Gold}, R., \& {Eichhorn}, A. 2019, JCAP, 2019, 029,
  \dodoi{10.1088/1475-7516/2019/06/029}

\bibitem[{{Herrnstein} {et~al.}(2004){Herrnstein}, {Zhao}, {Bower}, \&
  {Goss}}]{2004AJ....127.3399H}
{Herrnstein}, R.~M., {Zhao}, J.-H., {Bower}, G.~C., \& {Goss}, W.~M. 2004, \aj,
  127, 3399, \dodoi{10.1086/420711}

\bibitem[{{Hora} {et~al.}(2014){Hora}, {Witzel}, {Ashby}, {Becklin}, {Carey},
  {Fazio}, {Ghez}, {Ingalls}, {Meyer}, {Morris}, {Smith}, \&
  {Willner}}]{Hora+2014}
{Hora}, J.~L., {Witzel}, G., {Ashby}, M.~L.~N., {et~al.} 2014, \apj, 793, 120,
  \dodoi{10.1088/0004-637X/793/2/120}

\bibitem[{{Hornstein} {et~al.}(2007){Hornstein}, {Matthews}, {Ghez}, {Lu},
  {Morris}, {Becklin}, {Rafelski}, \& {Baganoff}}]{2007ApJ...667..900H}
{Hornstein}, S.~D., {Matthews}, K., {Ghez}, A.~M., {et~al.} 2007, \apj, 667,
  900, \dodoi{10.1086/520762}

\bibitem[{{Israel}(1967)}]{Israel1967}
{Israel}, W. 1967, Physical Review, 164, 1776, \dodoi{10.1103/PhysRev.164.1776}

\bibitem[{{Israel}(1968)}]{Israel1968}
---. 1968, Communications in Mathematical Physics, 8, 245,
  \dodoi{10.1007/BF01645859}

\bibitem[{{Janis} {et~al.}(1968){Janis}, {Newman}, \& {Winicour}}]{Janis+1968}
{Janis}, A.~I., {Newman}, E.~T., \& {Winicour}, J. 1968, Phys. Rev. Lett., 20,
  878, \dodoi{10.1103/PhysRevLett.20.878}

\bibitem[{{Janssen} {et~al.}(2019){Janssen}, {Goddi}, {van Bemmel}, {Kettenis},
  {Small}, {Liuzzo}, {Rygl}, {Mart{\'\i}-Vidal}, {Blackburn}, {Wielgus}, \&
  {Falcke}}]{2019Janssen}
{Janssen}, M., {Goddi}, C., {van Bemmel}, I.~M., {et~al.} 2019, \aap, 626, A75,
  \dodoi{10.1051/0004-6361/201935181}

\bibitem[{{Jaroszynski} \& {Kurpiewski}(1997)}]{Jaroszynski1997}
{Jaroszynski}, M., \& {Kurpiewski}, A. 1997, \aap, 326, 419.
\newblock \doarXiv{astro-ph/9705044}

\bibitem[{{Jia} {et~al.}(2019){Jia}, {Lu}, {Sakai}, {Gautam}, {Do}, {Hosek},
  {Service}, {Ghez}, {Gallego-Cano}, {Sch{\"o}del}, {Hees}, {Morris},
  {Becklin}, \& {Matthews}}]{2019ApJ...873....9J}
{Jia}, S., {Lu}, J.~R., {Sakai}, S., {et~al.} 2019, \apj, 873, 9,
  \dodoi{10.3847/1538-4357/ab01de}

\bibitem[{{Johannsen}(2013{\natexlab{a}})}]{Johannsen2013b}
{Johannsen}, T. 2013{\natexlab{a}}, \prd, 88, 044002,
  \dodoi{10.1103/PhysRevD.88.044002}

\bibitem[{{Johannsen}(2013{\natexlab{b}})}]{Johannsen2013a}
---. 2013{\natexlab{b}}, \prd, 87, 124017, \dodoi{10.1103/PhysRevD.87.124017}

\bibitem[{{Johannsen} \& {Psaltis}(2010)}]{Johannsen2010}
{Johannsen}, T., \& {Psaltis}, D. 2010, \apj, 718, 446,
  \dodoi{10.1088/0004-637X/718/1/446}

\bibitem[{{Johannsen} \& {Psaltis}(2011)}]{Johannsen2011}
---. 2011, \prd, 83, 124015, \dodoi{10.1103/PhysRevD.83.124015}

\bibitem[{{Johnson} {et~al.}(2020){Johnson}, {Lupsasca}, {Strominger}, {Wong},
  {Hadar}, {Kapec}, {Narayan}, {Chael}, {Gammie}, {Galison}, {Palumbo},
  {Doeleman}, {Blackburn}, {Wielgus}, {Pesce}, {Farah}, \&
  {Moran}}]{Johnson+2020}
{Johnson}, M.~D., {Lupsasca}, A., {Strominger}, A., {et~al.} 2020, Science
  Advances, 6, eaaz1310, \dodoi{10.1126/sciadv.aaz1310}

\bibitem[{{Joshi} {et~al.}(2011){Joshi}, {Malafarina}, \&
  {Narayan}}]{Joshi+2011}
{Joshi}, P.~S., {Malafarina}, D., \& {Narayan}, R. 2011, Classical and Quantum
  Gravity, 28, 235018, \dodoi{10.1088/0264-9381/28/23/235018}

\bibitem[{{Joshi} {et~al.}(2014){Joshi}, {Malafarina}, \&
  {Narayan}}]{Joshi+2014}
---. 2014, Classical and Quantum Gravity, 31, 015002,
  \dodoi{10.1088/0264-9381/31/1/015002}

\bibitem[{{Kallosh} {et~al.}(1992){Kallosh}, {Linde}, {Ort{\'\i}n}, {Peet}, \&
  {van Proeyen}}]{Kallosh+1992}
{Kallosh}, R., {Linde}, A., {Ort{\'\i}n}, T., {Peet}, A., \& {van Proeyen}, A.
  1992, Phys. Rev. D, 46, 5278, \dodoi{10.1103/PhysRevD.46.5278}

\bibitem[{{Kazakov} \& {Solodukhin}(1994)}]{Kazakov+1994}
{Kazakov}, D.~I., \& {Solodukhin}, S.~N. 1994, Nuclear Physics B, 429, 153,
  \dodoi{10.1016/S0550-3213(94)80045-6}

\bibitem[{{Kerr}(1963)}]{Kerr1963}
{Kerr}, R.~P. 1963, \prl, 11, 237, \dodoi{10.1103/PhysRevLett.11.237}

\bibitem[{{Khan} {et~al.}(2016){Khan}, {Husa}, {Hannam}, {Ohme}, {P{\"u}rrer},
  {Forteza}, \& {Boh{\'e}}}]{Khan2016}
{Khan}, S., {Husa}, S., {Hannam}, M., {et~al.} 2016, \prd, 93, 044007,
  \dodoi{10.1103/PhysRevD.93.044007}

\bibitem[{{Kocherlakota} \& {Rezzolla}(2020)}]{Kocherlakota+2020}
{Kocherlakota}, P., \& {Rezzolla}, L. 2020, Phys. Rev. D, 102, 064058,
  \dodoi{10.1103/PhysRevD.102.064058}

\bibitem[{{Kocherlakota} \& {Rezzolla}(2022)}]{Kocherlakota+2022}
---. 2022, arXiv e-prints, arXiv:2201.05641.
\newblock \doarXiv{2201.05641}

\bibitem[{{Kocherlakota} {et~al.}(2021){Kocherlakota}, {Rezzolla}, {Falcke},
  {Fromm}, {Kramer}, {Mizuno}, {Nathanail}, {Olivares}, {Younsi}, {Akiyama},
  {Alberdi}, {Alef}, {Algaba}, {Anantua}, {Asada}, {Azulay}, {Baczko}, {Ball},
  {Balokovi{\'c}}, {Barrett}, {Benson}, {Bintley}, {Blackburn}, {Blundell},
  {Boland}, {Bouman}, {Bower}, {Boyce}, {Bremer}, {Brinkerink}, {Brissenden},
  {Britzen}, {Broderick}, {Broguiere}, {Bronzwaer}, {Byun}, {Carlstrom},
  {Chael}, {Chan}, {Chatterjee}, {Chatterjee}, {Chen}, {Chen}, {Chesler},
  {Cho}, {Christian}, {Conway}, {Cordes}, {Crawford}, {Crew}, {Cruz-Osorio},
  {Cui}, {Davelaar}, {De Laurentis}, {Deane}, {Dempsey}, {Desvignes},
  {Doeleman}, {Eatough}, {Farah}, {Fish}, {Fomalont}, {Fraga-Encinas},
  {Friberg}, {Ford}, {Fuentes}, {Galison}, {Gammie}, {Garc{\'\i}a}, {Gentaz},
  {Georgiev}, {Goddi}, {Gold}, {G{\'o}mez}, {G{\'o}mez-Ruiz}, {Gu}, {Gurwell},
  {Hada}, {Haggard}, {Hecht}, {Hesper}, {Ho}, {Ho}, {Honma}, {Huang}, {Huang},
  {Hughes}, {Ikeda}, {Inoue}, {Issaoun}, {James}, {Jannuzi}, {Janssen},
  {Jeter}, {Jiang}, {Jimenez-Rosales}, {Johnson}, {Jorstad}, {Jung}, {Karami},
  {Karuppusamy}, {Kawashima}, {Keating}, {Kettenis}, {Kim}, {Kim}, {Kim},
  {Kim}, {Kino}, {Koay}, {Kofuji}, {Koch}, {Koyama}, {Kramer}, {Krichbaum},
  {Kuo}, {Lauer}, {Lee}, {Levis}, {Li}, {Li}, {Lindqvist}, {Lico}, {Lindahl},
  {Liu}, {Liu}, {Liuzzo}, {Lo}, {Lobanov}, {Loinard}, {Lonsdale}, {Lu},
  {MacDonald}, {Mao}, {Marchili}, {Markoff}, {Marrone}, {Marscher},
  {Mart{\'\i}-Vidal}, {Matsushita}, {Matthews}, {Medeiros}, {Menten}, {Mizuno},
  {Moran}, {Moriyama}, {Moscibrodzka}, {M{\"u}ller}, {Musoke}, {Mej{\'\i}as},
  {Nagai}, {Nagar}, {Nakamura}, {Narayan}, {Narayanan}, {Natarajan}, {Neilsen},
  {Neri}, {Ni}, {Noutsos}, {Nowak}, {Okino}, {Ortiz-Le{\'o}n}, {Oyama},
  {{\"O}zel}, {Palumbo}, {Park}, {Patel}, {Pen}, {Pesce}, {Pi{\'e}tu},
  {Plambeck}, {PopStefanija}, {Porth}, {P{\"o}tzl}, {Prather},
  {Preciado-L{\'o}pez}, {Psaltis}, {Pu}, {Ramakrishnan}, {Rao}, {Rawlings},
  {Raymond}, {Ricarte}, {Ripperda}, {Roelofs}, {Rogers}, {Ros}, {Rose},
  {Roshanineshat}, {Rottmann}, {Roy}, {Ruszczyk}, {Rygl}, {S{\'a}nchez},
  {S{\'a}nchez-Arguelles}, {Sasada}, {Savolainen}, {Schloerb}, {Schuster},
  {Shao}, {Shen}, {Small}, {Sohn}, {SooHoo}, {Sun}, {Tazaki}, {Tetarenko},
  {Tiede}, {Tilanus}, {Titus}, {Toma}, {Torne}, {Trent}, {Traianou}, {Trippe},
  {van Bemmel}, {van Langevelde}, {van Rossum}, {Wagner}, {Ward-Thompson},
  {Wardle}, {Weintroub}, {Wex}, {Wharton}, {Wielgus}, {Wong}, {Wu}, {Yoon},
  {Young}, {Young}, {Yuan}, {Yuan}, {Zensus}, {Zhao}, {Zhao}, \& {EHT
  Collaboration}}]{Kocherlakota2021}
{Kocherlakota}, P., {Rezzolla}, L., {Falcke}, H., {et~al.} 2021, \prd, 103,
  104047, \dodoi{10.1103/PhysRevD.103.104047}

\bibitem[{{Konoplya} {et~al.}(2016){Konoplya}, {Rezzolla}, \&
  {Zhidenko}}]{Konoplya2016}
{Konoplya}, R., {Rezzolla}, L., \& {Zhidenko}, A. 2016, \prd, 93, 064015,
  \dodoi{10.1103/PhysRevD.93.064015}

\bibitem[{Kramer {et~al.}(2021)Kramer, Stairs, Manchester, Wex, Deller, Coles,
  Ali, Burgay, Camilo, Cognard, Damour, Desvignes, Ferdman, Freire, Grondin,
  Guillemot, Hobbs, Janssen, Karuppusamy, Lorimer, Lyne, McKee, McLaughlin,
  M\"unch, Perera, Pol, Possenti, Sarkissian, Stappers, \&
  Theureau}]{Kramer2021}
Kramer, M., Stairs, I.~H., Manchester, R.~N., {et~al.} 2021, Phys. Rev. X, 11,
  041050, \dodoi{10.1103/PhysRevX.11.041050}

\bibitem[{{Kudriashov} {et~al.}(2021){Kudriashov}, {Martin-Neira}, {Roelofs},
  {Falcke}, {Brinkerink}, {Baryshev}, {Hogerheijde}, {Young}, {Pourshaghaghi},
  {Klein-Wolt}, {Moscibrodzka}, {Davelaar}, {Barat}, {Duesmann}, {Valenta},
  {Perdigues Armengol}, {De Wilde}, {Iglesias}, {Alagha}, \& {Van Der
  Vorst}}]{Kudriashov+2021}
{Kudriashov}, V., {Martin-Neira}, M., {Roelofs}, F., {et~al.} 2021, Chinese
  Journal of Space Science, 41, 211, \dodoi{10.11728/cjss2021.02.211}

\bibitem[{{Lambert} \& {Le Poncin-Lafitte}(2011)}]{Lambert2011}
{Lambert}, S.~B., \& {Le Poncin-Lafitte}, C. 2011, \aap, 529, A70,
  \dodoi{10.1051/0004-6361/201016370}

\bibitem[{{Liu} {et~al.}(2016){Liu}, {Wright}, {Zhao}, {Brinkerink}, {Ho},
  {Mills}, {Mart{\'\i}n}, {Falcke}, {Matsushita}, \&
  {Mart{\'\i}-Vidal}}]{2016A&A...593A.107L}
{Liu}, H.~B., {Wright}, M. C.~H., {Zhao}, J.-H., {et~al.} 2016, \aap, 593,
  A107, \dodoi{10.1051/0004-6361/201628731}

\bibitem[{{Lu} {et~al.}(2017){Lu}, {Kumar}, \& {Narayan}}]{Lu+2017}
{Lu}, W., {Kumar}, P., \& {Narayan}, R. 2017, \mnras, 468, 910,
  \dodoi{10.1093/mnras/stx542}

\bibitem[{{Luminet}(1979)}]{Luminet1979}
{Luminet}, J.~P. 1979, \aap, 75, 228

\bibitem[{{Magueijo}(2003)}]{Magueijo2003}
{Magueijo}, J. 2003, Reports on Progress in Physics, 66, 2025,
  \dodoi{10.1088/0034-4885/66/11/R04}

\bibitem[{{Markoff} {et~al.}(2001){Markoff}, {Falcke}, {Yuan}, \&
  {Biermann}}]{Markoff+2001}
{Markoff}, S., {Falcke}, H., {Yuan}, F., \& {Biermann}, P.~L. 2001, \aap, 379,
  L13, \dodoi{10.1051/0004-6361:20011346}

\bibitem[{{Marrone} {et~al.}(2008){Marrone}, {Baganoff}, {Morris}, {Moran},
  {Ghez}, {Hornstein}, {Dowell}, {Mu{\~n}oz}, {Bautz}, {Ricker}, {Brandt},
  {Garmire}, {Lu}, {Matthews}, {Zhao}, {Rao}, \& {Bower}}]{2008ApJ...682..373M}
{Marrone}, D.~P., {Baganoff}, F.~K., {Morris}, M.~R., {et~al.} 2008, \apj, 682,
  373, \dodoi{10.1086/588806}

\bibitem[{{Martins} {et~al.}(2008){Martins}, {Gillessen}, {Eisenhauer},
  {Genzel}, {Ott}, \& {Trippe}}]{2008ApJ...672L.119M}
{Martins}, F., {Gillessen}, S., {Eisenhauer}, F., {et~al.} 2008, \apjl, 672,
  L119, \dodoi{10.1086/526768}

\bibitem[{{Mazur} \& {Mottola}(2001)}]{Mazur_Mottola_2001}
{Mazur}, P.~O., \& {Mottola}, E. 2001, arXiv e-prints, gr.
\newblock \doarXiv{gr-qc/0109035}

\bibitem[{{McClintock} {et~al.}(2004){McClintock}, {Narayan}, \&
  {Rybicki}}]{McClintock+2004}
{McClintock}, J.~E., {Narayan}, R., \& {Rybicki}, G.~B. 2004, \apj, 615, 402,
  \dodoi{10.1086/424474}

\bibitem[{{Medeiros} {et~al.}(2020){Medeiros}, {Psaltis}, \&
  {{\"O}zel}}]{Medeiros2020}
{Medeiros}, L., {Psaltis}, D., \& {{\"O}zel}, F. 2020, \apj, 896, 7,
  \dodoi{10.3847/1538-4357/ab8bd1}

\bibitem[{{Menten} {et~al.}(1997){Menten}, {Reid}, {Eckart}, \&
  {Genzel}}]{1997ApJ...475L.111M}
{Menten}, K.~M., {Reid}, M.~J., {Eckart}, A., \& {Genzel}, R. 1997, \apjl, 475,
  L111, \dodoi{10.1086/310472}

\bibitem[{{Meyer} {et~al.}(2012){Meyer}, {Ghez}, {Sch{\"o}del}, {Yelda},
  {Boehle}, {Lu}, {Do}, {Morris}, {Becklin}, \&
  {Matthews}}]{2012Sci...338...84M}
{Meyer}, L., {Ghez}, A.~M., {Sch{\"o}del}, R., {et~al.} 2012, Science, 338, 84,
  \dodoi{10.1126/science.1225506}

\bibitem[{{Mizuno} {et~al.}(2018){Mizuno}, {Younsi}, {Fromm}, {Porth}, {De
  Laurentis}, {Olivares}, {Falcke}, {Kramer}, \& {Rezzolla}}]{Mizuno+2018}
{Mizuno}, Y., {Younsi}, Z., {Fromm}, C.~M., {et~al.} 2018, Nature Astronomy, 2,
  585, \dodoi{10.1038/s41550-018-0449-5}

\bibitem[{{Morris} {et~al.}(2012){Morris}, {Meyer}, \&
  {Ghez}}]{2012RAA....12..995M}
{Morris}, M.~R., {Meyer}, L., \& {Ghez}, A.~M. 2012, Research in Astronomy and
  Astrophysics, 12, 995, \dodoi{10.1088/1674-4527/12/8/007}

\bibitem[{{Morris} \& {Thorne}(1988)}]{Morris+1988}
{Morris}, M.~S., \& {Thorne}, K.~S. 1988, American Journal of Physics, 56, 395,
  \dodoi{10.1119/1.15620}

\bibitem[{{Morris} {et~al.}(1988){Morris}, {Thorne}, \&
  {Yurtsever}}]{Morris+1988b}
{Morris}, M.~S., {Thorne}, K.~S., \& {Yurtsever}, U. 1988, \prl, 61, 1446,
  \dodoi{10.1103/PhysRevLett.61.1446}

\bibitem[{{Mo{\'s}cibrodzka} {et~al.}(2016){Mo{\'s}cibrodzka}, {Falcke}, \&
  {Shiokawa}}]{Moscibrodzka+2016}
{Mo{\'s}cibrodzka}, M., {Falcke}, H., \& {Shiokawa}, H. 2016, \aap, 586, A38,
  \dodoi{10.1051/0004-6361/201526630}

\bibitem[{{Mo{\'s}cibrodzka} \& {Gammie}(2018)}]{Moscibrodzka+2018}
{Mo{\'s}cibrodzka}, M., \& {Gammie}, C.~F. 2018, \mnras, 475, 43,
  \dodoi{10.1093/mnras/stx3162}

\bibitem[{{Mou} {et~al.}(2014){Mou}, {Yuan}, {Bu}, {Sun}, \& {Su}}]{Mou+2014}
{Mou}, G., {Yuan}, F., {Bu}, D., {Sun}, M., \& {Su}, M. 2014, \apj, 790, 109,
  \dodoi{10.1088/0004-637X/790/2/109}

\bibitem[{{Murchikova} \& {Witzel}(2021)}]{2021ApJ...920L...7M}
{Murchikova}, L., \& {Witzel}, G. 2021, \apjl, 920, L7,
  \dodoi{10.3847/2041-8213/ac2308}

\bibitem[{{Narayan}(2002)}]{Narayan_2002}
{Narayan}, R. 2002, in Lighthouses of the Universe: The Most Luminous Celestial
  Objects and Their Use for Cosmology, ed. M.~{Gilfanov}, R.~{Sunyeav}, \&
  E.~{Churazov}, 405, \dodoi{10.1007/10856495\_60}

\bibitem[{{Narayan} {et~al.}(2021){Narayan}, {Chael}, {Chatterjee}, {Ricarte},
  \& {Curd}}]{Narayan+2021}
{Narayan}, R., {Chael}, A., {Chatterjee}, K., {Ricarte}, A., \& {Curd}, B.
  2021, arXiv e-prints, arXiv:2108.12380.
\newblock \doarXiv{2108.12380}

\bibitem[{{Narayan} {et~al.}(1997){Narayan}, {Garcia}, \&
  {McClintock}}]{Narayan+1997}
{Narayan}, R., {Garcia}, M.~R., \& {McClintock}, J.~E. 1997, \apjl, 478, L79,
  \dodoi{10.1086/310554}

\bibitem[{{Narayan} {et~al.}(2019){Narayan}, {Johnson}, \&
  {Gammie}}]{Narayan2019}
{Narayan}, R., {Johnson}, M.~D., \& {Gammie}, C.~F. 2019, \apjl, 885, L33,
  \dodoi{10.3847/2041-8213/ab518c}

\bibitem[{{Narayan} {et~al.}(1998){Narayan}, {Mahadevan}, {Grindlay}, {Popham},
  \& {Gammie}}]{Narayan+1998}
{Narayan}, R., {Mahadevan}, R., {Grindlay}, J.~E., {Popham}, R.~G., \&
  {Gammie}, C. 1998, \apj, 492, 554, \dodoi{10.1086/305070}

\bibitem[{{Narayan} \& {McClintock}(2008)}]{Narayan_McClintock_2008}
{Narayan}, R., \& {McClintock}, J.~E. 2008, \nar, 51, 733,
  \dodoi{10.1016/j.newar.2008.03.002}

\bibitem[{{Narayan} \& {Yi}(1995)}]{Narayan_Yi_1995}
{Narayan}, R., \& {Yi}, I. 1995, \apj, 452, 710, \dodoi{10.1086/176343}

\bibitem[{{Narayan} {et~al.}(1995){Narayan}, {Yi}, \&
  {Mahadevan}}]{Narayan+1995}
{Narayan}, R., {Yi}, I., \& {Mahadevan}, R. 1995, \nat, 374, 623,
  \dodoi{10.1038/374623a0}

\bibitem[{{Narayan} {et~al.}(2016){Narayan}, {Zhu}, {Psaltis}, \&
  {S\k{a}dowski}}]{Narayan+2016}
{Narayan}, R., {Zhu}, Y., {Psaltis}, D., \& {S\k{a}dowski}, A. 2016, \mnras,
  457, 608, \dodoi{10.1093/mnras/stv2979}

\bibitem[{{Natarajan} {et~al.}(2022){Natarajan}, {Deane}, {Mart{\'\i}-Vidal},
  {Roelofs}, {Janssen}, {Wielgus}, {Blackburn}, {Blecher}, {Perkins},
  {Smirnov}, {Davelaar}, {Moscibrodzka}, {Chael}, {Bouman}, {Kim}, {Bernardi},
  {van Bemmel}, {Falcke}, {{\"O}zel}, \& {Psaltis}}]{natarajan2022}
{Natarajan}, I., {Deane}, R., {Mart{\'\i}-Vidal}, I., {et~al.} 2022, \mnras,
  512, 490, \dodoi{10.1093/mnras/stac531}

\bibitem[{{Nathanail} {et~al.}(2020){Nathanail}, {Fromm}, {Porth}, {Olivares},
  {Younsi}, {Mizuno}, \& {Rezzolla}}]{Nathanail+2020}
{Nathanail}, A., {Fromm}, C.~M., {Porth}, O., {et~al.} 2020, \mnras, 495, 1549,
  \dodoi{10.1093/mnras/staa1165}

\bibitem[{{Nathanail} {et~al.}(2021){Nathanail}, {Mpisketzis}, {Porth},
  {Fromm}, \& {Rezzolla}}]{Nathanail+2021}
{Nathanail}, A., {Mpisketzis}, V., {Porth}, O., {Fromm}, C.~M., \& {Rezzolla},
  L. 2021, arXiv e-prints, arXiv:2111.03689.
\newblock \doarXiv{2111.03689}

\bibitem[{{Neilsen} {et~al.}(2013){Neilsen}, {Nowak}, {Gammie}, {Dexter},
  {Markoff}, {Haggard}, {Nayakshin}, {Wang}, {Grosso}, {Porquet}, {Tomsick},
  {Degenaar}, {Fragile}, {Houck}, {Wijnands}, {Miller}, \&
  {Baganoff}}]{2013ApJ...774...42N}
{Neilsen}, J., {Nowak}, M.~A., {Gammie}, C., {et~al.} 2013, \apj, 774, 42,
  \dodoi{10.1088/0004-637X/774/1/42}

\bibitem[{{Neilsen} {et~al.}(2015){Neilsen}, {Markoff}, {Nowak}, {Dexter},
  {Witzel}, {Barri{\`e}re}, {Li}, {Baganoff}, {Degenaar}, {Fragile}, {Gammie},
  {Goldwurm}, {Grosso}, \& {Haggard}}]{2015ApJ...799..199N}
{Neilsen}, J., {Markoff}, S., {Nowak}, M.~A., {et~al.} 2015, \apj, 799, 199,
  \dodoi{10.1088/0004-637X/799/2/199}

\bibitem[{{Newman} {et~al.}(1965){Newman}, {Couch}, {Chinnapared}, {Exton},
  {Prakash}, \& {Torrence}}]{Newman+1965}
{Newman}, E.~T., {Couch}, E., {Chinnapared}, K., {et~al.} 1965, Journal of
  Mathematical Physics, 6, 918, \dodoi{10.1063/1.1704351}

\bibitem[{Nordstr{\"o}m(1918)}]{Nordstrom1918}
Nordstr{\"o}m, G. 1918, Proc. Kon. Ned. Akad. Wet., 20, 1238

\bibitem[{{Nowak} {et~al.}(2012){Nowak}, {Neilsen}, {Markoff}, {Baganoff},
  {Porquet}, {Grosso}, {Levin}, {Houck}, {Eckart}, {Falcke}, {Ji}, {Miller}, \&
  {Wang}}]{2012ApJ...759...95N}
{Nowak}, M.~A., {Neilsen}, J., {Markoff}, S.~B., {et~al.} 2012, \apj, 759, 95,
  \dodoi{10.1088/0004-637X/759/2/95}

\bibitem[{{Olivares} {et~al.}(2020){Olivares}, {Younsi}, {Fromm}, {De
  Laurentis}, {Porth}, {Mizuno}, {Falcke}, {Kramer}, \&
  {Rezzolla}}]{Olivares+2020}
{Olivares}, H., {Younsi}, Z., {Fromm}, C.~M., {et~al.} 2020, \mnras, 497, 521,
  \dodoi{10.1093/mnras/staa1878}

\bibitem[{{O'Neil} {et~al.}(2019){O'Neil}, {Martinez}, {Hees}, {Ghez}, {Do},
  {Witzel}, {Konopacky}, {Becklin}, {Chu}, {Lu}, {Matthews}, \&
  {Sakai}}]{2019AJ....158....4O}
{O'Neil}, K.~K., {Martinez}, G.~D., {Hees}, A., {et~al.} 2019, \aj, 158, 4,
  \dodoi{10.3847/1538-3881/ab1d66}

\bibitem[{{{\"O}zel} {et~al.}(2000){{\"O}zel}, {Psaltis}, \&
  {Narayan}}]{Ozel2000}
{{\"O}zel}, F., {Psaltis}, D., \& {Narayan}, R. 2000, \apj, 541, 234,
  \dodoi{10.1086/309396}

\bibitem[{{\"Ozel} {et~al.}(2021){\"Ozel}, {Psaltis}, \& {Younsi}}]{Ozel2021}
{\"Ozel}, F., {Psaltis}, D., \& {Younsi}, Z. 2021, arXiv e-prints,
  arXiv:2111.01123.
\newblock \doarXiv{2111.01123}

\bibitem[{{Palumbo} {et~al.}(2019){Palumbo}, {Doeleman}, {Johnson}, {Bouman},
  \& {Chael}}]{Palumbo+2019}
{Palumbo}, D. C.~M., {Doeleman}, S.~S., {Johnson}, M.~D., {Bouman}, K.~L., \&
  {Chael}, A.~A. 2019, \apj, 881, 62, \dodoi{10.3847/1538-4357/ab2bed}

\bibitem[{Peters(1964)}]{Peters1964}
Peters, P.~C. 1964, Phys. Rev., 136, B1224, \dodoi{10.1103/PhysRev.136.B1224}

\bibitem[{{Plewa} {et~al.}(2015){Plewa}, {Gillessen}, {Eisenhauer}, {Ott},
  {Pfuhl}, {George}, {Dexter}, {Habibi}, {Genzel}, {Reid}, \&
  {Menten}}]{2015MNRAS.453.3234P}
{Plewa}, P.~M., {Gillessen}, S., {Eisenhauer}, F., {et~al.} 2015, \mnras, 453,
  3234, \dodoi{10.1093/mnras/stv1910}

\bibitem[{{Ponti} {et~al.}(2010){Ponti}, {Terrier}, {Goldwurm}, {Belanger}, \&
  {Trap}}]{Ponti+2010}
{Ponti}, G., {Terrier}, R., {Goldwurm}, A., {Belanger}, G., \& {Trap}, G. 2010,
  \apj, 714, 732, \dodoi{10.1088/0004-637X/714/1/732}

\bibitem[{{Ponti} {et~al.}(2017){Ponti}, {George}, {Scaringi}, {Zhang}, {Jin},
  {Dexter}, {Terrier}, {Clavel}, {Degenaar}, {Eisenhauer}, {Genzel},
  {Gillessen}, {Goldwurm}, {Habibi}, {Haggard}, {Hailey}, {Harrison},
  {Merloni}, {Mori}, {Nandra}, {Ott}, {Pfuhl}, {Plewa}, \&
  {Waisberg}}]{2017MNRAS.468.2447P}
{Ponti}, G., {George}, E., {Scaringi}, S., {et~al.} 2017, \mnras, 468, 2447,
  \dodoi{10.1093/mnras/stx596}

\bibitem[{{Porth} {et~al.}(2021){Porth}, {Mizuno}, {Younsi}, \&
  {Fromm}}]{Porth+2021}
{Porth}, O., {Mizuno}, Y., {Younsi}, Z., \& {Fromm}, C.~M. 2021, \mnras, 502,
  2023, \dodoi{10.1093/mnras/stab163}

\bibitem[{{Porth} {et~al.}(2017){Porth}, {Olivares}, {Mizuno}, {Younsi},
  {Rezzolla}, {Moscibrodzka}, {Falcke}, \& {Kramer}}]{Porth2017}
{Porth}, O., {Olivares}, H., {Mizuno}, Y., {et~al.} 2017, Computational
  Astrophysics and Cosmology, 4, 1, \dodoi{10.1186/s40668-017-0020-2}

\bibitem[{{Prather} {et~al.}(2021){Prather}, {Wong}, {Dhruv}, {Ryan},
  {Dolence}, {Ressler}, \& {Gammie}}]{Prather2021}
{Prather}, B., {Wong}, G., {Dhruv}, V., {et~al.} 2021, The Journal of Open
  Source Software, 6, 3336, \dodoi{10.21105/joss.03336}

\bibitem[{{Price}(1972{\natexlab{a}})}]{Price1972a}
{Price}, R.~H. 1972{\natexlab{a}}, \prd, 5, 2419,
  \dodoi{10.1103/PhysRevD.5.2419}

\bibitem[{{Price}(1972{\natexlab{b}})}]{Price1972b}
---. 1972{\natexlab{b}}, \prd, 5, 2439, \dodoi{10.1103/PhysRevD.5.2439}

\bibitem[{{Psaltis}(2019)}]{Psaltis2019}
{Psaltis}, D. 2019, General Relativity and Gravitation, 51, 137,
  \dodoi{10.1007/s10714-019-2611-5}

\bibitem[{{Psaltis} {et~al.}(2020{\natexlab{a}}){Psaltis}, {et al}, \& {the EHT
  Collaboration}}]{Psaltis2020}
{Psaltis}, D., {et al}, \& {the EHT Collaboration}. 2020{\natexlab{a}}, \prl

\bibitem[{{Psaltis} \& {Johannsen}(2011)}]{Psaltis2011}
{Psaltis}, D., \& {Johannsen}, T. 2011, in Journal of Physics Conference
  Series, Vol. 283, Journal of Physics Conference Series, 012030,
  \dodoi{10.1088/1742-6596/283/1/012030}

\bibitem[{{Psaltis} {et~al.}(2015){Psaltis}, {{\"O}zel}, {Chan}, \&
  {Marrone}}]{Psaltis2015}
{Psaltis}, D., {{\"O}zel}, F., {Chan}, C.-K., \& {Marrone}, D.~P. 2015, \apj,
  814, 115, \dodoi{10.1088/0004-637X/814/2/115}

\bibitem[{{Psaltis} {et~al.}(2008){Psaltis}, {Perrodin}, {Dienes}, \&
  {Mocioiu}}]{Psaltis2008}
{Psaltis}, D., {Perrodin}, D., {Dienes}, K.~R., \& {Mocioiu}, I. 2008, \prl,
  100, 091101, \dodoi{10.1103/PhysRevLett.100.091101}

\bibitem[{{Psaltis} {et~al.}(2021){Psaltis}, {Talbot}, {Payne}, \&
  {Mandel}}]{Psaltis2021}
{Psaltis}, D., {Talbot}, C., {Payne}, E., \& {Mandel}, I. 2021, \prd, 103,
  104036, \dodoi{10.1103/PhysRevD.103.104036}

\bibitem[{{Psaltis} {et~al.}(2016){Psaltis}, {Wex}, \& {Kramer}}]{Psaltis2016}
{Psaltis}, D., {Wex}, N., \& {Kramer}, M. 2016, \apj, 818, 121,
  \dodoi{10.3847/0004-637X/818/2/121}

\bibitem[{{Psaltis} {et~al.}(2020{\natexlab{b}}){Psaltis}, {Ozel}, {Medeiros},
  {Christian}, {Kim}, {Chan}, {Conway}, {Raithel}, {Marrone}, \&
  {Lauer}}]{Psaltis2020b}
{Psaltis}, D., {Ozel}, F., {Medeiros}, L., {et~al.} 2020{\natexlab{b}}, arXiv
  e-prints, arXiv:2005.09632.
\newblock \doarXiv{2005.09632}

\bibitem[{{Ransom} {et~al.}(2014){Ransom}, {Stairs}, {Archibald}, {Hessels},
  {Kaplan}, {van Kerkwijk}, {Boyles}, {Deller}, {Chatterjee},
  {Schechtman-Rook}, {Berndsen}, {Lynch}, {Lorimer}, {Karako-Argaman}, {Kaspi},
  {Kondratiev}, {McLaughlin}, {van Leeuwen}, {Rosen}, {Roberts}, \&
  {Stovall}}]{triple}
{Ransom}, S.~M., {Stairs}, I.~H., {Archibald}, A.~M., {et~al.} 2014, \nat, 505,
  520, \dodoi{10.1038/nature12917}

\bibitem[{{Raymond} {et~al.}(2021){Raymond}, {Palumbo}, {Paine}, {Blackburn},
  {C{\'o}rdova Rosado}, {Doeleman}, {Farah}, {Johnson}, {Roelofs}, {Tilanus},
  \& {Weintroub}}]{Raymond+2021}
{Raymond}, A.~W., {Palumbo}, D., {Paine}, S.~N., {et~al.} 2021, \apjs, 253, 5,
  \dodoi{10.3847/1538-3881/abc3c3}

\bibitem[{{Rees}(1982)}]{Rees_1982}
{Rees}, M.~J. 1982, in American Institute of Physics Conference Series,
  Vol.~83, The Galactic Center, ed. G.~R. {Riegler} \& R.~D. {Blandford},
  166--176, \dodoi{10.1063/1.33482}

\bibitem[{{Reid} {et~al.}(2009){Reid}, {Menten}, {Zheng}, {Brunthaler},
  {Moscadelli}, {Xu}, {Zhang}, {Sato}, {Honma}, {Hirota}, {Hachisuka}, {Choi},
  {Moellenbrock}, \& {Bartkiewicz}}]{2009ApJ...700..137R}
{Reid}, M.~J., {Menten}, K.~M., {Zheng}, X.~W., {et~al.} 2009, \apj, 700, 137,
  \dodoi{10.1088/0004-637X/700/1/137}

\bibitem[{{Reid} {et~al.}(2014){Reid}, {Menten}, {Brunthaler}, {Zheng}, {Dame},
  {Xu}, {Wu}, {Zhang}, {Sanna}, {Sato}, {Hachisuka}, {Choi}, {Immer},
  {Moscadelli}, {Rygl}, \& {Bartkiewicz}}]{2014ApJ...783..130R}
{Reid}, M.~J., {Menten}, K.~M., {Brunthaler}, A., {et~al.} 2014, \apj, 783,
  130, \dodoi{10.1088/0004-637X/783/2/130}

\bibitem[{{Reid} {et~al.}(2019){Reid}, {Menten}, {Brunthaler}, {Zheng}, {Dame},
  {Xu}, {Li}, {Sakai}, {Wu}, {Immer}, {Zhang}, {Sanna}, {Moscadelli}, {Rygl},
  {Bartkiewicz}, {Hu}, {Quiroga-Nu{\~n}ez}, \& {van
  Langevelde}}]{2019ApJ...885..131R}
---. 2019, \apj, 885, 131, \dodoi{10.3847/1538-4357/ab4a11}

\bibitem[{Reissner(1916)}]{Reissner1916}
Reissner, H. 1916, Annalen der Physik, 50, 106

\bibitem[{{Ressler} {et~al.}(2017){Ressler}, {Tchekhovskoy}, {Quataert}, \&
  {Gammie}}]{Ressler+2017}
{Ressler}, S.~M., {Tchekhovskoy}, A., {Quataert}, E., \& {Gammie}, C.~F. 2017,
  \mnras, 467, 3604, \dodoi{10.1093/mnras/stx364}

\bibitem[{{Ressler} {et~al.}(2020){Ressler}, {White}, {Quataert}, \&
  {Stone}}]{Ressler+2020}
{Ressler}, S.~M., {White}, C.~J., {Quataert}, E., \& {Stone}, J.~M. 2020,
  \apjl, 896, L6, \dodoi{10.3847/2041-8213/ab9532}

\bibitem[{{Rezzolla} \& {Zhidenko}(2014)}]{Rezzolla2014}
{Rezzolla}, L., \& {Zhidenko}, A. 2014, \prd, 90, 084009,
  \dodoi{10.1103/PhysRevD.90.084009}

\bibitem[{{Ripperda} {et~al.}(2021){Ripperda}, {Liska}, {Chatterjee}, {Musoke},
  {Philippov}, {Markoff}, {Tchekhovskoy}, \& {Younsi}}]{Ripperda+2021}
{Ripperda}, B., {Liska}, M., {Chatterjee}, K., {et~al.} 2021, arXiv e-prints,
  arXiv:2109.15115.
\newblock \doarXiv{2109.15115}

\bibitem[{{Robinson}(1975)}]{Robinson1975}
{Robinson}, D.~C. 1975, \prl, 34, 905, \dodoi{10.1103/PhysRevLett.34.905}

\bibitem[{{Roelofs} {et~al.}(2020{\natexlab{a}}){Roelofs}, {Janssen},
  {Natarajan}, {Deane}, {Davelaar}, {Olivares}, {Porth}, {Paine}, {Bouman},
  {Tilanus}, {van Bemmel}, {Falcke}, {Akiyama}, {Alberdi}, {Alef}, {Asada},
  {Azulay}, {Baczko}, {Ball}, {Balokovi{\'c}}, {Barrett}, {Bintley},
  {Blackburn}, {Boland}, {Bower}, {Bremer}, {Brinkerink}, {Brissenden},
  {Britzen}, {Broderick}, {Broguiere}, {Bronzwaer}, {Byun}, {Carlstrom},
  {Chael}, {Chan}, {Chatterjee}, {Chatterjee}, {Chen}, {Chen}, {Cho},
  {Christian}, {Conway}, {Cordes}, {Crew}, {Cui}, {De Laurentis}, {Dempsey},
  {Desvignes}, {Dexter}, {Doeleman}, {Eatough}, {Fish}, {Fomalont},
  {Fraga-Encinas}, {Friberg}, {Fromm}, {G{\'o}mez}, {Galison}, {Gammie},
  {Garc{\'\i}a}, {Gentaz}, {Georgiev}, {Goddi}, {Gold}, {Gu}, {Gurwell},
  {Hada}, {Hecht}, {Hesper}, {Ho}, {Ho}, {Honma}, {Huang}, {Huang}, {Hughes},
  {Ikeda}, {Inoue}, {Issaoun}, {James}, {Jannuzi}, {Jeter}, {Jiang}, {Johnson},
  {Jorstad}, {Jung}, {Karami}, {Karuppusamy}, {Kawashima}, {Keating},
  {Kettenis}, {Kim}, {Kim}, {Kim}, {Kino}, {Koay}, {Koch}, {Koyama}, {Kramer},
  {Kramer}, {Krichbaum}, {Kuo}, {Lauer}, {Lee}, {Li}, {Li}, {Lindqvist},
  {Lico}, {Liu}, {Liuzzo}, {Lo}, {Lobanov}, {Loinard}, {Lonsdale}, {Lu},
  {MacDonald}, {Mao}, {Markoff}, {Marrone}, {Marscher}, {Mart{\'\i}-Vidal},
  {Matsushita}, {Matthews}, {Medeiros}, {Menten}, {Mizuno}, {Mizuno}, {Moran},
  {Moriyama}, {Moscibrodzka}, {M{\"u}ller}, {Nagai}, {Nagar}, {Nakamura},
  {Narayan}, {Narayanan}, {Neri}, {Ni}, {Noutsos}, {Okino}, {Olivares},
  {Ortiz-Le{\'o}n}, {Oyama}, {{\"O}zel}, {Palumbo}, {Patel}, {Pen}, {Pesce},
  {Pi{\'e}tu}, {Plambeck}, {PopStefanija}, {Prather}, {Preciado-L{\'o}pez},
  {Psaltis}, {Pu}, {Ramakrishnan}, {Rao}, {Rawlings}, {Raymond}, {Rezzolla},
  {Ripperda}, {Rogers}, {Ros}, {Rose}, {Roshanineshat}, {Rottmann}, {Roy},
  {Ruszczyk}, {Ryan}, {Rygl}, {S{\'a}nchez}, {S{\'a}nchez-Arguelles}, {Sasada},
  {Savolainen}, {Schloerb}, {Schuster}, {Shao}, {Shen}, {Small}, {Won Sohn},
  {SooHoo}, {Tazaki}, {Tiede}, {Titus}, {Toma}, {Torne}, {Traianou}, {Trent},
  {Trippe}, {Tsuda}, {van Langevelde}, {van Rossum}, {Wagner}, {Wardle},
  {Weintroub}, {Wex}, {Wharton}, {Wielgus}, {Wong}, {Wu}, {Young}, {Young},
  {Younsi}, {Yuan}, {Yuan}, {Zensus}, {Zhao}, {Zhao}, \& {Zhu}}]{Roelofs2020}
{Roelofs}, F., {Janssen}, M., {Natarajan}, I., {et~al.} 2020{\natexlab{a}},
  \aap, 636, A5, \dodoi{10.1051/0004-6361/201936622}

\bibitem[{{Roelofs} {et~al.}(2020{\natexlab{b}}){Roelofs}, {Falcke},
  {Brinkerink}, {Moscibrodzka}, {Gurvits}, {Martin-Neira}, {Kudriashov},
  {Klein-Wolt}, {Tilanus}, {Kramer}, \& {Rezzolla}}]{Roelofs+2020}
{Roelofs}, F., {Falcke}, H., {Brinkerink}, C., {et~al.} 2020{\natexlab{b}}, in
  Perseus in Sicily: From Black Hole to Cluster Outskirts, ed. K.~{Asada},
  E.~{de Gouveia Dal Pino}, M.~{Giroletti}, H.~{Nagai}, \& R.~{Nemmen}, Vol.
  342, 24--28, \dodoi{10.1017/S1743921318007676}

\bibitem[{{Ryan}(1995)}]{Ryan1995}
{Ryan}, F.~D. 1995, \prd, 52, 5707, \dodoi{10.1103/PhysRevD.52.5707}

\bibitem[{{Sakai} {et~al.}(2019){Sakai}, {Lu}, {Ghez}, {Jia}, {Do}, {Witzel},
  {Gautam}, {Hees}, {Becklin}, {Matthews}, \& {Hosek}}]{2019ApJ...873...65S}
{Sakai}, S., {Lu}, J.~R., {Ghez}, A., {et~al.} 2019, \apj, 873, 65,
  \dodoi{10.3847/1538-4357/ab0361}

\bibitem[{{Sch{\"o}del} {et~al.}(2011){Sch{\"o}del}, {Morris}, {Muzic},
  {Alberdi}, {Meyer}, {Eckart}, \& {Gezari}}]{2011A&A...532A..83S}
{Sch{\"o}del}, R., {Morris}, M.~R., {Muzic}, K., {et~al.} 2011, \aap, 532, A83,
  \dodoi{10.1051/0004-6361/201116994}

\bibitem[{{Sch{\"o}del} {et~al.}(2003){Sch{\"o}del}, {Ott}, {Genzel}, {Eckart},
  {Mouawad}, \& {Alexander}}]{2003ApJ...596.1015S}
{Sch{\"o}del}, R., {Ott}, T., {Genzel}, R., {et~al.} 2003, \apj, 596, 1015,
  \dodoi{10.1086/378122}

\bibitem[{{Sch{\"o}del} {et~al.}(2007){Sch{\"o}del}, {Eckart}, {Alexander},
  {Merritt}, {Genzel}, {Sternberg}, {Meyer}, {Kul}, {Moultaka}, {Ott}, \&
  {Straubmeier}}]{2007A&A...469..125S}
{Sch{\"o}del}, R., {Eckart}, A., {Alexander}, T., {et~al.} 2007, \aap, 469,
  125, \dodoi{10.1051/0004-6361:20065089}

\bibitem[{{Sen}(1992)}]{Sen1992}
{Sen}, A. 1992, Phys. Rev. Lett., 69, 1006, \dodoi{10.1103/PhysRevLett.69.1006}

\bibitem[{{Shaikh}(2018)}]{Shaikh2018}
{Shaikh}, R. 2018, \prd, 98, 024044, \dodoi{10.1103/PhysRevD.98.024044}

\bibitem[{{Shaikh} {et~al.}(2019){Shaikh}, {Kocherlakota}, {Narayan}, \&
  {Joshi}}]{Shaikh+2019}
{Shaikh}, R., {Kocherlakota}, P., {Narayan}, R., \& {Joshi}, P.~S. 2019, Mon.
  Not. R. Astron. Soc., 482, 52, \dodoi{10.1093/mnras/sty2624}

\bibitem[{{Shao} \& {Wex}(2016)}]{ShaoWex2016}
{Shao}, L., \& {Wex}, N. 2016, Science China Physics, Mechanics, and Astronomy,
  59, 699501, \dodoi{10.1007/s11433-016-0087-6}

\bibitem[{{Shapiro} \& {Salpeter}(1975)}]{Shapiro_Salpeter_1975}
{Shapiro}, S.~L., \& {Salpeter}, E.~E. 1975, \apj, 198, 671,
  \dodoi{10.1086/153645}

\bibitem[{{Stairs}(2003)}]{Stairs2003}
{Stairs}, I.~H. 2003, Living Reviews in Relativity, 6, 5,
  \dodoi{10.12942/lrr-2003-5}

\bibitem[{{Stone} {et~al.}(2016){Stone}, {Marrone}, {Dowell}, {Schulz},
  {Heinke}, \& {Yusef-Zadeh}}]{2016ApJ...825...32S}
{Stone}, J.~M., {Marrone}, D.~P., {Dowell}, C.~D., {et~al.} 2016, \apj, 825,
  32, \dodoi{10.3847/0004-637X/825/1/32}

\bibitem[{{Subroweit} {et~al.}(2017){Subroweit}, {Garc{\'{\i}}a-Mar{\'{\i}}n},
  {Eckart}, {Borkar}, {Valencia-S.}, {Witzel}, {Shahzamanian}, \&
  {Straubmeier}}]{2017A&A...601A..80S}
{Subroweit}, M., {Garc{\'{\i}}a-Mar{\'{\i}}n}, M., {Eckart}, A., {et~al.} 2017,
  \aap, 601, A80, \dodoi{10.1051/0004-6361/201628530}

\bibitem[{{Suvorov} \& {V{\"o}lkel}(2021)}]{Suvorov2021}
{Suvorov}, A.~G., \& {V{\"o}lkel}, S.~H. 2021, \prd, 103, 044027,
  \dodoi{10.1103/PhysRevD.103.044027}

\bibitem[{{Takahashi}(2004)}]{Takahashi2004}
{Takahashi}, R. 2004, \apj, 611, 996, \dodoi{10.1086/422403}

\bibitem[{{Telesco} {et~al.}(1996){Telesco}, {Davidson}, \&
  {Werner}}]{1996ApJ...456..541T}
{Telesco}, C.~M., {Davidson}, J.~A., \& {Werner}, M.~W. 1996, \apj, 456, 541,
  \dodoi{10.1086/176678}

\bibitem[{{Teo}(1998)}]{Teo1998}
{Teo}, E. 1998, \prd, 58, 024014, \dodoi{10.1103/PhysRevD.58.024014}

\bibitem[{{Tiede}(2022)}]{tiede2022}
{Tiede}, P. 2022, Journal of Open Source Science (submitted)

\bibitem[{{Verma} {et~al.}(2014){Verma}, {Fienga}, {Laskar}, {Manche}, \&
  {Gastineau}}]{Verma2014}
{Verma}, A.~K., {Fienga}, A., {Laskar}, J., {Manche}, H., \& {Gastineau}, M.
  2014, \aap, 561, A115, \dodoi{10.1051/0004-6361/201322124}

\bibitem[{{Vigeland} {et~al.}(2011){Vigeland}, {Yunes}, \&
  {Stein}}]{Vigeland2011}
{Vigeland}, S., {Yunes}, N., \& {Stein}, L.~C. 2011, \prd, 83, 104027,
  \dodoi{10.1103/PhysRevD.83.104027}

\bibitem[{{Vincent} {et~al.}(2021){Vincent}, {Wielgus}, {Abramowicz},
  {Gourgoulhon}, {Lasota}, {Paumard}, \& {Perrin}}]{Vincent+2021}
{Vincent}, F.~H., {Wielgus}, M., {Abramowicz}, M.~A., {et~al.} 2021, \aap, 646,
  A37, \dodoi{10.1051/0004-6361/202037787}

\bibitem[{{V{\"o}lkel} \& {Barausse}(2020)}]{Volkel2020}
{V{\"o}lkel}, S.~H., \& {Barausse}, E. 2020, \prd, 102, 084025,
  \dodoi{10.1103/PhysRevD.102.084025}

\bibitem[{{von Fellenberg} {et~al.}(2018){von Fellenberg}, {Gillessen},
  {Graci{\'a}-Carpio}, {Fritz}, {Dexter}, {Baub{\"o}ck}, {Ponti}, {Gao},
  {Habibi}, {Plewa}, {Pfuhl}, {Jimenez-Rosales}, {Waisberg}, {Widmann}, {Ott},
  {Eisenhauer}, \& {Genzel}}]{2018ApJ...862..129V}
{von Fellenberg}, S.~D., {Gillessen}, S., {Graci{\'a}-Carpio}, J., {et~al.}
  2018, \apj, 862, 129, \dodoi{10.3847/1538-4357/aacd4b}

\bibitem[{{Wald}(1984)}]{Wald1984}
{Wald}, R.~M. 1984, {General relativity} (University of Chicago Press)

\bibitem[{{Walsh} {et~al.}(2013){Walsh}, {Barth}, {Ho}, \& {Sarzi}}]{Walsh2013}
{Walsh}, J.~L., {Barth}, A.~J., {Ho}, L.~C., \& {Sarzi}, M. 2013, \apj, 770,
  86, \dodoi{10.1088/0004-637X/770/2/86}

\bibitem[{{Westerweck} {et~al.}(2018){Westerweck}, {Nielsen},
  {Fischer-Birnholtz}, {Cabero}, {Capano}, {Dent}, {Krishnan}, {Meadors}, \&
  {Nitz}}]{Westerweck+2018}
{Westerweck}, J., {Nielsen}, A.~B., {Fischer-Birnholtz}, O., {et~al.} 2018,
  \prd, 97, 124037, \dodoi{10.1103/PhysRevD.97.124037}

\bibitem[{{Wex} \& {Kramer}(2020)}]{Wex2020}
{Wex}, N., \& {Kramer}, M. 2020, Universe, 6, 156,
  \dodoi{10.3390/universe6090156}

\bibitem[{{Wielgus} {et~al.}(2020){Wielgus}, {Hor{\'a}k}, {Vincent}, \&
  {Abramowicz}}]{Wielgus+2020}
{Wielgus}, M., {Hor{\'a}k}, J., {Vincent}, F., \& {Abramowicz}, M. 2020, \prd,
  102, 084044, \dodoi{10.1103/PhysRevD.102.084044}

\bibitem[{{Wielgus} {et~al.}(2022){Wielgus}, {Marchili}, {Mart{\'i}-Vidal},
  {Keating}, {Ramakrishnan}, \& {Tiede}}]{Wielgus2022}
{Wielgus}, M., {Marchili}, N., {Mart{\'i}-Vidal}, I., {et~al.} 2022, in prep.,
  00, 0, \dodoi{TBD}

\bibitem[{{Will}(2014)}]{Will2014}
{Will}, C.~M. 2014, Living Reviews in Relativity, 17, 4,
  \dodoi{10.12942/lrr-2014-4}

\bibitem[{{Will}(2018)}]{Will2018}
---. 2018, \prl, 120, 191101, \dodoi{10.1103/PhysRevLett.120.191101}

\bibitem[{{Witzel} {et~al.}(2012){Witzel}, {Eckart}, {Bremer}, {Zamaninasab},
  {Shahzamanian}, {Valencia-S.}, {Sch{\"o}del}, {Karas}, {Lenzen}, {Marchili},
  {Sabha}, {Garcia-Marin}, {Buchholz}, {Kunneriath}, \&
  {Straubmeier}}]{2012ApJS..203...18W}
{Witzel}, G., {Eckart}, A., {Bremer}, M., {et~al.} 2012, \apjs, 203, 18,
  \dodoi{10.1088/0067-0049/203/2/18}

\bibitem[{{Witzel} {et~al.}(2018){Witzel}, {Martinez}, {Hora}, {Willner},
  {Morris}, {Gammie}, {Becklin}, {Ashby}, {Baganoff}, {Carey}, {Do}, {Fazio},
  {Ghez}, {Glaccum}, {Haggard}, {Herrero-Illana}, {Ingalls}, {Narayan}, \&
  {Smith}}]{2018ApJ...863...15W}
{Witzel}, G., {Martinez}, G., {Hora}, J., {et~al.} 2018, \apj, 863, 15,
  \dodoi{10.3847/1538-4357/aace62}

\bibitem[{{Yagi} {et~al.}(2016){Yagi}, {Stein}, \& {Yunes}}]{Yagi2016}
{Yagi}, K., {Stein}, L.~C., \& {Yunes}, N. 2016, \prd, 93, 024010,
  \dodoi{10.1103/PhysRevD.93.024010}

\bibitem[{{Yelda} {et~al.}(2010){Yelda}, {Lu}, {Ghez}, {Clarkson}, {Anderson},
  {Do}, \& {Matthews}}]{2010ApJ...725..331Y}
{Yelda}, S., {Lu}, J.~R., {Ghez}, A.~M., {et~al.} 2010, \apj, 725, 331,
  \dodoi{10.1088/0004-637X/725/1/331}

\bibitem[{{Younsi} {et~al.}(2021){Younsi}, {Psaltis}, \&
  {{\"O}zel}}]{Younsi2021}
{Younsi}, Z., {Psaltis}, D., \& {{\"O}zel}, F. 2021, arXiv e-prints,
  arXiv:2111.01752.
\newblock \doarXiv{2111.01752}

\bibitem[{{Younsi} {et~al.}(2012){Younsi}, {Wu}, \& {Fuerst}}]{Younsi+2012}
{Younsi}, Z., {Wu}, K., \& {Fuerst}, S.~V. 2012, \aap, 545, A13,
  \dodoi{10.1051/0004-6361/201219599}

\bibitem[{{Younsi} {et~al.}(2016){Younsi}, {Zhidenko}, {Rezzolla}, {Konoplya},
  \& {Mizuno}}]{Younsi+2016}
{Younsi}, Z., {Zhidenko}, A., {Rezzolla}, L., {Konoplya}, R., \& {Mizuno}, Y.
  2016, \prd, 94, 084025, \dodoi{10.1103/PhysRevD.94.084025}

\bibitem[{{Yuan} \& {Narayan}(2014)}]{Yuan_Narayan_2014}
{Yuan}, F., \& {Narayan}, R. 2014, \araa, 52, 529,
  \dodoi{10.1146/annurev-astro-082812-141003}

\bibitem[{{Yuan} {et~al.}(2003){Yuan}, {Quataert}, \& {Narayan}}]{Yuan+2003}
{Yuan}, F., {Quataert}, E., \& {Narayan}, R. 2003, \apj, 598, 301,
  \dodoi{10.1086/378716}

\bibitem[{{Yuan} {et~al.}(2004){Yuan}, {Quataert}, \& {Narayan}}]{Yuan+2004}
---. 2004, \apj, 606, 894, \dodoi{10.1086/383117}

\bibitem[{{Yunes} \& {Pretorius}(2009)}]{Yunes2009}
{Yunes}, N., \& {Pretorius}, F. 2009, \prd, 80, 122003,
  \dodoi{10.1103/PhysRevD.80.122003}

\bibitem[{{Zhang} {et~al.}(2017){Zhang}, {Baganoff}, {Ponti}, {Neilsen},
  {Tomsick}, {Dexter}, {Clavel}, {Markoff}, {Hailey}, {Mori}, {Barri{\`e}re},
  {Nowak}, {Boggs}, {Christensen}, {Craig}, {Grefenstette}, {Harrison},
  {Madsen}, {Stern}, \& {Zhang}}]{2017ApJ...843...96Z}
{Zhang}, S., {Baganoff}, F.~K., {Ponti}, G., {et~al.} 2017, \apj, 843, 96,
  \dodoi{10.3847/1538-4357/aa74e8}

\bibitem[{{Zhu} {et~al.}(2015){Zhu}, {Narayan}, {Sadowski}, \&
  {Psaltis}}]{Zhu+2015}
{Zhu}, Y., {Narayan}, R., {Sadowski}, A., \& {Psaltis}, D. 2015, \mnras, 451,
  1661, \dodoi{10.1093/mnras/stv1046}

\bibitem[{{Zylka} {et~al.}(1992){Zylka}, {Mezger}, \&
  {Lesch}}]{1992A&A...261..119Z}
{Zylka}, R., {Mezger}, P.~G., \& {Lesch}, H. 1992, \aap, 261, 119

\end{thebibliography}

\allauthors

\end{document}